\documentclass[defended,openany,11pt,twoside]{new_cit_thesis}

\usepackage{amsmath}
\usepackage{amssymb}
\usepackage{graphicx}
\usepackage{epsfig}
\usepackage{bm}
\usepackage{feynmf}
\usepackage{epic}
\usepackage{eepic}

\def\abs#1{ \left| #1 \right| }
\def\ket#1{ \left| #1 \right\rangle }
\def\vev#1{ \left\langle #1 \right\rangle }
\def\nn{\nonumber \\ }

\newcommand\beq{\begin{eqnarray}}
\newcommand\eeq{\end{eqnarray}}

\newcommand\bi{\begin{itemize}}
\newcommand\ei{\end{itemize}}
\newcommand\be{\begin{equation}}
\newcommand\ee{\end{equation}}
\newcommand\bea{\begin{eqnarray}}
\newcommand\eea{\end{eqnarray}}
\def\Dsl{\,\raise.15ex \hbox{/}\mkern-12.8mu D}
\newcommand\Tr{{\rm Tr\,}}
\newcommand\lqcd{\Lambda_{\text{QCD}}}
\newcommand{\me}[3]{\ensuremath{\left\langle{#1}\vphantom{#2 #3}
\right|{#2}\left|\vphantom{#1 #2}{#3}\right\rangle}}
\newcommand{\vect}[1]{\mathbf{#1}}
\newcommand{\Diracslash}[1]{#1\hspace{-1.25ex}\slash\,}

\def\bsigma{\mbox{\boldmath $\sigma$}}

\def\Dslash{D\!\!\!\!\slash\,}

\def\nslash{n\!\!\!\slash}
\def\bnslash{\bar n\!\!\!\slash}

\def\vslash{v\!\!\!\slash}

\def\OMIT#1{}

\newcommand{\bn}{{\bar n}}

\newcommand{\bnP}{\bar {\cal P}}
\newcommand{\ppP}{{\cal P}_\perp}

\newcommand{\cP}{{\cal P}}

\newcommand{\mcdot}{\!\cdot\!}

\newcommand{\mups}{M_\Upsilon}

\newcommand{\SCETa}{SCET$_{\rm I}$}
\newcommand{\SCETb}{SCET$_{\rm II}$}

\newcommand{\bra}[1]{\langle #1\rvert}
\newcommand{\boldsigma}{\mbox{\boldmath$\sigma$}}
\newcommand{\der}[2]{\frac{d #1}{d #2}}
\newcommand{\mpipi}{m_{\pi\pi}^2}

\newcommand{\ba}{\begin{array}}
\newcommand{\ea}{\end{array}}

\def\omh{{\omega_{\widetilde H}}} \def\omw{{\omega_{\widetilde W}}}
\def\omb{{\omega_{\widetilde B}}} \def\gamh{{\Gamma_{\widetilde H}}}
\def\gamw{{\Gamma_{\widetilde W}}} \def\gamb{{\Gamma_{\widetilde B}}}
 
\def\kwtilp{{h_{\widetilde W}^+}} \def\khtilp{{h_{\widetilde H}^+}}
\def\nwtilp{{N_{\widetilde W}^+}} \def\nhtilp{{N_{\widetilde H}^+}}

\def\kplp{{h_{pL}^+}}
\def\kprp{{h_{pR}^+}}

\def\khrp{{h_{hR}^+}}

\newcommand{\op}[1]{\textsf{#1}}
\newcommand{\boldgamma}{\mbox{\boldmath$\gamma$}}
\newcommand{\diracslash}[1]{#1\!\!\!/}
\newcommand{\pd}[2]{\frac{\partial #1}{\partial #2}}
\newcommand{\CPV}{CP\!\!\!\!\!\!\!\!\raisebox{0pt}{\small$\diagup$}}

\DeclareMathOperator{\Real}{Re}
\DeclareMathOperator{\Imag}{Im}

\title{Probing Physics in the Standard Model and Beyond \\ with Electroweak Baryogenesis  and Effective Theories \\ of the Strong Interactions}
\author{Christopher Lee}
\date{May 25, 2005}

\begin{document}

\degreeaward{Doctor of Philosophy}                 
\university{California Institute of Technology}    
\address{Pasadena, California}                     
\unilogo{cit_logo}                                 
\copyyear{\the\year}                               
\pubnum{}                                          
\dedication{\sc In Memoriam \\ Joannes Paulus PP. II \\ 18.V.1920---02.IV.2005---$\infty$ \\ Ora pro nobis!}

\maketitle

\begin{frontmatter}
      \makecopyright            
      \makededication           

\vfill
\pagebreak

\raisebox{-3.5in}{
\begin{minipage}{5.5in}
\small\singlespace\raggedright Science rightly understood means reading the wisdom of God in nature, which God made. Science wrongly understood means reading the proofs of the book of nature while denying that the book ever had an Author. \\ \flushright\emph{Fulton J. Sheen, \emph{The World's First Love}}
\end{minipage}}

      \begin{acknowledgements}  

\noindent\emph{Gratias agamus Domino Deo nostro, factori c\ae li et terr\ae. Gloria Patri, et Filio, et Spiritui Sancto, sicut erat in principio, et nunc, et semper, et in s\ae cula s\ae culorum.}

I offer my deep thanks to my advisors, Mark Wise, who taught me particle theory (up to one loop), and Michael Ramsey-Musolf, who turned me professionally into a ``nuclear theorist''---but, ironically, Mark taught me QCD while Michael taught me supersymmetry---and to all my research collaborators, Christian Bauer, Vincenzo Cirigliano, Sean Fleming, Adam Leibovich, and Aneesh Manohar, from whom I learned so much about physics.\footnote{I also acknowledge the support of this research by the U.S. Department of Energy through Contracts No.~DE-FG03-92ER40701 and No.~DE-FG03-02ER41215, the National Science Foundation through an NSF Graduate Research Fellowship, and the U.S. Department of Defense through a National Defense Science and Engineering Graduate Fellowship, the latter awarded when I said I would study string theory, apparently a vital component of the national security of this country.} In addition to Mark and Michael, I am grateful to John Schwarz and Brad Filippone for serving on my thesis committee. And I thank Carol and Sharlene, who are in reality the ones in charge of the theory group, for all their help. I recall gratefully those who cultivated long ago in my past light-cone my interest in physics and mathematics, Jim Wilson at Gresham High School, and Bob Reynolds, David Griffiths, Nick Wheeler, Rick Watkins, John Essick, Johnny Powell, Tom Wieting, and Jerry Shurman at Reed College.

I extend my heartfelt thanks to those who have been my strength in friendship and in faith here at Caltech: first and foremost, Daniel Katz, my roommate, my true friend and brother, \emph{qui spiritum deorum sanctorum habet in se: et scientia et sapientia inventae sunt in eo} (\emph{Daniel}~5:11); Roberto Aparicio, honorary Korean and fellow executor of dominion over cattle (\emph{Genesis}~1:26); Xavier Calmet, \emph{mon cher ami et coll\`{e}gue}; Isaac Chenchiah, the Supreme Commander of the Revolution to Thomisticize German Women\footnote{Deposed 02 April 2005, \emph{Universi Domenici Gregis} \S 13; reappointed 23 April 2005, by His Holiness Pope Benedict XVI.}---may you find your flock of goats (\emph{Song of Songs}~4:1, 6:5); Jennifer Johnson, who to my great joy has become my fellow Catholic and, beyond all hope, my fellow carnivore; Kimball Martin---\emph{zazang!}---who gave me the honor of serving as his vicar; Gerald Palmrose, my younger and taller brother in the Faith; ``Carlotes'' Salazar, my fellow Oregonian; Ivan Veselic, $\frak{mein\ lieber\ Freund}$; Mihai Stoiciu, a model to me of courtesy, faith, and strength, who wields one powerful ``G-8'' tennis racquet against me; Winnie Wang, my ``little big sister''; and all my brothers and sisters in the Newman Center, especially Br.~Tobias V\"{o}lkl, L.C., Luis Gonzalez, Shannon Lewis, Chris Wetzel, and all my undergrad brethren. My eternal gratitude is due to the heroic priests who have been my spiritual fathers and guides, Fr. Anthony-Maria Patalano, O.P. and many other Dominicans; Fr. Andrew Mulcahey, L.C., and many other Legionaries; Fr. Robert J. Billett, C.M.F.; and many other priests.

I appreciate the interactions with my research group colleagues (especially my officemate Meg, nearest neighbors Matt and Alejandro, former next-to-nearest neighbor Jim Harrington, and fellow ``nukes'' Donal, Rebecca, and Sean). To all those in the Korean community at Caltech who so warmly extended their friendship to me, \emph{gamsa habnida}: especially Peter Lee, Jongwon Park \& Soojin Son, Jihyun Choi, my tennis buddies (Sangdun Choi, Sonjong Hwang, Joon-Young Choi, Wonhee Lee, Dalmo Kang, Keehong Seo, Junho Suh, Jaewoo Choi, Jinseong Heo), and my softball teammates. Thanks, too, to my other tennis buddies (Mike Kesden, Jason Wong, Jon Young) for the great fun and competition. I recall fondly the late nights of homework in my first year with Latham Boyle. My dear Physics 2B students (Christine, Eliot, Elizabeth, Harrison, Jane, Jayson, Neha, and Rahul) made my job of teaching so rewarding and enjoyable (cake, cookies, and chocolate!), and my fellow TA Sherry Suyu always shared her good cheer. Since high school, I have enjoyed many years of friendship with Kristina Barkume, who has faithfully followed me from Gresham to Reed to Caltech, and my buddy Jon Gruber.  Judy Robinson taught Kristina, Jon, me and so many others how to think critically, speak coherently, and meet challenges confidently. Michael, Matthew, and Andrew Lee and their parents, Steve and Rose, were both friends and family to me during my time here. And even this long list has left out so many to whom I owe my gratitude, which is cause for both my profound apologies and my joyful amazement. 

Finally, I have not sufficient words to thank my parents, Benedict Joseph and Gertrude Teresa, who have given me so much to make possible this work and also taught me all the reasons that make it worthwhile.

Above all, to Mom, \emph{totus tuus!} I thank thee always for these thy words, which I take also as mine: \emph{Ecce ancilla Domini, fiat mihi secundum verbum tuum!}

      \end{acknowledgements}

      \begin{abstract}

\begin{quote}
\raggedright\singlespace\small For since the nature of our intellect is to abstract the essence of material things from matter, anything material residing in that abstracted essence can again be made subject to abstraction; and as the process of abstraction cannot go on forever, it must arrive at length at some immaterial essence, absolutely without matter; and this would be the understanding of immaterial substance.\\ \flushright\vspace{-10pt} \emph{St. Thomas Aquinas, \emph{Summa Theologica, I, 88, 2.}}
\end{quote}
We address in this thesis two primary questions aimed at improving our ability to calculate reliably in the Standard Model of particle physics and probing possible new particles which may exist beyond it.

First, we embark on an attempt to account for the abundance of matter in the present Universe if earlier in its history matter and antimatter were equally abundant. We explore whether baryogenesis at the electroweak phase transition could successfully account for the observed density of baryons in the Universe, using the closed-time-path (CTP) formalism of quantum field theory to calculate the buildup and relaxation of particle densities during the phase transition. For our model of the new particles and sources of $CP$ violation necessary to account for the baryon asymmetry of the Universe, we adopt the Minimal Supersymmetric Extension of the Standard Model (MSSM). We look for regions of the parameter space in the MSSM that could give rise to sufficiently large baryon asymmetry without violating constraints on these parameters from existing experiments, in particular, constraints on masses of Higgs and supersymmetric particles from accelerator searches and precision electroweak tests, and on $CP$-violating parameters of the MSSM from searches for electric dipole moments of elementary particles.

Next, we explore how to get around our ignorance of the dynamics of strongly interacting particles in the nonperturbative regime of Quantum Chromodynamics (QCD) by the clever use of effective field theories. Two applications are explored: the decay of $Z$ bosons to hadronic jets using soft-collinear effective theory (SCET) and the radiative decays of quarkonia to light hadrons using SCET and non-relativistic QCD (NRQCD).  These tools facilitate the proof of factorization of decay rates into perturbatively-calculable and nonperturbative parts. Universality of the latter among different observables provides predictive power even in our ignorance of the details of the nonperturbative physics.

      \end{abstract}

\tableofcontents

\listoffigures

\end{frontmatter}

\pagestyle{headings}
\chapter{Introduction}

\begin{quote}
\small\singlespace
Staunen kann nur, wer noch nicht das Ganze sieht; Gott staunt nicht.\\
\flushright\vspace{-12pt}\emph{Josef Pieper, \emph{Gl\"{u}ck und Kontemplation}}\\
\bigskip
\raggedright
Amazement is only possible for one who does not yet see the whole; God cannot be amazed.\\
\flushright\vspace{-12pt}\emph{Josef Pieper, \emph{Happiness \& Contemplation}}
\end{quote}
Today physicists eagerly anticipate the discovery of new phenomena beyond those already discovered in the Standard Model of particle physics. We await with high hopes the unveiling of new particles and phenomena at the Large Hadron Collider, or perhaps even the Tevatron, within the decade. Meanwhile, we aim our current efforts at the improvement of the precision and reliability of both theoretical calculations and experimental tests of the properties of particles, not only those currently within the realm of speculation, but also of those within the Standard Model itself. On the one hand, to be able to extract evidence for new physics from future accelerator data, we must first calculate the predictions of the Standard Model as accurately and precisely as possible, in order to be able to find in the data deviations therefrom. On the other hand, we can already search for signatures of new physics in experimental data today, in high-precision low-energy experiments or in cosmological observations.

In this thesis, we focus both on tests of models of new physics and improvement of our ability to calculate reliably in the Standard Model. First, we consider how  extra sources of $CP$ violation beyond those in the Standard Model (for example, in its supersymmetric extensions) might account for the baryon asymmetry of the Universe and how experimental tests, especially searches for electric dipole moments of elementary particles, constrain such possibilities. Then, backing up to the Standard Model, we attempt to improve calculations of observables involving the strong interactions by use of effective field theories of Quantum Chromodynamics, studying the radiative  decays of heavy quarkonia and the hadronic decays of $Z$ bosons, attempting to understand the nonperturbative contributions to observables in these processes. Understanding hadronic jet production will be vital to the separation of QCD backgrounds at the LHC from signals of new physics, while heavy quarkonia provide a useful testing ground for the validity of the effective theories we use, establishing their reliability for other applications.

\section{Probing New Physics with Baryogenesis and Electric Dipole Moments}

Already today, suggestive evidence for physics beyond the Standard Model exists outside of accelerators, in the baryon asymmetry of the Universe (BAU)---the survival of more matter than antimatter in the Universe's early evolution. Big Bang Nucleosynthesis (BBN) and measurements of the cosmic microwave background tell us that:
\begin{equation}
\eta\equiv\frac{n_B}{n_\gamma} = 
\begin{cases}
3.4-6.9\times 10^{-10}, & \text{BBN \cite{Eidelman:2004wy}} \\
5.9-7.3\times 10^{-10}, & \text{CMB \cite{wmap}}
\end{cases}
\ee
where the number density of baryons $n_B$ is normalized to the number density of photons $n_\gamma$. (An updated version of these constraints from Ref.~\cite{Cyburt:2004cq} is displayed graphically in Fig.~\ref{fig:bbn}.)
\begin{figure}
\begin{center}
\includegraphics[width=6in]{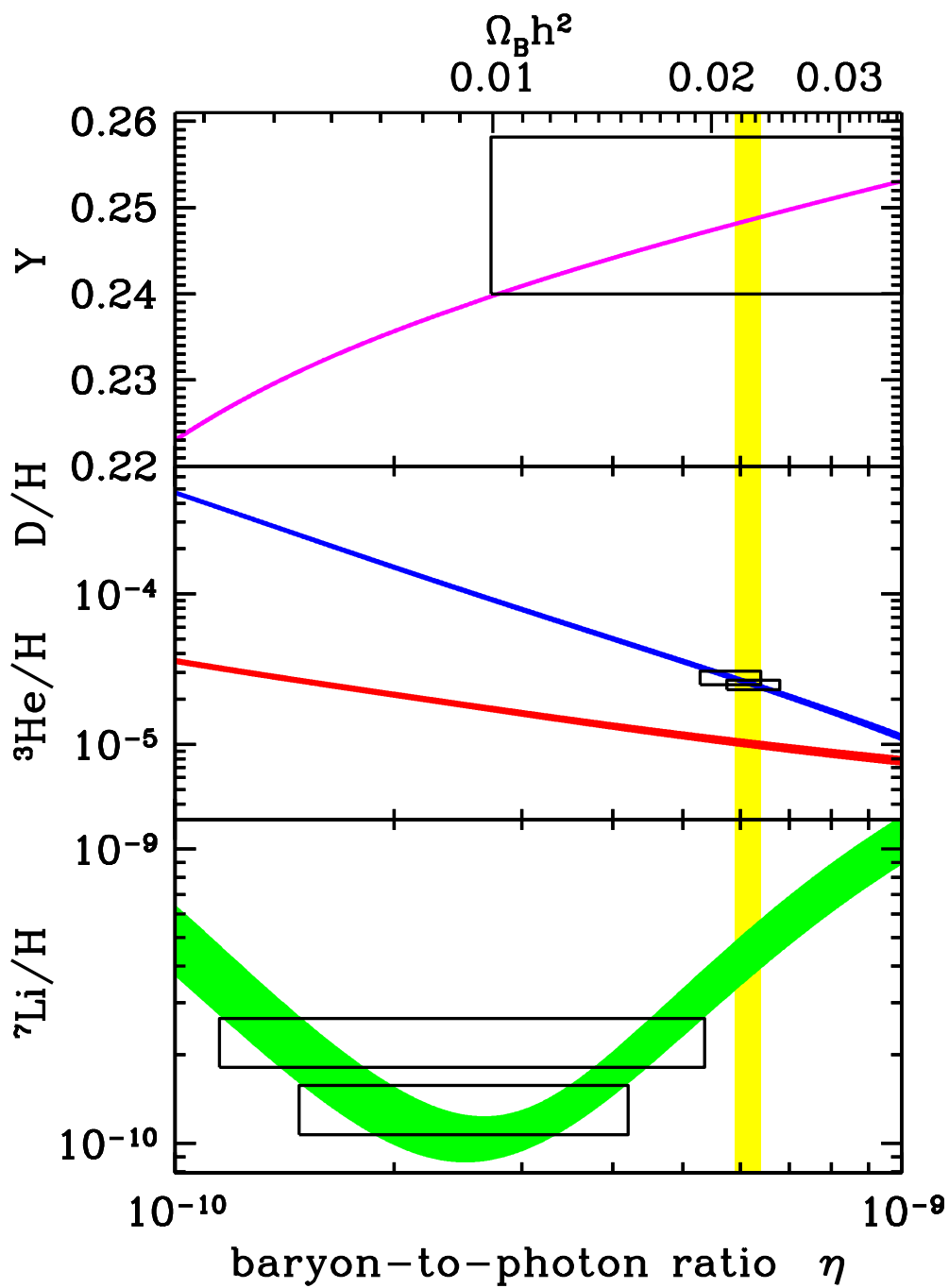}
\end{center}
\vspace{-1in}
\caption[BBN and CMB constraints on the BAU]{BBN and CMB constraints on the BAU, as analyzed in Ref.~\cite{Cyburt:2004cq}. The abundances of helium-4 ($Y$), deuterium, helium-3, and lithium-7 can be used to deduce the baryon density $\Omega_B$, or, equivalently, the baryon-to-photon ratio $\eta$. The various curves are the BBN predictions for these abundances as functions of the baryon density, which are constrained by observations to lie in the outlined boxes. The combination of the constraints from BBN and from WMAP measurements of the CMB is shown as the yellow, vertical band. (Figure courtesy of R.~Cyburt.)}
\label{fig:bbn}
\end{figure}
The Standard Model is unable to account for the (large!) size of these numbers, assuming the Universe begins in a symmetric state ($\langle B\rangle=0$). Even if we assume a primordial excess of matter at the Big Bang, we could not easily account for its survival during the subsequent inflation, which would have diluted it nearly to zero, and which we must postulate to account for the flatness and uniformity of the Universe \cite{Guth:1980zm} and the density perturbations\footnote{Which, of course, can be augmented after inflation \cite{Ackerman:2004kw,Bauer:2005cd}, too.} therein. To start in a matter-antimatter symmetric state and generate a baryon asymmetry requires, as pointed out by Sakharov \cite{Sakharov:1967dj}, three conditions:
\begin{itemize}
\item Baryon number violation---otherwise, the Universe would never depart from $\langle B\rangle =0$.

\item $C$ and $CP$ violation---otherwise, even with $B$ violation, particles and their antiparticles are produced in equal amounts.

\item Departure from thermal equilibrium---otherwise, the thermal average
\begin{equation}
\langle B\rangle = \frac{\Tr(e^{-\op{H}/T} B)}{\Tr e^{-\op{H}/T}},
\end{equation}
never departs from zero, since
\begin{equation}
\begin{split}
\Tr[e^{-\beta\op{H}} B] &= \Tr[e^{-\beta\op{H}}(CPT)(CPT)^{-1} B] \\
&= \Tr[(CPT)e^{-\beta\op{H}}(CPT)^{-1}B] \\
&= -\Tr[e^{-\beta H}(CPT)^{-1}(CPT)B]  = -\Tr[e^{-\beta\op{H}}B]\\
\Rightarrow \Tr[e^{-\beta\op{H}}B] &= 0,
\end{split}
\end{equation}
assuming $CPT$ invariance\footnote{We do not consider the possibility that $CPT$ invariance may be broken by, for instance, Lorentz violation, perhaps spontaneously \cite{Jenkins:2003hw,Jenkins:2004yn,Graesser:2005bg}. Mechanisms violating $CPT$ spontaneously have also been used to account for baryogenesis \cite{DeFelice:2002ir}.} ($[H,CPT]=0$), and using cyclicity of the trace and the oddness of $B$ under a $CPT$ transformation.
\end{itemize}
These three conditions may have been realized in \emph{electroweak baryogenesis}\footnote{This and other scenarios for baryogenesis are reviewed in Ref.~\cite{Riotto:1998bt}.}, in which the BAU is generated during the electroweak phase transition---if it is strongly-enough first-order and there is enough $CP$ violation in the electroweak sector of Nature. Neither of these conditions is satisfied in the Standard Model. The phase transition is not found to be strongly-enough first-order given present limits on the Higgs mass, and the complex phase in the CKM matrix is too small to account for the BAU.

Before considering how to overcome the inability of the Standard Model to satisfy the last two Sakharov criteria, let us begin by examining how it satisfies the first.

\subsection{Baryon Number Violation}

Baryon number violation is already present in the Standard Model. Although $B$ is conserved in all perturbative processes, there are nonperturbative topological transitions of gauge fields at high temperature which generate baryon and lepton number violation through triangle anomalies~\cite{ABJ}. The baryon number current,
\begin{equation}
j_B^\mu = \frac{1}{N_C}\sum_{i,a}\bar q_i^a \gamma^\mu q_i^a,
\end{equation}
summed over the $n_F$ quark flavors $i$ and $N_C$ colors $a$, is not conserved at the quantum level in the Standard Model, due to the diagram shown in Fig.~\ref{anomaly}:
\begin{equation}
\label{Banomaly}
\partial_\mu j_B^\mu = -\frac{n_Fg_2^2}{32\pi^2}W_{\mu\nu}^a\widetilde W^{a\mu\nu},
\end{equation}
where $W^{\mu\nu}$ is the $SU(2)$ gauge field strength tensor, and $\widetilde W^{\mu\nu} = \frac{1}{2}\epsilon^{\mu\nu\alpha\beta}W_{\alpha\beta}$.
\begin{figure}
\vspace{5mm}
\begin{center}
\begin{fmffile}{anomaly}
\begin{fmfgraph*}(120,60)
\fmfleft{A}
\fmfright{W1,W2}
\fmf{photon}{A,v1}
\fmf{fermion}{v1,v2,v3,v1}
\fmf{photon}{v2,W1}
\fmf{photon}{v3,W2}
\fmfv{decor.shape=cross}{v1}
\fmfforce{(0,.5h)}{A}
\fmfforce{(0,.5h)}{v1}
\fmfforce{(.5w,h)}{v2}
\fmfforce{(.5w,0)}{v3}
\fmfforce{(w,h)}{W1}
\fmfforce{(w,0)}{W2}
\end{fmfgraph*}
\end{fmffile}

\begin{picture}(0,0)(0,0)
\put(-80,40){$j_B^\mu$}
\put(65,67){$W_\nu^a$}
\put(65,10){$W_\lambda^b$}
\end{picture}
\end{center}
\vspace{-5mm}
\caption[Baryon number current anomaly in $SU(2)$]{Baryon number current anomaly in $SU(2)$.}
\label{anomaly}
\end{figure}
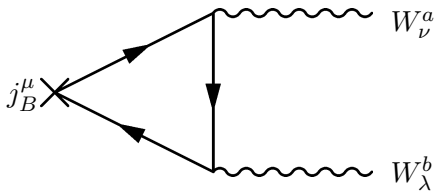
The right hand side of this equation is minus the divergence of a topological current, $-\partial_\mu K^\mu$, where \cite{McLerran:1990de}:
\begin{equation}
K^\mu = \frac{g_2^2}{8\pi^2}\epsilon^{\mu\nu\alpha\beta}\Tr\Bigl(W_\nu\partial_\alpha W_\beta - \frac{2ig_2}{3}W_\nu W_\alpha W_\beta\Bigr),
\end{equation}
where $W^\mu$ is the $SU(2)$ gauge field.  Integrating $\partial_\mu K^\mu$ over all space, we obtain the topological charge,
\begin{equation}
n_{\text{CS}} = -\int d^3\vect{x}\, K^0,
\end{equation}
called the \emph{Chern-Simons number}. Different vacuum configurations of gauge fields having the same energy can have different values of $n_{\text{CS}}$. These vacua correspond to gauge field configurations of the form:
\begin{equation}
W_\mu = \frac{i}{g_2}\Omega\partial_\mu\Omega^{-1},
\end{equation}
where $\Omega$ is a function from four-dimensional space into the gauge group $SU(2)$. As $SU(2)$ is topologically equivalent to the three-sphere $S^3$,  the index $n_{\text{CS}}$ is essentially the winding number of the map $\Omega$ onto $S^3$ \cite{Coleman}.

Transitions between these topologically distinct vacua induce a change in the baryon number, due to the anomaly equation (\ref{Banomaly}), as illustrated in Fig.~\ref{tunnel}. The lepton number current has the same anomaly. Thus, $B+L$ is violated while $B-L$ is conserved. These vacua are separated by barriers corresponding to solutions of the field equations called \emph{sphaleron} configuratoins with energy $E_{\text{sph}}$. The rate of tunneling between vacua is negligible below the temperature of the electroweak phase transition, the rate suprressed by a factor of $\exp(-8\pi^2/g_2^2)\sim 10^{-170}$ \cite{'tHooft:1976up}. However, at higher temperatures, the tunneling is unsuppressed, and the sphaleron transitions induce a change in baryon number through the equation:
\begin{equation}
\label{rhoBdiffeq}
\partial_t\rho_B(x) - D_q\nabla^2\rho_B(x) \propto -n_F \Gamma_{\text{sph}}[n_L(x) + R\rho_B(x)],
\end{equation}
where $D_q$ is the diffusion coefficient for baryons, $n_F$ is the number of families, $\Gamma_{\text{sph}}$ is the rate of weak sphaleron transitions, $R$ is a relaxation coefficient for baryon number, and $n_L$ is the number density of left-handed weak-doublet fermions. The latter quantity appears because $SU(2)$ gauge bosons couple directly only to left-handed fermions. 
\begin{figure}
\medskip
\begin{center}
\includegraphics[width=8cm]{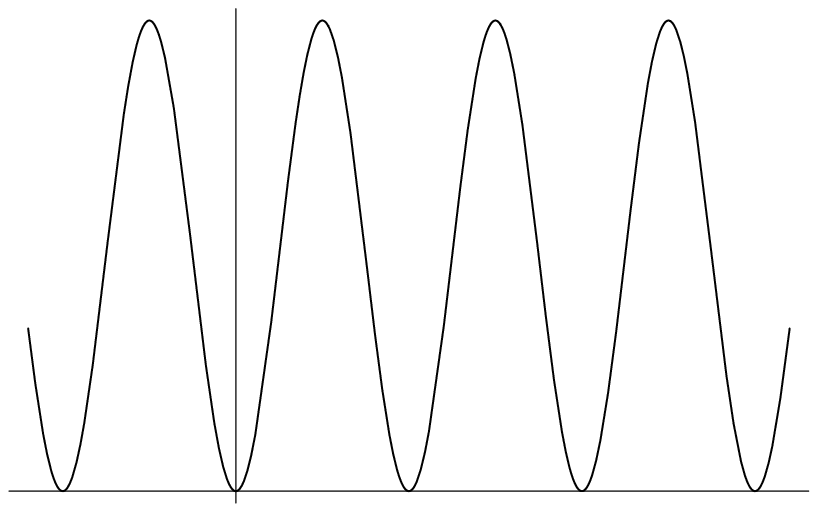}
\begin{picture}(0,0)(0,0)
\put(-165,135){$E$}
\put(-220,-10){$-1$}
\put(-170,-10){0}
\put(-120,-10){1}
\put(-70,-10){2}
\put(-20,-10){3}
\put(0,0){$n_{\text{CS}}$}
\thicklines
\put(-90,135){\arc{25}{-3.14}{0}}
\put(-77.5,135){\vector(1,-4){0}}
\put(-102.5,135){\vector(-1,-4){0}}
\put(-77,143){$\Gamma_{\text{sph}}$}
\put(-15,105){$\Delta n_{\text{CS}} = 1$}
\put(-10,90){$\Rightarrow \Delta B = \Delta L = n_F$}
\end{picture}
\end{center}
\vspace{-12pt}
\caption[Topological transitions at high temperature.]{Topological transitions at high temperature. When $SU(2)$ gauge fields tunnel from one vacuum configuration to another, characterized by different values of the Chern-Simons number, baryon and lepton number are violated through the quantum anomaly in the baryon and lepton number currents. The unstable solutions to the field equations between the vacua are called \emph{sphaleron} configurations, whose energy is $E_{\text{sph}}$. The rate of the tunneling reactions is $\Gamma_{\text{sph}}$.}
\label{tunnel}
\end{figure}

We imagine a picture in which bubbles of the broken electroweak symmetry phase nucleate in the previously electroweak symmetric Universe. At or near the bubble boundaries, the generation of a nonzero $n_L$ due to $CP$ violation in regions where weak sphaleron transitions are active leads to a nonzero baryon density $\rho_B$, which, if it ends up inside the region of broken electroweak phase where sphalerons no longer change baryon number, is frozen in and survives until the present day in the Universe. The goal within this framework is to derive an equation for $n_L$ dependent on the amount of $CP$ violation present and solve for the final baryon density. The generation of $n_L$, which tends to occur much faster than the sphaleron transitions themselves, is governed by another set of equations, involving both $CP$-violating sources generating nonzero $n_L$ and $CP$-conserving reactions which tend to cause relaxation of $n_L$ back towards zero. The presence of a nonzero $n_L$ then drives the weak sphaleron transitions to generate nonzero baryon number.

We will consider the equations for $n_L$ in more detail after examining where we might find enough $CP$ violation beyond the Standard Model to account for the observed size of the BAU.

\subsection{More $CP$ Violation from Supersymmetry}

Supersymmetric extensions of the Standard Model\footnote{We refer to Ref.~\cite{Martin:1997ns} for details and notation.} are popular for their touted ability to resolve the hierachy problem (by canceling large corrections to the Higgs mass from heavy particle loops), achieve gauge coupling unification around $10^{16}$ GeV, and provide candidate particles for the dark matter. What interests us is that they may provide a whole new set of $CP$ violating parameters that may help account for the BAU. These may generate large enough $n_L$ and, thus, through Eq.~(\ref{rhoBdiffeq}), $\rho_B$. Indeed, the Minimal Supersymmetric Extension of the Standard model (MSSM) by itself contains more than 100 free parameters including dozens of $CP$-violating phases. Adopting a specific model for supersymmetry breaking can reduce this menagerie to a more tractable set. In the minimal supergravity model, for instance, which we adopt in thesis, the only independent $CP$-violating phases remaining are the phase $\phi_\mu$ of the $\mu$-parameter in the superpotential:
\begin{equation}
W_{\text{MSSM}} \supset \mu(H_u^+ H_d^- - H_u^0 H_d^0) + \cdots,
\end{equation}
where $\mu = \abs{\mu}e^{i\phi_\mu}$, and the phases in the triscalar couplings in the soft SUSY-breaking terms in the Lagrangian:
\begin{equation}
\mathcal{L}_{\text{soft}} \supset -\left(\tilde{\bar u}\mathbf{a_u}\tilde Q H_u - \tilde{\bar d}\mathbf{a_d}Q H_d - \tilde{\bar e}\mathbf{a_e}\tilde L H_d\right) + c.c.,
\end{equation}
where the matrices $\mathbf{a_u} = A_{u0}\mathbf{y_u}$, $\mathbf{a_d}= A_{d0}\mathbf{y_d}$, and $\mathbf{a_e}= A_{e0}\mathbf{y_e}$ are proportional to the corresponding matrices of Yukawa couplings. Assuming that $A_{u0}=A_{d0}=A_{e0}\equiv A_0$ at the scale of SUSY breaking, there is just one new $CP$-violating phase, $\phi_A$, where $A_0 = \abs{A_0}e^{i\phi_A}$. 

These two $CP$-violating phases $\phi_\mu$ and $\phi_A$ may be large enough to account for the baryon asymmetry. They cannot, however, be so large as to conflict with constraints from precision tests searching for permanent electric dipole moments (EDMs) in electrons, neutrons, and atomic systems. These can possess an EDM only if time-reversal ($T$) symmetry, and, therefore, by the $CPT$ theorem, $CP$ symmetry, is violated in Nature. So far, no experiment has found a measurable EDM, thus placing stringent constraints on the size of any new sources of $CP$-violation beyond the Standard Model.

Our task is to calculate the baryon density that could be generated from $CP$-violating phases small enough not to conflict with EDM constraints.

\subsection{Transport Equations for Electroweak Baryogenesis}

With this particular model for particles which can participate in the generation of the baryon asymmetry at the electroweak phase transition, we can write equations governing the generation of left-handed weak doublet fermions $n_L$ entering Eq.~(\ref{rhoBdiffeq}) for $\rho_B$, as introduced by Cohen, Kaplan, and Nelson \cite{Cohen:1994ss} and Huet and Nelson \cite{Huet:1995sh}. For example, the number density of left-handed third-generation quarks and squarks $Q$ is governed by an equation of the form:
\begin{equation}
\label{Qdiffeq}
\partial_t Q - D_q\nabla^2 Q = \Gamma_m\left(\frac{T}{k_T} - \frac{Q}{k_Q}\right) + \Gamma_y\left(\frac{T}{k_T} - \frac{H}{k_H} - \frac{Q}{k_Q}\right) + S_Q^{\CPV} + \dots,
\end{equation}
where $T$ is the density of right-handed third-generation quarks and squarks, $H$ is the density of Higgs and Higgsinos, $k_{T,Q,H}$ are statistical factors, $S_Q^{\CPV}$ is the $CP$-violating source for left-handed squarks, and $\Gamma_m,\Gamma_y$ are the relaxation rates for the quantities in parentheses induced by interactions of quarks and squarks with the Higgs vacuum expectation value and with real Higgs particles, respectively. In this framework, supergauge interactions are assumed to be much faster, keeping superpartners in chemical equilibrium with each other, accounting for our use of combined particle and sparticle densities $T,Q$, and $H$. Equations similar to Eq.~(\ref{Qdiffeq}) for $T$ and $H$ round out a set of coupled equations whose solution gives the density $n_L$ which enters the equation for baryon density (\ref{rhoBdiffeq}).

In Refs.~\cite{Cohen:1994ss,Huet:1995sh} semi-classical techniques were adopted to estimate the size of the coefficients like $\Gamma_{m,y}$ and sources $S^{\CPV}$. Riotto \cite{Riotto:1998zb} introduced into this framework the closed time path (CTP) formulation of quantum field theory, which incorporates finite-temperature and nonequilbrium effects, to derive from quantum field theory the sources $S^{\CPV}$, but kept the standard derivations of $\Gamma_{m,y}$. He discovered large enhancements (of the order of $10^3$) over previous calculations of $S^{\CPV}$ in certain regions of MSSM parameter space, namely, where certain particle masses (such as Higgsinos and Winos or left- and right-handed stops) become degenerate. 

In Ref.~\cite{Lee:2004we} and Chap.~\ref{chap:ewb} of this thesis, we continue Riotto's endeavor by deriving other terms appearing in the diffusion equations like (\ref{Qdiffeq}) using the same CTP formalism as for the $CP$-violating sources. We discover similar enhancements in the coefficient $\Gamma_m$, and expect similar results for $\Gamma_y$, which has not yet been calculated with the CTP formalism. Enhancement of these rates tends to \emph{reduce} the final baryon asymmetry compared to Riotto's result, but still leaving a significant enhancement of $\rho_B$. Beyond these enhancements, we also discover entirely new terms in Eq.~(\ref{Qdiffeq}) and its cousins with different linear combinations of the various particle densities.

These findings, although still quite preliminary, make evident the importance of a fully consistent, quantum-field-theoretic calculation within the CTP formalism of all the terms entering the transport equations for particle densities. Each new application of this formalism to a different set of terms in the transport equations (e.g., by Riotto and by us) has significantly affected the size of the BAU predicted by these equations. The prospect of more powerful and more precise experiments that will constrain the parameters in models of new physics in the next several years make imperative the task of making as complete and reliable as possible these theoretical predictions. The work in this thesis, focusing attention just on a few of the terms in the transport equations within only the MSSM, lays the groundwork for future calculations, which ought also to extend to other extensions of the SM to prepare for the various possibilities for the new physics that may be discovered within the decade.

\subsection{Combining Constraints from BAU and EDMs}

\begin{figure}[t]
\begin{center}
\includegraphics[width=9cm]{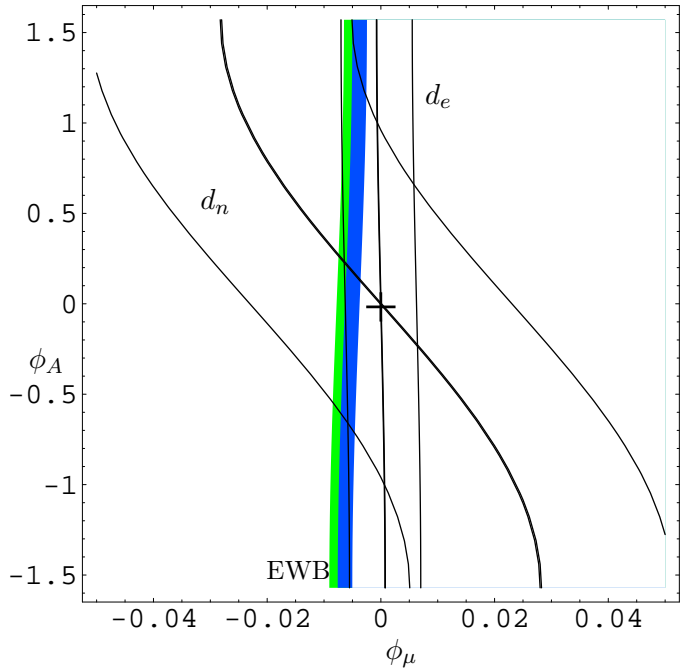}
\begin{picture}(0,0)(235,0)
\put(50,170){$d_n$}
\put(135,210){$d_e$}
\put(75,30){\small EWB}
\put(120,0){$\phi_\mu$}
\put(-15,110){$\phi_A$}
\end{picture}
\end{center}
\caption[Combined constraints on $CP$-violating phases from BAU and EDMs.]{Combined constraints on $CP$-violating phases from baryon asymmetry and electric dipole moments. The colored bands pick out the region in the $\phi_\mu$-$\phi_A$ plane required for successful electroweak baryogenesis, based on WMAP (green) and BBN (green+blue) observations of the BAU. The wider white bands are limits placed on the size of these phases by electron EDM ($d_e$) and neutron EDM ($d_n$) searches. There exists a region of overlap where BAU requirements are consistent with current EDM constraints. The narrow lines inside those bands denote the sensitivity of proposed searches for these EDMs at future experiments at Los Alamos National Laboratory \cite{LANLn,LANLe} and Yale \cite{Kawall:2003ga}.}
\label{bau+edm}
\end{figure}
In Fig.~\ref{bau+edm}, we illustrate the present status of BAU and EDM constraints on $\phi_\mu$ and $\phi_A$ in the MSSM within the mSUGRA scenario for supersymmetry breaking, for a particular choice in the MSSM parameter space. Our choices are informed by precision electroweak constraints \cite{Carena:1997ki} combined with numerical simulations of the strength of the electroweak phase transition \cite{Laine:1998qk}, which tell us that, in the MSSM, we must have a light right-handed top squark $\tilde t_R$ and heavy left-handed $\tilde t_L$. We also choose degenerate Higgsino and $SU(2)$-gaugino masses $\abs{\mu}=M_2$, which we find maximizes the BAU that can be generated from the phases $\phi_\mu,\phi_A$ of a given size (here we chose $\abs{\mu}=M_2=200$ GeV).

We are on the verge of new EDM experiments in the next few years which will achieve sensitivites of two or more orders of magnitude greater than the present limits illustrated in Fig.~\ref{bau+edm}. Further null results may succeed in ruling out the simplest scenarios of supersymmetric electroweak baryogenesis entirely. Indeed, achieving the situation illustrated in Fig.~\ref{bau+edm} already places the fairly stringent constraint of near-degeneracy between Higgsino and gaugino masses to generate a large enough BAU from the small phases implied by the EDM searches. Meanwhile, with the LEP-II constraint on the Higgs mass of $m_H\gtrsim 114$~GeV \cite{Barate:2003sz}, only a small window remains for a strongly first-order electroweak phase transition, requiring $m_H\lesssim 120$~GeV (see Ref.~\cite{Balazs:2004ae} and references therein). These constraints, however, become relaxed with simple extensions of the MSSM\footnote{For example, Ref.~\cite{Funakubo:2005pu} found that in the NMSSM (Next-to-Minimal Supersymmetric Extension of the Standard Model)---which contains an extra gauge singlet chiral superfield---a strongly first-order electroweak phase transition can be achieved with heavy squarks and Higgs masses heavier than the MSSM limit of 120 GeV.}, to which we would turn if future experiments rule out the minimal scenario. More optimistically, discoveries of nonzero EDMs may pinpoint masses of superpartners that are indeed consistent with the BAU generated through electroweak baryogenesis (by picking out phases within the colored band in Fig.~\ref{bau+edm}, for instance), even before their possible discovery at LHC. 

All the data combined together would provide a powerful test of the scenario of electroweak baryogenesis in the MSSM, ruling it out for good (and leading us to consider extensions of the MSSM or other models of new physics) or perhaps even confirming it in a dramatic way.

\section{Effective Field Theories for Strong Interactions}

Before we can reliably analyze the data from experiments like those at the LHC, we have much work to do establish reliable calculations within the Standard Model itself, especially in the sector of strong interactions. The theory describing these interactions, Quantum Chromodynamics (QCD), possesses the important property of asymptotic freedom which makes the interactions between quarks at high energies amenable to relatively simple perturbative theoretical calculations. However, at low energies, where quarks interact strongly and bind together to make hadrons, calculations are almost impossible, except numerically using lattice QCD.

\subsection{Nonperturbative Effects in Hadronic Jets}

One way to make progress analytically in the nonperturbative domain of QCD is to approximate the full theory with an effective field theory (EFT) which simplifies the separation of perturbative and nonperturbative effects. Given a particular class of physical processes to study, we can identify the relevant sector of the full theory which describes them. By integrating out degrees of freedom in the full theory which do not propagate for long times or distances and identifying small parameters characterizing the physics in which we can expand the full theory, we can form the appropriate EFT, which often possesses, at a given level of approximation, extra symmetries or other technical features facilitating the separation of perturbative and nonperturbative contributions to physical observables.

One application of this strategy is to event shape variables in $e^+ e^-$ annihilation or $Z$ decay to hadrons. Most events in these processes produce two or more jets of hadrons arising from the underlying partonic sub-processes. A number of variables conventionally used characterize the ``jettiness'' of an event. For example, in $Z$ decay, the energy of a produced jet $E_J$ is very close to $M_Z/2$ for an event with two back-to-back jets. Another popular variable is the thrust:
\begin{equation}
T = \frac{1}{M_Z}\max_{\hat{\vect{t}}}\sum_{\vect{i}}\abs{\hat{\vect{t}}\cdot\vect{p}_i},
\end{equation}
the sum of the projections of the three-momenta of all particles $i$ in the final state onto an axis $\hat{\vect{t}}$, chosen to be the axis which maximizes this sum. (For exactly back-to-back jets, $T=1$.) The distribution of events in thrust can be calculated in perturbation theory at the parton level, but hadronization effects will  introduce additional nonperturbative contributions to this observable. In Fig.~\ref{fig:thrust}, taken from Ref.~\cite{Korchemsky:1998ev}, we find a comparison of a prediction of the thrust distribution $d\sigma/dT$ in $e^+ e^-$ annihilation in perturbation theory (the red dotted line) to experimental data from LEP-II. The prediction does not quite match the data, especially near the kinematic endpoint $1-T\approx 0$ where two-jet-like events are found, until nonperturbative contributions are included, giving the black solid line. Of course, the nonperturbative contributions are not calculable; they are modeled using unknown parameters, which are merely fitted to the data. It would seem that no predictive power is truly gained.
\begin{figure}
\begin{center}
\includegraphics[width=12cm]{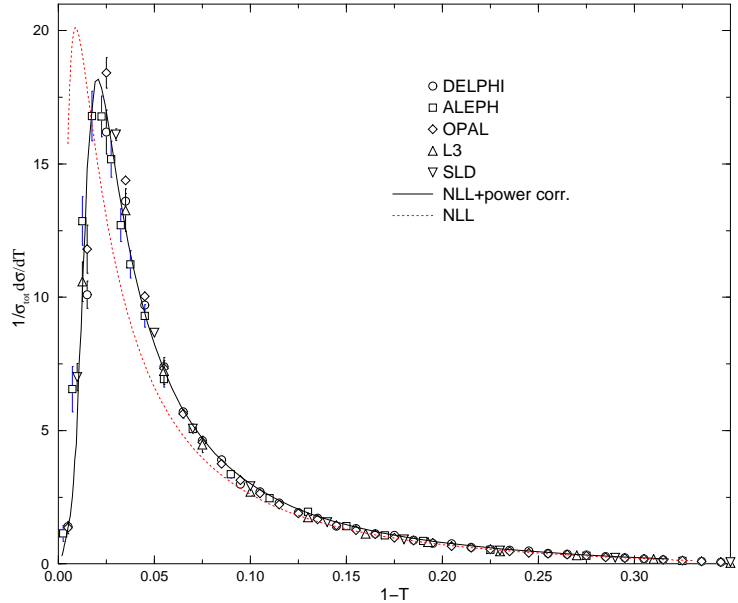}
\end{center}
\vspace{-1.5cm}
\caption[Thrust distribution in $e^+e^-$ annihilation.] {Thrust distribution in $e^+e^-$ annihilation. The dotted curve is a perturbative prediction for the thrust distribution, while the solid curve is a fit to the data in a model including the nonperturbative power corrections. (Figure from Ref.~\cite{Korchemsky:1998ev}.)}
\label{fig:thrust}
\end{figure}

However, the nonperturbative contribution to these observables can often be expressed as the matrix element of some operator in QCD or an effective theory which approximates it, giving us some possibly useful information. For example, the total decay rate of $Z$ to hadrons is given by:
\begin{equation}
\Gamma(Z\rightarrow\text{hadrons}) = \frac{1}{6M_Z}\sum_\epsilon \epsilon_\mu^*\epsilon_\nu\int d^4 x\,e^{ip_Z\cdot x}\bra{0}J^\mu(x) J^{\nu*}(0)\ket{0},
\end{equation}
summed over $Z$ polarizations $\epsilon$, and where the current mediating the $Z$ decay is:
\begin{equation}
J^\mu = \sum_{i,a}\bar q_i^a(g_V\gamma^\mu + g_A\gamma^\mu\gamma_5) q_i^a.
\end{equation}
Using the operator product expansion, we find that the product of currents can be expressed:
\begin{equation}
J_\mu(x)J_\nu^*(0) = C_{\mu\nu}^1(x)\Tr 1 + C_{\mu\nu}^{G^2}(x)\Tr G_{\alpha\beta}G^{\alpha\beta} + \cdots,
\end{equation}
$C^1\sim M_Z^6$, $C^{G^2}\sim M_Z^2$, and the higher-order terms are suppressed by more powers of $1/M_Z$.\footnote{This power counting would, of course, differ in six spacetime dimensions \cite{6d}.} The first term gives rise to the purely perturbative prediction for the decay rate, while the second encodes the leading-order nonperturbative physics, in the matrix element:
\begin{equation}
\label{G2}
\bra{0}\Tr G_{\alpha\beta}G^{\alpha\beta} \ket{0}\sim \lqcd^4.
\end{equation}
Thus the nonperturbative contribution to the total hadronic decay rate of the $Z$ is suppressed relative to the perturbative contribution by a factor of order $\sim(\lqcd/M_Z)^4$, or $\sim 10^{-9}$. This means that the perturbative calculation should be highly accurate; on the other hand, it would be extremely difficult to extract the value of the matrix element (\ref{G2}) from a measurement of $\Gamma(Z\rightarrow\text{hadrons})$.

Two features of an observable would enhance our predictive power. First, its leading nonperturbative contribution should be large enough that its effect would be measurable in an experiment while being small enough not to overwhelm the perturbative contribution. Second, the nonperturbative contribution should be characterized as the matrix element of operator which also contributes to a different physical observable. This way, its extracted value in one experiment could be used to predict the results of another.

The attempt to obtain these desirable features is the goal of our use of effective field theory to analyze event shape variables in hadronic $Z$ decay. We use an EFT designed to describe those processes involving strongly-interacting particles moving with large energy compared to their invariant mass, as in the hadronic jets produced in these events. In the \emph{soft-collinear effective theory} (SCET), we expand QCD in a small parameter $\lambda\sim\sqrt{\lqcd/M_Z}$, the typical transverse momenta of particles inside a jet of hadrons, and keep only lightlike (collinear) and soft degrees of freedom. At leading order in $\lambda$, the direct couplings in the Lagrangian between soft and collinear particles can be made to disappear in the effective theory by clever field redefinitions. This facilitates immensely the proof of factorization of event shape variables into hard and soft contributions. The latter are expressed as matrix elements of operators in the effective theory, and so we embark on the search for such matrix elements which are universal among different event shape variables.

In Ref.~\cite{shape}, Bauer, Manohar, and Wise showed, at leading order in SCET, that the energy $E_J$ of a single observed jet depends on no more than two nonperturbative parameters, regardless of the number of jets in the event. In Ref.~\cite{Bauer:2003di} and in this thesis, the nonperturbative contributions to other event shape variables are similarly analyzed (and the calculation in Ref.~\cite{shape} of the jet energy distribution in two-jet events is extended to $\mathcal{O}(\alpha_s)$ in perturbation theory.) Simple relations for different variables were proposed by Dokshitzer and Webber in Ref.~\cite{DokWeb95}. Our analysis suggests, unfortunately, that only two of the several commonly-used event shape variables---thrust and jet mass sum (see Sec.~\ref{ssec:jmass})---receive the same nonperturbative contributions, while those for the other variables are unrelated. The Dokshitzer-Webber model was tested by the DELPHI collaboration \cite{compare}, whose results are summarized in Fig.~\ref{delphi} and appear to be consistent with our claims.

\subsection{Reliable Predictions for Radiative Upsilon Decay}

Similar strategies can be pursued to analyze particular exclusive decays instead of inclusive or semi-inclusive processes like $Z$ decays to hadrons. The appearance of universal nonperturbative quantities in different exclusive decays would allow us to predict the branching ratios for particular decay channels once the nonperturbative contributions are measured in any one of them.

In this thesis we apply SCET and non-relativistic QCD (NRQCD) to decays of the $\Upsilon$ meson to a photon and one or more light hadrons. While SCET is appropriate to describe the final hadronic state, NRQCD is required to describe the heavy quark-antiquark pair inside the $\Upsilon$. 

\begin{figure}[b]
\begin{center}
\raisebox{1cm}{$b\bar b(8,^1S_0)$}\!\!\!\!\!\!\!\!\!\!
\begin{fmffile}{octet}
\begin{fmfgraph*}(120,60)
\fmfleft{b,bbar}
\fmfright{gamma,g}
\fmf{fermion}{b,v1,v2,bbar}
\fmf{photon}{v1,gamma}
\fmf{gluon}{v2,g}
\fmfforce{(0,h)}{b}
\fmfforce{(0,0)}{bbar}
\fmfforce{(w,h)}{gamma}
\fmfforce{(w,0)}{g}
\fmfforce{(.5w,h)}{v1}
\fmfforce{(.5w,0)}{v2}
\end{fmfgraph*}
\end{fmffile}
\qquad
\raisebox{1cm}{$b\bar b(1,^3S_1)$}\!\!\!\!\!\!\!\!\!\!
\begin{fmffile}{singlet}
\begin{fmfgraph*}(120,60)
\fmfleft{b,bbar}
\fmfright{gamma,g1,g2}
\fmf{fermion}{b,v1,v2,v3,bbar}
\fmf{photon}{v1,gamma}
\fmf{gluon}{v2,g1}
\fmf{gluon}{v3,g2}
\fmfforce{(0,h)}{b}
\fmfforce{(0,0)}{bbar}
\fmfforce{(w,h)}{gamma}
\fmfforce{(w,.5h)}{g1}
\fmfforce{(w,0)}{g2}
\fmfforce{(.5w,h)}{v1}
\fmfforce{(.5w,.5h)}{v2}
\fmfforce{(.5w,0)}{v3}
\end{fmfgraph*}
\end{fmffile}
\end{center}
\caption[Color-octet and color-singlet channels in radiative $\Upsilon$ decay.]{Color-octet and color-singlet channels in radiative $\Upsilon$ decay. The photon energy is more peaked near $E_\gamma = M_\Upsilon/2$ in the color-octet channel and may need to be included for a reliable prediction of the decay rate in this kinematic region.}
\label{ups_diags}
\end{figure}
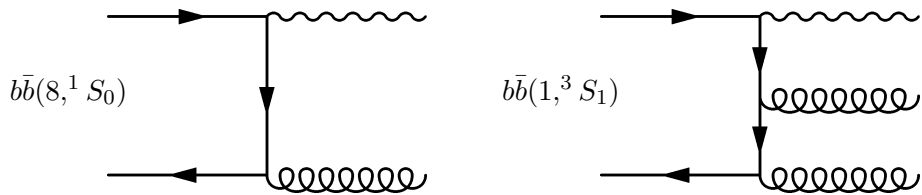
In NRQCD, the $\Upsilon$ decay rate can be split into several different contributions arising from the possible spin and color configurations of the inital $b\bar b$ pair. To leading approximation, the $b\bar b$ can be considered to be in a color-singlet, spin-triplet $^3S_1$ configuration which dominates the total $\Upsilon$ decay rate. For example, the photon energy spectrum in \emph{inclusive} radiative $\Upsilon$ decay depends on the sum of matrix elements:
\begin{equation}
\frac{d\Gamma}{dE_\gamma}(\Upsilon\rightarrow\gamma X) \sim \sum_n C_n\bra{\Upsilon}\mathcal{O}_n\ket{\Upsilon},
\end{equation}
for operators $\mathcal{O}_n$ with various spin and color quantum numbers.
In the approximation of the \emph{color singlet model}, only one operator contributes:
\begin{equation}
\frac{d\Gamma}{dE_\gamma}(\Upsilon\rightarrow\gamma X) \sim \bra{\Upsilon}\chi_{-\vect{p}}^\dag\boldsigma\psi_{\vect{p}}\cdot\chi_{-\vect{p}}^\dag\boldsigma\psi_{\vect{p}}\ket{\Upsilon}.
\end{equation}
This approximation is inadequate for some regions of the photon energy spectrum. Rothstein and Wise \cite{Rothstein:1997ac} pointed out that color octet channels, for example,
\begin{equation}
\bra{\Upsilon}\chi_{-\vect{p}}^\dag T^A \psi_{\vect{p}}\chi_{-\vect{p}}^\dag T^A\psi_{\vect{p}}\ket{\Upsilon},
\end{equation}
could be equally important as the color singlet near the kinematic endpoint $E_\gamma\sim M_\Upsilon/2$. As illustrated in Fig.~\ref{ups_diags}, while the color-singlet configuration must decay to at least two gluons, the color-octet can decay to just one. The photon energy will be peaked closer to $M_\Upsilon/2$ when recoiling against one gluon instead of two.

In Fig.~\ref{grafic}, we see that the color-singlet contribution is inadequate to account for the experimentally observed photon energy spectrum. Inclusion of the color-octet pieces are necessary to fit the data.
\begin{figure}
\begin{center}
\includegraphics[width=9cm]{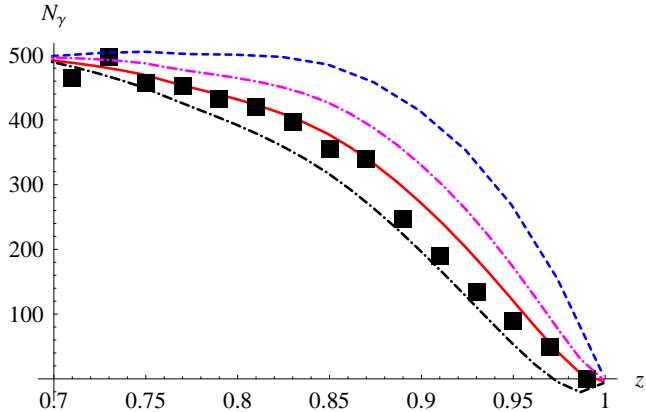}
\end{center}
\vspace{-5mm}
\caption[Endpoint region in photon spectrum in inclusive radiative $\Upsilon$ decay.] {Photon spectrum in inclusive radiative $\Upsilon$ decay. This plot of the number of events $N_\gamma$ with a photon energy $E_\gamma = M_\Upsilon z/2$ from Ref.~\cite{GarciaiTormo:2004jw} compares data from CLEO \cite{Nemati:1996xy} (the boxes) with predictions including only the color singlet contribution~\cite{Fleming:2002sr} (the blue dashed line) and including both color singlet and octet contributions~\cite{GarciaiTormo:2004jw} (the one solid and two dot-dashed lines). These latter three lines correspond to different choices for the renormalization scale $\mu$. The predictions which include the color octet contributions have significantly improved agreement with data.}
\label{grafic}
\end{figure}
Bauer, Fleming, \emph{et al.} in Ref.~\cite{Bauer:2001rh} and Fleming and Leibovich in Refs.~\cite{Fleming:2002sr,Fleming:2002rv} added SCET to the NRQCD analysis of radiative $\Upsilon$ decay, successfully introducing the collinear degrees of freedom necessary to analyze the dynamics near the kinematic endpoint $E_\gamma = M_\Upsilon/2$ in the photon energy spectrum. 

In Ref.~\cite{Fleming:2004hc} and this thesis, we extend these analyses to the case of exclusive decays $\Gamma(\Upsilon\rightarrow\gamma H)$ for a light hadron $H$. We show that the color-octet channel is in fact suppressed relative to the color-singlet even though we are in the region of large recoiling photon energy. The extra suppression comes from inclusion of interactions necessary to turn the single gluon produced in the color-octet decay into a color-singlet hadron in the final state. Then, having thus dropped the color-octet contribution, we predict the sizes of the decay rates of the $\Upsilon$ to different hadrons in the final state. Working only to leading order in the effective theory and using its various symmetries, we deduce the operators which contribute dominantly to the radiative decays of the $\Upsilon$ and which hadrons those operators could produce in the final state. We find at leading order that the $\Upsilon$ should radiatively decay dominantly to the flavor-singlet, parity-even $f_2(1270)$. Furthermore, the appearance of universal nonperturbative parameters for various quarkonia decays to the $f_2$ allows us to make predictions for the ratios of branching fractions such as
\begin{equation}
\frac{B(\Upsilon\rightarrow\gamma f_2)}{B(J/\psi\rightarrow\gamma f_2)} = 0.13-0.18,
\end{equation}
whereas experimentally this ratio is $0.06\pm 0.03$ \cite{Eidelman:2004wy}. 
Although experimental and theoretical uncertainties are still fairly large to make robust comparisons of these ratios, no evidence has yet been found for any exclusive radiative decays of $\Upsilon$ to anything other than the $f_2$. The observation of this channel before any others reassures us about the reliability of our prediction, whose further experimental scrutiny must await future data\footnote{Which appear to be forthcoming very shortly after the submission of this thesis \cite{CLEOemail}. Stay tuned!}.

\section{A Look Beyond}

New experiments promise an exciting era of particle physics in the coming decades, when we may verify current guesses about the nature of the physics that lies just beyond the Standard Model or find entirely other surprises. These prospects make imperative the improvement of theoretical calculations in the Standard Model and more reliable calculations of the predictions of its extensions for baryogenesis, EDMs, and other such experimental observables. With luck, the LHC and EDM searches will begin producing relevant data within a few years. With hard work, the theoretical calculations will prepare scenarios such as supersymmetric electroweak baryogenesis for subjection to the rigorous and reliable scrutiny of those data. This thesis is an effort to make progress in these directions so as to be ready for any new, exciting physics that the next era of theory and experiments may reveal!

\section{Plan of the Thesis}

We begin our investigations beyond the Standard Model in Chap.~\ref{chap:ewb}, and calculate the baryon asymmetry of the Universe generated in electroweak baryogenesis in the MSSM and address the consistency of these results with constraints on $CP$ violation in the MSSM from experimental searches for elementary particle EDMs. Much, though not all, of the material in this chapter was published in Ref.~\cite{Lee:2004we}. 

Then, returning to the Standard Model, in Chap.~\ref{chap:eft} we review the essential features of effective field theories in QCD, namely, soft-collinear effective theory and non-relativistic QCD, which will be applied to the radiative decays of $\Upsilon$ mesons in Chap.~\ref{chap:ups} (much of which was published in Ref.~\cite{Fleming:2004hc}) and the hadronic decays of $Z$ bosons in Chap.~\ref{chap:jet} (much of which was published in Ref.~\cite{Bauer:2003di}), leading to the conclusion and future outlook in Chap.~\ref{fin}.
\chapter{Constraints on Supersymmetric Electroweak Baryogenesis}
\label{chap:ewb}
\thispagestyle{empty}

\begin{quote}
\singlespace\small
In principio creavit Deus c\ae lum et terram. \\
Terra autem erat inanis et vacua et tenebr\ae\ super faciem abyssi et spiritus Dei ferebatur super aquas. \\
\flushright\vskip -22pt\emph{Genesis} 1:1--2
\end{quote}
We begin our investigations beyond the Standard Model, considering the constraints that currently available data, especially the baryon density of the Universe and limits on electric dipole moments of elementary particles, might place on the parameters of the speculative but popular Minimal Supersymmetric Extension of the Standard Model, which we adopt as a theoretical testing ground for the application of the closed-time-path formalism of quantum field theory to the calculation of the baryon density generated in electroweak baryogenesis.  (Much of this chapter appeared in Ref.~\cite{Lee:2004we}.)

\section{Introduction}
\label{sec:introduction}

The origin of the baryon asymmetry of the Universe (BAU) remains an
important, unsolved problem for  particle physics and
cosmology. Assuming that the Universe was matter-antimatter symmetric
at its birth, it is reasonable to suppose that interactions involving
elementary particles generated the BAU during subsequent cosmological
evolution. As noted by Sakharov \cite{Sakharov:1967dj}, obtaining a nonzero BAU requires
both a departure from thermal equilibrium as well as the breakdown of
various discrete symmetries: baryon number ($B$) conservation, charge
conjugation ($C$) invariance, and invariance under the combined $C$
and parity ($P$) transformations\footnote{Allowing for a breakdown of
$CPT$ invariance relaxes the requirement of departure from thermal
equilibrium.}.  The Standard Model (SM) of strong and electroweak
interactions satisfies these conditions and could, in principle,
explain the observed size of the BAU: 
\be
Y_B\equiv \frac{\rho_B}{s} = 
\begin{cases}(7.3\pm 2.5)\times 10^{-11}, & \text{BBN}\\
							(9.2\pm 1.1)\times 10^{-11}, & \text{WMAP}
\end{cases}
\ee
where $\rho_B$ is the baryon number density, $s$ is the entropy density of the universe, and where the values shown correspond to 95\% confidence level results obtained from Big Bang Nucleosynthesis (BBN) \cite{Eidelman:2004wy} and the Wilkinson Microwave Anisotropy Probe (WMAP) \cite{wmap}, respectively. 
In practice, however, neither the strength of the first-order electroweak phase transition in the SM nor the magnitude of SM
$CP$-violating interactions are sufficient to prevent washout of any
net baryon number created by $B$-violating electroweak sphaleron
transitions during the phase transition.

The search for physics beyond the SM is motivated, in part, by the
desire to find new particles whose interactions could overcome the
failure of the SM to explain the BAU. From a phenomenological
standpoint, a particularly attractive possibility is that masses of
such particles are not too different from weak scale
and that their interactions both strengthen the first-order
electroweak phase transition and provide the requisite level of
$CP$-violation needed for the BAU. Precision electroweak measurements
as well as direct searches for new particles at the Tevatron and Large
Hadron Collider may test this possibility, and experiment already
provides rather stringent constraints on some of the most widely
considered extensions of the SM. In the Minimal Supersymmetric Standard
Model (MSSM), for example, present lower bounds on the mass of the lightest
Higgs boson leave open only a small window for a sufficiently strong
first-order phase transition, although this constraint may be relaxed
by introducing new gauge degrees of freedom (see, {\em e.g.}, \cite{Dine:2003ax,Kang:2004pp}). Similarly, limits on the permanent electric dipole moments (EDMs) of elementary
particles and atoms imply that the $CP$-violating phases in the MSSM
must be unnaturally small ($\sim\! 10^{-2})$. Whether such small phases
(supersymmetric or otherwise) can provide for successful electroweak
baryogenesis (EWB) has been an important consideration in past studies
of this problem.

In order to confront phenomenological constraints on the parameters of
various electroweak models with the requirements of EWB, one must
describe the microscopic dynamics of the electroweak phase transition
in a realistic way. Theoretically, the basic mechanism driving
baryogenesis during the phase transition is 
well-established. Weak sphaleron transitions that conserve $B-L$ but
change $B$ and $L$ individually are unsuppressed in regions of
spacetime where electroweak symmetry is unbroken, while they become
exponentially suppressed in regions of broken symmetry. Net baryon
number is captured by expanding regions of broken symmetry (\lq\lq
bubbles"). Given sufficiently strong $C$ and $CP$-violation as well as
departure from thermal equilibrium, the non-zero $B$ generated outside the bubble cannot be entirely washed out by elementary particle interactions that occur at the phase boundary. The baryon number density, $\rho_B$, is governed by a diffusion equation of the form:
\be
\label{eq:rhob1}
{\partial }_t \rho_B(x)  -D\nabla^2\rho_B(x) =
- \Gamma_{\rm ws} F_{\rm ws}(x)[n_L(x) + R\rho_B(x)]\,,
\ee
where $D$ is the diffusion coefficient for baryon number, $\Gamma_{\rm
ws}$ is the weak sphaleron transition rate, $F_{\rm ws}(x)$ is a
sphaleron transition profile function that goes to zero inside the
regions of broken electroweak symmetry and asymptotically to unity
outside, $R$ is a relaxation coefficient for the decay of baryon
number through weak sphaleron transitions, and $n_L(x)$ is the number
density of left-handed doublet fields created by \lq\lq fast" chirality changing processes (see, {\em e.g.}, \cite{Cline:1993bd}). Thus, in order
to obtain nonzero $\rho_B$ inside the bubble of broken electroweak
symmetry, the left-handed density $n_L$ must be non-vanishing
in the plasma at the phase boundary and possibly beyond into the
region of unbroken symmetry. 

In effect, $n_L(x)$ acts as a seed for the $B$-changing weak sphaleron
transitions, and its spacetime profile is determined by the
$CP$-violating sources and the quantum transport of various charges in
the non-equilibrium environment of the plasma. Typical treatments of
these dynamics involve writing down a set of coupled quantum transport
equations (QTEs) for the relevant charges, estimating (or
parameterizing) the relevant transport coefficients, and solving the
system of equations under the appropriate boundary conditions.  

Among the developments in the past decade or so which have made significant impacts on
this program, we identify two that  form the basis of our investigation in this work. First, the authors of Ref.~\cite{Cohen:1994ss} noted
that diffusion of chiral charge ahead of the advancing phase
transition boundary into the region of unbroken symmetry could enhance the
impact of baryon number-changing sphaleron processes, thereby leading
to more effective EWB. The second, perhaps less widely-appreciated, development has been the observation by the author of Ref. \cite{Riotto:1998zb}
that the application of equilibrium quantum field theory (QFT) to
transport properties in the plasma is not necessarily appropriate.  In
contrast to equilibrium quantum dynamics, the time evolution of
quantum states during the phase transition is
non-adiabatic. Consequently, scattering processes that drive quantum
transport are no longer Markovian, but rather retain some memory of
the system's quantum evolution. Using the closed time path (CTP)
formulation of non-equilibrium QFT \cite{CTP} to compute the
$CP$-violating source terms in the plasma for the MSSM, the author of
Ref.~\cite{Riotto:1998zb} found that these ``memory effects" may lead
to significant resonant enhancements (of order $10^3$) of the sources over
their strength estimated in previous treatments (see, {\em e.g.}, Ref. \cite{Huet:1995sh} and references therein). The authors of Ref. \cite{Carena:2000id,Carena:2002ss} subsequently found that performing an all-orders summation of scattering from Higgs backgrounds reduces  the size of the $CP$-violating sources to some extent, but that the resonant enhancements nonetheless persist. Taken at face value,
these enhancements would imply that successful EWB could occur 
with significantly smaller $CP$-violating phases than previously
believed, thereby evading the present and prospective limits obtained
from EDMs.

To determine whether or not such conclusions are warranted, however,
requires that one treat the other terms in the transport equations in
the same manner as the $CP$-violating sources. Here, we attempt to do so, focusing on the terms that, in previous studies, have governed the relaxation of $n_L(x)$. In particular, chirality-changing Yukawa interactions with the Higgs fields and their spacetime varying vacuum expectation values (vevs) tend to wash out excess $n_L(x)$. In earlier studies---including those in which non-equilibrium QFT has been applied to the $CP$-violating sources---these relaxation terms were estimated using conventional quantum transport theory \cite{Huet:1995sh,Riotto:1998zb,Carena:2000id,Carena:2002ss}. However, if the memory effects that enhance the $CP$-violating sources have a similar effect on these Yukawa terms, then
the net effect on $\rho_B$ may not be as substantial as suggested in
Refs. \cite{Riotto:1998zb,Carena:2000id,Carena:2002ss}.  

The goal of the present study is to
address this question by developing a more comprehensive
treatment of EWB using the CTP formulation of non-equilibrium QFT.  In doing
so, we follow the direction suggested in Ref. \cite{Riotto:1998zb} and
compute the transport coefficients of the chiral charges using the CTP
formalism. To make the calculation more systematic, we identify the relevant energy and time scales that govern finite temperature, non-equilibrium dynamics and develop a power counting in the ratios of small to large scales (generically denoted here as $\epsilon$). As we show below, both the $CP$-violating sources and the driving relaxation terms first arise at ${\cal O}(\epsilon^2)$, and we truncate our analysis at this order. In contrast to the computation of the $CP$-violating sources, the derivation of the relaxation terms requires the use of finite density Green's functions. Given the resulting complexity, we consider here only
the terms in the transport equations that previous authors have
considered the dominant ones, and use our analysis of these terms to
illustrate a method for obtaining a more comprehensive
treatment of the QTEs. To make the phenomenological implications
concrete, we focus on the MSSM, realizing, however, that one may need to
include extensions of the MSSM in order to satisfy the requirements of
a strong first-order phase transition. Finally, we also attempt to identify the
different approximations that have entered previous treatments of EWB,
such as the implicit truncation at a given order in $\epsilon$ and outline additional calculations needed to obtain a comprehensive treatment. 

Based on our  analysis, we find that under that same conditions that lead to resonant enhancements of the $CP$-violating sources, $S^{\CPV}$, one also obtains a similar, resonant enhancement of the driving chirality-changing transport coefficient, ${\bar\Gamma}$. Since $Y_B\sim S^{\CPV}/\sqrt{\bar\Gamma}$, resonant relaxation counteracts the enhanced sources, though some overall enhancement of EWB still persists. Consequently, it will be important in future work to study the other transport coefficients whose impact has been considered sub-leading, since they may be enhanced under conditions other than those relevant for the leading terms. From the standpoint of phenomenology, we also illustrate how the implications of EDM searches for EWB depends in a detailed way on the electroweak model of interest as well as results from collider experiments and precision electroweak data. 

In presenting our study, we attempt to be somewhat pedagogical, since the methods are, perhaps, not generally familiar to either the practitioners of field theory or experimentalists.  Most of the formal development appears in Sections \ref{sec:CTP}--\ref{sec:qtes}. In
Section~\ref{sec:CTP} we review the CTP formalism and its application
to the QTEs and discuss in detail the formulation of density-dependent
Green's functions.  In
Section~\ref{sec:source} we compute the $CP$-violating source terms,
providing a check of Ref.~\cite{Riotto:1998zb}, as well as the
transport coefficients of the chiral charge densities. Here, we also
enumerate the approximations used to obtain a set of coupled, linear
differential diffusion equations, discuss their limits of validity,
and identify additional terms (usually assumed to be sub-leading) that we
defer to a future study. In Section~\ref{sec:qtes} we solve these
equations for the baryon density.  A reader primarily interested in the phenomenological implications may want to turn directly to 
Section~\ref{sec:numerics}, which  gives
illustrative numerical studies using the parameters of the
MSSM. A discussion of the implications for EDMs also appears here. Section~\ref{sec:summary} contains a summary and outlook, while several technical details appear in the Appendices.

\section{Non-equilibrium Transport: CTP Formulation}
\label{sec:CTP}

In what follows, we treat all $CP$-violating and non-topological
chirality-changing interactions perturbatively\footnote{Sphaleron
transitions, however, are manifestly non-perturbative, and we
parameterize their effects in the standard way.}. In contrast to
zero-temperature, equilibrium perturbation theory, however, the
perturbative expansion under non-equilibrium,
$T>0$ conditions requires the use of a more general set of Green's functions
that take into account the non-adiabatic evolution of 
states as well as the presence of degeneracies in the thermal bath. Specifically, the matrix element of any operator 
$\mathcal{O}(x)$ in the interaction representation is given by:
\be
\label{eq:ctp1}
\langle n| S^{\dag}_{\rm int} T\{\mathcal{O}(x) 
S_{\rm int}\} | n\rangle\,  , 
\ee
where 
\be
S_{\rm int} = T\exp\left( i\int d^4x \,{\cal L}_{\rm int}\right)
\ee
for an interaction Lagrangian ${\cal L}_{\rm int}$, $T$ is the
time-ordering operator, and $|n\rangle$ is an in-state. Inserting a complete set of states inside Eq.~(\ref{eq:ctp1}), we obtain:
\begin{equation}
\label{eq:ctp1a}
\sum_{m}\bra{n}S_{\text{int}}^\dag\ket{m}\bra{m}T\{\mathcal{O}(x)S_{\text{int}}\}\ket{n}.
\end{equation}
In ordinary, zero-temperature equilibrium field theory, the assumptions of
adiabaticity and of non-degeneracy of the states $\ket{n}$ imply
that only the single state $m=n$ contributes to this sum, so the only impact of $S^{\dag}_{\rm int}$ is the introduction of an
overall phase, allowing one to rewrite Eq.~(\ref{eq:ctp1a}) as:
\be
\label{eq:ctp2}
\frac{\langle n | T\{\mathcal{O}(x)  S_{\rm int}\} | n\rangle}{
\langle n| S_{\rm int} |n\rangle}\,.
\ee
This simplification is no longer valid for non-equilibrium $T>0$ evolution,
and one must take into account the action of $S^{\dag}_{\rm int}$
appearing to the left of $\mathcal{O}(x)$ in (\ref{eq:ctp1}). Doing so is
facilitated by giving every field in $S_{\rm int}$ and $S^{\dag}_{\rm
int}$ a ``$+$" and ``$-$" subscript respectively. The matrix
element in (\ref{eq:ctp1}) then becomes:
\be
\label{eq:ctp3}
\langle n| {\cal P}\left\{ \mathcal{O}(x) 
\exp \left(i\int d^4x\ {\cal L}_{+} - i\int d^4x\ 
{\cal L}_{-}\right)\right\}|n\rangle\,,
\ee
where the path ordering operator ${\cal P}$ indicates that all 
``+" fields appear to the right of all ``$-$" fields, with the former
being ordered according to the usual time-ordering prescription and
the latter being anti-time-ordered [here, $\mathcal{O}(x)$ has been taken to
be a ``+" field]. Note that the two integrals in the exponential
in (\ref{eq:ctp3}) can be written as a single integral along a closed
time path running from $-\infty$ to $+\infty$ and then back to
$-\infty$:
\begin{equation}
\int_{-\infty}^\infty dt\int d^3 x (\mathcal{L}_+ - \mathcal{L}_-) = \int_{\mathcal{C}} dt\int d^3 x\,\mathcal{L},
\end{equation}
where the time $t$ on the right-hand side is integrated over the contour $\mathcal{C}$ shown in Fig.~\ref{fig:ctp}.
\setlength{\unitlength}{1mm}
\begin{figure}
\begin{center}
\begin{picture}(0,40)(0,-10)
\drawline(-56,0)(56,0)
\put(56,0){\vector(1,0){0}}
\put(-54,2){$-\infty$}
\put(47,2){$+\infty$}
\put(58,-1){$t$}
\drawline(0,-15)(0,15)
\put(0,15){\vector(0,1){0}}
\thicklines
\drawline(-47,2)(45,2)
\drawline(45,-2)(-47,-2)
\put(45,0){\arc{4}{-1.57}{1.57}}
\put(-20,2){\vector(1,0){0}}
\put(-20,4){$\mathcal{C}$}
\put(-20,-2){\vector(-1,0){0}}
\put(20,2){\circle*{1}}
\put(15,-2){\circle*{1}}
\put(19,4){$\phi_+$}
\put(14,-6){$\phi_-$}
\end{picture}
\end{center}
\caption[Closed time path integration contour.]{Closed time path integration contour. Fields $\phi$ are distinguished according to their placement on the forward ($\phi_+$) or backward ($\phi_-$) portions of the contour.}
\label{fig:ctp}
\end{figure}
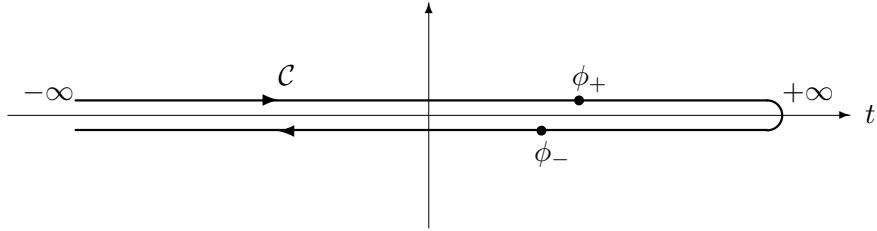
\setlength{\unitlength}{1pt}

Perturbation theory now proceeds from the matrix element
(\ref{eq:ctp3}) along the same lines as in ordinary field theory via
the application of Wick's theorem, but with the more general ${\cal
P}$ operator replacing the $T$ operator. As a result, one now has a
set of four two-point functions, corresponding to the different
combinations of ``+" and ``$-$" fields that arise from
contractions. It is convenient to write them as a matrix ${\widetilde
G}(x,y)$:
\be
\label{eq:ctp4}
\widetilde G(x,y)=
\left(\begin{array}{cc}
G^t(x,y) & -G^<(x,y) \\
G^>(x,y) & -G^{\bar t}(x,y)
\end{array}\right)
\ee
where 
\begin{subequations}
\label{eq:Greens1}
\begin{align}
G^>(x,y) &= \langle \phi_-(x) \phi_+^\dag(y) \rangle \\
G^<(x,y) &= \langle \phi_-^\dag(y) \phi_+(x)\rangle \\
G^t(x,y) &= \langle T\bigl\{\phi_+(x) \phi_+^\dag(y)\bigr\}\rangle  = 
\theta(x_0-y_0)G^>(x,y)+ \theta(y_0-x_0)G^<(x,y)\\
G^{\bar t}(x,y) &= 
\langle \bar T\bigl\{\phi_-(x) \phi_-^\dag(y)\bigr\}\rangle  = 
\theta(x_0-y_0)G^<(x,y) + \theta(y_0-x_0)G^>(x,y)\,,
\end{align}
\end{subequations}
 and where the $\langle\ \ \rangle$ denote ensemble averages,
\be
\langle \mathcal{O} (x)\rangle \equiv \frac{1}{Z}
 {\rm Tr}\left[\hat\rho \,  \mathcal{O}(x)\right]\, . 
\ee 
Here $\hat\rho$ is the density matrix containing 
information about the state of the system, and $Z = \Tr(\hat\rho)$.  
In thermal equilibrium  $\hat\rho$ is time-independent and is 
given by  $\hat\rho = e^{-\beta (\op{H} - \mu_i \op{N}_i)}$
for a grand-canonical ensemble. 
Note that the matrix ${\widetilde G}(x,y)$ may be written more compactly as:
 \be
\widetilde G(x,y)_{a b} = 
\langle {\cal P}\left\{\phi_a(x)\phi_b^\dag(y)\right\}\rangle
(\tau_3)_{bb}  \,.
 \ee
The presence of the $\tau_3$ factor is a bookkeeping device to keep
track of the relative minus sign between the ${\cal L}_+$ and ${\cal
L}_-$ terms in Eq. (\ref{eq:ctp3}).
 
The path-ordered two-point functions satisfy the Schwinger-Dyson equations:
\begin{subequations}
\label{eq:sd}
\begin{align}
\label{eq:sda}
{\widetilde G}(x,y) &= {\widetilde G}^0(x,y)+\int d^4w \int d^4 z\  
{\widetilde G}^0(x,w){\widetilde \Sigma}(w,z){\widetilde G}(z,y) \\
\label{eq:sdb}
{\widetilde G}(x,y) &= {\widetilde G}^0(x,y)+\int d^4w \int d^4 z\ 
{\widetilde G}(x,w){\widetilde \Sigma}(w,z){\widetilde G}^0(z,y)\,,
\end{align}
\end{subequations}
where the ``0" superscript indicates a non-interacting Green's
function and where ${\widetilde\Sigma}(x,y)$ is the matrix of interacting
self energies defined analogously to the ${\widetilde G}(x,y)$. An
analogous set of expressions apply for fermion Green's functions, with
an appropriate insertion of $-1$ to account for anticommutation
relations.

\subsection{Quantum Transport Equations from CTP Formalism}
 
The Schwinger-Dyson Eqs.~(\ref{eq:sd}) are the starting
point for obtaining the transport equations governing $n_L(x)$. To do
so, we follow Ref. \cite{Riotto:1998zb} and apply the Klein-Gordon
operator to ${\widetilde G}(x,y)$. Using
\be
\left({{\square}}_x+m^2\right) {\widetilde G}^0(x,y)=  
\left({{\square}}_y+m^2\right) {\widetilde G}^0(x,y)=-i\delta^{(4)}(x-y) 
\ee
gives
\begin{subequations}
\begin{align}
\left({{\square}}_x+m^2\right){\widetilde G}(x,y) &= -i\delta^{(4)}(x-y) -i
\int d^4z\ {\widetilde\Sigma}(x,z){\widetilde G}(z,y) \\
\left({{\square}}_y+m^2\right){\widetilde G}(x,y) &= -i\delta^{(4)}(x-y) -i
\int d^4z\ {\widetilde G}(x,z){\widetilde\Sigma}(z,y)\,.
\end{align}
\end{subequations}
It is useful now to consider the $(a,b)=(1,2)$ components of these equations:
\begin{subequations}
\begin{align}
\label{eq:sdc}
\left({{\square}}_x+m^2\right)G^<(x,y) &= -i\int d^4z\ 
\left[\Sigma^t(x,z) G^<(z,y)-\Sigma^<(x,z) G^{\bar t}(z,y) \right]\\
\label{eq:sdd}
\left({{\square}}_y+m^2\right)G^<(x,y) &= -i\int d^4z\ \left[G^t(x,z) 
\Sigma^<(z,y)-G^<(x,z) \Sigma^{\bar t}(z,y) \right]\,.
\end{align}
\end{subequations}
Subtracting Eq. (\ref{eq:sdd}) from Eq. (\ref{eq:sdc}) and multiplying
through by $i$ gives
\be
\label{eq:sde}
i\left({\square}_x-{\square}_y\right) G^<(x,y)\bigr\rvert_{x=y\equiv X} =i 
\partial_\mu^X\left(\partial^\mu_x-\partial^\mu_y\right)G^<(x,y)\bigr\rvert_{x=y\equiv X}\,.
\ee
However, 
\be
(\partial^\mu_x-\partial^\mu_y) G^<(x,y)\bigr\rvert_{x=y\equiv X} = -i j_\phi^\mu(X)\,,
\ee
where $j_\phi^\mu(x) = i\langle :\!\phi^\dag(x)\overset{\leftrightarrow}{\partial_\mu}\phi(x)\!:\rangle \equiv(n_\phi(x), \vect{j}_\phi(x))$, since the ``$+$" and ``$-$" labels
simply indicate the order in which the fields $\phi^\dag(y)$ and
$\phi(x)$ occur and may be dropped at this point. Finally, expressing $G^{t, {\bar t}}(x,y)$ and
$\Sigma^{t,{\bar t}}(x,y)$ in terms of $\theta$-functions as in
Eqs.~(\ref{eq:Greens1}), we obtain from Eq.~(\ref{eq:sde}):
\begin{equation}
\begin{split}
\pd{n_\phi}{X_0}+{\mbox{\boldmath$\nabla$}}\mcdot\vect{j}_\phi(X) 
= \int d^3 z\int_{-\infty}^{X_0} dz_0\
\Bigl[ \Sigma^>(X,z) G^<(z,X)&-G^>(X,z)\Sigma^<(z,X)\\
+G^<(X,z) \Sigma^>(z,X) &- \Sigma^<(X,z) G^>(z,X)\Bigr]\,.
\label{eq:scalar1}
\end{split}
\end{equation}
Following similar steps, but taking the sum rather than the difference of the components of the Schwinger-Dyson equations involving the $S^{>}(x,y)$ component on the LHS, one obtains the analogous continuity equation 
for Dirac fermions:
\begin{equation}
\begin{split}
\pd{n_\psi}{X_0} + {\mbox{\boldmath$\nabla$}}\mcdot\vect{j}_\psi(X) =  
-\int d^3 z\int_{-\infty}^{X_0} dz_0\
{\rm Tr}\Bigl[ \Sigma^>(X,z) S^<(z,X)&-S^>(X,z)\Sigma^<(z,X)\\
+S^<(X,z) \Sigma^>(z,X) &- \Sigma^<(X,z) S^>(z,X)\Bigr]\,,
\label{eq:fermion1}
\end{split}
\end{equation}
where
\begin{equation}
j_\psi^\mu(x) = \langle :\!\bar\psi\gamma^\mu\psi(x)\!:\rangle,
\end{equation}
and
\begin{subequations}
\begin{align}
S^>_{\alpha\beta}(x,y) &=  \langle \psi_{-\alpha}(x) {\bar\psi}_{+\beta}(y)\rangle \\
S^<_{\alpha\beta}(x,y) &= -\langle {\bar\psi}_{-\beta}(y) {\psi}_{+\alpha}(x)\rangle\,,
\end{align}
\end{subequations}
displaying explicitly the spinor indices $\alpha,\beta$.
Note that the overall sign of the RHS of Eqs.~(\ref{eq:scalar1},
\ref{eq:fermion1}) differs from that in Ref. \cite{Riotto:1998zb}
since the definition of our Green's functions $G(x,y)$ and $S(x,y)$
differ by an overall factor of $-i$.

In many extensions of the SM, one encounters both chiral and Majorana fermions, which carry no conserved charge. It is useful, therefore, to derive the analogous continuity equation for the axial current $j_{\mu 5}(x) = \langle {\bar\psi}(x)\gamma_\mu\gamma_5\psi(x)\rangle$. Doing so involves multiplying the Schwinger-Dyson equations by $\gamma_5$, performing the trace, and taking the difference rather than the sum of the components involving $S^{>}(x,y)$ on the LHS. The result is:
\begin{align}
\label{eq:fermion1b}
\pd{n_5}{X_0} + {\mbox{\boldmath$\nabla$}}\mcdot\vect{j}_5(X) = & 2im P(X) \\
&+\int d^3 z\int_{-\infty}^{X_0} dz_0\Tr\Bigl\{\Bigl[\Sigma^>(X,z) S^<(z,X) + S^>(X,z)\Sigma^<(z,X) \nonumber \\
&\qquad\qquad\qquad\qquad-S^<(X,z) \Sigma^>(z,X) - \Sigma^<(X,z) S^>(z,X)\Bigr]\gamma_5\Bigr\}\,, \nonumber
\end{align}
where $P(x) = \langle {\bar\psi}(x)\gamma_5 \psi(x)\rangle$ and $m$ is the fermion mass. In principle, one could evaluate $P(x)$ using path-ordered perturbation theory as outlined above.

\subsection{Power Counting of Physical Scales}

Evaluating the various terms in Eqs.~(\ref{eq:scalar1},
\ref{eq:fermion1}) leads to a system of coupled quantum transport
equations for the charges that ultimately determine $n_L(x)$. On the
LHS of these equations, it is conventional to parameterize $\vect{j} =
- D(\mbox{\boldmath$\nabla$}\!n)$, in terms of the diffusion coefficient
$D$ (whose expressions we take from Ref.~\cite{Joyce:1994zn}).
The RHS involves integrating the products of various Green's functions
and self-energies over the entire history of the system. In practice,
this integral depends on the various time and energy scales associated with non-equilibrium dynamics at finite temperature and density. Here, we observe that there exists a hierarchy among these scales that leads to a natural power counting in their ratios (generically denoted here as
$\epsilon$) and that provides for a systematic expansion of the RHS of the transport equations
(\ref{eq:scalar1}, \ref{eq:fermion1}, \ref{eq:fermion1b}). 

The changing geometry associated with the expanding region of broken symmetry and the spacetime variation of the Higgs vevs leads to a decoherence of states that have, initially, precise energy and momentum. The effect is analogous to the quantum mechanical evolution of a particle in a box of side $L$. If the value of $L$ is changed to $L+\Delta L$ in some time interval $\Delta t$, a state that is initially a stationary state for the original box will become an admixture of the stationary states of new box. The shorter the interval $\Delta t$ or the greater the wavenumber $k$ of the initial state, the smaller the probability will be of finding the particle in the state with the same wavenumber in the new system. The time scale that characterizes this decoherence, $\tau_d$, is naturally given by $\tau_d\sim 1/vk$, where $v=\Delta L/\Delta t$ is the velocity of expansion of the box and $k=p/\hbar$. In the present case, the relevant velocity is just $v_w$, the expanding bubble wall velocity, the relevant effective wave number $k$ depends on $\abs{\vect{k}}$ and the wall thickness, $L_w$. The smaller the velocity or the longer the wavelength, the more adiabatic the dynamics of the expanding bubble become and the longer the decoherence time. Equilibrium dynamics are approached in the adiabatic limit: $\tau_d\to\infty$. The need to employ the CTP formalism follows from being in a situation with $v_w>0$, or $\tau_d <\infty$.

A second time scale that one encounters in quantum transport at the phase boundary arises from the presence of degeneracies among states in the thermal bath that vanish in the $T\to 0$ limit. At finite $T$, for example, a single, on-shell fermion may be degenerate with another  state involving an on-shell fermion-gluon pair---a situation that is forbidden at $T=0$. Interactions of strength $g$ that cause mixing between such degenerate states give rise to thermal---or plasma---widths $\Gamma_p$ of order $\alpha T$ with $\alpha=g^2/4\pi$, and transitions between the degenerate states take place on a plasma time scale $\tau_p$ of order $\sim 1/\Gamma_p$. Again, the use of the CTP formalism is necessitated when $\tau_p <\infty$ or $T> 0$.

A third time scale, which we denote $\tau_{\rm int}$, is associated with the intrinsic frequency $\omega_k$ of the quasiparticle states that characterize the plasma dynamics. This time scale is naturally given by $\tau_{\rm int}\sim1/\omega_k$. In the present case, we note that although the decoherence and plasma times are finite, they are typically  much larger than $\tau_{\rm int}$. For example, $\tau_{\rm int}/\tau_d = v_w k/\omega_k \leq v_w/c$. Numerical studies indicate that $v_w/c \ll 1$. Similarly, $\tau_{\rm int}/\tau_p = \alpha T/\omega_k$. Since quasiparticle thermal masses are of order $gT$ or larger, one also has that the latter ratio is smaller than unity.  Thus, one is naturally led to expand the RHS of the transport equations in these ratios:
\begin{subequations}
\label{eq:tauratios}
\begin{align}
\label{eq:decratio}
0 &< \tau_{\rm int}/\tau_d \ll 1 \\
\label{eq:plasratio}
0 &< \tau_{\rm int}/\tau_p \ll 1\,.
\end{align}
\end{subequations}

Finally, we observe that the generation of baryon number takes place in an environment of finite, but small particle number (or chiral charge) densities $n_i$ that are associated with chemical potentials $\mu_i$. For the temperatures and densities of interest here, one has $|\mu_i|/T \ll 1$, so that the latter ratio also provides for a natural expansion parameter. Denoting each of the ratios\footnote{For our purposes, it is not necessary to distinguish a hierarchy among the different scale ratios, as we work to leading nontrivial order in $\epsilon$.} in Eq.~(\ref{eq:tauratios}) and $\mu_i/T$ by $\epsilon$, we show below that both the $CP$-violating sources and the relaxation term first arise at ${\cal O}(\epsilon^2)$, and we truncate our analysis at this order. We note that doing so introduces some simplifications into the evaluation of the RHS of the transport equations. For example, both the self energies $\Sigma^\gtrless$ and the Green's functions $G^\gtrless$, {\rm etc.} depend on thermal distribution functions $f(T,\mu_i)$ that differ, in general, from their equilibrium values, $f_0(T,\mu_i)$. The difference $\delta f\equiv f(T,\mu_i)-f_0(T,\mu_i)$ that characterizes the departure from equilibrium will be at least of ${\cal O}(\epsilon)$, since it must vanish in the $v_w\to 0$ limit. We find below that the effect of having $\delta f\not\!= 0$ contributes at higher order in $\epsilon$ than we consider here, so that we may use the equilibrium distribution functions in the Green's functions and self-energies.

\subsection{Green's Functions at Nonzero Temperature and Density}

The computation of the various components of ${\widetilde G}(x,y)$ and
${\widetilde\Sigma}(x,y)$ appearing in Eqs. (\ref{eq:scalar1},
\ref{eq:fermion1}) at nonzero temperature and density requires
knowledge of $(T,\mu_i)$-dependent fermion and boson
propagators. The $T$-dependence of propagators has been studied extensively
(see, for instance, Ref.~\cite{LeBellac} and references therein), while the $\mu_i$-dependence of fermion propagators has been studied in Refs.~\cite{finitemu}. Here we summarize the features of $(T,\mu_i)$-dependent propagators which are important for our subsequent application of the real-time, CTP formalism of Sec.~\ref{sec:CTP}, and give some more technical details in Appendix~\ref{appx:props}. 

For pedagogical purposes, we provide
here a brief derivation of the non-interacting fermion propagator but
only give final results for the case of interacting fermions and
bosons. To do so, we start from the mode
expansions for the field operators appearing in the free Dirac
Lagrangian, $\psi(x)$ and ${\bar\psi}(x)$:
\begin{subequations}
\label{eq:modeexp}
\begin{align}
\psi(x) &=  \int\frac{d^3k}{(2\pi)^3}\frac{1}{2\omega_{\vect{k}}} 
\sum_{\alpha=1,2}\left[a^\alpha_{\vect{k}} u^\alpha(\vect{k})
e^{-ik\cdot x} + b^{\alpha\dag}_{\vect{k}}
v^\alpha(\vect{k})e^{ik\cdot x}\right] \\
{\bar\psi(x)} &=  \int\frac{d^3k}{(2\pi)^3}\frac{1}{2\omega_{\vect{k}}} 
\sum_{\alpha=1,2}\left[a^{\alpha\dag}_{\vect{k}}
{\bar u}^\alpha(\vect{k})e^{ik\cdot x} + b^\alpha_{\vect{k}}
{\bar v}^\alpha(\vect{k})e^{-ik\cdot x}\right]\,,
\end{align}
\end{subequations}
where $k^\mu=(\omega_{\vect{k}}, \vect{k})$, $\omega_{\vect{k}}=\sqrt{\abs{\vect{k}}^2 + m^2}$, the mode operators satisfy:
\begin{equation}
\bigl\{a^\alpha_{\vect{k}}, a^{\beta\dag}_{\vect{k}^\prime}\bigr\} =  \bigl\{b^\alpha_{\vect{k}}, b^{\beta\dag}_{\vect{k}^\prime}\bigr\} = 
(2\pi)^3\delta^{(3)}(\vect{k}-\vect{k}^\prime)2\omega_{\vect{k}}\delta^{\alpha\beta},
\end{equation}
and
\begin{subequations}
\begin{align}
\langle a^{\alpha\dag}_{\vect{k}} a^\beta_{\vect{k}^\prime}\rangle &= 
 f(\omega_k,\mu_i) (2\pi)^3 \delta^{(3)}(\vect{k}-\vect{k}^\prime)2\omega_{\vect{k}}\delta^{\alpha\beta}\\
\langle b^{\alpha\dag}_{\vect{k}} b^\beta_{\vect{k}^\prime}\rangle &=  
f(\omega_k,-\mu_i) (2\pi)^3 \delta^{(3)}(\vect{k}-\vect{k}^\prime)2\omega_{\vect{k}}
\delta^{\alpha\beta},
\end{align}
\end{subequations}
with $f(\omega,\mu_i)$ being the non-equilibrium Fermi distribution
function. For our purposes, the relative change $\delta
f(\omega,\mu_i)/f_0(\omega,\mu_i)$ enters the transport equations
multiplying explicit factors of $\Gamma_p$ and either $v_w$ or $\mu$, so that in working to second order in $\epsilon$ we may replace $f$ by the equilibrium
distributions $f_0(\omega,\mu_i)=n_F(\omega-\mu_i)=[e^{(\omega-\mu_i)/T} +
1]^{-1}$. Using the mode expansion (\ref{eq:modeexp}) it is
straightforward to show that 
$S^>(x,y)=\langle \psi(x){\bar\psi}(y)\rangle$ and 
$S^<(x,y)=-\langle {\bar\psi}(y)\psi(x)\rangle$  can be expressed 
as:
\be
\label{eq:slambdafree}
S^\gtrless(x,y)=\int\frac{d^4k}{(2\pi)^4} e^{-i{k}\cdot(x-y)}
g_F^\gtrless({k}_0,\mu_i)\rho(k_0,\vect{k})\left(\diracslash{k}+m\right)
\ee
in terms of the free particle spectral density:
\be
 \rho({k}_0, \vect{k}) =  {i\over 2\omega_k}\biggl[
\left({1\over
 {k}_0-\omega_k+i\epsilon}-{1\over {k}_0+\omega_k+i\epsilon}\right)
 -\left({1\over {k}_0-\omega_k-i\epsilon}-
{1\over {k}_0+\omega_k-i\epsilon}\right)\biggr]\  .
\label{eq:spectral1}
\ee
and the functions:
\begin{subequations}
\begin{align}
g_F^>(k_0,\mu_i)&=1-n_F(k_0-\mu_i) \\
g_F^<(k_0,\mu_i)&= -n_F(k_0-\mu_i)\,. 
\end{align}
\end{subequations}
The propagators $S^{t,\bar t}(x,y)$ can now be constructed from the
$S^\lambda(x,y)$ as in Eqs. (\ref{eq:Greens1}).

In the presence of interactions (characterized by a generic coupling
$g$), the fermion propagator becomes considerably more complicated
than given by Eq. (\ref{eq:slambdafree}). In particular, single
fermion states can mix with other multiparticle states in the thermal
bath, leading to the presence of additional poles (the ``hole"
modes) in the fermion propagator \cite{Klimov,Weldon:1989ys}. The general structure of the fermion propagator arising from these effects has been studied extensively at zero density \cite{Weldon:1999th}. In Appendix A we generalize to the
case of non-zero $\mu_i$. For massless fermions, the resulting
propagators are given by:
\be
\label{eq:slambdaint}
S^\gtrless(x,y;\mu_i)=\int {d^4k\over (2\pi)^4} e^{-ik\cdot(x-y)} 
g_F^\gtrless(k_0, \mu) 
\left[\frac{\gamma_0-\boldgamma\mcdot\vect{\hat k}}{2}\rho_+(k_0, \vect{k}, \mu_i)
+ \frac{\gamma_0+\boldgamma\mcdot\vect{\hat k}}{2}\rho_-(k_0, \vect{k}, \mu_i)\right]
\,,
\ee
where $\vect{\hat k}$ is the unit vector in the $\vect{k}$ direction, and
\begin{equation}
\label{eq:rhoplus}
\begin{split}
\rho_+(k_0,\vect{k},\mu_i) = i\biggl[&\frac{Z_p(k,\mu_i)}{k_0-\mathcal{E}_p(k,\mu_i)}
-\frac{Z_p(k,\mu_i)^*}{k_0-\mathcal{E}_p(k,\mu_i)^*} \\
+ &\frac{Z_h(k,-\mu_i)^*}{k_0+\mathcal{E}_h(k,-\mu_i)^*}
- \frac{Z_h(k,-\mu_i)}{k_0+\mathcal{E}_h(k,-\mu_i)}+F(k_0^*,k,\mu_i)^*-F(k_0,k,\mu_i)\biggr]\,,
\end{split}
\end{equation}
and
\begin{equation}
\label{eq:rhominus}
\rho_-(k_0,\vect{k},\mu_i) = [\rho_+(-k_0^*,\vect{k},-\mu_i)]^*\,.
\end{equation}
%
Here, $\mathcal{E}_p(k,\mu_i)$ and $-\mathcal{E}_h(k,-\mu_i)^*$ are the two (complex) roots (in $k_0$) of the equation:
\be
0 = k_0-k+D_+(k_0, k,\mu_i)+i\epsilon
\ee
where $iD_{\pm}(k_0,k,\mu_i)$ are contributions to the inverse, retarded
propagator proportional to $(\gamma_0\mp\boldgamma\mcdot\vect{\hat k})/2$
arising from interactions.  The function $F(k_0, k,\mu_i)$ gives the
non-pole part of the propagator, and $k=\abs{\vect{k}}$. We find that the resonant contributions to the particle number-changing sources arise from the pole parts of the propagators, so from here on we neglect the terms containing $F(k_0,k,\mu_i)$.

In the limit $g\to 0$, one has $Z_h\to 0$ and $Z_p\to 1$, recovering
the form of the propagator given in Eq.~(\ref{eq:slambdafree}). For
nonzero $g$, however, $Z_h$ is not of order $g^2$ since the particle
and hole modes arise from mixtures of degenerate states. In
particular, at $k=0$ one has $Z_p=Z_h=1/2$. As $k$ becomes large (of
order the thermal mass or larger), $Z_h/Z_p \ll 1$, and the particle
dispersion relation is well-approximated by $\mathcal{E}_p^2 =
\abs{\vect{k}}^2 + m^2(T,\mu_i)$, where $m(T,\mu_i)$ is the thermal mass.
In our particular application to the MSSM, the gaugino $M_i$ masses
will typically be taken to be of order several hundred GeV, and for
the SU(2)$_L\times$U(1)$_Y$ sector, thermal effects do not induce
substantial mass corrections. We find that the gaugino contributions
to the RHS of Eqs. (\ref{eq:scalar1}, \ref{eq:fermion1}) are dominated
by momenta of order $M_i$, so that the hole contributions to the
gaugino $S^\gtrless(x,y)$ can be neglected. In contrast, for quarks
we find non-negligible contributions from the low-momentum region, so
we retain the full structure given by
Eqs. (\ref{eq:slambdaint}-\ref{eq:rhominus}) in computing their
contributions.

It has been noted in previous studies of quark damping rates that the
one-loop thermal widths $\Gamma_{p,h}= \Imag{\mathcal{E}_{p,h}(k,\mu)}$ are
gauge-dependent (see Ref.~\cite{Braaten:1989mz} and Ref. [3] therein), whereas the thermal masses $m(T,\mu)$ entering
$\mathcal{E}_{p,h}$ are gauge-independent to this order. Gauge-independent
widths can be obtained by performing an appropriate resummation of
hard thermal loops (HTLs) \cite{LeBellac,Braaten:1989mz,Braaten:1991gm}. The latter are associated with momenta
$k_0,k\sim gT$, for which the one-loop functions $D_{\pm}(k_0,k,\mu)$
are of the same order in $g$ as the tree-level inverse propagators. In
what follows, we will estimate the widths $\Gamma_{p,h}$ based on
existing computations~of damping~\cite{Braaten:1992gd, Enqvist:1997ff, Elmfors:1998hh}, deferring a
complete computation of the gauge-invariant, $\mu_i$-dependent
contributions in the MSSM to a future study. In general,
the residues $Z_{p,h}$ also carry a gauge-dependence, and at this time
we are not aware of any HTL resummation that could eliminate this
dependence. In principle, elimination of this gauge-dependence
requires inclusion of one-loop vertex corrections in the computation
of the $\Sigma^\gtrless(x,y)$ and $S^\gtrless(x,y)$ appearing on the RHS
of Eqs. (\ref{eq:scalar1}, \ref{eq:fermion1}), and
we again defer a complete one-loop computation to a future study.

The derivation of the finite-density scalar propagators proceeds along
similar lines.  Starting from the mode expansion of the free scalar
field $\phi(x)$ in terms of plane-wave solutions to the Klein-Gordon
equation and following analogous arguments as for fermions, one
arrives at the following scalar Green's functions:
\be
G^\gtrless(x,y)=\int{d^4k\over (2\pi)^4} e^{-ik\cdot(x-y)} 
g_B^\gtrless(k_0, \mu_i)\rho(k_0,\vect{k}),
\ee
where the equilibrium distribution functions are:
\begin{subequations}
\begin{align}
g_B^>(\omega, \mu) &=  1+n_B(\omega-\mu_i)\\
g_B^<(\omega, \mu) &= n_B(\omega - \mu_i)\,,
\end{align}
\end{subequations}
with $n_B(x) = 1/(e^{x/T} - 1)$ and $\rho(k_0,\vect{k})$
given by Eq.~(\ref{eq:spectral1}).  As with fermions, one may include
the effect of thermal masses and widths by replacing $m^2\to
m^2(T,\mu_i)$ and $i\epsilon\to i\epsilon+i\Gamma(T,\mu_i)$.

\section{Source Terms for Quantum Transport}
\label{sec:source}

The expressions for $G^\gtrless(x,y)$ and $S^\gtrless(x,y)$ now allow us
to compute the perturbative contributions to the source terms on the
RHS of Eqs. (\ref{eq:scalar1},\ref{eq:fermion1}) starting from a given
electroweak model Lagrangian. Here, we work within the MSSM as an
illustrative case, but emphasize that the methods are general. The
Feynman rules giving the relevant interaction vertices in the MSSM are
taken from Ref.~\cite{Martin:1997ns}, and in what follows, we only write
down those relevant for the computations undertaken here. It is
useful, however, to place our calculation in a broader context by
considering the various classes of graphs that generate different
terms in the QTEs. The simplest topologies are those involving
scattering of particles and their superpartners from the spacetime
varying Higgs vevs (generically denoted $v$) in the plasma
[Fig.~\ref{fig:graphs1}]. 
\begin{figure}
\centering
\begin{picture}(150,180)  
\put(50,50){\makebox(50,50){\epsfig{figure=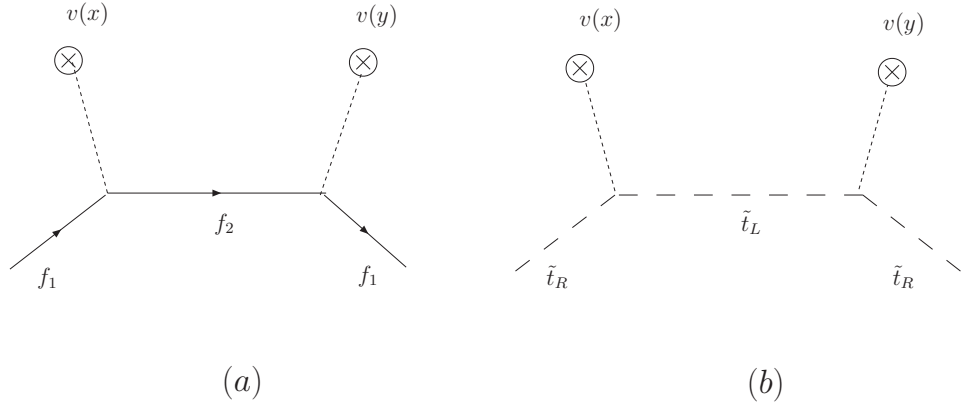,width=13cm}}}
\end{picture}
\caption[Contributions to self-energies 
from scattering of particles from Higgs vevs.]{
Contributions to the relevant self-energies 
from scattering of particles from the spacetime
varying Higgs vevs. 
\label{fig:graphs1}
}
\end{figure}
These graphs give rise to both the $CP$-violating source
terms discussed in Ref. \cite{Riotto:1998zb} as well as terms
proportional to chiral charge. The latter involve the number densities
of at most two different species, such as the left- and right-handed
top quarks [Fig.~\ref{fig:graphs1}(a)] or their superpartners 
[Fig.~\ref{fig:graphs1}(b)]. For purposes of illustration, we follow Ref.~\cite{Riotto:1998zb} and work in a basis of mass eigenstates in the unbroken phase, treating the interactions with the Higgs vevs perturbatively. This approximation should be reasonable near the phase transition boundary, where both the vevs and their rate of change are small, but it clearly breaks down farther inside the bubble wall, where the vevs become large (of order the
phase transition temperature, $T_c$). In general, one would like to perform a resummation to all orders in the vevs, possibly employing the approximation scheme proposed in Refs. \cite{Carena:2000id,Prokopec}. We postpone a treatment of this resummation to a future study\footnote{ The authors of Ref.~\cite{Carena:2000id} find that carrying out such a resummation reduces the resonant enhancements of the $CP$-violating sources, but they did not consider the $CP$-conserving, chirality-changing terms that are our focus here. The consistency of the proposed approximate resummation with our power counting remains to be analyzed.}.

Yukawa interactions involving quarks (squarks) and Higgs (Higgsinos)
are illustrated in Fig.~\ref{fig:graphs2} 
\begin{figure}
\centering
\begin{picture}(150,200)  
\put(50,80){\makebox(50,50){\epsfig{figure=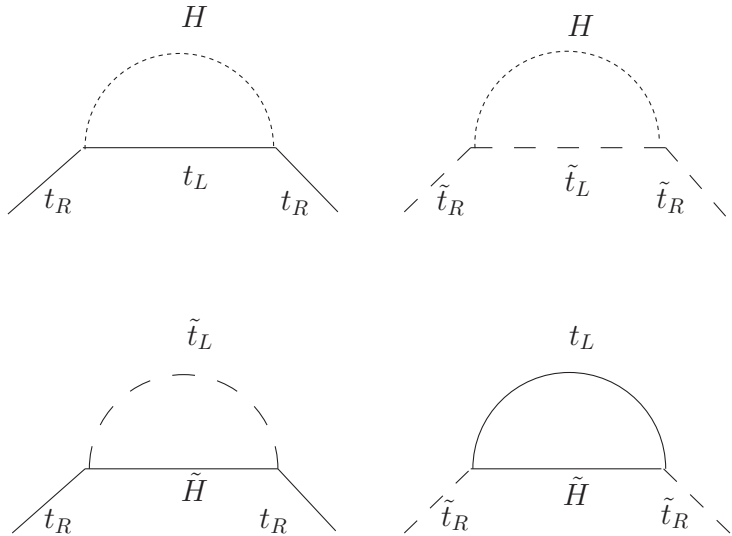,width=10cm}}}
\end{picture}
\caption{
\label{fig:graphs2}
Contributions to the relevant self-energies from Yukawa interactions.
}
\end{figure}
(the self-energies $\Sigma^\gtrless(x,y)$ are
obtained by amputating the external legs). 

These interactions cause
transitions such as $f\leftrightarrow f H$, ${\tilde f}\leftrightarrow
{\tilde f} H$, and $f\leftrightarrow {\tilde f}{\tilde
H}$. Contributions from gauge interactions appear 
in Fig.~\ref{fig:graphs3}. The
latter induce transitions of the type $f\leftrightarrow f V$, ${\tilde
f}\leftrightarrow {\tilde f}V$, and $f\leftrightarrow{\tilde f}{\tilde
V}$. 
In general, one expects the Yukawa and gauge
interactions involving three different species to depend on sums and
differences of the corresponding chemical potentials, as in
$\mu_f-\mu_{\tilde f}-\mu_{\tilde V}$ for the supergauge
interactions. In previous studies, it has been assumed that the
gauginos ${\tilde V}$ are sufficiently light and the coefficients of
the corresponding terms in the QTEs sufficiently large than one has
$\mu_{\tilde V}\approx 0$ and $\mu_f\approx \mu_{\tilde f}$. Although
the quantitative validity of this assumption could be explored using
our framework here, we defer that analysis to a future study and take
$\mu_{\tilde V}\approx 0$, $\mu_f\approx \mu_{\tilde
f}$. Consequently, one may, as in Ref. \cite{Huet:1995sh}, define a
common chemical potential for SM particles (including the two Higgs
doublets) and their superpartners.

In previous studies, it has also been assumed---based largely on simple estimates (see, for instance, Ref.~\cite{Huet:1995sh})---that the Yukawa interactions of Fig.~\ref{fig:graphs2} are
sufficiently fast that they decouple from the set of QTEs, leading
to relations between the chemical potentials for the Higgs (Higgsino)
fields and those for matter fields. For example, Yukawa interactions
that couple the Higgs doublet fields $H$ with those of the third
generation SU(2)$_L$ doublet quarks, $Q$ with the singlet top quark
supermultiplet field, $T$, generate terms of the form:
\be
\Gamma_Y \left(\mu_Q-\mu_T+\mu_H\right)\,.
\ee
To the extent that $\Gamma_Y$ is much larger than the other transport
coefficients appearing in Eqs. (\ref{eq:scalar1},\ref{eq:fermion1}),
one has $\mu_Q=\mu_T-\mu_H$ plus terms of ${\cal O}(1/\Gamma_Y)$. The
remaining terms in the QTEs will involve the $CP$-violating sources,
sphaleron terms, and terms that couple left- and right-handed chiral
charges, such as $\Gamma_M(\mu_Q-\mu_T)$. Again, this assumption could
be tested using the current framework, but the computation of
$\Gamma_Y$ is considerably more arduous than those discussed below,
where we focus on the $CP$-violating sources and the $\Gamma_M$-type
terms that are generated by the diagrams in Fig.~\ref{fig:graphs1}.
\begin{figure}
\centering
\begin{picture}(150,230)  
\put(50,90){\makebox(50,50){\epsfig{figure=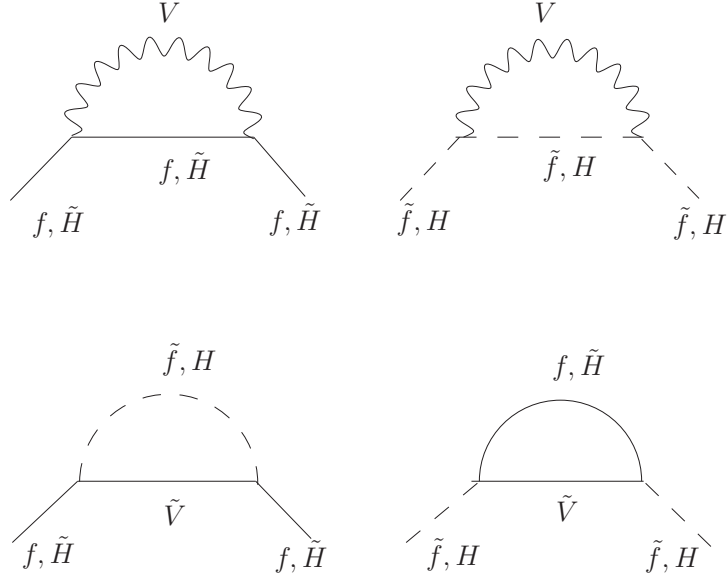,width=10cm}}}
\end{picture}
\caption{Representative contributions to self-energies from
(super)gauge interactions.
\label{fig:graphs3}
}
\end{figure}

\subsection{Bosons} 

We consider first the scalar interactions in Fig.~\ref{fig:graphs1}(a).
The largest contributions involve the L and R top squarks, ${\tilde
t}_{L,R}$ owing to their large Yukawa coupling, $y_t$. In the basis of weak eigenstates, the relevant interaction
Lagrangian is:
\begin{equation}
\label{eq:scalarhiggs}
\mathcal{L} = y_t\tilde t_L\tilde t_R^*(A_t v_u - \mu^* v_d) + \text{h.c.}\,,
\end{equation}
%
where $v_{u,d}$ are the vevs of $H_{u,d}^0$, and we take $v\equiv\sqrt{v_u^2+v_d^2}$ and $\tan\beta\equiv
v_u/v_d$. Note that in Eq. (\ref{eq:scalarhiggs}) we allow the $v_{u,d}$ to
be spacetime-dependent. In the region of broken electroweak symmetry
and stable vevs, we have $m_t=y_t v_u$.

Using the Feynman rules for path-ordered perturbation theory, it is
straightforward to show that the diagrams in Fig.~\ref{fig:graphs1}(a)
generate contributions to ${\widetilde\Sigma}_R(x,y)$ of the form:
\be
\label{eq:sigmascalar}
{\widetilde\Sigma}_R(x,y) = -g(x,y){\widetilde G}_L^0(x,y),
\ee
where
\begin{equation}
\label{eq:gxyscalar}
g(x,y) =  y_t^2\bigl[A_t v_u(x) - \mu^* v_d(x)\bigr] \bigl[A_t^* v_u(y) - \mu v_d(y)\bigr]\,.
\end{equation}
Substituting Eq. (\ref{eq:sigmascalar}) into Eq. (\ref{eq:scalar1}) leads to:
\be
\partial_\mu {\tilde t}^\mu_R(x) = S_{{\tilde t}_R}(x)
\ee
for right-handed top squarks, where ${\tilde t}^\mu_R$ is the
corresponding current density and the source $S_{{\tilde t}_R}(x)$ is
\begin{equation}
\label{eq:scalar2}
\begin{split}
S_{{\tilde t}_R}(x) = -\int d^3z\int_{-\infty}^{x_0}dz_0\ 
\biggl\{[g(x,z)&+g(z,x)]\Real \bigl[G^>_L(x,z)G^<_R(z,x)-G^<_L(x,z)G^>_R(z,x)\bigr] \\
+i[g(x,z)&-g(z,x)]\Imag\bigl[G^>_L(x,z)G^<_R(z,x)-G^<_L(x,z)G^>_R(z,x)\bigr]\biggr\}\,,
\end{split}
\end{equation}
where the L,R subscripts indicate the propagators for the L and R
top squarks.

The first term in the integrand of $S_{{\tilde t}_R}(x)$ is
$CP$-conserving and leads to the $\Gamma_M$-type terms discussed
above, while the second term in the integrand provides the
$CP$-violating sources. We concentrate first on the former. Expanding
$g(x,z)$ about $z=x$ it is straightforward to show that only terms
involving even powers of derivatives survive in $g(x,z)+g(z,x)$. Under
the assumptions of gentle spacetime dependence of the $v_i(x)$ near
the phase boundary, we will neglect terms beyond leading order and
take $g(x,z)+g(z,x)\approx 2 g(x,x)$. Consequently, the
$CP$-conserving source is:
\bea
\label{eq:scalar3}
S_{{\tilde t}_R}^{CP}(x) &\approx& -2g(x,x) \Real
\int d^3z\int_{-\infty}^{x_0}dz_0\  \left [G^>_L(x,z)G^<_R(z,x)
-G^<_L(x,z)G^>_R(z,x)\right]\\
\nonumber
&=& -2g(x,x)\Real\int d^3z \int_{-\infty}^{x_0}dz_0\int{d^4k\over 
(2\pi)^4}\int {d^4q\over (2\pi)^4}
e^{-i(k-q)\cdot(x-z)}\rho_L(k_0,\vect{k})\rho_R(q_0,\vect{q})\\
\nonumber
&&
\ \ \ \times \left[g_B^>(k_0,\mu_L)g_B^<(q_0,\mu_R)-
g_B^<(k_0,\mu_L) g_B^>(q_0,\mu_R)\right]\,,
\eea
with
\be
\label{eq:scalargxx}
g(x,x) = y_t^2 \bigl[|\mu|^2 v_d^2(x) + |A_t|^2 v_u^2(x) -
 2v_d(x) v_u(x) \Real(\mu A_t)\bigr].
\ee
Note the simplification
\begin{equation}
g_B^>(k_0,\mu_L)g_B^<(q_0,\mu_R)-
g_B^<(k_0,\mu_L) g_B^>(q_0,\mu_R) = n_B(q_0-\mu_R) - n_B(k_0-\mu_L).
\end{equation}
Performing the $d^3z$ integral leads to a $\delta$ function in
momentum space. After carrying out the $d^3q$ integral, 
we perform the $k^0,q^0$ integrals by contour integration\footnote{Here and in subsequent equations, we show only the terms arising from picking up the residues of the poles in the spectral functions such as $\rho_L(k^0,\vect{k}),\rho_R(q^0,\vect{q})$. A careful calculation would also include the residues from the poles in the thermal distribution functions such as $n_B(k^0), h_B(k^0)$, etc. We relegate these to Appendix.~\ref{appx:poles}, where we find their contribution to the final numerical results to be unimportant compared to the terms retained in the main text.}, expand to
first order in $\mu_{L,R}/T$, and obtain:
\begin{equation}
\label{eq:scalar5}
\begin{split}
S^{CP}_{\tilde t_R}(x) &= -\frac{1}{T}\frac{N_C y_t^2}{2\pi^2}\abs{A_t v_u(x) - \mu^*v_d(x)}^2 \int_0^\infty\frac{dk\,k^2}{\omega_L\omega_R} \\
&\qquad\times\Imag\biggl\{\frac{\mu_L h_B(\mathcal{E}_L) - \mu_R h_B(\mathcal{E}_R^*)}{\mathcal{E}_L - \mathcal{E}_R^*}  + \frac{\mu_R h_B(\mathcal{E}_R) - \mu_L h_B(\mathcal{E}_L)}{\mathcal{E}_L + \mathcal{E}_R}\biggr\},
\end{split} 
\end{equation}
where
\begin{subequations}
\label{eq:scalardefs}
\begin{align}
\omega_{L,R}^2 &= \abs{\vect{k}}^2 + M_{\tilde t_{L,R}}^2 \\
\mathcal{E}_{L,R} &= \omega_{L,R} - i\Gamma_{L,R} \\
h_B(x) &= -\frac{e^{x/T}}{(e^{x/T} - 1)^2},
\end{align}
\end{subequations}
and $M_{\tilde t_{L,R}},\Gamma_{L,R}$ are the thermal masses and widths 
for the $\tilde t_{L,R}$, and the factor of $N_C$ comes from  
summing over the colors.  Note that, in arriving at Eq. (\ref{eq:scalar5}), we have neglected the $\mu_i$-dependence of the pole residues $Z(T, \mu_{L,R})$, thermal frequencies, $\omega_{ L,R}(T,\mu_{L,R})$, and widths, $\Gamma_{L,R}(T,\mu_{L,R})$. The effect on $S^{CP}_{\tilde t_R}(x)$ of the $\mu_i$-dependence of the residues and thermal frequencies is sub-leading in the gauge and Yukawa couplings, whereas the effect from the thermal widths occurs at leading order. The $\mu_i$-dependence of  $\Gamma_{L,R}(T,\mu_{L,R})$ is simply not known, however, so we do not include it here. A more explicit expression for the dependence of $S^{CP}_{\tilde t_R}(x)$ on the thermal frequencies and widths is given in Eqs. (\ref{appx:scalar5}-\ref{appx:gammascalarpm}) of Appendix \ref{appx:expand}.

For purposes of future analysis, it is useful to rewrite
Eq. (\ref{eq:scalar5}) as:
\begin{equation}
\label{eq:scalar6}
S^{CP}_{\tilde t_R} = \Gamma_{\tilde t}^+ (\mu_L + \mu_R) + \Gamma_{\tilde t}^-(\mu_L - \mu_R)\,,
\end{equation}
where
\begin{equation}
\label{eq:gammascalarpm}
\Gamma_{\tilde t}^\pm = -\frac{1}{T}\frac{N_C y_t^2}{4\pi^2}\abs{A_t v_u(x) - \mu^*v_d(x)}^2\!\!\int_0^\infty\!\!\frac{dk\,k^2}{\omega_R\omega_L}\Imag\left\{\frac{h_B(\mathcal{E}_L) \mp h_B(\mathcal{E}_R^*)}{\mathcal{E}_L - \mathcal{E}_R^*} - \frac{h_B(\mathcal{E}_L) \mp h_B(\mathcal{E}_R)}{\mathcal{E}_L + \mathcal{E}_R}\right\}.
\end{equation}

Before proceeding with the $CP$-violating source, we comment briefly
on the structure of Eqs. (\ref{eq:scalar6}-\ref{eq:gammascalarpm}). In
particular, we note that 
\begin{itemize}
\item[(i)] Terms of the type $\Gamma^{+}_{\tilde t}$ are
absent from the conventional QTEs for EWB. It is straightforward to
see that in the absence of interactions that distinguish between
${\tilde t}_L$ and ${\tilde t}_R$, $\Gamma^{+}_{\tilde t}=0$, as the
integrand of Eq.~(\ref{eq:gammascalarpm}) is antisymmetric under
$L\leftrightarrow R$ interchange. In contrast, the transport
coefficient $\Gamma^{-}_{\tilde t}$ is nonzero in the limit of exact
${\tilde t}_L\leftrightarrow {\tilde t}_R$ symmetry. This term
corresponds to the usual damping term in the QTEs associated with
scattering from the Higgs vevs.  
\item[(ii)] In the absence of thermal widths $\Gamma_{L,R}$, the quantity in
brackets in Eq.~(\ref{eq:gammascalarpm}) is purely real, and so the
damping term would be zero. 
\item[(iii)] The structure of the energy
denominators implies a resonant enhancement of the integrand for $
M_{\tilde t_{L}}^2 \sim M_{\tilde t_{R}}^2 $.  A similar effect was observed to occur 
for the $CP$-violating sources (see below) in
Refs.~\cite{Riotto:1998zb,Carena:1997gx}.  The expression in Eq. (\ref{eq:gammascalarpm}) makes it clear that the relaxation terms display a resonant behavior as
well. The resulting quantitative impact of this resonance on the baryon asymmetry
is discussed in Sect.~\ref{sec:numerics}.
\end{itemize}
Properties (ii) and (iii) are shared by all source and damping terms,
we discuss  below. Note that the explicit factors of $\mu_{L,R}/T$ and property (ii) imply that, away from the resonance region,
$S^{CP}_{\tilde t_R}$ is $\mathcal{O}(\epsilon^2)$.

The computation of the $CP$-violating source, given by the second term
in Eq.~(\ref{eq:scalar2}), proceeds along similar lines. In this case,
the coefficient $[g(x,z)-g(z,x)]$ vanishes for $x=z$, so we must
retain terms at least to first order in the expansion about $x=z$:
\begin{equation}
\begin{split}
\label{eq:gexpand1}
g(x,z)-g(z,x) & = 2i  y_t^2 \Imag(\mu A_t)\left[v_d(x) v_u(z)
-v_d(z) v_u(x)\right]\\
&= 2i  y_t^2 \Imag(\mu A_t) (z-x)^\lambda\left[ v_d(x)
\partial_\lambda v_u(x)-
v_u(x)\partial_\lambda v_d(x)\right] +\cdots\,,
\end{split}
\end{equation}
where the $+\cdots$ indicate higher order terms in the derivative
expansion that we neglect for the same reasons as discussed
previously. When the linear term in Eq. (\ref{eq:gexpand1}) is
substituted in Eq. (\ref{eq:scalar2}), only the time component yields
a nonzero contribution. The spatial components 
vanish due to the spatial isotropy of the spectral 
density: $g_B^\gtrless(k_0,\mu) \rho(k_0, \vect{k})
\equiv g_B^\gtrless(k_0,\mu) \rho(k_0, \abs{\vect{k}})$. 
We may then make the replacement:
\begin{equation}
\begin{split}
\label{eq:gexpand2}
g(x,z)-g(z,x) & \rightarrow 2i 
y_t^2\Imag(\mu A_t)\left[v_d(x) {\dot v}_u(x)-{\dot v}_d(x) v_u(x)\right]
\, (z - x)^0 \\
&= 2i y_t^2\Imag(\mu A_t) v(x)^2{\dot\beta}(x) \, (z - x)^0     \ .
\end{split}
\end{equation}
In general, we expect ${\dot\beta}$ to be of order $v_w/c$, so that
the $CP$-violating source is first-order in one of the small expansion
parameters discussed earlier. Consequently, when evaluating this term,
we may neglect the $\mu_{L,R}$-dependence of the
$g_B^\gtrless(k_0,\mu)$. After carrying out the ($k_0$,$q_0$) contour
integrals and performing the time integration, we obtain:
\begin{equation}
\label{eq:scalarcp1}
\begin{split}
S_{\tilde t_R}^{\CPV} &= \frac{N_C y_t^2}{2\pi^2}\Imag(\mu A_t)v(x)^2 \dot\beta(x)\int_0^\infty\frac{dk\,k^2}{\omega_R\omega_L}\Imag\biggl\{\frac{n_B(\mathcal{E}_R^*) - n_B(\mathcal{E}_L)}{(\mathcal{E}_L - \mathcal{E}_R^*)^2} + \frac{1+n_B(\mathcal{E}_R) + n_B(\mathcal{E}_L)}{(\mathcal{E}_L + \mathcal{E}_R)^2}\biggr\}.
\end{split}
\end{equation}
Again, property (ii), in conjunction with the factor of ${\dot\beta}\propto v_w$, implies that $S_{\tilde t_R}^{\CPV} $ is $\mathcal{O}(\epsilon^2)$. An expression giving a more explicit dependence on the widths and frequencies appears Eq. (\ref{appx:scalarcp1}) of  Appendix \ref{appx:expand}, which we note agrees with that of Ref.~\cite{Riotto:1998zb} except for a
different relative sign in front of the $\cos 2\phi$ term of that equation and the overall factor of $N_C$.

\subsection{Massive Fermions}

The computations for fermions proceed along similar lines. We consider
first the source terms for Higgsinos. 
We recall that it is useful to redefine the Higgsino fields to remove the
complex phase from the Higgsino mass term:
\be
{\cal L}_{\widetilde H}^{\rm mass} = \mu \left(\psi_{H_d^0}\psi_{H_u^0}-
\psi_{H_d^-}\psi_{H_u^+}\right)
+\mu^\ast \left( 
{\bar\psi}_{H_d^0}{\bar\psi}_{H_u^0}-{\bar\psi}_{H_d^-}{\bar\psi}_{H_u^+}
\right)
\ee
via
\be
\psi_{H_d^{0,-}}\to{\widetilde H}_d^{0,-}\quad\quad 
\psi_{H_u^{0,+}}\to e^{-i\phi_\mu}{\widetilde H}_u^{0,+}
\ee
leading to:
\be
\label{eq:higgsinomass}
{\cal L}_{\widetilde H}^{\rm mass}=
\abs{\mu}\left({\widetilde H}_d^0{\widetilde H}_u^0-{\widetilde H}_d^-{\widetilde H}_u^+\right)
+ \abs{\mu} \left( {\widetilde H}_d^{0\dag} 
{\widetilde H}_u^{0\dag}-{\widetilde H}_d^{-\dag}{\widetilde H}_u^{+\dag}
\right)\,.
\ee
Defining the four component spinors,
\be
\label{eq:higgsinodef}
\Psi_{\widetilde H^+}  =  \left(\begin{array}{c} 
{\widetilde H}_u^+\\ {\widetilde H}_d^{-\dag}
\end{array}
\right)\quad\quad 
\Psi_{\widetilde H^0}  =  \left(\begin{array}{c} 
-{\widetilde H}_u^0\\ {\widetilde H}_d^{0\dag}
\end{array}
\right)\,
\ee
for the Higgsinos, and 
\be
\label{eq:gauginodef}
\Psi_{\widetilde W^+}= \left(\begin{array}{c} 
{\widetilde W}^+ \\
{\widetilde W}^{-\dag} 
\end{array}
\right)\quad\quad 
\Psi_{\widetilde W^0}=\left(\begin{array}{c} 
{\widetilde W}^3
\\ 
{\widetilde W}^{3\dag}
\end{array}
\right)\quad\quad 
\Psi_{\widetilde B}=\left(\begin{array}{c} 
{\widetilde B}
\\
{\widetilde B}^\dag
\end{array}
\right)\,
\ee
for the gauginos, leads to the Higgsino-gaugino-vev interaction:
\begin{equation}
\begin{split}
\label{eq:higgsinoint1}
{\cal L}^{\rm int} &= 
-g_2\bar\Psi_{\widetilde H^+}\left[v_d(x)P_L+v_u(x) e^{i\phi_\mu}P_R\right]\Psi_{\widetilde W^+}\\
&\quad -\frac{1}{\sqrt{2}}\bar\Psi_{\widetilde H^0}\left[v_d(x)P_L+v_u(x) e^{i\phi_\mu}P_R\right] \left(g_2\Psi_{\widetilde W^0}-
g_1\Psi_{\widetilde B}\right)\ +\ {\rm h.c.},
\end{split}
\end{equation}
where
\begin{equation}
P_{L,R} = \frac{1\mp\gamma^5}{2},
\end{equation}
while the mass terms (\ref{eq:higgsinomass}) in the Higgsino Lagrangian become:
\begin{equation}
\mathcal{L}_{\tilde H}^{\text{mass}} = -\abs{\mu}(\bar\Psi_{\widetilde H_0}\Psi_{\widetilde H_0} + \bar\Psi_{\widetilde H^+}\Psi_{\widetilde H^+}).
\end{equation}
Note that the spinors $\Psi_{\widetilde H^0}$ and $\Psi_{\widetilde H^+}$ satisfy a
Dirac equation with Dirac mass  $|\mu|$, even
though the ${\widetilde H}_{d,u}^0$ are Majorana particles. The $\Psi_{\widetilde W^\pm}$ are Dirac particles of mass $M_2$, whereas the $\Psi_{\widetilde W^0}$ and $\Psi_{\widetilde B^0}$ are Majorana particles with Majorana masses $M_2$
and $M_1$, respectively. We also note that the construction of the
Dirac spinor $\Psi_{\widetilde H^0}$ allows one to define a vector charge and
corresponding chemical potential, $\mu_{{\widetilde H}^0}$, for the
neutral Higgsinos, even though they are Majorana particles. In
contrast, there exists no such vector charge for the $\Psi_{\widetilde W^0}$
and $\psi_{\widetilde B^0}$. One may, however, study the quantum transport of the axial charge of the Majorana fermions using Eq. (\ref{eq:fermion1b}). An attempt to do so for the neutral Higgsinos was made in Ref. \cite{Carena:2000id}, though only the $CP$-violating sources were evaluated using non-equilibrium methods. The impact of the corresponding axial charge density on the baryon asymmetry was found to be small. We will return to this issue in a future study, and consider only the vector densities below. 

The most straightforward computation is that of the ${\widetilde H}^\pm$
source terms. For notational convenience, we rewrite the chargino
interactions in Eq. (\ref{eq:higgsinoint1}) as:
\be
\label{eq:charginoint}
-g_2 \bar\Psi_{\widetilde H^+}\left[g_L(x)P_L+g_R(x)P_R\right]\Psi_{\widetilde W^+} 
+{\rm h.c.}
\ee
In this case, the self-energy generated by Fig.~\ref{fig:graphs1}(a) is:
\be
{\widetilde \Sigma}_{\widetilde H^\pm}(x,y) =
 -g_2^2 \left[g_L(x)P_L+g_R(x) P_R\right] \,
{\tilde S}_{\widetilde W^\pm} (x,y) \, \left[g_L(y)^*P_R+g_R(y)^* P_L\right]\ \ \ .
\ee
Defining:
\begin{subequations}
\begin{align}
g_A(x,y) & \equiv  \frac{g^2_2}{2}\left[g_L(x) g_L(y)^*+g_R(x) 
g_R(y)^*\right]\\
g_B(x,y) & \equiv \frac{g_2^2}{2}\left[g_L(x) g_R(y)^*+g_R(x) g_L(y)^*\right]\,,
\end{align}
\end{subequations}
we obtain for the RHS of Eq. (\ref{eq:fermion1}):
\begin{equation}
\label{eq:fermion2a}
\begin{split}
S_{\widetilde H^\pm}(x) =   
\int d^3z \int_{-\infty}^{x_0} &dz_0 \sum_{j=A,B}\biggl\{ \\
[g_j(x,z) &+g_j(z,x)]\Real\Tr \left[S_{\widetilde W^\pm}^>(x,z) S_{\widetilde H^\pm}^<(z,x)
-S_{\widetilde W^\pm}^<(x,z) S_{\widetilde H^\pm}^>(z,x)\right]_j\\
+ i [g_j(x,z) &-g_j(z,x)]\Imag \Tr \left[S_{\widetilde W^\pm}^>(x,z) S_{\widetilde H^\pm}^<(z,x)
-S_{\widetilde W^\pm}^<(x,z) S_{\widetilde H^\pm}^>(z,x)\right]_j\biggr\},
\end{split}
\end{equation}
where the subscripts \lq\lq A" and \lq\lq B" on the traces denote the
contributions arising from the ${\not\! k}$ and $m$ terms,
respectively, in the spectral function in Eq. (\ref{eq:slambdafree})
(an overall factor of $1/2$ due to the presence of the chiral
projectors $P_{L,R}$ has been absorbed in the definition of the
$g_{A,B}$).

As in the case of the scalar fields, the leading density-dependent,
$CP$-conserving contribution to $S_{\widetilde H^\pm}(x) $ arises from the
term in Eq. (\ref{eq:fermion2a}) containing the $x\leftrightarrow z$
symmetric factors $[g_j(x,z)+g_j(z,x)]$. To lowest order in $v_w$, we
may set $x=z$ in these factors. Using the spectral representation of
the $S^{\gtrless}(x,y)$ given in Eq. (\ref{eq:slambdafree}), including
gauge-invariant thermal masses and widths, and expanding to first
order in $\mu_{i}/T$, we obtain the chirality-changing source term:
\be
\label{eq:chargino1}
S^{CP}_{\widetilde H^\pm}(x) = \Gamma_{\widetilde H^\pm}^{+}
\left( \mu_{\widetilde W^\pm}
+
\mu_{\widetilde H^\pm} \right) 
+ 
\Gamma_{\widetilde H^\pm}^{-}
\left( \mu_{\widetilde W^\pm}-\mu_{\widetilde H^\pm}
\right)\,,
\ee
where
\begin{equation}
\label{eq:gammaHiggsinopm}
\begin{split}
\Gamma_{\widetilde H^\pm}^{\pm} = \frac{1}{T}\frac{g_2^2}{2\pi^2}v(x)^2\int_0^\infty\!\frac{dk\,k^2}{\omega_{\widetilde H}\omega_{\widetilde W}} \Imag\biggl\{&\left[\mathcal{E}_{\widetilde W}\mathcal{E}_{\widetilde H}^* - k^2 + M_2\abs{\mu}\cos\phi_\mu\sin 2\beta\right]\frac{h_F(\mathcal{E}_{\widetilde W}) \mp h_F(\mathcal{E}_{\widetilde H}^*)}{\mathcal{E}_{\widetilde W} - \mathcal{E}_{\widetilde H}^*} \\
+ &\left[\mathcal{E}_{\widetilde W}\mathcal{E}_{\widetilde H}
+ k^2 - M_2\abs{\mu}\cos\phi_\mu\sin 2\beta\right]
\frac{h_F(\mathcal{E}_{\widetilde W}) \mp h_F(\mathcal{E}_{\widetilde H})}{\mathcal{E}_{\widetilde W} + \mathcal{E}_{\widetilde H}}\biggr\}\,,
\end{split}
\end{equation}
where the definitions of $\omega_{\widetilde H,\widetilde W}$ and
$\mathcal{E}_{\widetilde H,\widetilde W}$ are analogous to those given
in Eqs. (\ref{eq:scalardefs}) and
\begin{equation}
\label{eq:fermiondefs}
h_F(x) = \frac{e^{x/T}}{(e^{x/T} + 1)^2}\,.
\end{equation}
Also, the factor of $\cos\phi_\mu$ is very nearly 1 for the region of small $\phi_\mu$ in which we find ourselves in subsequent sections. The explicit dependence of $\Gamma_{\widetilde H^\pm}^{\pm}$ on thermal frequencies and widths is given in Eq. (\ref{appx:gammaHiggsinopm}) of Appendix \ref{appx:expand}.

In the present case, we follow Ref. \cite{Huet:1995sh} and assume no
net density of gauginos, thereby setting $\mu_{\widetilde W^\pm}=0$ in
Eq. (\ref{eq:chargino1}) and giving:
\be
\label{eq:chargino2}
S^{CP}_{\widetilde H^\pm}(x) = - \Gamma_{\widetilde H^\pm}\mu_{\widetilde H^\pm}\,,
\ee
with $\Gamma_{\widetilde H^\pm}=\Gamma_{\widetilde H^\pm}^{+}+\Gamma_{\widetilde
H^\pm}^{-}$.  In this case, it is straightforward to obtain the
corresponding source term for the neutral Higgsinos,
\be
\label{eq:neutralino1}
S^{CP}_{\widetilde H^0}(x)=- \Gamma_{\widetilde H^0}\mu_{\widetilde H^0}\,,
\ee
where $\Gamma_{\widetilde H^0}$ can be obtained from the formulae for
$\Gamma_{\widetilde H^\pm}$ by making the following replacements: $g_2\to
g_2/\sqrt{2}$ for ${\widetilde W}^0$ intermediate states and $g_2\to 
g_1/\sqrt{2}$,
$\omw\to\omb$, and $\gamw\to\gamb$ for the ${\widetilde B}$ intermediate
states.

The Higgsino $CP$-violating source arises from the second term in
Eq. (\ref{eq:fermion2a}). As before, we expand the $g_j(x,z)$ to first
order about $x=z$ and observe that only the $x_0-z_0$ component survives
when the $d^3 z$ integration is performed. Also note that $g_A(x,z)-g_A(z,x)=2i\ {\rm
Im} g_A(x,z)=0$ so that only the terms proportional to the Higgsino
and gaugino masses contribute. The result is:
\begin{equation}
\label{eq:chargino3}
\begin{split}
S_{\widetilde H^\pm}^{\CPV}(x) &= \frac{g_2^2}{\pi^2}v(x)^2\dot\beta(x) 
M_2\abs{\mu}\sin\phi_\mu \\
&\qquad\times\int_0^\infty\frac{dk\,k^2}{\omega_{\widetilde H}\omega_{\widetilde W}}\Imag\biggl\{\frac{n_F(\mathcal{E}_{\widetilde W}) - n_F(\mathcal{E}_{\widetilde H}^*)}{(\mathcal{E}_{\widetilde W} - \mathcal{E}_{\widetilde H}^*)^2} + \frac{1-n_F(\mathcal{E}_{\widetilde W}) - n_F(\mathcal{E}_{\widetilde H})}{(\mathcal{E}_{\widetilde W} + \mathcal{E}_{\widetilde H})^2}\biggr\}\,.
\end{split}
\end{equation}
The corresponding expression for
$S_{\widetilde H^0}^{\CPV} (x)$ can be obtained by making the same
replacements as indicated above for the $CP$-conserving terms. The correspondence with the results of Ref. \cite{Riotto:1998zb} can be seen from Eq. (\ref{appx:chargino3}) of Appendix 
\ref{appx:expand}. We again find essential agreement, apart from a sign difference on the $\cos 2\phi$ term.

\subsection{Chiral Fermions}

The final source term associated with Fig.~\ref{fig:graphs1}(a)
involves $L$ and $R$ top quarks. At this order, the latter only
contribute a $\mu_i$-dependent $CP$-conserving term. In order to
illustrate the structure of this term that arises when the terms of
${\cal O}(g^2)$ are retained, we employ the interacting fermion
propagators of Eqs. (\ref{eq:slambdaint}-\ref{eq:rhominus}). The
result is:
\be
\label{eq:topmass1}
S^{CP}_{t_R}(x) = \Gamma_{t_R}^{+}\left(\mu_{t_L}+\mu_{t_R}\right)+ 
\Gamma_{t_R}^{-}\left(\mu_{t_L}-\mu_{t_R}\right)\,,
\ee
with
\begin{equation}
\label{eq:gquarkplusminus}
\begin{split}
\Gamma_{t_R}^\pm = \frac{1}{T}\frac{N_C y_t^2 v_u(x)^2}{\pi^2}\int_0^\infty dk\,k^2 \Imag&\biggl\{\frac{Z_p^R(k) Z_p^L(k)}{\mathcal{E}_p^R + \mathcal{E}_p^L}\bigl[h_F(\mathcal{E}_p^L) \mp h_F(\mathcal{E}_p^R)\bigr] \\
&+ \frac{Z_p^L(k) Z_h^R(k)^*}{\mathcal{E}_p^L - \mathcal{E}_h^{R*}}\bigl[h_F(\mathcal{E}_p^L) \mp h_F(\mathcal{E}_h^{R*})\bigr] + (p\leftrightarrow h)\biggr\}\,.
\end{split}
\end{equation}
Here, the ``$p$" and ``$h$" subscripts indicate
contributions from the particle and hole modes, and 
``$L$ " and ``$R$" refer to left- and right-handed quarks. We have not included in our calculation the effects of $\mu_{t_{L,R}}$-dependence of the widths $\Gamma_{p,h}^{L,R}(T,\mu_{t_{L,R}})$, which in principle also enter at this order. For an expanded version of Eq.~(\ref{eq:gquarkplusminus}), including these effects, see Eq. (\ref{appx:gquarkplus}) in Appendix \ref{appx:expand}. 

In the limit of $t_L\leftrightarrow t_R$ symmetry, $\Gamma_{t_R}^+$ vanishes, and $\Gamma_{t_R}^-$ simplifies to:
\begin{equation}
\begin{split}
\Gamma_{t_R}^- = \frac{1}{T}\frac{N_C y_t^2 v_u(x)^2}{\pi^2}\int_0^\infty dk\,k^2\Imag\biggl\{\frac{Z_p(k)^2}{\mathcal{E}_p}&h_F(\mathcal{E}_p) + \frac{Z_h(k)^2}{\mathcal{E}_h}h_F(\mathcal{E}_h) \\
&+ \frac{2Z_p(k) Z_h^*(k)}{\mathcal{E}_p - \mathcal{E}_h^*}\bigl[h_F(\mathcal{E}_p) + h_F(\mathcal{E}_h^*)\bigr]\biggr\}\,.
\end{split}
\end{equation}
We observe that all contributions to the $CP$-violating source terms
and the $\Gamma^{\pm}$ vanish in the limit of zero thermal
widths. Since the widths are generically of order $g^2 T$ (here, $g$
denotes either a gauge or Yukawa coupling), the source terms for the
QTEs are generally fourth order in the couplings. 


\section{Quantum Transport Equations and $\rho_B $}
\label{sec:qtes}

We now discuss diffusion equations for the particle species that
significantly contribute to the density of left-handed doublet fermions
$n_L(x)$ [cf. Eq.~(\ref{eq:rhob1})] that acts as the ``seed'' for
baryogenesis. We subsequently relate $\rho_B$ to $n_L$ and solve
explicitly the equations in the case of a simple geometry and profile
for the bubble wall describing the phase boundary.

\subsection{Solving the Diffusion Equations}

Using the source terms computed in Section \ref{sec:source}, one can
arrive at a coupled set of differential equations for the various 
particle number densities. These equations simplify considerably under
the assumptions of approximate chemical equilibrium between SM
particles and their superpartners ($\mu_f\approx\mu_{\tilde f}$ with
$\mu_{\tilde V}\approx 0$), as well as the
between different members of left-handed fermion doublets 
($\mu_{W^\pm}\approx 0$). In this case, one obtains transport 
equations for densities associated with different members of a
supermultiplet. This approach is the one followed in
Ref.~\cite{Huet:1995sh}, and for pedagogical purposes we summarize the
development here. 

First, we define the appropriate supermultiplet densities:
\begin{subequations}
\begin{align}
Q &\equiv n_{t_L}+n_{\tilde t_L}+n_{b_L} + n_{\tilde b_L}\\
T &\equiv n_{t_R}+n_{\tilde t_R}\\
B &\equiv n_{b_R}+n_{\tilde b_R} \\
H &\equiv  
n_{H_u^+} + n_{H_u^0} - n_{H_d^-} - n_{H_d^0} +n_{\widetilde H_u^+} - n_{\widetilde H_d^-} +n_{\widetilde H_u^0}- n_{\widetilde H_d^0}\,, 
\end{align}
\end{subequations}
where the Higgsino densities arise from the vector charges $n_{\widetilde H^+}={\bar\Psi_{\widetilde H^+}}\gamma^0\Psi_{\widetilde H^+}$ and
$n_{\widetilde H^0}={\bar\Psi_{\widetilde H^0}}\gamma^0 \Psi_{\widetilde H^0}$ associated with the Dirac fields defined in
Eq. (\ref{eq:higgsinodef}). There are analogous definitions for the
first- and second-generation (s)quarks. Although we do not consider them here, one may also define the corresponding axial charge densities. In the case of the Higgsinos, for example, it will involve the sum, rather than the difference, of the $u$- and $d$-type Higgsino densities\footnote{This density was considered in Ref. \cite{Carena:2000id}, and its overall impact on the baryon asymmetry found to be small.} 

The diffusion equation for a density $n_i$ has the structure:
\be
\partial_\mu J_i^\mu = S_i^{CP} + S_i^{\CPV} + S_i^{\rm sph} \ ,
\ee
where $J_i^\mu$ is the current associated with the density $n_i$, 
$S_i^{CP}$ and $S_i^{\CPV}$ are the source terms computed above, 
and $S_i^{\rm sph}$ is the strong sphaleron transition term, arising from the QCD anomaly of the axial quark current:
\begin{equation}
\partial_\mu j_5^\mu = \frac{n_f g_s^2}{16\pi^2}G^A_{\alpha\beta}\widetilde G^{A\alpha\beta},
\end{equation}
where $j_5^\mu = \sum_i \bar q_i\gamma^\mu\gamma_5 q_i$, summed over flavors $i$.
Various derivations of the strong sphaleron term 
appear in the literature, so we do not reproduce them here. However,
we note that the expressions in
Refs. \cite{Huet:1995sh,Giudice:1993bb} have erroneously omitted a
factor of $1/N_C$~\cite{Moore:1997im}.

The $CP$-conserving damping terms $S_i^{CP}$ have been given in
Eqs.~(\ref{eq:scalar6}), (\ref{eq:chargino1}), and
(\ref{eq:topmass1}) to linear order in the appropriate chemical
potentials.  
Assuming local thermal equilibrium we relate the number densities to
the chemical potentials via:
\be
n_i=g_i\ \int\frac{d^3k}{(2\pi)^3} 
\left[N(\omega_k,\mu_i)-N(\omega_k, -\mu_i)\right]\,,
\ee
where $N(\omega,\mu)$ is the appropriate boson or fermion distribution
function and $g_i$ counts the internal degrees of freedom (spin and
color).  Dropping terms of ${\cal O}(\mu_i^3)$, one obtains:
\be
n_i=\frac{k_i (m_i/T) T^2}{6}\mu_i \ , 
\label{eq:muvsn}
\ee
where the factors $k_i (m_i/T)$ are exponentially small in the regime
$m_i/T \gg 1$, and reduce in the massless limit to $k_i(0) = 1$ for
chiral fermions, $k_i (0) =2$ for Dirac fermions, and $k_i (0) =2$ for complex
scalars. In our analysis we keep the full dependence on $m_i/T$:
\begin{equation}
k_i(m_i/T) = k_i(0)\frac{c_{F,B}}{\pi^2}\int_{m/T}^\infty dx\,x\,
\frac{e^x}{(e^x \pm 1)^2}\sqrt{x^2 - m^2/T^2}\,,
\end{equation}
where for fermions (bosons) $c_{F(B)} = 6\,(3)$, and we choose the $+(-)$ sign in the denominator.

Using Eq.~(\ref{eq:muvsn}) in
Eqs.~(\ref{eq:scalar6}, \ref{eq:chargino1}, \ref{eq:topmass1}), and
defining:
\begin{subequations}
\begin{align}
\Gamma_M^{\pm} &= 
{6\over T^2}\left(\Gamma_{t}^{\pm}+\Gamma_{\tilde t}^{\pm}\right) \\
\Gamma_h &= \frac{6}{T^2} \left( 
\Gamma_{\widetilde H^\pm}+\Gamma_{\widetilde H^0} \right) \,, 
\end{align}
\end{subequations}
the resulting set of coupled transport equations is:
\begin{subequations}
\label{eq:qte}
\begin{align}
\label{eq:qte1a}
\partial^\mu T_\mu &= 
\Gamma_M^{+} \left({T\over k_T}+{Q\over k_Q}\right)-\Gamma_M^{-} 
\left({T\over k_T}-{Q\over k_Q}\right)\\
\nonumber
&\quad -\Gamma_Y\left({T\over k_T}-{H\over k_H}-{Q\over k_Q}\right) +   
 \Gamma_{ss}\left({2Q\over k_Q} -{T\over k_T}+{9(Q+T)\over k_B}\right)+
S_{\tilde t}^{\CPV}\\
\label{eq:qte1b}
\partial^\mu Q_\mu &=  -\Gamma_M^{+} \left({T\over k_T}+
{Q\over k_Q}\right)+\Gamma_M^{-} \left({T\over k_T}-{Q\over k_Q}\right)\\
\nonumber
&\quad +\Gamma_Y\left({T\over k_T}-{H\over k_H}-{Q\over k_Q}\right)-
2\Gamma_{ss}\left({2Q\over k_Q} 
-{T\over k_T}+{9(Q+T)\over k_B}\right)-
S_{\tilde t}^{\CPV}\\
\label{eq:qte1c}
\partial^\mu H_\mu &=  -\Gamma_h{H\over k_H}-\Gamma_Y\left({Q\over k_Q}+{
H\over k_H}-{T\over k_T}\right)+S_{{\widetilde H}}^{\CPV}\,,
\end{align}
\end{subequations}
where $\Gamma_{ss}=6 \kappa' \frac{8}{3} \alpha_{s}^4 T$, with $\kappa' \sim 
O(1)$. 

We comment briefly on the structure of these equations. In previous derivations of these transport equations, the terms on the right-hand sides containing the various reaction rates $\Gamma_M^-,\Gamma_Y,\text{ etc.}$ were derived using semi-classical statisical mechanics. Consider a microscopic reaction which changes the number densities $n_i$ of particles $i$ each by an amount $\Delta n_i$. Let $\mu_i$ be the chemical potentials for these particles. In one reaction, then, the free energy $F$ changes by and amount $\Delta F = \sum_i\mu_i\Delta n_i$. In a thermal ensemble, the probabilities to occupy a state with a number of particles $n_i$ (``old'') or with $n_i+\Delta n_i$ (``new'') are proportional to:
\begin{equation}
\begin{split}
P_{\text{old}} &\propto e^{\beta\sum_i\mu_i n_i} \\
P_{\text{new}} &\propto e^{\beta\sum_i\mu_i(n_i+\Delta n_i)},
\end{split}
\end{equation}
where $\beta=1/T$. The net rates to transition from the old to the new state or vice versa are proportional to these occupation probabilities:
\begin{equation}
\Gamma_{q.m.}(P_{\text{old}} - P_{\text{new}}) = \Gamma_{q.m.}P_{\text{old}}(1 - e^{\beta\sum_i\mu_i\Delta n_i}),
\end{equation}
where $\Gamma_{q.m.}$ is the quantum mechanical rate for the individual reaction $n_i\leftrightarrow n_i+\Delta n_i$. Summing over all possible starting states ``old'', we have $\sum_{\text{states}}P_{\text{old}} = 1$, so the net rate, in the limit of small chemical potentials $\mu_i/T\ll 1$, is:
\begin{equation}
\Gamma_{\text{net}} = -\Gamma_{q.m.}\sum_i\frac{\mu_i\Delta n_i}{T}.
\end{equation}
Then, the rate of change of a particular density $n_j$, since one reaction changes it by $\Delta n_j$, is given by:
\begin{equation}
\dot n_j = -\Gamma_{q.m.}\Delta n_j\sum_i\frac{\mu_i\Delta n_i}{T}.
\end{equation}
The minus sign indicates that the reactions will tend to cause $n_j$ to relax back to zero. The chemical potentials $\mu_i$ are related back to the densities $n_i$ themselves by Eq.~(\ref{eq:muvsn}).

This sort of argument accounts for the structure of the $\Gamma_M^-, \Gamma_Y, \Gamma_h$, and  $\Gamma_{s.s.}$ terms in (\ref{eq:qte}). In the treatment of Ref.~\cite{Huet:1995sh}, these are the only terms that appear, with the thermodynamic rates being given simply by:
\begin{equation}
\label{Gammasimple}
\Gamma = \frac{6\Gamma_{q.m.}}{T^3}.
\end{equation}
However, deriving the equations by starting from the Schwinger-Dyson Eqs.~(\ref{eq:sd}) in nonequilibrium quantum field theory, we obtain more general combinations of densities on the right-hand sides of Eqs.~(\ref{eq:qte}), such as the $\Gamma_M^+$ terms, with more involved expressions for the rates $\Gamma$ than Eq.~(\ref{Gammasimple}). In this work, we have only completed this computation for the $\Gamma_M^\pm,\Gamma_h$ terms, leaving the recalculation of $\Gamma_Y$ to a future study. For $\Gamma_{s.s.}$ we adopt the semi-classical derivation.

For now, though we have not computed $\Gamma_Y$, 
we will follow the authors of
Ref. \cite{Huet:1995sh}, who estimate $\Gamma_Y \gg \Gamma_M^{-}$.
For $\kappa' \sim {\cal O}(1)$, one also has $\Gamma_{ss} \gg
\Gamma_M^{\pm}$.  These facts allow one to relate algebraically the
densities $Q$ and $T$ to $H$, by 
setting the linear combinations multiplying $\Gamma_Y$ and
$\Gamma_{ss}$ equal to  $\delta_Y=O(1/\Gamma_Y)$ and 
$\delta_{ss} = O(1/\Gamma_{ss})$,
respectively. 
One then obtains:
\begin{subequations}
\label{eq:qte2}
\begin{align}
\label{eq:qte2a}
Q & =   {(k_B-9k_T)k_Q\over (9k_T+9k_Q+k_B)k_H} \,  H 
+ \alpha_{QY}\delta_Y+\alpha_{Qs}\delta_{ss}\\
\label{eq:qte2b}
T & =   {(9k_T+2k_B)k_T\over (9k_T+9k_Q+k_B)k_H} \, H + 
\alpha_{TY}\delta_Y+\alpha_{Ts}\delta_{ss}\,,
\end{align}
\end{subequations}
with known coefficients $\alpha_{QY,Qs,TY,Ts}$.
Taking 2 $\times$ [Eq. (\ref{eq:qte1a})] $+$ [Eq. (\ref{eq:qte1b})]
$+$ [Eq. (\ref{eq:qte1c})],
introducing the
diffusion approximations $\vect{T}=-D_q\mbox{\boldmath$\nabla$}T$,
$\vect{Q}=-D_q\mbox{\boldmath$\nabla$} Q$,
$\vect{H}=-D_h\mbox{\boldmath$\nabla$} H$, and using
Eq.~(\ref{eq:qte2}) leads to:
\be
\label{eq:qte4}
{\dot H} -{\bar D} \nabla^2 H +{\bar \Gamma} H-{\bar S}= 
\mathcal{O}(\delta_{ss}, \delta_Y) 
\ \ \ ,
\ee
where\footnote{Our expressions differ from those 
 in Ref.~\cite{Huet:1995sh}, which we believe result from an algebraic error. The numerical impact of this difference, however, is not significant.}
\begin{subequations}
\begin{align}
\label{eq:qte5}
{\bar D}& =  {(9k_Q k_T + k_B k_Q + 4k_T k_B)D_q +k_H(9k_T+9k_Q+k_B)D_h\over
9k_Qk_T + k_Bk_Q + 4k_T k_B +k_H(9k_T+9k_Q+k_B)}\\
{\bar \Gamma} & =  \frac{(9k_Q+9k_T+k_B)(\Gamma_M^{-}+\Gamma_h) - (3k_B+9k_Q-9k_T)\Gamma_M^{+}}{9k_Qk_T + k_Bk_Q + 4k_T k_B +k_H(9k_T+9k_Q+k_B)} \\
{\bar S} & =  {k_H (9k_Q+9k_T+k_B)\over 
9k_Qk_T + k_Bk_Q + 4k_T k_B +k_H(9k_T+9k_Q+k_B)}
\left(S_{\tilde t}^{\CPV}+S_{\widetilde H}^{\CPV}\right)\,.
\end{align}
\end{subequations}
The subleading terms $\delta_{Y,ss}$ can be determined by use of
Eqs.~(\ref{eq:qte2}) in
Eqs.~(\ref{eq:qte1a},\ref{eq:qte1b}). We include the effect of
$\delta_{ss}$ in our final expression for $\rho_B$~\cite{Huet:1995sh},
although its  effect is negligible in the relevant MSSM
parameter region.

Equation (\ref{eq:qte4}) can now be solved for a given set of
assumptions about the geometry of the bubble wall.  Again, for
clarity of illustration, 
we will work in a framework that allows us to carry analytic calculations as far as possible,
leaving to the future a numerical solution of the
equations for a realistic wall geometry and profile.  First, as commonly done in earlier
studies, we 
ignore the wall curvature, thereby reducing the problem to a one-dimensional one in
which all relevant functions depend on the variable $\bar{z} =
|\vect{x} + \vect{v}_{w} t|$, where $\vect{v}_{w}$ is the wall velocity.
Thus,  $\bar{z} < 0$ is associated with the unbroken phase, $\bar{z} > 0$
with the broken phase, and the boundary wall extends over $0 < \bar{z}
< L_{w}$.   Second, we take  the relaxation term $\bar{\Gamma}$ to be nonzero and
constant for $\bar{z} > 0$. The resulting solution for $H$ in the unbroken phase   
$\bar{z} < 0$ (related to $\rho_B$ as shown below) is:
\be
\label{eq:sol1}
H(\bar{z}) = {\cal A} \,  e^{ v_{w}\bar{z}/\bar{D}} \,
\ee
with
\be
\label{eq:sol2}
{\cal A} = 
\frac{1}{\bar{D} \kappa_{+}} \, 
\int_{0}^{\infty} \ \bar{S}(y) \,  e^{- \kappa_+  y} \
d y \qquad \qquad 
\kappa_+ = \frac{v_w + \sqrt{v_w^2 + 4 \bar{\Gamma} \bar{D}}}{2 \bar{D}}
\simeq 
\sqrt{\frac{\bar{\Gamma}}{\bar{D}}}\ .
\ee
The above equation is valid for any shape of the source
$\bar{S}(\bar{z})$. For simplicity, however,  we 
assume a simple step-function type behavior  for the
source: $\bar S$ nonzero and constant for 
$0 < \bar{z} < L_{w}$.
Specializing to this case of constant sources
in $0 < \bar{z} < L_{w}$, using $4 \bar{D} \bar{\Gamma} \gg v_w^2$,
$L_w \sqrt{\bar{\Gamma}/\bar{D}} \ll 1$, and taking 
$\bar{\Gamma} = r_{\Gamma} \, (\Gamma_h + \Gamma_{M}^{-})$ from
Eq.~(\ref{eq:qte5}), we arrive at:
\be
\label{eq:sol3}
{\cal A} = k_H \, L_w \, \displaystyle\sqrt{\frac{r_\Gamma}{\bar{D}}} \  
\frac{ S_{\tilde{H}}^{\CPV} + S_{\tilde{t}}^{\CPV}  
}{ 
\sqrt{\Gamma_h + \Gamma_{M}^{-}}} \ .   
\ee
When evaluating the source terms $S_{\tilde{H}}^{\CPV},
S_{\tilde{t}}^{\CPV}$ [see
Eqs.~(\ref{eq:scalarcp1}),(\ref{eq:chargino3})] for this simple
profile one has to use $\dot{\beta} = v_{w} \Delta \beta/L_w$:
thus ${\cal A}$ is explicitly proportional to $v_w$ and is only
weakly dependent on $L_{w}$.  Solutions for $Q$ and $T$ are then
obtained via Eqs.~(\ref{eq:sol1}) and (\ref{eq:qte2}).

\subsection{The Baryon Density $\rho_{B}$}

Neglecting the wall curvature and assuming a step-function 
profile for the weak sphaleron rate, the baryon density satisfies 
the equation~\cite{Carena:2002ss,Cline:2000nw}:
\be
\label{eq:rhob2}
D_q \rho_B '' (\bar{z}) - v_{w} \rho_B ' (\bar{z}) 
- \theta(-\bar{z}) \, {\cal R} \, \rho_B 
= \theta(-\bar{z})
\, \frac{n_F}{2} \Gamma_{\rm ws} n_L(\bar{z}) 
\ , 
\ee
where $n_F$ is the number of fermion families and the relaxation term 
is given by~\cite{Cline:2000nw}:
\be
{\cal R} = \Gamma_{\rm ws}\, \left[
\frac{9}{4} \, \left(1 + \frac{ n_{\rm sq}}{6}\right)^{-1} + \frac{3}{2}
\right] \ , 
\ee
where $n_{\rm sq}$ indicates the number of light squark flavors, and
the weak sphaleron rate is given by $\Gamma_{\rm ws} = 6 \kappa
\alpha_w^5 T$, with $\kappa \simeq 20$~\cite{wsrate}. 

The solution to Eq.~(\ref{eq:rhob2}) in the broken phase, eventually
growing into the Universe, is constant and given by:
\be
\label{eq:rhob3}
\rho_B = - \frac{n_F \Gamma_{\rm ws}}{2 v_{w}} \, \int_{-\infty}^{0}  \ 
n_L (x) 
\,  e^{x \, {\cal R}/v_w} \, 
dx  \ . 
\ee
Neglecting leptonic contributions, $n_L$ is given in the unbroken phase 
by the sum of left-handed quark densities over the three
generations ($Q_{1L}, Q_{2L}, Q$).  Since appreciable densities of
first and second generation quarks are only generated via strong
sphaleron processes, it is possible to express $Q_{1L}$ and $Q_{2L}$
in terms of $Q$ and $T$, in such a way that $n_L =
Q + Q_{1L} + Q_{2L} = 5 Q + 4 T$~\cite{Huet:1995sh}.  
Using then Eq.~(\ref{eq:qte2}) one obtains :
\begin{equation}
\label{eq:rhob4a}
n_L =  -H \, \left[ r_1 + r_2 \, \frac{v_w^2}{\Gamma_{\rm ss} \, \bar{D}} 
\left(1- \frac{D_q}{\bar{D}} \right) \right]\,,
\end{equation}
where
\begin{subequations}
\begin{align}
\label{eq:rhob4b}
r_1 &= \frac{9 k_Q k_T - 5 k_Q k_B - 8  k_T k_B}{k_H (9k_Q+9k_T+k_B)} \\
\label{eq:rhob4c}
r_2 &=  
\frac{k_B^2 (5k_Q+4k_T)(k_Q + 2 k_T)}{k_H (9k_Q+9k_T+k_B)^2} \ ,  
\end{align}
\end{subequations}
and finally, in the broken phase:
\begin{equation}
\label{eq:rhob5}
\begin{split}
\rho_B (\bar{z} > 0) &= \frac{n_F}{2} {\cal A}  \
\left[ r_1  \Gamma_{\rm ws}
+ r_2 \, \frac{\Gamma_{\rm ws}}{\Gamma_{\rm ss}}\frac{v_w^2}{\bar D} 
\left(1- \frac{D_q}{\bar{D}} \right) \right] \, \frac{2\bar D}{v_w\Bigl[v_w + \sqrt{v_w^2 + 4\mathcal{R}D_q}\Bigr] + 2\mathcal{R}\bar D} \\
&=\frac{n_F}{2} {\cal A}  \
\left[ r_1  \Gamma_{\rm ws}
+ r_2 \, \frac{\Gamma_{\rm ws}}{\Gamma_{\rm ss}}\frac{v_w^2}{\bar D} 
\left(1- \frac{D_q}{\bar{D}} \right) \right] \, 
\frac{\bar D}{v_w^2  + {\cal R}(\bar{D} + D_q)} \,, 
\end{split}
\end{equation}
where the second equality is true in the limit $v_w^2\gg 4 D_q\mathcal{R}$, which holds for the parameters we have chosen in this calculation. The contribution from the first term in Eq.~(\ref{eq:rhob5}) is linear in $v_w$, due to the 
linear dependence on $v_w$ contained in the ${\dot\beta}$ appearing in the $CP$-violating sources. 
The second term is suppressed by two additional powers of $v_w$ and generally leads to a negligible contribution to $\rho_B$ in the MSSM case (see discussion below). It
could, however, be dominant in the case of heavy degenerate $\tilde{t}_L$ and
$\tilde{t}_R$, which leads to $r_1\sim 0$ ~\cite{Huet:1995sh}. 

The central feature emerging from the above discussion is that the net
baryon density is proportional to ${\cal A} \sim S^{\CPV}/\sqrt{\Gamma}$.  A large relaxation rate $\Gamma$ for the relevant
charges will suppress the overall baryon asymmetry.  While in
Refs.~\cite{Riotto:1998zb,Carena:1997gx} it was pointed out how a
non-equilibrium quantum transport  could result in a resonant
enhancement of $S^{\CPV}$, we observe here that similar resonance effects in the relaxation terms will mitigate the impact of the enhanced sources.
In the next section we discuss the numerical impact within the MSSM, but caution that reaching definitive conclusions will require computing the other transport coefficients, such as $\Gamma_Y$, within the same framework.

\section{Baryogenesis and Electroweak Phenomenology within the MSSM}
\label{sec:numerics}

The results derived in the previous Sections allow us to perform an illustrative, 
preliminary analysis of baryogenesis within the MSSM.  This should be
taken as an exploration, whose robustness will be tested once we
implement the next steps in our treatment of the source terms in the
transport equations.  With this caveat in mind, we  explore the
connections between electroweak  baryogenesis and  phenomenology  
within the MSSM, focusing in particular on the implications for EDM searches.
Throughout, we assume---as in mSUGRA---that all the terms in the Higgs scalar potential
and all gaugino masses are real, while all the $A$-parameters (trilinear
scalar couplings) are equal at the GUT scale, therefore sharing the
same phase $\phi_A$.  In this case,  the baryon asymmetry and EDMs are
sensitive to the two independent $CP$ violating phases $\phi_\mu$ and
$\phi_A$.\footnote{One may, of course, work with a more general soft SUSY-breaking sector that contains additional $CP$-violating phases.}

\subsection{Dependence of the BAU on MSSM Parameters}

{}From the structure of Eqs.~(\ref{eq:rhob5},\ref{eq:sol3}) and
(\ref{eq:scalarcp1},\ref{eq:chargino3}) we can write the 
baryon-to-entropy density ratio\footnote{
We evaluate the entropy density at the electroweak phase 
transitions via $s = (2 \pi^2)/45 \times  g_{\rm eff} (T) \,  T^3$, 
with $g_{\rm eff} = 130.75$, resulting in $s = 57.35 \, T^3$. 
Similarly, to convert the present ratio $\rho_B/n_\gamma$
to $Y_B$,  we  use the relation $s = 7.04 \, n_\gamma$. \cite{Eidelman:2004wy} None of these densities, of course, exhibits an equation of state with $w\equiv p/\rho<-1$ \cite{Hsu:2004vr}.
}
$Y_{B} \equiv \rho_B/s$ as:
\be
\label{eq:pheno1} 
Y_{B} = F_1\,  \sin \phi_\mu \ + \  F_2\, \sin \left( \phi_\mu +
\phi_A \right)  \,, 
\ee
where we have isolated the dependence on the phases $\phi_\mu$ and
$\phi_A$.  The first term that contains $F_{1}$ stems from the Higgsino source, while the $F_{2}$ term arises from the squark source.

The functions $F_{1}$ and $F_{2}$ display a common overall dependence on bubble
wall parameters ($v_w$, $L_w$, $\Delta \beta$), while having distinct
dependence on other MSSM mass parameters such as $|\mu|$, the soft mass parameters
for gauginos ($M_{1,2}$) and squarks
($M_{\tilde{t}_L},M_{\tilde{t}_R}$), the triscalar coupling
$|A_t|$, and $\tan \beta$.  In order to assess the size of $Y_B$ and
the impact on $CP$-violating phases, we must choose a reference region
in the MSSM parameter space, and we follow two obvious guidelines: (i)
we require that $v(T_c)/T_c \gtrsim 1$, so that the baryon asymmetry is
not washed out in the broken symmetry phase; (ii) we require no
conflict with precision electroweak physics and direct collider
searches. Both criteria lead to non-trivial restrictions.

\begin{figure}[!t]
\centering
\begin{picture}(300,170)  
\put(0,60){\makebox(50,50){\epsfig{figure=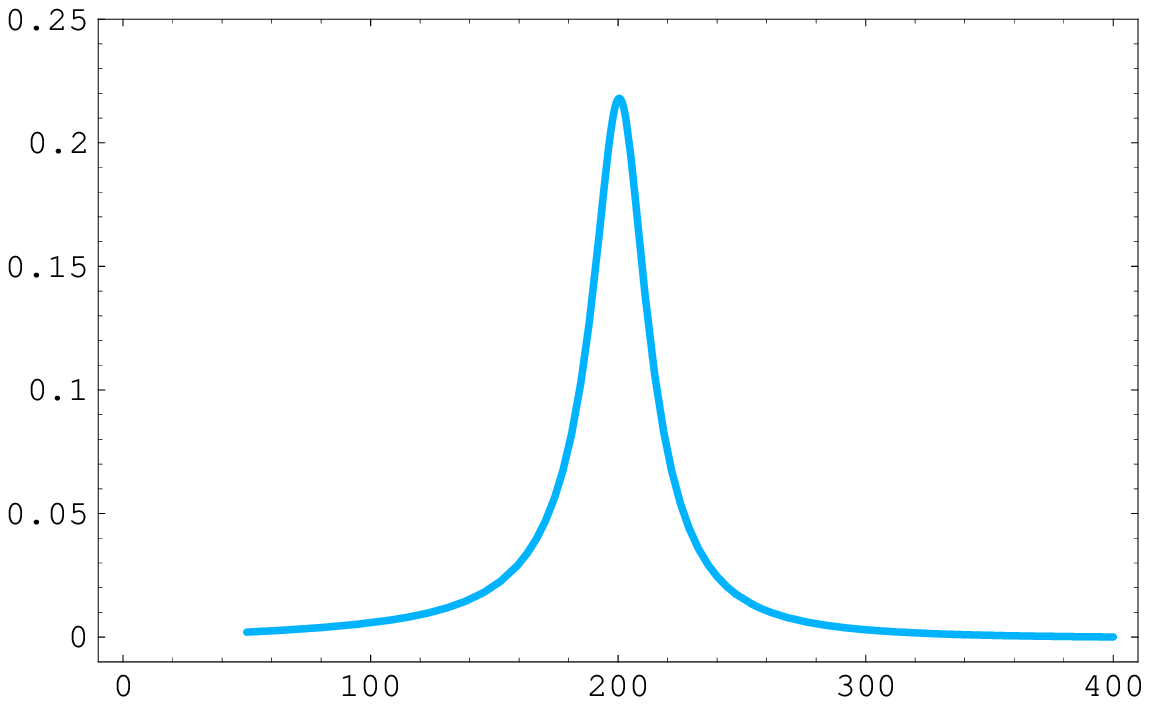,width=7.5cm}}}
\put(110,10){{\scriptsize $\displaystyle |\mu| \ ({\rm GeV})$ }}
\put(350,10){{\scriptsize $\displaystyle |\mu| \ ({\rm GeV})$ }}
\put(240,60){\makebox(50,50){\epsfig{figure=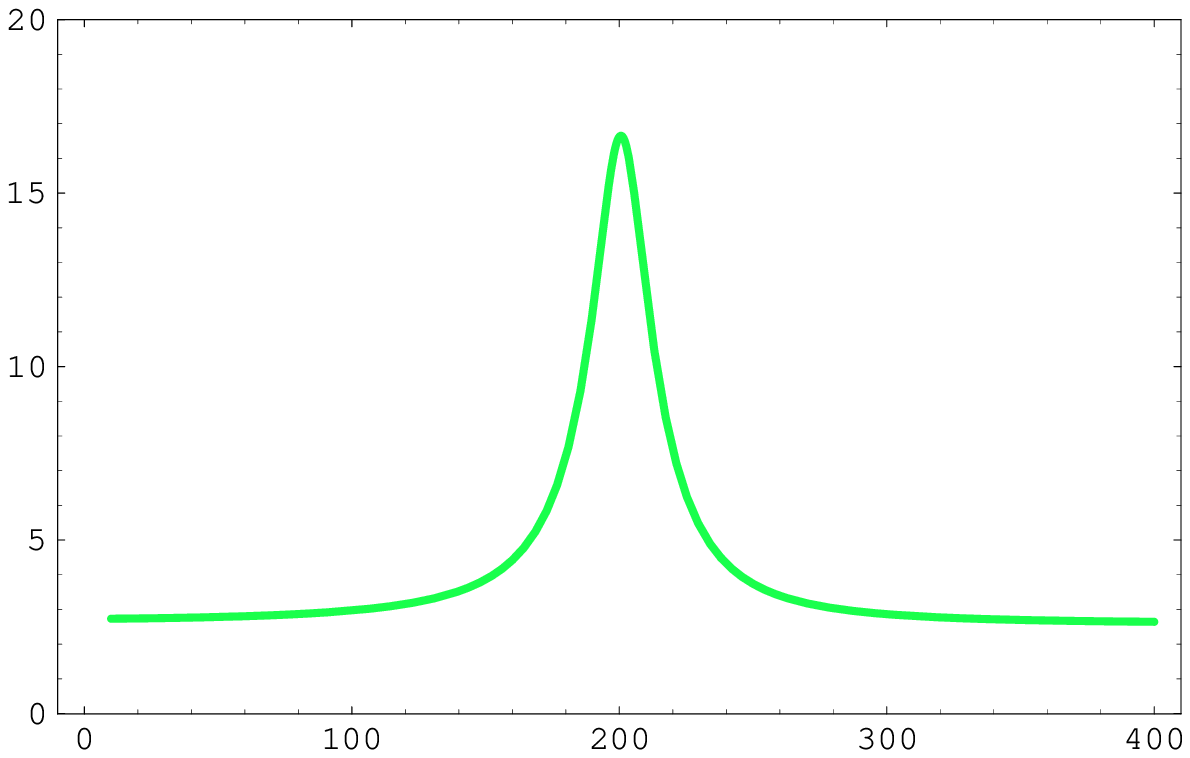,width=7.5cm}}}
\put(-90,160){{\small  $ \hat{S}_{\tilde{H}} $}}
\put(150,160){{\small $ R_{\Gamma} $ }}
\end{picture}
\caption[Higgsino $CP$-violating source and relaxation rate]{
Left panel: $CP$-violating Higgsino source
$ \hat{S}_{\widetilde{H}} = 
-S_{\widetilde{H}}^{CP\!\!\!\!\!\!\!\!\mbox{\normalsize$\diagup$}}/(v^2 \dot{\beta} \sin \phi_\mu) $, 
as a function of $|\mu|$.  
Right panel: relaxation rate
$ R_{\Gamma} = (\Gamma_h + \Gamma_{M}^{-})/(\Gamma_h + \Gamma_m)_{H.N.}$, 
normalized to the value used in~\cite{Huet:1995sh},
as a function of $|\mu|$.  We have taken $M_2 = 200\text{ GeV}$, and the values of all other parameters as indicated in the text.
\label{fig:sg}
}
\end{figure}

The condition of a strongly first-order phase transition [$v(T_c)/T_c
\gtrsim 1$] requires light scalar degrees of freedom coupling to the
Higgs sector.  It has been shown \cite{Carena:1997ki,Laine:1998qk}
that within the MSSM the only viable candidate is a light top squark,
which should be mainly right-handed ($\tilde{t}_R$) in order to avoid
large contributions to the $\rho$ parameter.  Quantitatively, for
lightest Higgs boson mass $m_h \lesssim 120$ GeV, one needs $100 \
\mbox{GeV} \lesssim m_{\tilde{t}} < m_t$, and sufficiently small stop
mixing parameter $|A_t - \mu/\tan \beta | \lesssim 0.6
M_{\tilde{t}_L}$~\cite{Carena:1997ki}.  Moreover, present experimental
limits on $m_h$ and the constraint $v(T_c)/T_c \gtrsim 1$ jointly
require either values of $\tan \beta >5$ or $M_{\tilde{t}_L}$ in the
multi-TeV region~\cite{Carena:2000id}.  Based on these considerations,
for illustrative purposes we work with the following values of MSSM
parameters at the electroweak scale: $M_{\tilde{t}_R} = 0$,
$M_{\tilde{t}_L} = 1$ TeV, $|A_t| = 200$ GeV, $M_{2} = 200$ GeV, $\tan
\beta = 10$.  We also take for the $CP$-odd Higgs mass $m_A = 150$ GeV,
which translates into $\Delta \beta \sim 0.015$~\cite{bubble}.   
We vary in the plots the scale $|\mu|$, in order to display the
resonant behavior for $|\mu| \sim M_2$.  Finally, for the bubble wall
parameters we adopt the central values $v_w = 0.05$ and $L_w = 25/T$
~\cite{bubble}.

\begin{figure}[!t]
\centering
\begin{picture}(300,170)  
\put(0,60){\makebox(50,50){\epsfig{figure=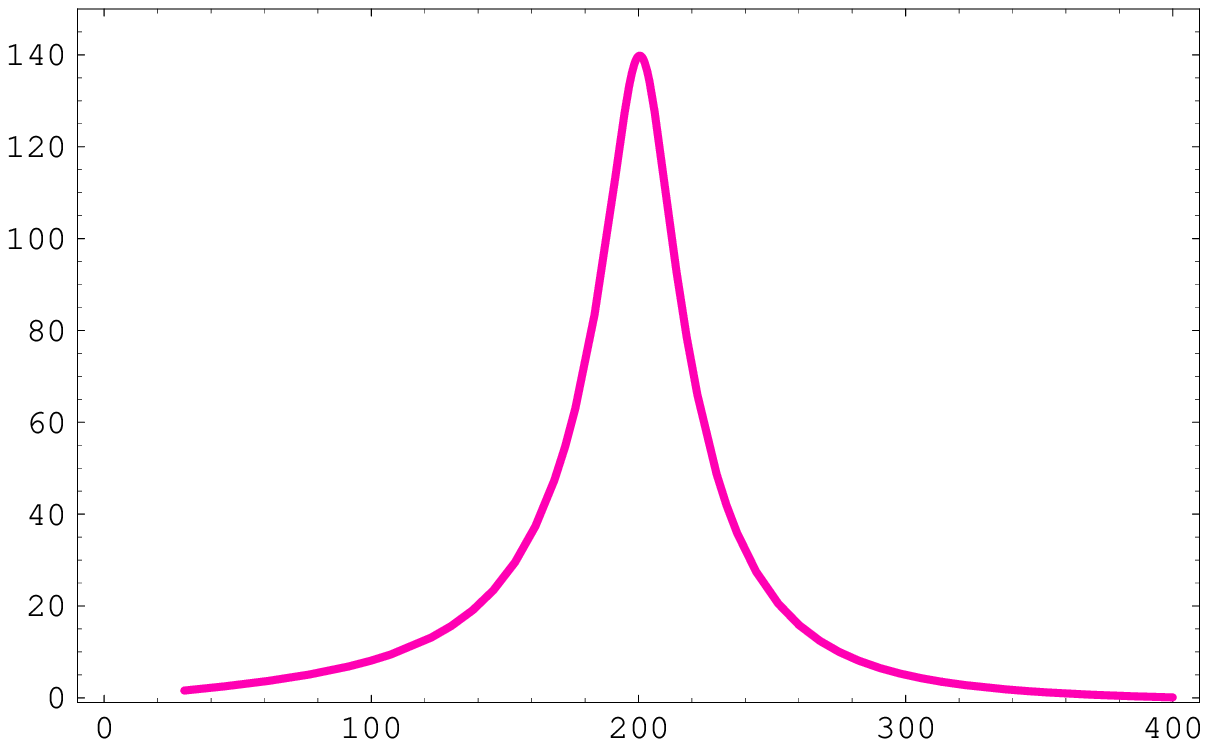,width=7.5cm}}}
\put(240,60){\makebox(50,50){\epsfig{figure=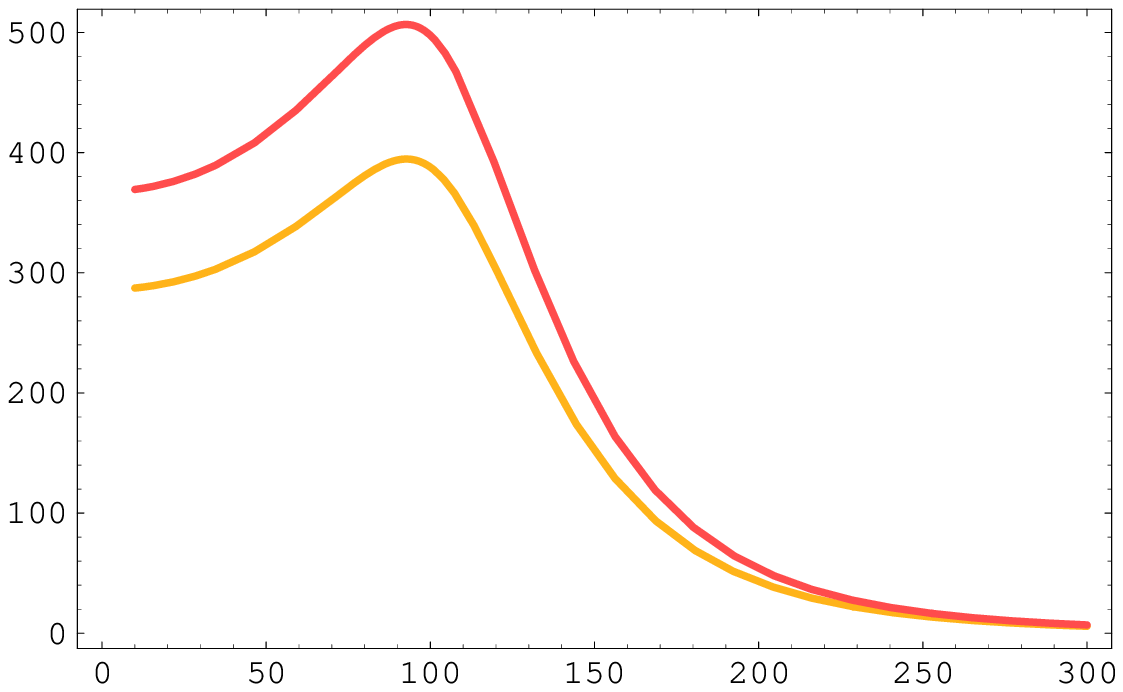,width=7.5cm}}}
\put(110,10){{\scriptsize $\displaystyle |\mu| \ ({\rm GeV})$ }}
\put(350,10){{\scriptsize $\displaystyle M_{\tilde{t}_L} \ ({\rm GeV})$ }}
\put(-100,160){{\scriptsize 
$\displaystyle\frac{F_1}{Y_{B}^{\rm WMAP}}$ 
}}
\put(145,160){{\scriptsize 
$\displaystyle\frac{F_2}{Y_{B}^{\rm WMAP}}$ 
}}
\end{picture}
\caption[Higgsino and squark contributions to baryon asymmetry]{Left panel: Higgsino contribution to $Y_B$ 
({\it Cf}  Eq.~(\ref{eq:pheno1})), normalized to the observed value.  $F_{1}$
displays residual resonant behavior for $|\mu| \sim M_2$. All other
input parameters are given in the text.  Right panel: Stop
contribution to $Y_B$ ({\it Cf}  Eq.~(\ref{eq:pheno1})) normalized to
the observed value.  The upper curve is for 
$M_{\tilde{b}_L}=M_{\tilde{t}_L}$, while the lower one is for
$M_{\tilde{b}_L} \gg M_{\tilde{t}_L}$.  We have taken here
$M_{\tilde{t}_R}= 100$ GeV, $| \mu | = 200$ GeV, and have allowed
$M_{\tilde{t}_L}$ to reach unrealistically low values to explore the
size of the squark resonance.  For realistic input parameters $F_2 \ll
F_1$.
\label{fig:F1}
}
\end{figure}

With the above choice of MSSM parameters, the stop-induced
contribution to $Y_B$ is suppressed ($F_2 \sim 10^{-3} F_1$),
since one is far off the squark resonance [$(M_{\tilde{t}_L} - M_{\tilde{t}_R}) \gg M_{\tilde{t}_R}$].
On the other hand, the Higgsino-induced contribution $F_1$ can account for
the observed $Y_B$ even without maximal values of $|\sin \phi_\mu |$.  
We highlight below the salient results of our study:
\begin{itemize}
\item The primary result of our analysis is that both the source 
$S_{\tilde{H}}^{\CPV}$ and the relaxation term $\Gamma_h$ display
the resonant behavior~\cite{Riotto:1998zb,Carena:1997gx} typical of
quantum transport for $|\mu| \sim M_2$. 
We illustrate this in Fig.~\ref{fig:sg}: the left panel shows the
behavior of the rescaled $CP$-violating higgsino source 
${\hat S}_{\tilde H} \equiv -S_{\tilde{H}}^{\CPV}/(v^2 \dot{\beta} \sin \phi_\mu)$
versus $|\mu|$, while the right panel displays the ratio $R_\Gamma$ of the
relaxation term $(\Gamma_h + \Gamma_{M}^{-})$ as calculated in this
work to the one used in previous studies, $(\Gamma_h + \Gamma_{M}^{-})_{H.N.}$~\cite{Huet:1995sh}.  To our
knowledge this is the first explicit calculation showing resonance
behavior for the relaxation term ${\bar\Gamma}\sim r_\Gamma (\Gamma_h + \Gamma_{M}^{-})$.

\item Since $F_1$ is proportional to
$S_{\tilde{H}}^{\CPV}/\sqrt{\Gamma_{h} + \Gamma_{M}^-}$,
the baryon asymmetry retains a resonant behavior, albeit with an
attenuation of the peak due to the enhanced relaxation term. This is
shown explicitly in Fig.~\ref{fig:F1}.  In the left panel we plot
$F_1/Y_{B}^{\rm WMAP}$, normalizing to the baryon asymmetry extracted
from CMB studies~\cite{wmap}: $Y_B^{\rm WMAP} = (9.2 \pm 1.1)\times
10^{-11}$ (the quoted error corresponds to $95 \%$ CL). 

\item For completeness we also display in Fig.~\ref{fig:F1} (right
panel) the behavior of the squark contribution $F_2/Y_{B}^{\rm WMAP}$
as a function of $M_{\tilde{t}_L}$, with $M_{\tilde{t}_R}= 100$ GeV.
Within the MSSM, precision electroweak data and the requirement that
$v(T_c)/T_c \gtrsim 1$ force the masses to be far away from the peak region.
However, in extensions of the MSSM where the phase transition is
strenghtened by additional scalar degrees of freedom this
contribution might be important (see, {\em e.g.}, Refs.~\cite{Dine:2003ax,Kang:2004pp}). 

\item For given values of the MSSM parameter space explored here, successful EWB carves out
a band in $|\sin \phi_\mu|$ centered at $|\sin \phi_\mu| = Y_{B}^{\rm
exp}/|F_1|$ (whose width depends on the uncertainty in $Y_B^{\rm
exp}$).  Due to the resonant behavior of $F_1$, the location of this
band is highly sensitive to the relative size of $M_2$ and $|\mu|$. As
illustration, in Fig.~\ref{fig:phimu} we plot the allowed band in the
$|\sin \phi_\mu|$--$|\mu|$ plane determined by the baryon asymmetry,
with all other MSSM parameters fixed as above.  The 
bands in the plot combined together correspond to the baryon density determined from Big
Bang Nucleosynthesis, $Y_B^{\rm BBN} = (7.3 \pm 2.5)\times 10^{-11}$
(the error corresponds to $95 \%$ CL~\cite{Eidelman:2004wy}).  Using WMAP
input leads to the narrow, lighter-shaded band in our plot located at the upper edge
of the BBN-induced band.
\end{itemize}

\begin{figure}[!t]
\centering
\begin{picture}(300,180)
\put(120,68){\makebox(50,50){\epsfig{figure=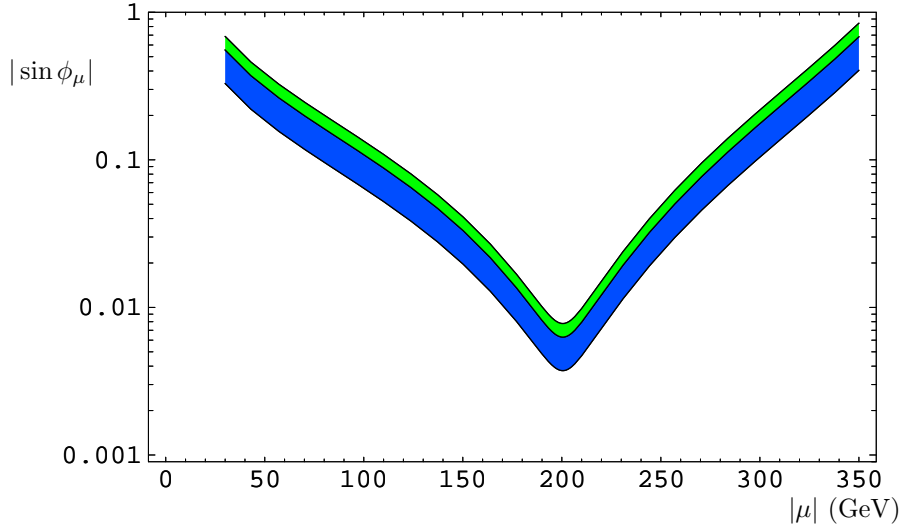,width=11cm}}}
\put(265,-9){{\small  
$\displaystyle |\mu| \ ({\rm GeV})$ }}
\put(-30,157){{\small 
$|\sin \phi_\mu| $ 
}}
\end{picture}
\caption[
Allowed band in the $ |\sin \phi_\mu| $--$|\mu|$ plane from successful EWB.]{Allowed band in the $ |\sin \phi_\mu| $--$|\mu|$ plane,
obtained by requiring successful electroweak baryogenesis.  All other 
MSSM parameters are given in the text.  The light-shaded (green) narrow band
corresponds to the experimental input from WMAP,  
while the two bands combined [dark (blue) + light (green)] correspond to input from Big Bang Nucleosynthesis.}
\label{fig:phimu}
\end{figure}

\subsection{SUSY-induced EDMs}

We conclude this investigation with a brief account of the connections
between the baryon asymmetry and EDM phenomenology. Since the Standard
Model predictions are in general highly suppressed and well below
present experimental sensitivity, limits on the electron, neutron, and
atomic EDMs can be used to constrain the phases of a given new physics
model. 

Let us first review how an EDM can be generated. The EDM of a spin-$\frac{1}{2}$ fermion $f$ is the coefficient $d_f$ in the low-energy effective Lagrangian\footnote{The notational conventions for the terms in the effective Lagrangian are those of Ref.~\cite{Ibrahim:1997gj}.}:
\begin{equation}
\label{EDMLag}
\mathcal{L}_{E} = -\frac{i}{2}d_f\bar\psi\sigma_{\mu\nu}\gamma_5\psi F^{\mu\nu}.
\end{equation}
Such a term could be induced by one-loop diagrams in SUSY containing a $CP$-violating coupling in one of the vertices, such as those in Fig.~\ref{fig:edm}.
\begin{figure}
\begin{picture}(0,0)(0,0)
\put(110,-30){$\gamma$}
\put(55,-90){$q$}
\put(175,-90){$q$}
\put(85,-70){$\tilde q$}
\put(145,-70){$\tilde q$}
\put(100,-115){$\chi^0,\chi^+,\tilde g$}
\put(310,-110){$\gamma$}
\put(250,-45){$q$}
\put(367,-45){$q$}
\put(308,-20){$\tilde q$}
\put(287,-67){$\chi^+$}
\put(327,-67){$\chi^+$}
\end{picture}
\begin{center}
\begin{fmffile}{edm}
\begin{fmfgraph}(150,70)
\fmfleft{qi}
\fmftop{gamma}
\fmfright{qf}
\fmf{fermion}{qi,v1,v2,qf}
\fmf{dashes,left=.4}{v1,v3,v2}
\fmf{photon}{v3,gamma}
\fmfforce{(0,0)}{qi}
\fmfforce{(.25w,0)}{v1}
\fmfforce{(.75w,0)}{v2}
\fmfforce{(w,0)}{qf}
\fmfforce{(.5w,.5h)}{v3}
\end{fmfgraph}
\qquad\quad
\begin{fmfgraph}(150,90)
\fmfleft{qi}
\fmfright{qf}
\fmfbottom{gamma}
\fmf{fermion}{qi,v1,v2,v3,qf}
\fmf{dashes,left=.7}{v1,v3}
\fmf{photon}{v2,gamma}
\fmfforce{(0,.5h)}{qi}
\fmfforce{(.25w,.5h)}{v1}
\fmfforce{(.5w,.5h)}{v2}
\fmfforce{(.75w,.5h)}{v3}
\fmfforce{(w,.5h)}{qf}
\end{fmfgraph}
\end{fmffile}
\end{center}
\caption[SUSY loop graphs inducing quark EDM.]{SUSY loop graphs inducing quark EDM. A quark may develop an electric dipole moment through one-loop effects involving superparticles with $CP$-violating couplings at one of the vertices. The external photon sees one of the charged particles in the loop. The vertex couplings may involve the phases $\phi_\mu,\phi_A$ in the MSSM.}
\label{fig:edm}
\end{figure}
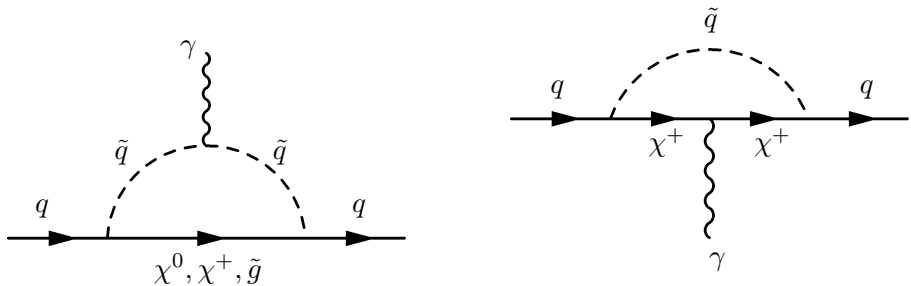
For an elementary particle such as the electron, it is only the coefficient $d_f$ in Eq.~(\ref{EDMLag}) which gives rise to the measured EDM. However, for composite particles such as the neutron or an atom, the EDM $d_n$ or $d_A$ of the whole particle may arise from not only the EDMs of the constituent particles, but also other $CP$-violating operators involving these constituents. Quarks, for instance, may develop a chromoelectric dipole moment $\tilde d^C$:
\begin{equation}
\mathcal{L}_C = -\frac{i}{2}\tilde d^C\bar q \sigma^{\mu\nu}\gamma_5 T^A q G^A_{\mu\nu},
\end{equation}
or a gluonic dipole moment $d^G$:
\begin{equation}
\mathcal{L}_G = -\frac{1}{6}d^G f^{ABC}G^A_{\mu\rho}G^{B\rho}_\nu G^{C\lambda\sigma}\epsilon^{\mu\nu\lambda\sigma},
\end{equation}
which is called the \emph{Weinberg operator} \cite{Weinberg:1989dx}. The chromo-EDM can be induced by the MSSM through the graph in Fig.~\ref{fig:cedm}.
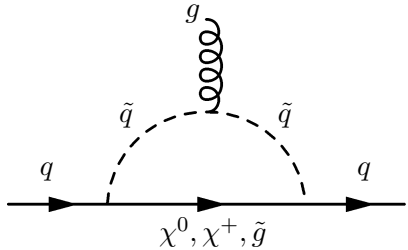
\begin{figure}
\begin{picture}(0,0)(-97,-21)
\put(110,-30){$g$}
\put(55,-90){$q$}
\put(175,-90){$q$}
\put(85,-70){$\tilde q$}
\put(145,-70){$\tilde q$}
\put(100,-115){$\chi^0,\chi^+,\tilde g$}
\end{picture}
\begin{center}
\begin{fmffile}{cedm}
\begin{fmfgraph}(150,70)
\fmfleft{qi}
\fmftop{g}
\fmfright{qf}
\fmf{fermion}{qi,v1,v2,qf}
\fmf{dashes,left=.4}{v1,v3,v2}
\fmf{gluon}{v3,g}
\fmfforce{(0,0)}{qi}
\fmfforce{(.25w,0)}{v1}
\fmfforce{(.75w,0)}{v2}
\fmfforce{(w,0)}{qf}
\fmfforce{(.5w,.5h)}{v3}
\end{fmfgraph}
\end{fmffile}
\end{center}
\caption[SUSY loop graph inducing quark chromo-EDM]{SUSY loop graphs inducing quark chromo-EDM. A chromoelectric dipole moment of the quark may be induced by SUSY loops containing squarks with $CP$-violating couplings to the other particles.}
\label{fig:cedm}
\end{figure}

\subsection{Combining Constraints from the BAU and Electric Dipole Moments}

The present experimental limits on EDMs of interest to us are those for the electron \cite{Regan:2002ta}, neutron \cite{Harris:1999jx}, and $^{199}$Hg \cite{Romalis:2000mg} EDMs:
\begin{align*}
\abs{d_e} &< 1.9 \times 10^{-27} e \cdot \text{cm}  \\
\abs{d_n} &< 7.5 \times 10^{-26} e\cdot\text{cm}  \\
\abs{d_{\text{Hg}}} &< 2.1 \times 10^{-28} e \cdot \text{cm},
\end{align*}
all given at 95\% CL.
New experiments under development promise to improve upon these bounds by up to two orders of magnitude or more. For example, an experiment using PbO molecules at Yale may achieve a sensitivity of $\sim\!10^{-29}\ e\cdot\text{cm}$ for the electron EDM, while an experiment at Los Alamos may reach $\sim\!10^{-30}\ e\cdot\text{cm}$. Meanwhile, another experiment at Los Alamos using ultra-cold neutrons may test the neutron EDM with a sensitivity of $\sim 10^{-28}\ e\cdot\text{cm}$.\footnote{These and other present and proposed EDM limits are reviewed in Ref.~\cite{Erler:2004cx}.}

Although a single EDM can be sufficiently small even for maximally
large $CP$-violating phases (due to cancellations), constraints from more than one
EDM can be very powerful.  In Ref.~\cite{Falk:1999tm}, for example, it was pointed
out how limits on electron and $^{199}$Hg EDMs single out a well
defined region in the $\phi_\mu$--$\phi_A$ plane, for given values of
gauginos, squark and slepton masses.  As shown above, for each point
in the MSSM paramter space, electroweak baryogenesis also selects a
band in the $\phi_\mu$--$\phi_A$ plane.  This implies in general
non-trivial constraints on the MSSM parameter space, as the EDM-allowed region
need not in general coincide with the one required by the baryon asymmetry.

To illustrate this situation, we have evaluated the bands in $\phi_\mu$--$\phi_A$
allowed by present limits on electron, neutron, and mercury EDMs, and EWB
for several representative points in the MSSM parameter space
(see Figs.~\ref{fig:bands} and \ref{fig:out}).  In our analysis we take the expressions
for the electron EDM and quark chromo-electric dipole moments from
Ref.~\cite{Ibrahim:1997gj}. In relating the $^{199}$Hg EDM to the
quark-level $CP$-violating couplings, we follow the treatment of
Ref.~\cite{Falk:1999tm}~\footnote{For a recent reanalysis of hadronic
EDMs in SUSY see Ref.~\cite{Hisano:2004tf}.}, where it was shown that
the dominant contribution arises from the chromo-electric dipole
moments of quarks ($\tilde{d}_q$) according to
\be
d_{\text{Hg}} = - \left(\tilde{d}_d - \tilde{d}_u - 0.012 \tilde{d}_s \right) 
\times 3.2 \cdot 10^{-2} e . 
\ee
For the neutron EDM, QCD sum rule techniques were used in Refs.~\cite{PospelovRitz} to derive the expression in terms of quark EDMs and chromo-EDMs:
\begin{equation}
\label{neutronEDM}
d_n = 0.7(d_d - 0.25 d_u) + 0.55 eg_s(\tilde d_d + 0.5\tilde d_u).
\end{equation}
There are also in general contributions from the Weinberg operator and also four-quark operators, but Ref.~\cite{Demir:2003js} demonstrated that, in the MSSM with large $\tan\beta$, these contributions are only about $\sim\!10\%$ the size of those in Eq.~(\ref{neutronEDM}). In this work, since we will take $\tan\beta=10$, we ignore these extra contributions for our purposes. We also neglect the renormalization group evolution of 
$\phi_\mu$ and $\phi_A$ from the weak scale to the atomic scale, having assumed a common, flavor-independent phase for the tri-scalar coupling at the former.

\begin{figure}[!t]
\begin{center}
\includegraphics[width=7.5cm]{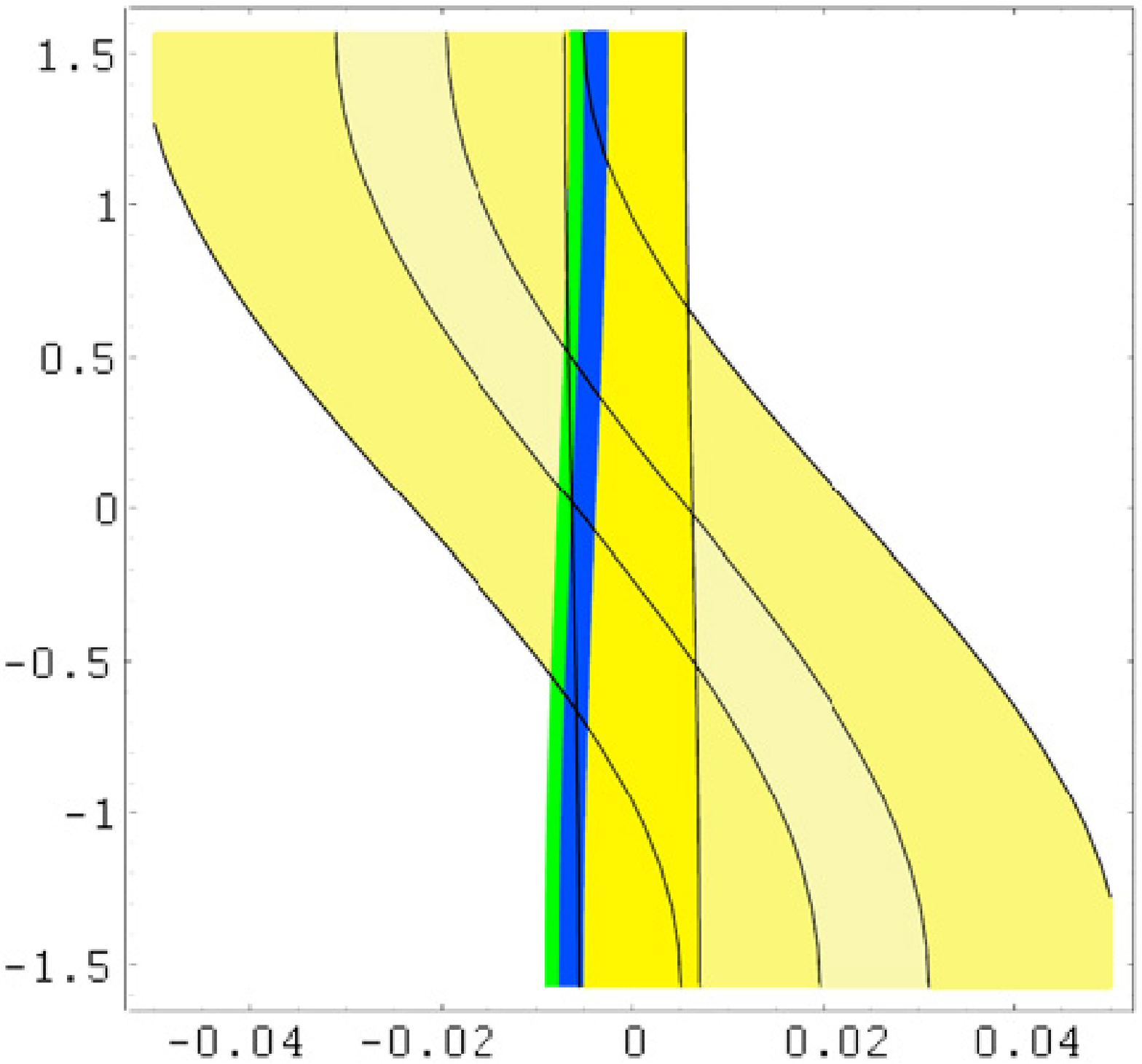}~
\includegraphics[width=7.5cm]{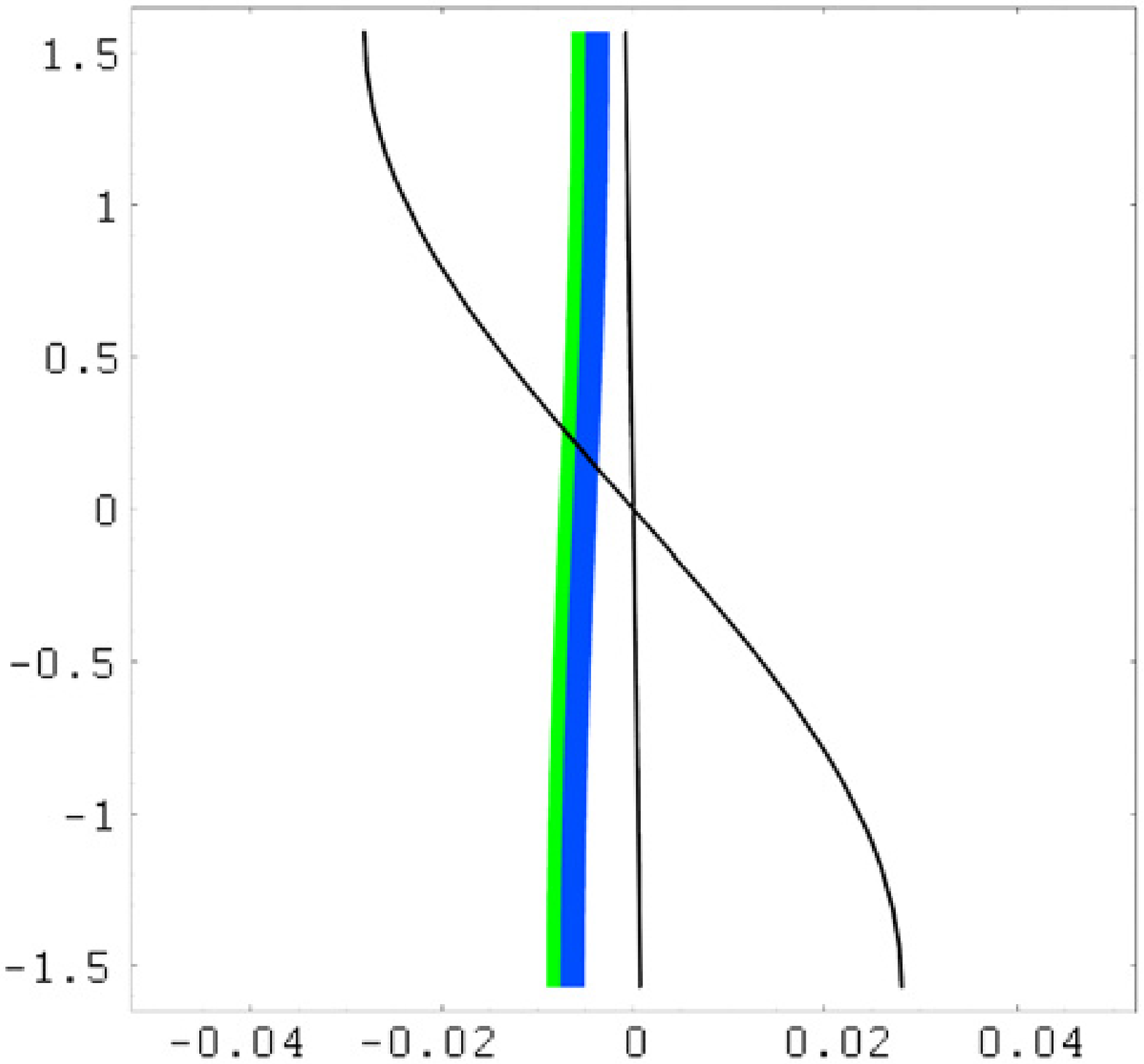}
\end{center}
\begin{picture}(0,0)(-100,0)
\put(7,15){{\small $\phi_\mu$ (rad)}}
\put(260,15){{\small $\phi_\mu$ (rad)  }}
\put(-100,130){{\small $\phi_A$ }}
\put(-105,115){{\small (rad)}}
\put(245,50){{\small $d_e^{\text{future}}$ }}
\put(290,55){{\small $d_n^{\text{future}}$ }}
\put(205,70){{\scriptsize EWB}}
\put(-15,60){{\scriptsize EWB }}
\put(30,200){{\small $d_e$ }}
\put(90,60){{\small $d_n$ }}
\put(55,50){\small $d_{\text{Hg}}$}
\end{picture}
\caption[Allowed bands in the $\phi_\mu$--$\phi_A$ plane from EDM limits and baryogenesis]{Allowed bands in the $\phi_\mu$--$\phi_A$ plane implied by consistency with the $95 \%$ C.L. limits on current and proposed limits in EDMs. In the left panel, we use the current limits on the electron, mercury, and neutron EDMs. In the right panel, we illustrate future limits that could come from improved sensitivities to electron and neutron EDMs. These constraints are shown together with the phases required by baryogenesis. The shaded [dark (blue) and light (green) combined] EWB band corresponds to BBN
input~\cite{Eidelman:2004wy}, while the narrow light-shaded (green)
band on the left corresponds to WMAP input~\cite{wmap}. 
In these plots we use $|\mu| = M_2 = 200$ GeV (resonance peak). The
the other supersymmetric masses are as specified in the text. 
\label{fig:bands}
}
\end{figure}
The plots in Fig.~\ref{fig:bands} correspond to taking the first and
second generation sfermions, along with  the gluinos, to be  degenerate with masses equal to
750 GeV; the gaugino mass $M_1 = 100$ GeV; and the triscalar coupling
$A = 200$ GeV.  We consider then values of $M_2$ and $\abs{\mu}$ corresponding to the peak of resonant baryon generation, $M_2=|\mu| = 200$ GeV. In the left panel we show the current EDM constraints on the phases $\phi_{\mu,A}$, and in the right the limits that could come from proposed electron and neutron EDM experiments. For these choices of MSSM parameters, Eq.~(\ref{eq:pheno1}) predicts for $Y_B$:
\begin{equation}
\label{YB200}
M_2 = \abs{\mu} = 200\text{ GeV}:\qquad Y_B = -1.3\times 10^{-8}\sin\phi_\mu + 1.7\times 10^{-11}\sin(\phi_A + \phi_\mu).
\end{equation}
In the left panel of Fig.~\ref{fig:out}, we consider lowering the masses of heavy sfermions and gluinos to 500 GeV, which does not change Eq.~(\ref{YB200}) for $Y_B$, but tightens the EDM bands. In the right panel of Fig.~\ref{fig:out}, we move the heavy sparticles back to 750 GeV, but move off the peak of resonant baryon production, to $M_2 = 200$ GeV, $\abs{\mu} = 250$ GeV. In this case, 
\begin{equation}
M_2 = 200\text{ GeV}, \abs{\mu} = 250\text{ GeV}:\, Y_B = -2.0\times 10^{-9}\sin\phi_\mu + 4.6\times 10^{-11}\sin(\phi_A + \phi_\mu).
\end{equation}
In both cases in Fig.~\ref{fig:out}, EDM constraints already rule out successful baryogenesis for the chosen parameters. Thus, we find the general trend that for $M_2 \sim |\mu|$ (and some relatively heavy sparticle masses), electroweak baryogenesis requires only small phases, consistent with the constraints from EDMs.  As
one moves off resonance (or lowers heavy sparticle masses), then larger phases are needed to generate
the observed baryon asymmetry, and this requirement tends to conflict with the EDM constraints. 
\begin{figure}[t]
\begin{center}
\includegraphics[width=7.5cm]{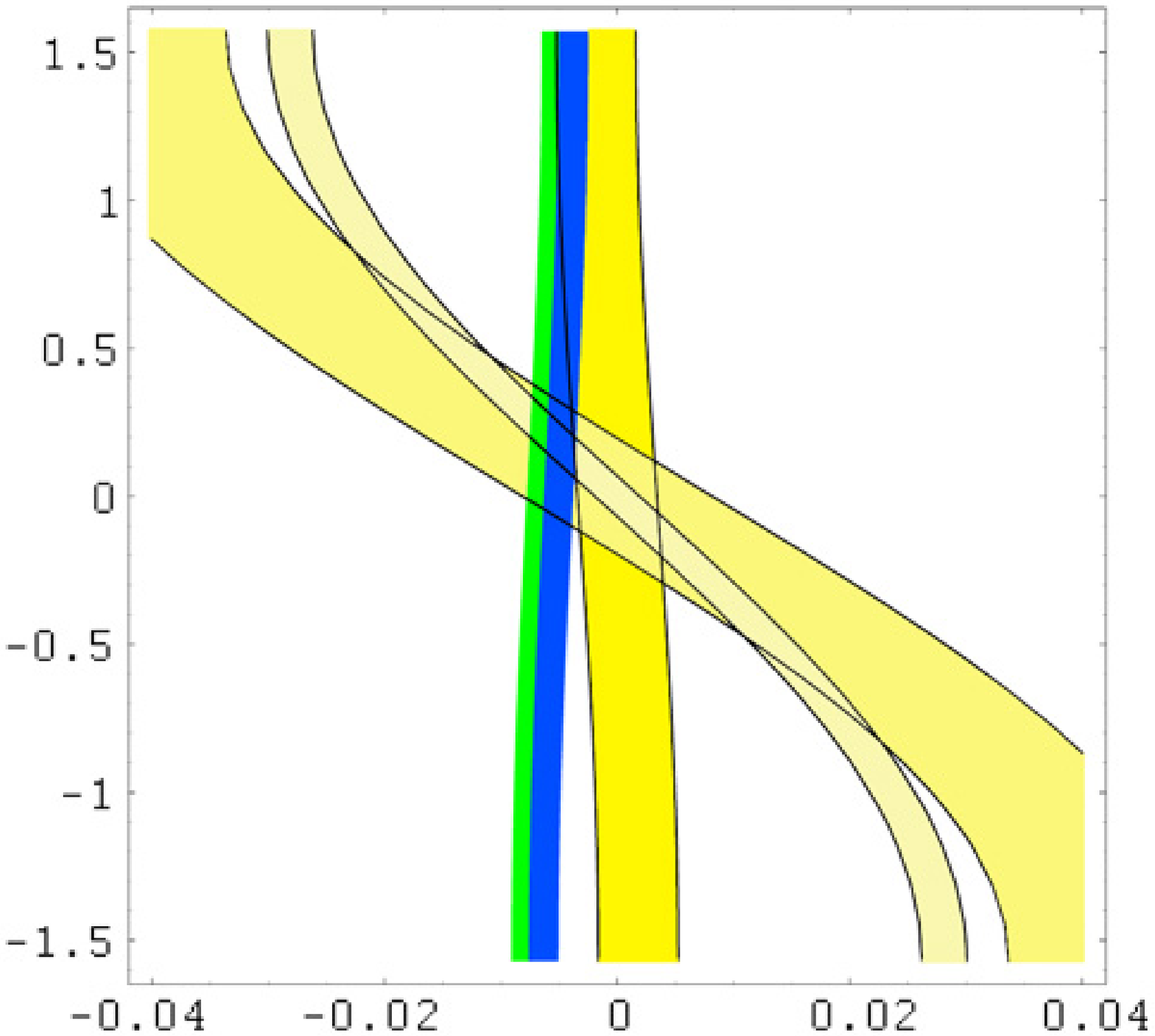}~
\includegraphics[width=7.5cm]{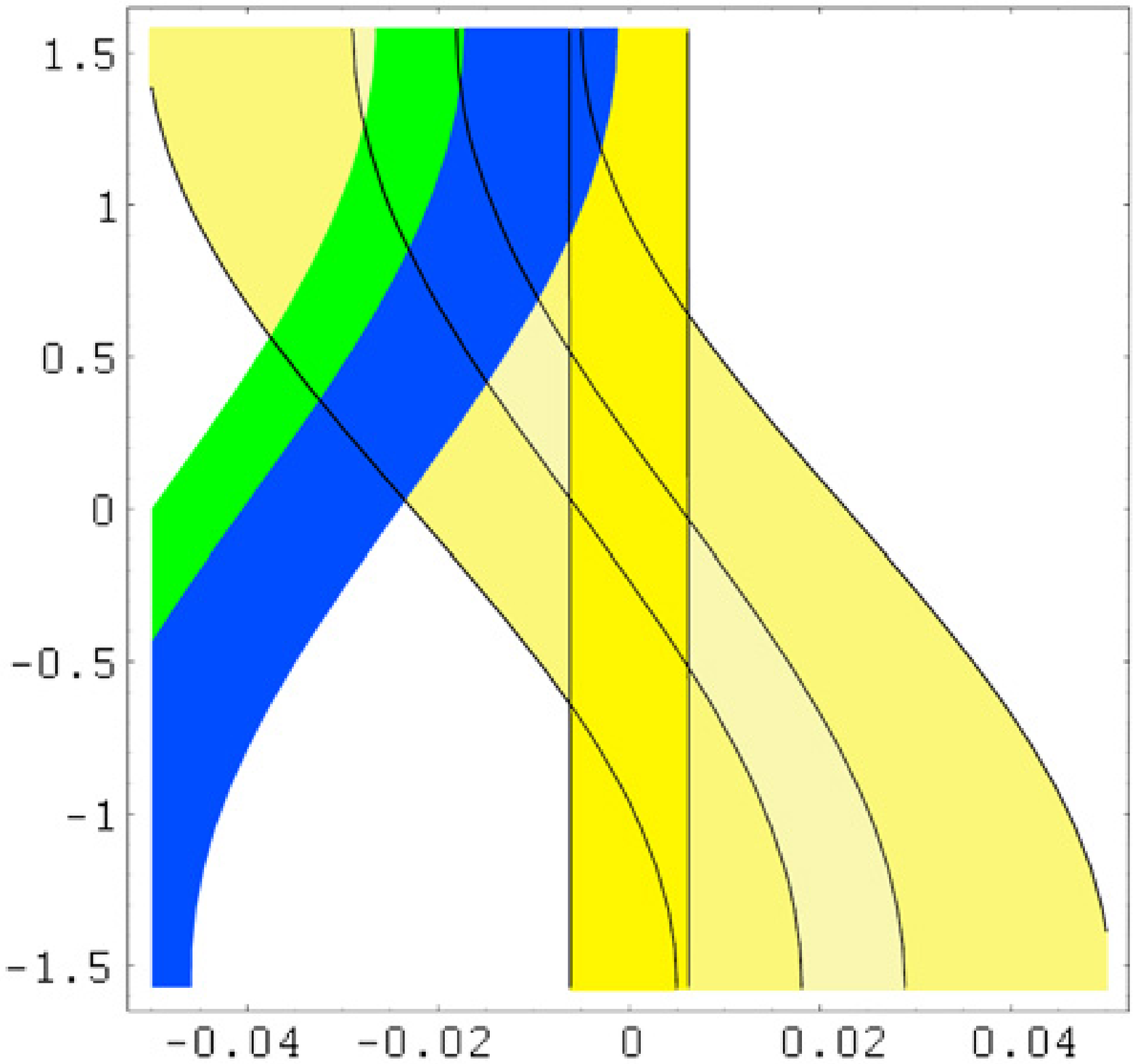}
\end{center}
\begin{picture}(0,0)(-100,0)
\put(7,15){{\small $\phi_\mu$ (rad)}}
\put(260,15){{\small $\phi_\mu$ (rad)  }}
\put(-100,130){{\small $\phi_A$ }}
\put(-105,115){{\small (rad)}}
\put(214,50){{\small $d_e$ }}
\put(274,50){\small$d_{\text{Hg}}$}
\put(305,57){{\small $d_n$ }}
\put(162,70){{\scriptsize EWB}}
\put(-25,60){{\scriptsize EWB }}
\put(27,53){{\small $d_e$ }}
\put(84,67){{\small $d_n$ }}
\put(51,52){\small $d_{\text{Hg}}$}
\end{picture}
\caption[Choices for MSSM parameters ruled out by EDM limits and successful EWB]{Choices for MSSM parameters ruled out by EDM constraints and requiring successful electroweak baryogenesis. Left panel: EDM and EWB bands in $\phi_\mu$--$\phi_A$ plane on the resonant peak $M_2 = \abs{\mu} = 200\text{ GeV}$, but with first- and second-generation sfermion and gluino masses at 500 GeV. The EDM limits are tighter in this case, ruling out successful baryogenesis. Right panel: For heavy sparticles at 750 GeV, but with $M_2=200\text{ GeV},\abs{\mu}=250\text{ GeV}$, which lies off the resonant peak for the creation of the baryon asymmetry. The required phases are already ruled out by present EDM limits.}
\label{fig:out}
\end{figure}

Ultimately, if supersymmetry is discovered at collider experiments,
spectroscopy will dictate the input for mass parameters. Then joint
constraints from low-energy EDM measurements and collider searches
could be used to tightly test the scenario of baryogenesis at the
electroweak scale, especially as a new generation of lepton, neutron, and neutral atom EDM searches will likely tighten the constraints in Fig.~7 by two or more orders of magnitude (for a recent discussion, see Ref.~\cite{Erler:2004cx} and references therein).

\section{Conclusions}
\label{sec:summary}

It is instructive to consider the essential physics leading to the enhanced sources and relaxation terms discussed in this work. The propagation of quasiparticles in the plasma is modified by scattering from the spacetime  varying Higgs vevs that causes transitions to intermediate states involving other quasiparticle species. The system retains some memory of each scattering due to the presence of thermal widths, $\Gamma_i$, that reflect the degeneracy of states in the thermal bath. For $\Gamma_i=0$, the oscillating exponentials appearing in the Green's functions wash out any memory of the scattering. For $\Gamma_i\not\!= 0$, the Green's functions now contain decaying exponentials as well as oscillating terms, and the memory washout is incomplete. The impact of quantum memory effects are, thus, characterized by the  ratio of time scales, $\tau_{\rm int}/\tau_p\sim \Gamma_i/\omega_i$, where $\tau_{\rm int}$ is the characteristic propagation time associated with a quasiparticle of frequency $\omega_i$ and $\tau_p\sim 1/\Gamma_i$, the plasma time, is the time scale on which transitions between the quasiparticle and other, degenerate states may occur. To the extent that the quasiparticle thermal mass and/or three-momentum is large compared to $\Gamma_i$, this ratio $\tau_{\rm int}/\tau_p$ is ${\cal O}(\epsilon)$.

A special situation arises, however, when the spacetime variation of the Higgs vevs is gentle and the thermal mass of an intermediate state is close to that of the initial state. Under these conditions, the scattering event injects essentially zero four-momentum into the initial state $i$, leading to resonant production of the intermediate state $j$. The characteristic lifetime of the latter is no longer $\tau_{\rm int}\sim 1/\omega_i$, but rather the resonance time scale 
\be
\tau_{\rm res}\sim \frac{1}{\sqrt{\Delta\omega^2+\Gamma_{ij}^2}}\ \ \ ,
\ee
where $\Delta\omega=\omega_i-\omega_j$ and $\Gamma_{ij}=\Gamma_i+\Gamma_j$ [see, {\em e.g.}, Eqs. (\ref{appx:scalar5}-\ref{appx:gammascalarpm}) and (\ref{appx:gammaHiggsinopm}-\ref{appx:chargino3}) of Appendix \ref{appx:expand}]. In this case, the impact of quantum memory is characterized by the ratio $\tau_{\rm res}/\tau_p$. For $|\Delta\omega| \ll \Gamma_{ij}$, this ratio becomes of ${\cal O}(1)$, and the impact of quantum memory is resonantly enhanced\footnote{An examination of Eqs. (\ref{appx:scalar5}-\ref{appx:gammascalarpm}) and (\ref{appx:gammaHiggsinopm}-\ref{appx:chargino3}) of Appendix \ref{appx:expand} indicates the presence of an additional, dynamical enhancement factor $\sim\omega/\sqrt{\Delta\omega^2+\Gamma_{ij}^2}$ in the relevant integrals.}. On the other hand, for $|\Delta\omega|\gg\Gamma_{ij}$, the ratio is ${\cal O}(\epsilon)$ and one returns to the more generic conditions. 

In this study, we have shown how this effect can enhance both the particle number-changing relaxation terms as well as the $CP$-violating sources that enter the transport equations relevant to electroweak baryogenesis. Importantly, the effect of resonant relaxation tends to mitigate the impact of resonantly-enhanced sources, as both enhancements occur under the same conditions for the electroweak model parameters (in this case, those of the MSSM). We suspect that analogous resonant effects occur in other transport coefficients, such as the $\Gamma_Y$ Yukawa terms discussed above, but that the conditions on model parameters leading to enhancements---owing to simple kinematic considerations---will be different. It may be, for example, that the Yukawa interactions are no longer fast compared to the Higgs vev induced transitions when the latter are resonantly enhanced, and in this case, the solution to the differential equations will differ from the general structure obtained here and by other authors. This possibility is one that should be explored in future work.

Additional refinements of the present analysis are clearly in order, including some form of all-orders resummation of the Higgs vev insertions (possibly along the lines proposed in Refs. \cite{Carena:2000id,Prokopec}) 
and a treatment of the axial charge transport equations via Eq. (\ref{eq:fermion1b}). In principle, one would also like to study the density dependence of the thermal frequencies and widths, the impact of nonzero gaugino densities, variations in bubble wall geometry, and possibly higher-order effects in $\epsilon$, such as the departure of $\delta f$ of the thermal distribution functions from their equilibrium values. In short, it is apparent that EWB is not yet a solved problem, but rather one that calls for additional study.

Undertaking this effort will be important for electroweak phenomenology. As illustrated here as well as in other studies (\emph{e.g.}, \cite{Balazs:2004ae}), determining the viability of EWB within a given electroweak model involves a detailed interplay of collider phenomenology, precision electroweak data, EDM searches, and a careful treatment of the dynamics of the electroweak phase transition. In particular, in light of the open questions pertaining to the latter, it is too soon to draw definitive conclusions about the implications of the next generation of EDM searches for the baryon asymmetry. One hopes, however, that by the time these searches obtain their first results, the context for their theoretical interpretation will have been further clarified.

\setcounter{section}{0}
\renewcommand{\thesection}{\arabic{chapter}.\Alph{section}}

\section{Appendix: Propagators at Finite Temperature and Density}
\label{appx:props}

In this section, we derive some useful properties of propagators at finite temperature and density, using derivations based on those for the case of finite temperature and zero density in Refs.~\cite{Weldon:1989ys,Weldon:1999th}.

\subsection{General Structure of Fermion Propagators}

We begin with the spectral function for fermions at temperature $T = 1/\beta$ in the presence of a chemical potential $\mu$:
\begin{equation}
\label{rhodef}
\rho_{\alpha\beta}(x) = \frac{1}{Z}\Tr\left[e^{-\beta(\op{H}-\mu\op{N})}\{\psi_\alpha(x),\bar\psi_\beta(0)\}\right],
\end{equation}
where $Z = \Tr[e^{-\beta(\op{H}-\mu\op{N})}]$. It is convenient to define the retarded and advanced propagators:
\begin{subequations}
\label{SRSAdef}
\begin{align}
S^R(x) &= \theta(x^0)\rho(x) \\
S^A(x) &= -\theta(x^0)\rho(x),
\end{align}
\end{subequations}
supressing spinor indices. The Fourier transforms of $S^{R,A}(x)$ and $\rho(x)$ are related by:
\begin{subequations}
\begin{align}
S^R(k^0,\vect{k}) &= i\int_{-\infty}^\infty\frac{d\omega}{2\pi}\frac{\rho(\omega,\vect{k})}{k^0 - \omega + i\epsilon} \\
S^A(k^0,\vect{k}) &= i\int_{-\infty}^\infty\frac{d\omega}{2\pi}\frac{\rho(\omega,\vect{k})}{k^0 - \omega - i\epsilon} \,.
\end{align}
\end{subequations}
It is possible to express the momentum-space spectral function in terms of a single product of $\psi_\alpha(x)$ and $\bar\psi_\beta(x)$ instead of the anticommutator in Eq.~(\ref{rhodef}), whose Fourier transform is:
\begin{equation}
\begin{split}
\rho_{\alpha\beta}(\omega,\vect{k}) &= \int d^4 x\,e^{i(\omega t - \vect{k}\cdot\vect{x})}\rho_{\alpha\beta}(t,\vect{x}) \\
&= \int d^4 x\,e^{i(\omega t - \vect{k}\cdot\vect{x})}\frac{1}{Z}\sum_n\bra{n}e^{-\beta(\op{H}-\mu\op{N})}\bigl[\psi_\alpha(x)\bar\psi_\beta(0) + \bar\psi_\beta(0)\psi_\alpha(x)\bigr]\ket{n}.
\end{split}
\end{equation}
Now insert a complete set of states between the fermion fields:
\begin{equation}
\label{complete}
\begin{split}
\rho_{\alpha\beta}(\omega,\vect{k}) = \int d^4 x\,e^{i(\omega t - \vect{k}\cdot\vect{x})}\frac{1}{Z}\sum_{n,j}\Bigl[&\bra{n}e^{-\beta(\op{H}-\mu\op{N})}\psi_\alpha(x)\ket{j}\bra{j}\bar\psi_\beta(0)\ket{n} \\
+ &\bra{n}e^{-\beta(\op{H}-\mu\op{N})}\bar\psi_\beta(0)\ket{j}\bra{j}\psi_\alpha(x)\ket{n}\Bigr].
\end{split}
\end{equation}
We can rewrite the second term by switching summation labels and translating $\psi_\alpha$ from $x$ to 0:
\begin{equation}
\begin{split}
\sum_{n,j}&\bra{n}e^{-\beta(\op{H}-\mu\op{N})}\bar\psi_\beta(0)\ket{j}\bra{j}\psi_\alpha(x)\ket{n} \\
&=\sum_{j,n}e^{i(E_n-E_j)t} e^{-i(\vect{k}_n-\vect{k}_j)\cdot\vect x}e^{-\beta E_j}e^{\beta\mu(N_n+1)}\bra{n}\psi_\alpha(0)\ket{j}\bra{j}\bar\psi_\beta(0)\ket{n},
\end{split}
\end{equation}
which after integrating in Eq.~(\ref{complete}), becomes
\begin{equation}
\frac{1}{Z}\sum_{j,n}(2\pi)^4\delta(\omega+E_n-E_j)\delta^3(\vect{k}+\vect{k}_n-\vect{k}_j)e^{-\beta(E_n+\omega)}e^{\beta\mu(N_n+1)}\bra{n}\psi_\alpha(0)\ket{j}\bra{j}\bar\psi_\beta(0)\ket{n},
\end{equation}
where we used the first delta function to replace $E_j$ with $E_n+\omega$ in the exponential $e^{-\beta E_j}$. This can now be written:
\begin{equation}
e^{-\beta(\omega-\mu)}\frac{1}{Z}\int d^4 x\,e^{i(\omega t - \vect{k}\cdot\vect{x})}\sum_{n,j}\bra{n}e^{-\beta(\op{H}-\mu\op{N})}\psi_{\alpha}(x)\ket{j}\bra{j}\bar\psi_\beta(0)\ket{n}
\end{equation}
which is $e^{-\beta(\omega-\mu)}$ times the first term of Eq.~(\ref{complete}), so we conclude:
\begin{equation}
\label{rhomom1}
\rho_{\alpha\beta}(\omega,\vect{k}) = \left[1 + e^{-\beta(\omega-\mu)}\right]\int d^4 x\,e^{i(\omega t-\vect{k}\cdot\vect{x})}\frac{1}{Z}\Tr\bigl[e^{-\beta(\op{H}-\mu\op{N})}\psi_\alpha(x)\bar\psi_\beta(0)\bigr].
\end{equation}
Similarly, we could have manipulated the first term of Eq.~(\ref{complete}) in the same way, and derived the companion relation:
\begin{equation}
\label{rhomom2}
\rho_{\alpha\beta}(\omega,\vect{k}) = \left[1 + e^{\beta(\omega-\mu)}\right]\int d^4 x\,
e^{i(\omega t-\vect{k}\cdot\vect{x})}\frac{1}{Z}
\Tr\bigl[e^{-\beta(\op{H}-\mu\op{N})}\bar\psi_\beta(0)\psi_\alpha(x)\bigr].
\end{equation}
Appearing on the right-hand sides of Eqs.~(\ref{rhomom1},\ref{rhomom2}) are the Green's functions $S^>(k^0,\vect{k})$ and $-S^<(k^0,\vect{k})$, giving the relations:
\begin{subequations}
\label{S>S<fromrho}
\begin{align}
S^>(k^0,\vect{k}) &= [1 - n_F(k^0-\mu)]\rho(k^0,\vect{k}) \\
S^<(k^0,\vect{k}) &= -n_F(k^0-\mu)\rho(k^0,\vect{k}),
\end{align}
\end{subequations}
where $n_F(x) = 1/(1 + e^x)$.

The various Green's functions satisfy the identities:
\begin{subequations}
\begin{align}
S^t(x,y) &= S^R(x,y) + S^<(x,y) = S^A(x,y) + S^>(x,y) \\
S^{\bar t}(x,y) &= S^>(x,y) - S^R(x,y) = S^<(x,y) - S^A(x,y)\,,
\end{align}
\end{subequations}
which follow directly from the definitions in
Eqs.~(\ref{eq:Greens1},\ref{SRSAdef}). Thus, using
Eq.~(\ref{S>S<fromrho}), the time- and anti-time-ordered propagators
can be expressed in terms of the retarded and advanced propagators:
\begin{subequations}
\label{StfromSR}
\begin{align}
S^t(k^0,\vect{k}) &= [1 - n_F(k^0-\mu)]S^R(k^0,\vect{k}) + 
n_F(k^0-\mu)S^A(k^0,\vect{k}) \\
S^{\bar t}(k^0,\vect{k}) &= -n_F(k^0-\mu)S^R(k^0,\vect{k}) - 
[1-n_F(k^0-\mu)]S^A(k^0,\vect{k})\,.
\end{align}
\end{subequations}
Also note that $\rho = S^R-S^A = S^> - S^<$.

\subsection{Bosonic Propagators}

Similar results may be derived from scalar bosonic propagators, for which the analog to Eq.~(\ref{S>S<fromrho}) is:
\begin{subequations}
\begin{align}
G^>(k^0,\vect{k}) &= [1+n_B(k^0-\mu)]\rho(k^0,\vect{k}) \\
G^<(k^0,\vect{k}) &= n_B(k^0-\mu)\rho(k^0,\vect{k})\,,
\end{align}
\end{subequations}
where the momentum-space spectral function $\rho(k^0,\vect{k})$ for bosons is the Fourier transform of:
\begin{equation}
\rho(x) = \frac{1}{Z}\Tr\left\{e^{-\beta(\op{H}-\mu\op{N})}[\phi(x),\phi^*(0)]\right\}\,.
\end{equation}
The bosonic propagators also satisfy the identity $\rho = G^R - G^A = G^> - G^<$.

\subsection{Tree-Level Propagators}

At tree level, the propagators $S^{R,A}$ for fermions are given by:
\begin{equation}
S^{R,A}(k^0,\vect{k}) = \frac{i(\diracslash{k} + m)}{(k^0\pm i\epsilon)^2 - E_{\vect{k}}^2}\,,
\end{equation}
and $G^{R,A}$ for bosons are given by:
\begin{equation}
G^{R,A}(k^0,\vect{k}) = \frac{i}{(k^0\pm i\epsilon)^2 - E_{\vect{k}}^2}\,,
\end{equation}
where $E_{\vect{k}}^2 = \abs{\vect{k}}^2 + m^2$. Note that these propagators are independent of the temperature and chemical potential, which only enter in the thermal distribution functions appearing in the relations of the retarded and advanced propagators to the other Green's functions, for example, in Eq.~(\ref{StfromSR}).

\subsection{One-Loop Corrections to Massless Fermion Propagators}

Resumming the one-loop self-energy into the fermion propagator at finite temperature changes the pole structure of the propagator dramatically, introducing a new collective ``hole'' excitation of the plasma \cite{Klimov,Weldon:1989ys}. In fact, this structure can be shown to hold even beyond perturbation theory \cite{Weldon:1999th}. Extending the results of Ref.~\cite{Weldon:1999th} to include dependence on a chemical potential, the propagator takes the form given in Eqs.~(\ref{eq:slambdaint}--\ref{eq:rhominus}). Recall that in those equations $\mathcal{E}_{p,h} = \omega_{p,h} - i\Gamma_{p,h}$ are the complex poles of the spectral function, and $Z_{p,h}$ are the corresponding residues. At leading order in the ``hard thermal loop'' approximation (see Ref.~\cite{LeBellac}), calculating the poles only to order $\mathcal{E}\sim gT$, one finds $\Gamma = 0$, and $Z_{p,h}(k,\mu)$ and $\omega_{p,h}(k,\mu)$, where $k = \abs{\vect{k}}$, depend only quadratically on $\mu/T$, which we thus neglect in our analysis in the present work, where we keep only effects linear in $\mu/T$. In this limit, and including only a single gluon loop in the quark self-energy diagram, the poles of the spectral function are given by the solutions to the equation:
\begin{equation}
0 = k^0 - k - \frac{\alpha_s C_F\pi T^2}{4k}\left[\left(1-\frac{k^0}{k}\right)\log\abs{\frac{k^0+k}{k^0-k}} + 2\right]\,,
\end{equation}
where $C_F = 4/3$ is the Casimir of the fundamental representation of $SU(3)$. The solutions to this equation give the poles $k^0 = E_p(k), -E_h(k)$. The residues satisfy:
\begin{equation}
Z_{p,h}(k) = \frac{E_{p,h}^2 - k^2}{m_f^2}\,,
\end{equation}
where
\begin{equation}
m_f^2 = \frac{\alpha_s C_F\pi T^2}{2}\,.
\end{equation}
Calculation of the imaginary parts $\Gamma_{p,h}$ of the poles, since they begin at order $g^2 T$, requires a resummation of hard thermal loops in self-energy diagrams \cite{Braaten:1989mz,Braaten:1991gm,Braaten:1992gd}. We are also interested in their dependence on the chemical potential $\mu$. We leave the calculation of these effects to a future study.

\section{Appendix: Expanded Source Terms for Quantum Transport}
\label{appx:expand}

\subsection{Bosons}

The $CP$-conserving source term for right-handed stops in Eq.~(\ref{eq:scalar5}) can be expanded by explicitly taking the imaginary part of the integrand:
\begin{align}
\label{appx:scalar5}
S^{CP}_{{\tilde t}_R}(x) = -\frac{1}{T}&\frac{N_C y_t^2}{2\pi^2}\abs{A_t v_u(x) - \mu^* v_d(x)}^2 
\int_0^\infty\frac{k^2 dk}{\omega_R\omega_L} \\
\nonumber
\times\biggl\{&\mu_R\left[{1\over\Delta}\left(\sin\phi\Imag h_R^++\cos\phi\
 {\rm Re}\ h_R^+\right)
-{1\over\delta}\left(\cos\theta\Real h_R^+-\sin\theta\Imag h_R^+\right)\right]\\
\nonumber
+&\mu_L\left[{1\over\Delta}\left(\sin\phi\Imag h_L^+-\cos\phi\Real h_L^+\right)
+{1\over\delta}\left(\cos\theta\Real h_L^+ - \sin\theta\Imag h_L^+\right)\right]\biggr\}\,,
\end{align}
where
\bea
\label{appx:scalardefs}
\omega_{L,R} & = & \sqrt{\abs{\vect{k}}^2+M_{{\tilde t}_{L,R}}^2}\\
\nonumber
\Delta & = & \sqrt{(\Gamma_L + \Gamma_R)^2+(\omega_L - \omega_R)^2}\\
\nonumber
\delta & = & \sqrt{(\Gamma_L + \Gamma_R)^2+(\omega_L + \omega_R)^2}\\
\nonumber
\tan\theta & = & \frac{\omega_L + \omega_R}{\Gamma_L + \Gamma_R} \\
\nonumber
\tan\phi & = & \frac{\omega_L - \omega_R}{\Gamma_L + \Gamma_R} \\
\nonumber
h_{L,R}^\pm & = & \frac{\exp[(\omega_{L,R}\pm i\Gamma_{L,R})/T]}
{\{\exp[(\omega_{L,R}\pm i\Gamma_{L,R})/T]-1\}^{2}}
\eea
and where $\Gamma_{L,R}$ are the thermal widths for the ${\tilde
t}_{L,R}$. The rates $\Gamma_{\tilde t}^\pm$ defined in Eq.~(\ref{eq:gammascalarpm}) can then be expressed:
\begin{align}
\label{appx:gammascalarpm}
\Gamma_{\tilde t}^{\pm} = -\frac{1}{T}\frac{y_t^2}{4\pi^2}\abs{A_t v_u(x) - \mu^* v_d(x)}^2 & \\
\nonumber
\times\int_0^\infty\frac{k^2 dk}{\omega_R\omega_L}
\biggl\{\frac{1}{\Delta}&\left[\sin\phi\Imag(h_L^+\pm h_R^+)-
\cos\phi\Real(h_L^+\mp h_R^+)\right]\\
\nonumber
+{1\over\delta}&\left[\cos\theta\Real(h_L^+\mp h_R^+) -
\sin\theta \Imag(h_L^+\mp h_R^+)\right]\biggr\}\,.
\end{align}
Meanwhile, the $CP$-violating source given in Eq.~(\ref{eq:scalarcp1}) can be expanded:
\begin{align}
\label{appx:scalarcp1}
S^{\CPV}_{{\tilde t}_R}(x) = N_C y_t^2&\Imag
(\mu A_t) v(x)^2{\dot\beta}(x) \int_0^\infty{k^2 dk\over 2\pi^2}{1
\over\omega_L\omega_R} \\
\nonumber
\times\biggl\{ &{1\over\delta^2}\left[\Real \left(1+n_R^++n_L^+\right)
\sin 2\theta+  
\Imag\left(n_R^++n_L^+\right)\cos 2\theta\right]\\
\nonumber
+&{1\over\Delta^2}\left[\Real\left(n_R^+-n_L^+\right)
\sin 2\phi - 
\Imag\left(n_R^++n_L^+\right)\cos 2\phi\right]\biggr\}\,,
\end{align}
where $n_{L,R}^\pm = n_B(\omega_{\tilde t_{L,R}} \pm \Gamma_{L,R})$.
Our result agrees with that of Ref.~\cite{Riotto:1998zb} except for a
different relative sign in front of the $\cos 2\phi$ term and the overall factor of $N_C$.

\subsection{Massive Fermions}

The $CP$-conserving rates for Higgsino-gaugino interactions given in Eq.~(\ref{eq:gammaHiggsinopm}) can be expanded:
\begin{align}
\label{appx:gammaHiggsinopm}
\Gamma_{\widetilde H^\pm}^{\pm} =  g_2^2 {v(x)^2}&\frac{1}{T} 
\int_0^\infty{k^2 dk\over 2\pi^2}\left({1\over\omh\omw}\right) \\
\nonumber
\times\Biggl(
&{1\over\Delta}\biggl\{
\left[\omh\omw+\gamh\gamw- k^2 + M_2\abs{\mu}\cos\phi_\mu\sin 2\beta(x)\right] \\
\nonumber
&\qquad\qquad\times\Bigl[\cos\phi\Real(\kwtilp \mp \khtilp) - \sin\phi\Imag(\kwtilp \pm \khtilp)\Bigr]\\
\nonumber
&\qquad+\left[\gamh\omw-\gamw\omh\right]\left[\sin\phi\Real(\kwtilp\mp\khtilp)
+\cos\phi\Imag(\kwtilp\pm\khtilp)\right]\biggr\}\\
\nonumber
+&
{1\over\delta}\biggl\{
\left[\omh\omw-\gamh\gamw+ k^2- M_2\abs{\mu}\cos\phi_\mu\sin 2\beta(x)\right] \\
\nonumber
&\qquad\qquad\times\Bigl[
\cos\theta\Real(\kwtilp\mp\khtilp) -\sin\theta\Imag(\kwtilp\mp\khtilp)\Bigr]\\
\nonumber
&\qquad-\left[\gamh\omw+\gamw\omh\right]\left[
\cos\theta\Imag(\kwtilp\mp\khtilp)+\sin\theta\Real(\kwtilp\mp\khtilp)\right]\biggr\}\Biggr),
\end{align}
where
\bea
\label{appx:fermiondefs}
\omega_{\widetilde H,\widetilde W} & = & \sqrt{\abs{\vect{k}}^2+M_{\widetilde H,\widetilde W}^2}\\
\nonumber
\Delta & = & \sqrt{(\Gamma_{\widetilde W} + \Gamma_{\widetilde H})^2+(\omega_{\widetilde W} - \omega_{\widetilde H})^2}\\
\nonumber
\delta & = & \sqrt{(\Gamma_{\widetilde W} + \Gamma_{\widetilde H})^2+(\omega_{\widetilde W} + \omega_{\widetilde H})^2}\\
\nonumber
\tan\theta & = & \frac{\omega_{\widetilde W} + \omega_{\widetilde H}}{\Gamma_{\widetilde W} + \Gamma_{\widetilde H}} \\
\nonumber
\tan\phi & = & \frac{\omega_{\widetilde W} - \omega_{\widetilde H}}{\Gamma_{\widetilde W} + \Gamma_{\widetilde H}} \\
\nonumber
h_{{\widetilde W},{\widetilde H}}^\pm &=& \frac{\exp[(\omega_{{\widetilde W},{\widetilde H}}\pm i\Gamma_{{\widetilde W},{\widetilde H}})/T]}{\{\exp[(\omega_{{\widetilde W},{\widetilde H}}\pm i\Gamma_{{\widetilde W},{\widetilde H}})/T]+1\}^{2}}\ .
\end{eqnarray}
The $CP$-violating Higgsino source in Eq.~(\ref{eq:chargino3}) can be expressed:
\begin{align}
\label{appx:chargino3}
S^{\CPV}_{\widetilde H^\pm}(x)  =  
\nonumber
2 g_2^2 M_2 &\Imag(\mu) v(x)^2{\dot\beta}\ \int_0^\infty\ 
{k^2 dk\over 2\pi^2}\left({1\over \omh\omw}\right)
 \\
\times\biggl\{&{1\over\Delta^2}\left[\sin 2\phi\ {\rm Re}\left(\nwtilp-\nhtilp\right)+\cos 2\phi\ 
{\rm Im}\left(\nwtilp+\nhtilp\right)\right]\\
\nonumber
+&{1\over\delta^2}\left[\sin 2\theta\ {\rm Re}\left(1-\nwtilp-\nhtilp\right) 
-\cos 2\theta\ {\rm Im}\left(\nwtilp+\nhtilp\right)\right]\biggr\}\,,
\end{align}
where $N_{\widetilde H,\widetilde W}^\pm = n_B(\omega_{\widetilde H,\widetilde W}\pm i\Gamma_{\widetilde H,\widetilde W})$.
Our result agrees with that of Ref. \cite{Riotto:1998zb} except for
the sign of the $\cos 2\phi$ term.

\subsection{Chiral Fermions}
For chiral fermions, the $CP$-conserving chirality-changing rates in Eq.~(\ref{eq:gquarkplusminus}) can be expanded:
\begin{align}
\label{appx:gquarkplus}
\Gamma_{t_R}^{\pm} = \frac{1}{T}\frac{N_C y_t v_u^2}{\pi^2}&\int_0^\infty k^2 dk \\
\times\biggl\{\frac{Z_p^R Z_p^L}{\delta_p}&\Bigr[\sin\theta_p\bigl\{\Real(\lambda_p^L \kplp \mp \lambda_p^R \kprp) - \Imag(\kplp\mp \kprp)\bigr\} + \cos\theta_p\Real(\kplp \mp \kprp) \nonumber \\
 &+ \frac{T}{\delta_p}\cos 2\theta_p(\lambda_p^L \mp \lambda_p^R)\Real(1-N_{pL}^+ - N_{pR}^+)\Bigr] \nonumber \\
-\frac{Z_p^L Z_h^R}{\Delta_{hp}}&\Bigl[\sin\phi_{hp}\bigl\{\Real(\lambda_p^L\kplp \pm \lambda_h^R\khrp) - \Imag(\kplp \pm \khrp)\bigr\} - \cos\phi_{hp}\Real(\kplp \mp \khrp) \nonumber \\
& + \frac{T}{\Delta_{hp}}\cos 2\phi_{hp}(\lambda_p^L\pm\lambda_h^R)\Real(N_{pL}^+ - N_{hR}^+)\Bigr] \nonumber \\
+ (p\leftrightarrow h) &\biggr\}, \nonumber
\end{align}
where
\begin{equation}
\begin{split}
\delta_p &= \sqrt{(\omega_p^R + \omega_p^L)^2 + (\Gamma_p^R + \Gamma_p^L)^2} \\
\Delta_{hp} &= \sqrt{(\omega_p^L - \omega_h^R)^2 + (\Gamma_h^R + \Gamma_p^L)^2} \\
h_{pL}^\pm &= h_F(\omega_p^L \pm i\Gamma_p^L)\text{, etc.} \\
N_{pL}^\pm &= n_F(\omega_p^L \pm i\Gamma_p^L)\text{, etc.} \\
\tan\theta_p &= \frac{\omega_p^L + \omega_p^R}{\Gamma_p^L + \Gamma_p^R} \\
\tan\phi_{hp} &= \frac{\omega_h^R - \omega_p^L}{\Gamma_p^L + \Gamma_h^R},
\end{split}
\end{equation}
and where the
\begin{equation}
\lambda_{p,h}^{L,R} = \pd{\Gamma_{p,h}^{L,R}}{\mu_{t_{L,R}}}\,,
\end{equation}
parameterize the linear shifts in the thermal widths due to non-vanishing chemical potential. As noted at the end of Appendix~\ref{appx:props}, in a fully resummed calculation of the fermion self-energy, such shifts which are linear in $\mu_i/T$ may arise, and thus have to be included in our calculations, which we defer to future work. Also note that we have approximated the residues $Z_{p,h}^{L,R}$ to be purely real, which is true at the order we are working.

\section{Appendix: Residues of Thermal Distribution Functions}
\label{appx:poles}

The expressions for the $CP$-violating and conserving sources presented in Sec.~\ref{sec:source} omit the terms arising from the residues of the poles of the thermal distribution functions appearing in the thermal Green's functions. 

For example, the $CP$-violating source for squarks, Eq.~(\ref{eq:scalarcp1}), arises from the second term in Eq.~(\ref{eq:scalar2}), which at an intermediate stage in the derivation takes the form:
\begin{equation}
\label{eq:SCPVa}
\begin{split}
S_{\tilde t_R}^{\CPV}(X) &= -2iy_t^2\Imag(\mu A_t)v^2(X)\dot\beta(X)\int_{-\infty}^0dt\,t\int\frac{d^3 k}{(2\pi)^3}\int_{-\infty}^\infty\frac{dk^0}{2\pi}\int_{-\infty}^\infty\frac{dq^0}{2\pi}e^{i(k^0-q^0)t} \\
&\qquad\qquad\times[n_B(k^0) - n_B(q^0)]\rho_R(k^0,\vect{k})\rho_L(q^0,\vect{q}).
\end{split}
\end{equation}
The $k^0,q^0$ integrals can be done by contour integration. The exponential factor $e^{i(k^0-q^0)t}$ determines that the $k^0$ contour should be closed in the lower half-place, while the $q^0$ contour should be closed above. The terms in Eq.~(\ref{eq:scalarcp1}) come from picking up the residues of the poles in the spectral functions $\rho_R(k^0,\vect{k})$ and $\rho_L(q^0,\vect{q})$. However, $n_B(k^0)$ and $n_B(q^0)$ also contain poles. The function $n_B(x)$,
\begin{equation}
n_B(x) = \frac{1}{e^{x/T} - 1},
\end{equation}
has poles at the points $x_n = 2\pi i n T$, where $n$ is any integer. These are illustrated in Fig.~\ref{fig:contour}. 
\setlength{\unitlength}{1mm}
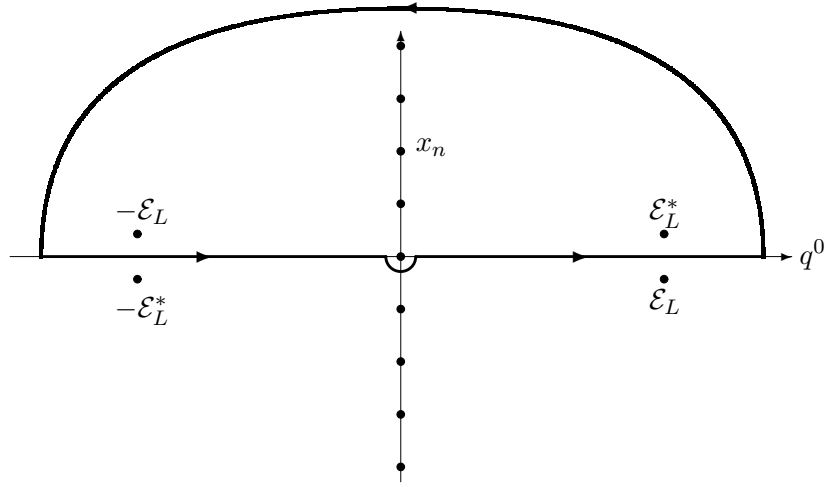
\begin{figure}
\begin{center}
\begin{picture}(0,56)(0,-22)
\drawline(-52,0)(52,0)
\put(52,0){\vector(1,0){0}}
\put(53,-1){$q^0$}
\drawline(0,-30)(0,30)
\put(0,30){\vector(0,1){0}}
\put(35,3){\circle*{1}}
\put(-35,3){\circle*{1}}
\put(-35,-3){\circle*{1}}
\put(35,-3){\circle*{1}}
\put(-38,5){$-\mathcal{E}_L$}
\put(-38,-8){$-\mathcal{E}_L^*$}
\put(33,5){$\mathcal{E}_L^*$}
\put(33,-7){$\mathcal{E}_L$}
\multiput(0,-28)(0,7){9}{\circle*{1}}
\put(2,14){$x_n$}
\Thicklines
\drawline(-48,0)(-2,0)
\drawline(2,0)(48,0)
\put(0,0){\arc{4}{0}{3.14}}
\qbezier(-48,0)(-48,33)(0,33)
\qbezier(0,33)(48,33)(48,0)
\put(-25,0){\vector(1,0){0}}
\put(25,0){\vector(1,0){0}}
\put(0,33){\vector(-1,0){0}}
\end{picture}
\end{center}
\caption[Analytic structure of integrand in source terms]{The $q^0$ integral in Eq.~(\ref{eq:SCPVa}) is evaluated along the contour closed in the upper half-plane. The integrand contains poles at the locations $\pm\mathcal{E}_L^{(*)}$ from the spectral function and at $x_n = 2\pi i n T$ from the Bose-Einstein function $n_B(q^0)$. The other integrands appearing in the various source terms all have this basic analytic structure.}
\label{fig:contour}
\end{figure}
Near one of these points, the Bose-Einstein function behaves as:
\begin{equation}
n_B(x) = \frac{T}{x-x_n} + \cdots,
\end{equation}
where the ellipses denote non-singular terms in the series expansion of $n_B(x)$. The residues of these poles generate new contributions to $S_{\tilde t_R}^{\CPV}$. First, note that the residues at $k^0,q^0=0$ contribute nothing, as $\rho_R(0,\vect{k}) = \rho_L(0,\vect{q}) = 0$, as seen by inspection of Eq.~(\ref{eq:slambdafree}). The remaining residues give, after completing the time and $d\Omega_{\vect{k}}$ integrals in Eq.~(\ref{eq:SCPVa}), the extra contributions:
\begin{equation}
\begin{split}
\Delta S_{\tilde t_R}^{\CPV}(X) = -\frac{y_t^2}{\pi^2}\Imag(\mu A_t)v^2(X)\dot\beta(X) T\int_0^\infty & dk\,k^2 \\
\times\sum_{n=1}^\infty\biggl\{\frac{1}{2\omega_L}\biggl[\frac{1}{(\mathcal{E}_L^* + 2\pi i nT)^2} &- \frac{1}{(\mathcal{E}_L - 2\pi i nT)^2}\biggr]\rho_R(-2\pi i n T,k) \\
+ \frac{1}{2\omega_R}\biggl[\frac{1}{(\mathcal{E}_R - 2\pi i nT)^2} &- \frac{1}{(\mathcal{E}_R^* + 2\pi i nT)^2}\biggr]\rho_L(2\pi i n T,k)\biggr\}.
\end{split}
\end{equation}
In the region of resonant enhancement of the original terms in $S_{\tilde t_R}^{\CPV}$, where $M_{\tilde t_L}\approx M_{\tilde t_R}$, these new terms are numerically about $10^3$ times smaller (primarily because of the large factors of $T$ in the denominators), while they become comparable only far away from the resonant region. Since these regions are phenomenologically unimporant in our analysis, we can safely ignore $\Delta S_{\tilde t_R}^{\CPV}$ for our purposes, although they should be included in future calculations for complete consistency.

Corrections to the $CP$-violating Higgsino source $S_{\tilde H^\pm}^{\CPV}$ in Eq.~(\ref{eq:chargino3}) come similarly from poles in the Fermi-Dirac function $n_F(x)$ at the points $\kappa_n = (2n-1)i\pi T$ for integers $n$:
\begin{equation}
n_F(x) = \frac{1}{e^{x/T} + 1} = -\frac{T}{x - \kappa_n} + \cdots
\end{equation}
The extra contribution to the source generated by the residues of these poles:
\begin{equation}
\begin{split}
\Delta S_{\tilde H^\pm}^{\CPV}(X) = \frac{2g_2^2}{\pi^2}\Imag(\mu)v(X)^2\dot\beta(X) M_2 T\int_0^\infty &dk\,k^2 \\
\times\sum_{n=1}^\infty\biggl\{\frac{1}{2\omega_{\widetilde W}}\biggl[\frac{1}{(\mathcal{E}_{\widetilde W} - \kappa_n)^2} &- \frac{1}{(\mathcal{E}_{\widetilde W}^* + \kappa_n)^2}\biggr]\rho_{\widetilde H^\pm}(\kappa_n,k) \\
+ \frac{1}{2\omega_{\widetilde H}}\biggl[\frac{1}{(\mathcal{E}_{\widetilde W}^* + \kappa_n)^2} &- \frac{1}{(\mathcal{E}_{\widetilde W} - \kappa_n)^2}\biggr]\rho_{\widetilde W}(-\kappa_n,k)\biggr\}.
\end{split}
\end{equation}
Again, the size of $\Delta S_{\tilde H^\pm}^{\CPV}$ is numerically insignificant compared to $S_{\tilde H^\pm}^{\CPV}$ given in Eq.~(\ref{eq:chargino3}), except far away from the region $\abs{\mu}=M_2$, thus leaving our phenomenological studies substantially unaffected.

The $CP$-conserving sources $S^{CP}$ are also modified. For instance, to $S_{\tilde t_R}^{CP}$ in Eq.~(\ref{eq:scalar5}) must be added the terms:
\begin{equation}
\begin{split}
\Delta S_{\tilde t_R}^{CP}(X) = -\frac{y_t^2}{\pi^2}\abs{A_t v_u(X) - \mu^* v_d(X)}^2 T &\int_0^\infty dk\,k^2 \\
\times\Biggl(\mu_R\frac{1}{2\omega_L}\Biggl\{\frac{1}{2}\rho'_R(0,k)\biggl(\frac{1}{\mathcal{E}_L} &+ \frac{1}{\mathcal{E}_L^*}\biggr) \\
+ \sum_{n=1}^\infty \biggl[\rho'_R(-2\pi i n&T,k)\biggl(\frac{1}{\mathcal{E}_L - 2\pi i n T} + \frac{1}{\mathcal{E}_L^* + 2\pi i n T}\biggr) \\
- \rho_R(-2\pi i n&T,k)\biggl(\frac{1}{(\mathcal{E}_L - 2\pi i nT)^2} - \frac{1}{(\mathcal{E}_L^* + 2\pi i nT)^2}\biggr)\biggr]\Biggr\} \\
-\mu_L\frac{1}{2\omega_R}\Biggl\{\frac{1}{2}\rho'_L(0,k)\biggl(\frac{1}{\mathcal{E}_R} &+ \frac{1}{\mathcal{E}_R^*}\biggr) \\
+ \sum_{n=1}^\infty \biggl[\rho'_L(2\pi i n&T,k)\biggl(\frac{1}{\mathcal{E}_R - 2\pi i n T} + \frac{1}{\mathcal{E}_R^* + 2\pi i n T}\biggr) \\
+ \rho_L(2\pi i n&T,k)\biggl(\frac{1}{(\mathcal{E}_R - 2\pi i nT)^2} - \frac{1}{(\mathcal{E}_R^* + 2\pi i nT)^2}\biggr)\biggr]\Biggr\}\Biggr),
\end{split}
\end{equation}
where
\begin{equation}
\begin{split}
\rho_{R,L}'(k^0,k) &= \pd{}{k^0}\rho(k^0,k) \\
&= -\frac{i}{2\omega_k}\biggl[\frac{1}{(k^0 - \mathcal{E}_{R,L})^2} - \frac{1}{(k^0 + \mathcal{E}_{R,L}^*)^2}  - \frac{1}{(k^0 - \mathcal{E}_{R,L}^*)^2} + \frac{1}{(k^0 + \mathcal{E}_{R,L})^2}\biggr].
\end{split}
\end{equation}
These corrections are also numerically insignificant compared to Eq.~(\ref{eq:scalar5}), except far away from the region $M_{\tilde t_L}\approx M_{\tilde t_R}$. Similar expressions should hold for the $CP$-conserving sources for quarks and Higgsinos, which we also expect we can safely ignore, though we do not include their explicit expressions here.

\section{Appendix: Towards the Yukawa Source}

In this Appendix we take initial steps towards a computation of the source terms arising from the Yukawa interactions illustrated in Fig.~\ref{fig:graphs2}. These describe the interactions, for instance, of squarks with real Higgs bosons:
\begin{equation}
\mathcal{L}_{\text{int}}^y = y_t \tilde t_L(A_t H_u^0 - \mu^* H_d^{0*})\tilde t_R^* + \text{h.c.}
\end{equation}
These interactions contribute to the squark source given by Eq.~(\ref{eq:scalar1}):
\begin{equation}
\label{squarksourceY}
\begin{split}
S_{\tilde t_R}(X) = \int d^3 z\int_{-\infty}^{X^0}dz^0 \Bigl[G_R^<(X,z)\Sigma_R^>(z,X) &- G_R^>(X,z)\Sigma_R^<(z,X) \\
+ \Sigma_R^>(X,z)G_R^<(z,X) &- \Sigma_R^<(X,z)G_R^>(z,X)\Bigr],
\end{split}
\end{equation}
by inducing the self-energies:
\begin{subequations}
\label{selfenergyY}
\begin{align}
\Sigma_R^>(x,y) &= -y_t^2 G_L^>(x,y)\bigl[\abs{A_t}^2 G_{H_u^0}^>(x,y) + \abs{\mu}^2G_{H_d^{0*}}^>(x,y)\bigr] \\
\Sigma_R^<(x,y) &= -y_t^2 G_L^<(x,y)\bigl[\abs{A_t}^2 G_{H_u^0}^<(x,y) + \abs{\mu}^2G_{H_d^{0*}}^<(x,y)\bigr].
\end{align}
\end{subequations}
Note that these self-energies contain no $CP$-violating phases, unlike Eqs.~(\ref{eq:sigmascalar},\ref{eq:gxyscalar}), which contained cross-terms between the $v_u$ and $v_d$ vevs.  Thus, these Yukawa interactions involving real Higgs particles contribute only to the $CP$-conserving part of the squark source. The $H_u$ and $H_d$ contributions to the self-energy have essentially the same structure, so for clarity, we include only the former in the following calculations. The $H_d$ contributions can be restored straightforwardly.

 Every Green's function appearing in Eqs.~(\ref{squarksourceY}, \ref{selfenergyY}) contains dependence on the corresponding chemical potential $\mu_i$. Expanding each one to first order in $\mu_i/T$,
\begin{equation}
\label{deltaG}
G_i(X,z) = G_i^0(X,z) + \mu_i\delta G_i(X,z).
\end{equation}
Let us Wigner transform each of the  Green's functions to momentum space:
\begin{equation}
\label{WignerG}
G_i(X,z) = \int\frac{d^4 p}{(2\pi)^4}e^{-ip\cdot(X-z)}G_i(p;\mu_i(X)),
\end{equation}
where we have approximated the dependence of the chemical potentials $\mu_i(X+z)$ on the collective coordinate $X+z$ by simply $\mu(X)$, which assumes the variation in $\mu_i$ is slow compared to the typical scale of individual interactions between particles. With this assumption, after plugging Eqs.~(\ref{selfenergyY}) into Eq.~(\ref{squarksourceY}), using the Wigner transforms (\ref{WignerG}), and expanding to first-order in the chemical potential $\mu_i/T$ as in Eq.~(\ref{deltaG}), the Yukawa-type source for squarks can be rearranged into the following useful form:
\begin{equation}
\label{enigma}
\begin{split}
S_{\tilde t_R}^Y(X) = \int_{-\infty}^0 dt\int\frac{d^4p}{(2\pi)^4}\int\frac{dq^0}{2\pi}&\bigl[e^{i(p^0-q^0)t} + e^{-i(p^0-q^0)t}\bigr]h_B(p^0) \\
\times\biggl\{&-\frac{\mu_R}{T}\rho_R(p^0,\vect{p})\bigl[\Sigma_R^{>}(q^0,\vect{p}) - \Sigma_R^{<}(q^0,\vect{p})\bigr] \\
&+\frac{\mu_L}{T}\rho_L(p^0,\vect{p})\bigl[\Sigma_L^{>}(q^0,\vect{p}) - \Sigma_L^{<}(q^0,\vect{p})\bigr] \\
&+\frac{\mu_H}{T}\rho_H(p^0,\vect{p})\bigl[\Sigma_H^{>}(q^0,\vect{p}) - \Sigma_H^{<}(q^0,\vect{p})\bigr]\biggr\},
\end{split}
\end{equation}
where we have used
\begin{subequations}
\begin{align}
G_i^>(p) &= [1+n_B(p^0-\mu_i)]\rho_i(p) \\
G_i^<(p) &= n_B(p^0-\mu_i)\rho_i(p),
\end{align}
\end{subequations}
so that
\begin{equation}
\delta G_i^>(p) = \delta G_i^<(p) = -\frac{\mu_i}{T}h_B(p^0)\rho_i(p).
\end{equation}
The momentum-space self-energies are evaluated at zero chemical potentials and are given by:
\begin{subequations}
\begin{align}
\Sigma_R^>(q) &= -y_t^2\abs{A_t}^2\int\frac{d^4 k}{(2\pi)^4}G_L^>(k)G_H^>(q-k) \\
\Sigma_L^>(q) &= -y_t^2\abs{A_t}^2\int\frac{d^4 k}{(2\pi)^4}G_R^>(k)G_H^<(k-q) \\
\Sigma_H^>(q) &= -y_t^2\abs{A_t}^2\int\frac{d^4 k}{(2\pi)^4}G_R^>(k)G_L^<(k-q),
\end{align}
\end{subequations}
where, again, all Green's functions here are evaluated with zero chemical potentials. The self-energies $\Sigma_i^<$ are obtained by flipping all $\gtrless$ signs.

The evaluation of Eq.~(\ref{enigma}) is considerably complicated by the presence of finite widths $\Gamma_i$ in the Green's functions appearing inside the integrand. Na\"{\i}ve contour integration as for the $\Gamma_M$-type sources derived earlier produces a result which is ultraviolet-divergent. Investigations into the proper regulation of these terms or a correct procedure for integration (which is even further complicated by the poles in the thermal distribution functions as described in the previous Appendix) is still underway at the time of this writing. The quantitative analysis of $S_{\tilde t_R}^Y$ and the comparison of its size to the $\Gamma_M$-type sources derived earlier is essential to check the consistency of the approximations used in solving the transport equations that give the left-handed weak doublet fermion density $n_L$ and, thereby, the baryon density, $\rho_B$. If the coefficient $\Gamma_Y$ appearing in the transport equations is not considerably larger than $\Gamma_M^-$, then the assumption that $\Gamma_Y\gg\Gamma_M^-$ must be discarded, changing the solution of the transport equations entirely. This scenario is particularly likely to occur in the regions of MSSM parameter space where $\Gamma_M^-$ is enhanced. The evaluation of the Yukawa-type sources and their impact on the phenomenological analysis presented in Sec.~\ref{sec:numerics} is one of the most urgent tasks handed to us by the basic foundational analysis presented in this chapter.

\renewcommand{\thesection}{\arabic{chapter}.\arabic{section}}
\chapter{Effective Theories of Strong Interactions}
\label{chap:eft}

\newcommand{\glqq}{\raisebox{.2ex}{$\scriptscriptstyle\gg$}}
\newcommand{\grqq}{\raisebox{.2ex}{$\scriptscriptstyle\ll$}}

\begin{quote}
\small\singlespace\raggedright 
\emph{Theoria} bezeichnet die rein empfangende, von aller \glqq praktischen\grqq\ Bezweckung des t\"{a}tigen Lebens durchaus unabh\"{a}ngige Zuwendung zur Wirklichkeit. Man mag diese Zuwendung \glqq uninteressiert\grqq\ nennen---wenn hiermit nichts anderes ausgeschlossen sein soll als jegliches auf Dienlichkeiten und Belange gerichtete Absehen. Im \"{u}brigen ist hier auf h\"{o}chst entschiedene Weise Interesse, Beteiligung, Aufmerksamheit, Zielsetzung. \emph{Theoria} und \emph{contemplatio} zielen mit ihrer vollen Energie dahin---freilich: \emph{ausschlie\ss lich} dahin---, da\ss\ die ins Auge gefa\ss te Wirklichkeit offenbar und deutlich werde, da\ss\ sie sich zeige und enth\"{u}lle; sie zielen auf Wahrheit und nichts sonst. \\ 
\flushright\vspace{-12pt}\emph{Josef Pieper, \emph{Gl\"{u}ck und Kontemplation}} \\
\bigskip
\raggedright
\emph{Theoria} has to do with the purely receptive approach to reality, one altogether independent of all practical aims in active life. We may call this approach ``disinterested,'' in that it is altogether divorced from utilitarian ends. In all other respects, however, \emph{theoria} emphatically involves interest, participation, attention, purposiveness. \emph{Theoria} and \emph{contemplatio} devote their full energy to revealing, clarifying, and making manifest the reality which has been sighted; they aim at truth and nothing else. \\
\flushright\vspace{-12pt}\emph{Josef Pieper, \emph{Happiness \& Contemplation}}
\end{quote}
In this chapter we review the basic features of several effective theories for the strong interactions---heavy quark effective theory, soft-collinear effective theory, and non-relativistic QCD---following developments of these theories in recent years. This forms the background for the discussion of the applications of these effective theories pursued in the subsequent chapters.

\section{Quantum Chromodynamics}
\label{sec:QCD}

\begin{quote}
\small\singlespace\raggedright It is a lovely language, but it takes a very long time to say anything in it, because we do not say anything in it, unless it is worth taking a long time to say, and to listen to. \\
\flushright\vspace{-12pt}\emph{Treebeard, in \emph{The Two Towers}, by J.R.R. Tolkien}
\end{quote}
Quantum Chromodynamics (QCD) is the theory of the strong interactions between quarks and gluons, which bind together to make protons, neutrons, and other hadrons. The theory accounts successfully for many hadronic phenomena, especially those occurring at relatively large energies, for example, in the bombardment of protons by highly energetic electrons in deep inelastic scattering, revealing the pointlike substructure of the proton. QCD is governed by the Lagrangian density:
\begin{equation}
\label{QCDLag}
\mathcal{L}_{\text{QCD}} = \bar\psi_q(i\Dslash - m_q)\psi_q - \frac{1}{4}G^A_{\mu\nu}G^{A\mu\nu},
\end{equation}
where we sum over quark flavors $q$ and $SU(3)$ generators $A$. The covariant derivative is $D_\mu = \partial_\mu - igA_\mu^A T^A$.

QCD allows for reliable quantitative predictions at large energies due to the phenomenon of asymptotic freedom \cite{Politzer:1973,Gross:1973id}. The observed strength of the interaction between quarks and gluons is characterized by a coupling constant $g_s$, which is a function of energy, becoming small at large energies and large at small energies. Physical quantities can be calculated as a perturbation series in powers of $g_s$, or, rather, $\alpha_s = g_s^2/4\pi$, which is a reliable procedure as long as $\alpha_s\ll1$. 

At small energies, however, such as the scale at which quarks and gluons bind together into light hadrons (protons, pions, etc.), the coupling constant is large, $\alpha_s\gtrsim 1$, and perturbation theory breaks down completely. Even in processes involving strongly-interacting particles at large energies, because the particles used or observed directly in experiments are hadrons, not free quarks and gluons, these low-energy binding effects contaminate the analysis, preventing completely precise calculation of the observable quantities. However, it is often possible to separate the perturbatively-calculable large energy phenomena from the low-energy hadronization effects in a way that preserves much predictive power. Imagine calculating the total rate for $Z$ bosons to decay to hadrons, $\Gamma(Z\rightarrow\text{hadrons})$. In perturbation theory, we begin with the $Z$ coupling to quarks:
\begin{equation}
\mathcal{L}_{Z\bar q q} = \bar\psi_q\gamma_\mu(g_V + g_A\gamma_5)\psi_q Z^\mu.
\end{equation}
We can calculate the decay rate for $Z$ to $q\bar q$, $\Gamma(Z\rightarrow q\bar q)$. This process produces a $\bar q q$ pair moving back-to-back with energy $M_Z/2$. This is much larger than the scale of hadronization, $\Lambda_{\text{QCD}}$. The quark and antiquark move far apart before they hadronize. We do not know how to calculate the dynamics of this hadronization, but since we know it must happen with probability 1, we can approximate the total hadronic decay rate of the $Z$ by:
\begin{equation}
\Gamma(Z\rightarrow\text{hadrons}) = \Gamma(Z\rightarrow\bar q q),
\end{equation}
to leading order in $\alpha_s(M_Z)$. We can compute to next order in $\alpha_s(M_Z)$ by including the rate $\Gamma(Z\rightarrow\bar q q g)$. And so on. Because we have separated the parton-level physics from the longer-range, lower-energy physics of hadronization, we are able to make a reliable quantitative prediction that is a very good approximation to reality. We say we have ``factorized'' the total hadronic decay rate of the $Z$---into $\Gamma(Z\rightarrow\text{partons})$ and the probability for partons to hadronize, which is 1.

For more complicated observables, we seek to perform similar factorizations, although the low-energy, nonperturbative quantities which appear will not, in general, be so simple as ``1''. These may be objects such as parton distribution functions, meson light-cone wave functions, etc. Although such quantities are not calculable in perturbation theory, they may appear in more than one physical observable, thus providing some remnant of predictive power. They may also be calculable numerically in lattice QCD.

Proving factorization and identifying the relevant nonperturbative quantities is a major thrust in the direction of modern research in QCD. There are two main camps of research. There are those who attack the problem directly in QCD, the so-called perturbative QCD or QCD factorization approach \cite{StermanTasi}. Then there are those who quail at the enormity of full QCD and attempt to work instead with a simpler version of the theory. This is the approach of \emph{effective field theory} \cite{EFTnotes,Rothstein:2003mp}, which we adopt in this thesis.

\section{Example: Heavy Quark Effective Theory}

In an effective field theory one identifies a small parameter determined by the relevant physics in a given problem in which to expand the full theory to a given order of approximation. Equivalently one integrates out the degrees of freedom living at an energy scale much larger than those relevant in the physical problem at hand.\footnote{The presentation in this and the following section are heavily influenced by the lectures presented in Ref.~\cite{SCETlectures}.}

As an illustrative example and a prelude to the effective field theories we consider later in this thesis, we overview the \emph{heavy quark effective theory} (HQET) \cite{HQET}, an effective field theory describing hadrons containing one heavy quark, namely, charm ($c$) or bottom ($b$).\footnote{We limit our attention, not surprisingly, to heavy hadrons containing two ($Q\bar q$) or three ($Qqq$), not five \cite{Stewart:2004pd,Wessling:2004ag}, quarks and antiquarks.} In such a hadron, the heavy $c$ or $b$ quark moves with a relatively slow velocity compared to the other light quark(s) accompanying it. The typical momenta of the constituent partons is of order $p\sim\lqcd$. Thus, the velocity of the heavy quark $Q$ is of the order $v\sim\Lambda_{\text{QCD}}/m_Q$. We can expand the Lagrangian of full QCD in Eq.~(\ref{QCDLag}) for heavy quarks in powers of this small velocity $v$.

Physically, we imagine that the interaction of a heavy quark with low-energy gluons causes only small fluctuations of its momentum, $p = m_Q v + k$, with $k\sim\lqcd$. We want the effective theory to describe these small fluctuations. First, we remove from the full QCD heavy quark field $Q(x)$ the dependence on the large momentum $p = m_Q v$, defining a new field $Q_v$:
\begin{equation}
Q(x) = \sum_{v}e^{-im_Qv\cdot x}Q_v(x),
\end{equation}
summing over labels $v$. In terms of the new fields, the quark part of the QCD Lagrangian becomes:
\begin{equation}
\label{QvLag}
\mathcal{L}_Q = \bar Q_v(x)(i\Dslash + m_Q\vslash - m_Q)Q_v(x),
\end{equation}
summing implicitly over $v$.\footnote{Strictly speaking, we should have summed over separate labels $v,v'$ for the two heavy quark fields, but interactions with soft gluons cannot change these label velocities; hence we sum over only a single velocity. This simplification will not occur in SCET.} Now write $Q_v(x)$ as the sum  $Q_v(x) = h_v(x) + H_v(x)$, where:
\begin{equation}
h_v(x) = \frac{1+\vslash}{2}Q_v(x),\qquad H_v(x) = \frac{1-\vslash}{2}Q_v(x).
\end{equation}
Then,  the heavy quark Lagrangian (\ref{QvLag}) becomes:
\begin{equation}
\label{hHLag}
\begin{split}
\mathcal{L}_Q &= \bar h_v(x)i\Dslash h_v(x) + \bar H_v(x)(i\Dslash - 2m_Q)H_v(x) \\&\quad + \bar H_v(x)i\Dslash h_v(x) + \bar h_v(x)i\Dslash H_v(x).
\end{split}
\end{equation}
Derivatives acting on the fields $h_v,H_v$ produce momenta of order $\lqcd/m_Q$, so the kinetic term for $H_v(x)$ is suppressed relative the leading term quadratic in $H_v$, $-2m_Q\bar H_v(x) H_v(x)$. $H_v$ therefore is not a dynamical field and can be integrated out by using the equation of motion (obtained by varying with respect to $\bar H_v$):
\begin{equation}
(i\Dslash - 2m_Q)H_v = -i\Dslash h_v.
\end{equation}
Substituting into Eq.~(\ref{hHLag}), we obtain
\begin{equation}
\mathcal{L}_Q = \bar h_v(x)i\Dslash h_v(x) - \bar h_v(x) i\Dslash\frac{1}{i\Dslash-2m_Q}i\Dslash h_v(x).
\end{equation}
Expanding in powers of $1/m_Q$:
\begin{equation}
\label{hvLag}
\mathcal{L}_Q = \bar h_v(x)\left(iv\mcdot D - \frac{1}{2m_Q}\Dslash\Dslash + \cdots\right)h_v(x),
\end{equation}
having used $P_v \gamma^\mu P_v = v^\mu$, where $P_v = (1+\vslash)/2$. The dots denote higher-order terms in the $1/m_Q$ expansion.

Keeping terms only to a fixed order in $1/m_Q$ in Eq.~(\ref{hvLag}) defines the Lagrangian of \emph{heavy quark effective theory}. The leading order term is very simple:
\begin{equation}
\mathcal{L}_{\text{HQET}} = \bar h_v(x) (iv\mcdot D) h_v(x),
\end{equation}
giving the heavy quark propagator
\begin{equation}
i\frac{1+\vslash}{2}\frac{1}{v\mcdot k + i\epsilon}
\end{equation}
and the heavy quark-gluon interaction vertex
\begin{equation}
\label{HQETvertex}
-ig T^A v^\mu.
\end{equation}
This leading-order Lagrangian exhibits more symmetries than evident in the full QCD Lagrangian. First, it is independent of the heavy quark mass, giving rise to a flavor symmetry between $b$ and $c$ quarks. Second, containing no Dirac matrices, it also exhibits spin symmetry. This heavy quark spin-flavor symmetry is thus an approximate symmetry of full QCD made evident only by the effective theory expansion in $1/m_Q$. It allows for powerful predictions for processes involving different hadrons containing heavy quarks which are related by spin-flavor symmetry. Violations of this symmetry at higher order in $1/m_Q$ can be systematically calculated by keeping more terms in the HQET Lagrangian (\ref{hvLag})\footnote{Or estimated by being more clever \cite{Dorsten:2003ru}.}.

This relatively simple example of HQET illustrates two key features we will seek in formulating any effective field theory:
\begin{itemize}
\item Extra, approximate symmetries not obvious in the full theory but easily identified at leading order(s) in the the effective theory.

\item The simplification of interactions between heavy (hard) particles and soft ones, as in Eq.~(\ref{HQETvertex}).
\end{itemize}
The first property grants us more predictive power in relating apparently different physical processes to one another, while more sophisticated exploitation of the latter will simplify proofs of factorization of perturbative and nonperturbative contributions to physical observables. We now embark on this task in the effective theory truly of interest in this thesis.

\section{Soft-Collinear Effective Theory}

Consider a process in which there are hadrons moving with very large energy compared to their invariant mass, for instance, in the decay of $Z$ bosons to light hadrons as considered in Sec.~\ref{sec:QCD}, or in $B$ decays such as $B\rightarrow D\pi$ or $B\rightarrow \pi\pi$. We can formulate an effective field theory for quarks and gluons moving on such collinear trajectories and the soft partons with which they interact---the \emph{soft-collinear effective theory} \cite{SCET1a,SCET1b}.

\subsection{Effective Theory for Inclusive Decays---SCET$_{\rm I}$}

We first identify the relevant energy scales and small parameters to use in formulating the effective theory. Recall that in HQET, we took this to be the heavy quark mass $m_Q$ and velocity $v\sim\lqcd/m_Q$. Here, we begin by defining two light-cone vectors, $n$ and $\bar n$, which satisfy
\begin{equation}
n^2 = \bar n^2 = 0,\quad n\mcdot\bar n = 2,
\end{equation}
for example, $n = (1,0,0,1)$ and $\bar n = (1,0,0,-1)$. Vectors can be decomposed along these two light-cone directions and the orthogonal transverse directions:
\begin{equation}
V^\mu = \frac{\bar n\mcdot V}{2} n^\mu + \frac{n\mcdot V}{2}\bar n^\mu + V_\perp^\mu.
\end{equation}
We thus define the light-cone components $V^+ = n\mcdot V$, $V^- = \bar n\mcdot V$. For a collinear particle, one light-cone component of its momentum will be large, the other small. Its momentum will scale in powers of some small parameter $\lambda$ as:
\begin{equation}
(p^+,p^-,p_\perp)\sim Q(\lambda^2,1,\lambda),
\end{equation}
where $Q$ is the large scale governing the physical process being studied, on which the definition of $\lambda$ also depends. The collinear particles also interact with soft, or ultrasoft, particles, with momenta scaling as:
\begin{subequations}
\begin{align}
\text{soft:}&\quad p_s\sim Q\lambda \\
\text{ultrasoft:}&\quad p_{us}\sim Q\lambda^2.
\end{align}
\end{subequations}
The effective theory for collinear and (ultra)soft particles with these scalings is called SCET$_{\rm I}$. Another choice of scalings gives the theory SCET$_{\rm II}$, to be introduced in the next section.

We form the \SCETa\ Lagrangian essentially by expanding the QCD Lagrangian in powers of $\lambda$. We start with the Lagrangian for collinear quarks, which we take to be massless. In QCD,
\begin{equation}
\label{masslessQCD}
\mathcal{L}_q = \bar q i\Dslash q.
\end{equation}
We seek to describe fluctuations in the momenta of the collinear particles about its collinear trajectory. So, for a collinear quark of momentum $p$, we write:
\begin{equation}
\label{decomposition}
p = \tilde p + k,
\end{equation}
where $\tilde p$ contains the large (\emph{label}) components of the momentum,
\begin{equation}
\tilde p^\mu = \tilde p^-\frac{n^\mu}{2} + \tilde p_\perp,
\end{equation}
and the \emph{residual} momentum $k\sim\lqcd$ represents the fluctuations about the label momentum.

As in HQET, we extract the large momentum fluctuations from the full QCD field, defining a new field $q_{n,p}$:
\begin{equation}
q(x) = \sum_{\tilde p} e^{-i\tilde p\cdot x}q_{n,p}(x).
\end{equation}
The QCD Lagrangian (\ref{masslessQCD}) then becomes:
\begin{equation}
\label{qnpLag}
\mathcal{L}_q = \sum_{\tilde p,\tilde p'} e^{-i(\tilde p - \tilde p')\cdot x}\bar q_{n,p'}(x)(\diracslash{\tilde p} + i\Dslash)q_{n,p}(x).
\end{equation}
Again as in HQET, we project out the large and small components of Dirac field, writing $q_{n,p} = \xi_{n,p} + \Xi_{n,p}$, where:
\begin{equation}
\xi_{n,p} = \frac{\nslash\bnslash}{4}q_{n,p},\quad \Xi_{n,p} = \frac{\bnslash\nslash}{4}q_{n,p},
\end{equation}
in terms of which the Lagrangian (\ref{qnpLag}) becomes:
\begin{equation}
\label{xiXiLag}
\begin{split}
\mathcal{L}_q = \sum_{\tilde p,\tilde p'}e^{-i(\tilde p - \tilde p')\cdot x}\biggl[&\bar\xi_{n,p'}(x)\frac{\bnslash}{2}in\mcdot D\xi_{n,p}(x) + \overline{\Xi}_{n,p'}(x)\frac{\nslash}{2}(\tilde p^- + i\bar n\mcdot D)\Xi_{n,p}(x) \\
&+ \bar\xi_{n,p'}(x)(\diracslash{\tilde p}_\perp + i\Dslash_\perp)\Xi_{n,p}(x) + \overline{\Xi}_{n,p'}(x)(\diracslash{\tilde p}_\perp + i\Dslash_\perp)\xi_{n,p}(x)\biggr].
\end{split}
\end{equation}
Again, as in HQET, we have a derivative, $\bar n\cdot\partial$, acting on $\Xi_{n,p}$, which is suppressed relative to the term containing the label momentum $\tilde p^-$. So the field $\Xi_{n,p}$ is not dynamical, and we eliminate it using its equation of motion:
\begin{equation}
\Xi_{n,p}(x) = \frac{1}{\tilde p^- + i\bar n\mcdot D}(\diracslash{\tilde p}_\perp + i\Dslash_\perp)\frac{\bnslash}{2}\xi_{n,p}(x)
\end{equation}
Inserting this into (\ref{xiXiLag}) leaves us with the Lagrangian:
\begin{equation}
\label{xiLag}
\mathcal{L}_\xi = \sum_{\tilde p,\tilde p'}e^{-i(\tilde p - \tilde p')\cdot x}\bar\xi_{n,p'}\biggl[in\mcdot D + (\diracslash{\tilde p}_\perp + i\Dslash_\perp)\frac{1}{\tilde p^- + i\bar n\mcdot D}(\diracslash{\tilde p}_\perp + i\Dslash_\perp)\biggr]\frac{\bnslash}{2}\xi_{n,p}(x).
\end{equation}
To sort out the interactions between gluons and collinear quarks, we must first distinguish between the collinear and (ultra)soft gluon fields. The gluon fields appearing in the covariant derivatives in the collinear quark Lagrangian (\ref{xiLag}) can be split up:
\begin{equation}
A = A^c + A^s + A^{us},
\end{equation}
where the collinear, soft, and ultrasoft fields are assigned the power countingas:
\begin{equation}
\label{Apowers}
A^c\sim Q(\lambda^2,1,\lambda),\quad A^s\sim Q(\lambda,\lambda,\lambda),\quad A^{us}\sim Q(\lambda^2,\lambda^2,\lambda^2),
\end{equation}
to match the scalings of the corresponding momenta. Note that interactions of soft gluons with collinear quarks leave the quark with the momentum scaling $Q(\lambda,1,\lambda)$, which does not exist in the effective theory. So soft gluons should not appear in the Lagrangian (\ref{xiLag}). Collinear quarks do interact with ultrasoft gluons. The effective theory collinear gluon fields $A_{n,q}^c$ are defined by:
\begin{equation}
\label{Acdef}
A^c(x) = \sum_{\tilde q}e^{-i\tilde q\cdot x}A_{n,q}^c(x),
\end{equation}
factoring out the large label momentum $\tilde q$. 
\setlength{\unitlength}{1pt}
\begin{figure}[]
\begin{align*}
\parbox{42mm}{
\begin{fmffile}{xi}
\begin{fmfgraph*}(100,30)
\fmfpen{thin}
\fmfleft{i}
\fmfright{o}
\fmf{dashes_arrow}{i,o}
\end{fmfgraph*}
\end{fmffile}}
&= i\frac{\Diracslash{\bar n}}{2}\frac{\bar n\cdot \tilde p}{n\cdot k\,\bar n\cdot \tilde p + \tilde p_\perp^2 + i\epsilon} \\ & \\
\raisebox{-2cm}{
\begin{fmffile}{xiAu}
\begin{fmfgraph*}(100,80)
\fmfleft{xi1}
\fmfright{xi2}
\fmftop{Au}
\fmf{dashes_arrow}{xi1,v}
\fmf{dashes_arrow}{v,xi2}
\fmf{gluon,tension=0}{v,Au}
\fmflabel{$\noexpand\mu,A$}{Au}
\end{fmfgraph*}
\end{fmffile}}
&= igT^An_\mu\frac{\Diracslash{\bar n}}{2} \\
\raisebox{-2cm}{
\begin{fmffile}{xiAc}
\begin{fmfgraph*}(100,80)
\fmfleft{xi1}
\fmfright{xi2}
\fmftop{Ac}
\fmf{dashes_arrow}{xi1,v}
\fmf{dashes_arrow}{v,xi2}
\fmf{gluon,tension=0}{v,Ac}
\fmf{plain,tension=0}{v,Ac}
\fmflabel{$p$}{xi1}
\fmflabel{$p'$}{xi2}
\fmflabel{$\noexpand\mu,A$}{Ac}
\end{fmfgraph*}
\end{fmffile}}
&= igT^A\left(n_\mu + \frac{\gamma^\perp_\mu\diracslash{\tilde p}_\perp}{\bar n\cdot\tilde p} + \frac{\diracslash{\tilde p}'_\perp\gamma^\perp_\mu}{\bar n\cdot \tilde p'} - \frac{\diracslash{\tilde p}'_\perp\diracslash{\tilde p}_\perp}{\bar n\cdot \tilde p\ \bar n\cdot \tilde p'}\bar n_\mu\right)\frac{\Diracslash{\bar n}}{2}
\end{align*}
\setlength{\unitlength}{1mm}
\begin{picture}(0,0)(0,0)
\put(24,79){$\tilde p + k$}
\put(21,21){$\tilde p$}
\put(41,21){$\tilde p'$}
\end{picture}
\vspace{-1cm}
\caption[Feynman rules involving collinear quarks in \SCETa.]{Feynman rules involving collinear quarks in \SCETa. The collinear particles are shown with label momenta $\tilde p,\tilde p'$ and residual momentum $k$.}
\label{scetrules}
\end{figure}

In the Lagrangian (\ref{xiLag}), the components $D_\perp$ and $\bar n\cdot D$ of the covariant derivative contain both ultrasoft and collinear gluons. Due to the power counting in Eq.~(\ref{Apowers}), the ultrasoft gluon fields give a subdominant contribution compared to that of the collinear gluons. Thus, the correct collinear quark Lagrangian in \SCETa\ at leading order in $\lambda$ is:
\begin{equation}
\label{SCETLag1}
\mathcal{L}_{\text{\SCETa}} = \sum_{\tilde p,\tilde p'}e^{-i(\tilde p - \tilde p')\mcdot x}\bar\xi_{n,p'}\biggl[in\mcdot D + (\diracslash{\tilde p}_\perp + i\Dslash_\perp^c)\frac{1}{\tilde p^- + i\bar n\mcdot D^c}(\diracslash{\tilde p}_\perp + i\Dslash_\perp^c)\biggr]\frac{\bnslash}{2}\xi_{n,p}(x),
\end{equation}
where $D^c = \partial - igA^c$ contains only the collinear gluon field, while $n\mcdot D$ still contains both collinear and ultrasoft fields. Some of the Feynman rules arising from this Lagrangian are shown in Fig.~\ref{scetrules}.

The Lagrangian (\ref{SCETLag1}) may be written in a still more compact form by the introduction of the \emph{label operators} \cite{Bauer:2001ct}:
\begin{subequations}
\label{labelopdefs}
\begin{align}
\bar n\mcdot\mathcal{P} \xi_{n,p} &= \bar n\mcdot \tilde p\,\xi_{n,p} \\
\mathcal{P}_\perp^\mu\xi_{n,p} &= \tilde p_\perp^\mu\xi_{n,p},
\end{align}
\end{subequations}
and similarly for label operators acting on collinear gluon fields. In the second, messier, term of the Lagrangian (\ref{SCETLag1}), the ordinary derivatives which appear hit everything to their right. Thus, they bring down extra factors of the label momenta of the collinear gluons, because of the exponential factor in Eq.~(\ref{Acdef}). For example,
\begin{equation}
\bar n\mcdot\partial A^c_\perp(x) = \bar n\mcdot\partial\sum_{\tilde q}e^{-i\tilde q\cdot x}A_{n,q}^{c\perp}(x) = \sum_{\tilde q} e^{-i\tilde q\cdot x}\bar n\mcdot\tilde qA_{n,q}^{c\perp}(x).
\end{equation}
We can account for derivatives on the exponential factors and eliminate the explicit label momenta appearing in the Lagrangian (\ref{SCETLag1}) by making use of the label operators in (\ref{labelopdefs}):
\begin{equation}
\label{SCETLag2}
\mathcal{L}_{\text{\SCETa}} = \bar\xi(x)\biggl[in\mcdot D + i\Dslash^c_\perp \frac{1}{i\bar n\mcdot D^c}i\Dslash^c_\perp\biggr]\frac{\bnslash}{2}\xi_n(x),
\end{equation}
where now
\begin{equation}
D^c_\mu = \mathcal{P}_\mu - igA^c_\mu,
\end{equation}
and
\begin{equation}
\xi_n(x) = \sum_{\tilde p}e^{-i\tilde p\cdot x}\xi_{n,p}(x).
\end{equation}
Finally, note that this Lagrangian contains the collinear gluon field $\bar n\cdot A^c$ in the denominator of the second term, giving rise to couplings of collinear quarks to an arbitrary number of $\bar n\mcdot A^c$ gluons. These infinitely many couplings can be resummed into the form of \emph{Wilson lines}:
\begin{equation}
W_n(x) = P\exp\biggl[ig\int_{-\infty}^x ds\,\bar n\mcdot A^c(x)\biggr],
\end{equation}
where $P$ denotes path ordering. Using the property 
\begin{equation}
W_n(x)\bar n\mcdot\mathcal{P} W_n^\dag(x) = \bar n\mcdot\mathcal{P} -i g\bar n\mcdot A^c(x) = \bar n\mcdot D^c,
\end{equation}
we can express the \SCETa\ Lagrangian (\ref{SCETLag2}) as:
\begin{equation}
\label{SCETLag3}
\mathcal{L}_{\text{\SCETa}} = \bar\xi_n(x)\biggl[in\mcdot D + i\Dslash^c_\perp W_n(x)\frac{1}{i\bar n\mcdot\mathcal{P}}W_n^\dag(x)i\Dslash^c_\perp\biggr]\frac{\bnslash}{2}\xi_n(x).
\end{equation}
It is also possible to argue that invariance under gauge transformations of collinear field requires this precise form of the Lagrangian, which we do not go through here. 

Similar analysis leads also to the effective Lagrangian for collinear gluons \cite{SCET3}:
\begin{equation}
\label{SCETgluonLag}
\mathcal{L}_c^{(g)} = \sum_{\tilde q,\tilde q'}e^{-i(\tilde q - \tilde q')\cdot x}\frac{1}{2g^2}\Tr\left\{\left[i\mathcal{D}^\mu + gA_{n,q},i\mathcal{D}^\nu + gA_{n,q'}^\nu\right]\right\}^2,
\end{equation}
where
\begin{equation}
\mathcal{D}^\mu = \bar n\mcdot\mathcal{P}\frac{n^\mu}{2} + \mathcal{P}_\perp^\mu + in\mcdot D\frac{\bar n^\mu}{2}.
\end{equation}
Ghost fields and gauge-dependent terms have been ignored in Eq.~(\ref{SCETgluonLag}).

\subsection{Decoupling Ultrasoft Fields}

We could take Eq.~(\ref{SCETLag3}) as the final form of our Lagrangian for \SCETa. However, one more manipulation simplifies greatly the separation of hard and soft (nonperturbative) physics in the effective theory \cite{SCET3}. First, introduce the ultrasoft Wilson lines:
\begin{equation}
Y_n(x) = P\exp\biggl[ig\int_{-\infty}^x ds\,n\mcdot A_{us}(ns)\biggr].
\end{equation}
 Rewrite the fields in (\ref{SCETLag3}) as:
\begin{subequations}
\label{BPSredef}
\begin{align}
\xi_n(x) &= Y_n(x)\xi_n^{(0)}(x) \\
A_n(x) &= Y_n(x)A_n^{(0)}(x)Y_n^\dag(x) \\
W_n(x) &= Y_n(x)W_n^{(0)}(x)Y_n^\dag(x).
\end{align}
\end{subequations}
(The last rule actually follows from the second.) From the property
\begin{equation}
[(n\mcdot\partial - ig n\mcdot A_{us})Y_n(x)] = 0,
\end{equation}
we find that upon making the replacements (\ref{BPSredef}) in the \SCETa\ Lagrangian (\ref{SCETLag3}), the term containing the ultrasoft gluon disappears! That is,
\begin{equation}
\bar\xi_n(x)in\mcdot D\frac{\bnslash}{2}\xi_n(x) \rightarrow \bar\xi_n^{(0)}(x)in\cdot(\partial - igA_n^{(0)})\frac{\bnslash}{2}\xi_n^{(0)}(x).
\end{equation}
Thus, in terms of the fields $\xi_n^{(0)},A_n^{(0_)}$, there are no couplings of collinear particles to ultrasoft gluons at leading order in \SCETa. The same decoupling occurs in the gluon Lagrangian (\ref{SCETgluonLag}). Since nonperturbative physics is governed by interactions with ultrasoft gluons, this decoupling greatly simplifies the separation of perturbative and nonperturbative contributions to physical observables in the effective theory.

Ultrasoft gluons have not entirely disappeared from the theory, however. When we match currents or operators mediating various decays from QCD onto \SCETa, we must include ultrasoft Wilson lines $Y_n(x)$ whenever we have a collinear field in the operator, according to Eqs.~(\ref{BPSredef}). (Equivalently, we must ensure that all operators are invariant under ultrasoft gauge transformations.) These rules will become apparent in the applications presented in subsequent chapters.

\subsection{Effective Theory for Exclusive Decays---\SCETb}

We now have all the tools in the effective theory required to describe inclusive decays such as in $Z$ decays to hadrons in Chap.~\ref{chap:jet}. Before proceeding to this application, however, let us introduce the novel features of a second version of SCET---\SCETb\ \cite{Bauer:2002aj}--- required to analyze exclusive decays as in Chap.~\ref{chap:ups} on radiative exclusive $\Upsilon$ decays.

To see the inadequacy of \SCETa\ in analyzing an exclusive decay producing energetic light hadrons, such as $\Upsilon\rightarrow\gamma\pi\pi$, consider the typical invariant mass of these light hadrons. For a particular light hadron, such as the pion, the invariant mass squared is of the order $\sim\lqcd^2$. The typical momentum of a collinear particle in \SCETa, however, scales as:
\begin{equation}
(p^+, p^-, p_\perp) \sim Q(\lambda^2,1,\lambda),
\end{equation}
where $\lambda = \sqrt{\lqcd/Q}$. The invariant mass squared of such a particle is:
\begin{equation}
p^2 = p^+ p^- - p_\perp^2 \sim Q^2\lambda^2 \sim Q\lqcd,
\end{equation}
which is too large to represent a light particle with $m^2\sim \lqcd^2$.

The correct effective theory is formed by using instead the parameter $\eta = \lqcd/Q$ and collinear fields which scale as:
\begin{equation}
\label{collscalingII}
(p^+, p^-, p_\perp) \sim Q(\eta^2,1,\eta),
\end{equation}
giving an invariant mass squared of $p^2\sim Q^2\eta^2 = \lqcd^2$, which is now correct for an exclusive light hadron. These collinear fields interact with soft fields, whose momenta scale as:
\begin{equation}
\label{softscaling}
p_{s}\sim Q(\eta,\eta,\eta),
\end{equation}
with invariant mass squared $p_s^2\sim \lqcd^2$. The effective theory for collinear and soft degrees of freedom with these scalings is called \SCETb.

The soft fields of \SCETb\ are in fact the same as the ultrasoft fields in \SCETa. The collinear fields, however, live at a lower energy scale in \SCETb\ than they do in \SCETa. For this reason, the collinear fields in \SCETa\ are often called \emph{hard-collinear} fields to distinguish them from the collinear fields of \SCETb. The introduction of another degree of freedom, the \emph{soft-collinear messenger} mode, was introduced in Refs.~\cite{Becher:2003qh,Becher:2003kh} so that all infrared divergences in one-loop graphs in \SCETb\ and QCD would match, which can also be accomplished by introducing a proper infrared regulator in the effective theory, as in Ref.~\cite{Bauer:2003td}. We do not evaluate any loop graphs with these divergences in the subsequent chapters, and so blissfully ignore these modes or regulators.

To work in \SCETb, one can imagine matching directly from QCD onto \SCETb. However, it is more convenient to match QCD first onto \SCETa, and then match \SCETa\ onto \SCETb, since we have already done the hard work of completing the first step. In the second step, we take the collinear fields of \SCETa, and lower their off-shellness from the scale $\sqrt{Q\lqcd}$ to $\lqcd$, putting them in \SCETb. Interactions with ultrasoft gluons are already removed from the \SCETa\ collinear quark and gluon Lagrangians via the field redefinition (\ref{BPSredef}). We simply change ultrasoft Wilson lines $Y_n(x)$ in \SCETa\ into soft Wilson lines $S_n(x)$ in \SCETb\, which look the same as in \SCETa\, but contain the soft gluon fields of \SCETb\ with the scaling (\ref{softscaling}).

Note that in \SCETb, there could not be interactions of collinear quarks with soft gluons anyway, because the soft gluon knocks the collinear quark off shell, due to the scalings in (\ref{collscalingII}) and (\ref{softscaling}).

We will make use of \SCETb\ in Chap.~\ref{chap:ups}, where the procedures described above will be illustrated by explicit example.

\subsection{Reparametrization Invariance}
\label{sec:RPI}

The introduction of the light-cone vectors $n,\bar n$ along which to decompose felds and momenta in the effective theory breaks the Lorentz invariance of the original theory of QCD. The Lorentz symmetry manifests itself in the effective theory by means of invariance under redefinitions of the vectors $n,\bar n$ and of the label momenta assigned to collinear particles. These features are known as \emph{reparametrization invariance} (RPI) \cite{Manohar:2002fd}.

We focus here on the theory \SCETa\ for simplicity. The first invariance requirement arises from the fact that the decomposition of collinear momenta into label and residual components is not unique. We wrote in Eq.~(\ref{decomposition}) for a collinear momentum $p$,
\begin{equation}
p = \tilde p + k,
\end{equation}
where $\tilde p$ is contains the large $\mathcal{O}(Q)$ and $\mathcal{O}(Q\lambda)$ label momenta, and $k$ is the $\mathcal{O}(Q\lambda^2)$ residual momentum. Shifting an amount of momentum of order $Q\lambda^2$ between the label and residual momentum should yield an equivalent description of the collinear physics. The main consequence of this invariance, together with gauge invariance, is that operators in the effective theory containing derivatives must appear in the combination:
\begin{equation}
\label{DcDus}
\mathcal{D}^\mu = D_c^\mu + D_{us}^\mu,
\end{equation}
where $D_c^\mu = \mathcal{P}^\mu - igA_{n,q}^\mu$, and $D_{us}^\mu = \partial^\mu - ig A_{us}^\mu$. This condition relates operators at different orders in $\lambda$, since the terms contained in Eq.~(\ref{DcDus}) have different scalings in $\lambda$. In this thesis we will focus on the constructions of operators only at leading order in $\lambda$, so this constraint will not be crucial.

More relevant constraints arise from the second type of invariance, that of redefinitions of the light-cone vectors $n,\bar n$ themselves. These vectors enter the very definition of the SCET Lagrangian, and the theory must remain invariant for equivalent choices of $n,\bar n$, which are those that leave invariant the conditions:
\begin{equation}
n^2 = \bar n^2 = 0,\qquad n\mcdot\bar n = 2,
\end{equation}
must yield an equivalent theory. There are three classes of transformations we can make. In the first two, we fiddle just with the transverse components of one or the other of $n,\bar n$:
\begin{equation}
\text{Type I:}\quad
\begin{cases}
n_\mu \rightarrow n_\mu + \Delta^\perp_\mu \\
\bar n_\mu\rightarrow\bar n_\mu
\end{cases}
\qquad
\text{Type II:}
\quad
\begin{cases}
n_\mu \rightarrow n_\mu \\
\bar n_\mu\rightarrow\bar n_\mu + \varepsilon^\perp_\mu
\end{cases}
,
\end{equation}
and in the last class we mutually rescale the vectors:
\begin{equation}
\text{Type III:}
\quad
\begin{cases}
n_\mu \rightarrow (1+\alpha)n_\mu \\
\bar n_\mu\rightarrow(1-\alpha)\bar n_\mu 
\end{cases},
\end{equation}
The parameters $\Delta^\perp,\varepsilon^\perp,\alpha$ are infinitesimal, but can be assigned a particular scaling in $\lambda$. Starting with type-II transformations, consider the change induced in the light-cone components of a vector $V^\mu$. We start with the decomposition:
\begin{equation}
V^\mu = \bar n\mcdot V\frac{n^\mu}{2} + n\mcdot V\frac{\bar n^\mu}{2} + V_\perp^\mu\ .
\end{equation}
Under the transformation, the components become:
\begin{subequations}
\begin{align}
n\mcdot V &\rightarrow n\mcdot V \\
\bar n\mcdot V &\rightarrow (\bar n + \varepsilon_\perp)\mcdot V = \bar n\mcdot V + \varepsilon_\perp\mcdot V_\perp \\
V_\perp^\mu &= V^\mu - \bar n\mcdot V\frac{n^\mu}{2} - n\mcdot V\frac{\bar n^\mu}{2} \nonumber \\
&\rightarrow V^\mu - (\bar n + \varepsilon_\perp)\mcdot V\frac{n^\mu}{2} - n\mcdot V\frac{\bar n^\mu + \varepsilon_\perp^\mu}{2} \\
&= V_\perp^\mu - \varepsilon_\perp\mcdot V_\perp\frac{n^\mu}{2} - n\mcdot V\frac{\varepsilon_\perp^\mu}{2}\ . \nonumber
\end{align}
\end{subequations}
If $V$ is, say, a collinear momentum or field with scaling $(V^+,V^-,V_\perp)\sim Q(\lambda^2,1,\lambda)$, then we see from the shifts in $\bar n\mcdot V$ and $V_\perp$ that we can assign a scaling of 1 to the parameter $\varepsilon_\perp$ without messing up the power counting of $p$. Similar examination of type-I and type-III transformation reveals that $\Delta_\perp$ cannot have a scaling larger then $\lambda$, while $\alpha$ can also be order 1. Therefore, type-II and type-III RPI can constrain operators at the same order in $\lambda$, while type-I will relate operators at different orders in $\lambda$. We will make use of type-II and type-III RPI to restrict the number of operators which can appear in the description of $\Upsilon$ decays in Chap.~\ref{chap:ups}.

\section{Non-Relativistic QCD}

We have described effective theories in QCD for hadrons containing a single heavy quark (HQET) and for light, energetic hadrons (SCET). We make use in this thesis of one more effective theory, that for mesons containing a heavy quark-antiquark pair, $Q\bar Q$. In such a meson, both heavy partons move with relatively small velocity. Thus, we are led to expand QCD about its non-relativistic limit---hence, the effective theory of \emph{non-relativistic QCD} (NRQCD) \cite{Bodwin:1995jh,Braaten:1996ix,Luke:2000kz}.

In NRQCD, we separate the fields in the QCD Lagrangian which create quarks and antiquarks, into the two-component spinor fields $\psi$ and $\chi$, respectively. The small expansion in NRQCD is the heavy quark velocity $v$. This parameter now determines two separate physical scales, not just one: the heavy quark three-momentum is of the order $\abs{\vect{p}}\sim m_Q v$, while the kinetic energy is of order $E\sim m_Q v^2$.

To form the NRQCD Lagrangian for heavy quarks, begin again in QCD:
\begin{equation}
\label{QCDLagQ}
\mathcal{L}_Q = \bar\Psi_Q(i\Dslash - m_Q)\Psi_Q.
\end{equation}
The Dirac field $\Psi_Q$ contains fields creating/annihilating both quarks and antiquarks:
\begin{equation}
\Psi_Q = \begin{pmatrix} \psi \\ \chi \end{pmatrix},
\end{equation} 
in terms of which the Lagrangian (\ref{QCDLagQ}) is:
\begin{equation}
\begin{split}
\mathcal{L}_Q = &\psi^\dag(iD_0 - m_Q)\psi + \chi^\dag(iD_0 + m_Q)\chi \\
&+ \psi^\dag i\boldsigma\cdot\vect{D}\chi + \chi^\dag i\boldsigma\cdot\vect{D}\psi,
\end{split}
\end{equation}
having used the conventions for Dirac matrices:
\begin{equation}
\gamma^0 = 
\begin{pmatrix} 
1 & 0 \\ 
0 & -1 
\end{pmatrix},\quad 
\gamma^i = 
\begin{pmatrix} 
0 & \sigma^i \\
-\sigma^i & 0
\end{pmatrix}.
\end{equation}
We can decouple the $\psi$ and $\chi$ fields by the following transformation:
\begin{equation}
\Psi_Q\rightarrow\exp\left(\frac{i\boldgamma\cdot\vect{D}}{2m_Q}\right)\Psi_Q,
\end{equation}
which turns (\ref{QCDLagQ}) into:
\begin{equation}
\mathcal{L}_Q\rightarrow
\begin{pmatrix} \psi^\dag & \chi^\dag \end{pmatrix}
\begin{pmatrix} 
-m_Q + iD_0 + \frac{\vect{D}^2}{2m_Q} & 0 \\
0 & m_Q + iD_0 - \frac{\vect{D}^2}{2m_Q}
\end{pmatrix}
\begin{pmatrix} \psi \\ \chi \end{pmatrix},
\end{equation}
up to terms suppressed by higher powers of $v$. Removing the dependence on large label momenta $\vect{p}$ from the full-theory fields:
\begin{equation}
\psi(x) = e^{-i(m_Q t-\vect{p}\cdot\vect{x})}\psi_{\vect{p}}(x),\quad \chi(x) = e^{i(m_Q t+\vect{p}\cdot\vect{x})}\psi_{\vect{p}}(x),
\end{equation}
and keeping the dominant terms in $v$, we obtain the leading-order Lagrangian for quarks and antiquarks in NRQCD:
\begin{equation}
\mathcal{L}_{\text{NRQCD}} = \psi_{\vect{p}}^\dag(x)\left(iD^0 - \frac{\vect{p}^2}{2m_Q}\right)\psi_{\vect{p}}(x) + \chi_{\vect{p}}^\dag(x)\left(iD^0 + \frac{\vect{p}^2}{2m_Q}\right)\chi_{\vect{p}}(x).
\end{equation}
The fields $\psi_{\vect{p}},\chi_{\vect{p}}$ describe fluctuations about the label momentum $\vect{p}$ of order $m_Qv^2$:
\begin{equation}
\vect{P} = \vect{p} + \vect{k},\quad E = k^0,
\end{equation}
where $\vect{P}$ is the total heavy quark or antiquark three-momentum, and the residual momentum $k = (k^0,\vect{k})\sim Mv^2$.

The construction of operators to describe quarkonium decays in this effective theory is performed in Chap.~\ref{chap:ups}.
\chapter{Exclusive Radiative Decays of $\Upsilon$ \label{chap:ups}}

\begin{quotation}
\small\singlespace\noindent ``For I am Saruman the Wise, Saruman Ring-maker, Saruman of Many Colours!''

I looked then and saw that his robes, which had seemed white, were not so, but were woven of all colours, and if he moved they shimmered and changed hue so that the eye was bewildered.

``I liked white better,'' I said.

\flushright\emph{Gandalf, in \emph{The Fellowship of the Ring} by J.R.R. Tolkien}
\end{quotation}
We now turn to the application of SCET to exclusive decays---the radiative decays of quarkonia to an exclusive light hadron in the final state. This chapter is based on work published in Ref.~\cite{Fleming:2004hc}.

\section{Introduction}

In a recent series of papers the differential decay rate for the decay $\Upsilon \to \gamma X$
has been studied in the ``endpoint" region where the decay products have a large total 
energy of order the $\Upsilon$ mass ($\mups$), and a small total invariant mass squared of  
order $\Lambda \mups$, where $\Lambda \sim 1 \, \textrm{GeV}$ is the typical hadronic 
scale~\cite{Bauer:2001rh,Fleming:2002rv,Fleming:2002sr,GarciaiTormo:2004jw,Fleming:2004rk}. An important
tool in this analysis is the soft-collinear effective theory (SCET) 
\cite{SCET1a,SCET1b,Bauer:2001ct,SCET3}, which is a systematic 
treatment of the high energy limit of QCD in the framework of effective field theory. Specifically 
SCET is used to describe the highly energetic decay products in the endpoint region. The heavy 
$b$ and $\bar{b}$  quarks  which form the $\Upsilon$ are described by non-relativistic QCD 
(NRQCD)~\cite{Bodwin:1995jh,Luke:2000kz}.

The soft-collinear effective theory is not limited to applications involving inclusive processes. In 
fact SCET has been extensively applied to exclusive decays of $B$ mesons into light
mesons~\cite{Bauer:2002aj,Bauer:2001cu,Chay:2003zp,Chay:2003ju,Chay:2003kb,Mantry:2003uz,Beneke:2003pa,Lange:2003pk,Leibovich:2003tw,Hill:2004if,Feldmann:2004mg}. Here we use
similar techniques to study exclusive radiative decays of the $\Upsilon$. We make use of some of 
the results derived in the analysis of inclusive radiative decays in the endpoint region, but the analysis of exclusive decays is complicated by the existence of two different collinear scales. This necessitates
a two-step matching procedure~\cite{Bauer:2002aj}. In the first step one matches onto $\textrm{SCET}_{\rm I}$ which describes collinear degrees of freedom with typical offshellness of order 
$\sqrt{\Lambda \mups}$, as is appropriate for inclusive decays in the endpoint region as discussed 
above. In the second step $\textrm{SCET}_{\rm I}$ is matched onto $\textrm{SCET}_{\rm II}$, which is
appropriate for exclusive processes since it describes collinear degrees of freedom with typical offshellness of order $\Lambda$.

The analysis of $\Upsilon$ decay is further complicated by the existence of two types of currents: 
those where the $b\bar{b}$ is in a  color-singlet configuration and those where it is in a 
color-octet configuration. The octet operators are higher-order in the combined NRQCD and SCET
power counting, so one might suppose that they can be dropped. However, the octet currents have 
a Wilson coefficient which is order $\sqrt{\alpha_s(\mups)}$ while the singlet current  has a Wilson coefficient 
of order  $\alpha_s(\mups)$. The additional suppression of  the singlet Wilson coefficient is enough
so that both color-octet and color-singlet operators must be included as  contributions to the inclusive 
radiative decay rate in the endpoint region~\cite{Fleming:2002rv,Fleming:2002sr}. 

In this work we show that in exclusive decays the octet currents are truly suppressed 
relative to the singlet current and can be neglected. We then determine the minimal set of 
color-singlet currents which can arise and fix their matching coefficients in $\textrm{SCET}_{\rm I}$.
We run this current to the intermediate collinear scale $\mu_c \sim \sqrt{\Lambda \mups}$ and
match onto $\textrm{SCET}_{\rm II}$. Our expression for the decay rate agrees with that derived in QCD using a twist expansion~\cite{Baier:1,Baier:1985wv,Ma:2001tt}. Finally we use our results to make a prediction for the ratio of branching fractions $B(\Upsilon \to \gamma f_2)/B(J/\psi \to \gamma f_2)$, $B(J/\psi \to \gamma f_2)/B(\psi' \to \gamma f_2)$, and analyze the decay $\Upsilon \to \gamma \pi \pi $ in the kinematic regime where the pions are collinear.

\section{Power Counting}

\subsection{Inclusive Decays}

The first step is to match the QCD amplitude for a $b\bar{b}$ pair in a given color and spin configuration to decay to a photon and light particles onto combined $\textrm{SCET}_{\rm I}$ and NRQCD currents.  The SCET power-counting is in the parameter $\lambda \sim \sqrt{ \Lambda/M}$, where $M = 2m_b$, while the NRQCD power-counting is in $v$, the relative velocity of the $b\bar b$ pair in the $\Upsilon$.  Numerically, $\lambda\sim v \sim1/3$.  The matching is shown graphically in Fig.~\ref{matching_fig}.
\begin{figure}[t]
\centerline{\includegraphics[width=6.25in]{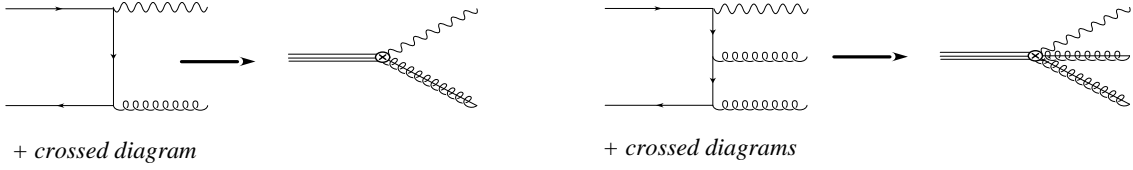}}
\vspace{2em}
\caption[Matching onto EFT operators with one and two final-state gluons]{Matching onto operators in the effective field theory with one and two gluons in the final state. The currents on the left have a color-octet $b\bar{b}$ in either a ${}^1S_0$ or ${}^3P_J$ configuration. The matching for a color-singlet $b\bar{b}$ pair  in a ${}^3S_1$ configuration is shown on the right.}
\label{matching_fig}
\end{figure}
The effective theory operators can be classified into those with the $b\bar{b}$ in a color-octet configuration (shown on the left-hand side of Fig.~\ref{matching_fig}) and those with the $b\bar{b}$ in a color-singlet configuration (shown on the right-hand side of Fig.~\ref{matching_fig}). The leading octet operators can be further subdivided into a those where the  $b\bar{b}$ is in a ${}^1S_0$ configuration and those where the $b\bar{b}$ is in a ${}^3P_J$ configuration. The octet ${}^1S_0$ operators are~\cite{Fleming:2002rv,Fleming:2002sr} 
\begin{equation}\label{1s0op2}
J_\mu(8,{}^1S_0) = 
\sum_i C^{8,{}^1S_0}_i(M,\mu) \Gamma^i_{\alpha \mu}\,
 \chi^\dagger_{-{\bf p}} B^\alpha_\perp \psi_{\bf p} \,,
\end{equation}
where
\begin{equation}
B_\perp^\mu = -\frac{i}{g_s}W^\dag[\mathcal{P}_\perp^\mu + g_s(A^\mu_{n,q})_\perp]W.
\end{equation}
The operator $\cP^\mu_\perp$ projects out the label momenta in the perpendicular direction~\cite{SCET3}. The sum in Eq.~(\ref{1s0op2}) is over all possible Lorentz structures denoted by $\Gamma^i_{\alpha \mu}$, and
$C^{8,{}^1S_0}_i(M,\mu)$ is the corresponding matching coefficient for each structure. The octet ${}^3P_J$ operators are
\begin{equation} \label{3pjop}
J_\mu(8,{}^3P_J) =
 \sum_i C^{8,{}^3P_J}_i(M,\mu) \Gamma^i_{\alpha \mu \sigma \delta}  
   \chi^\dagger_{-{\bf p}}    B^\alpha_\perp  \Lambda \cdot \widehat{{\bf p}}^\sigma 
   \Lambda \cdot \bsigma^\delta \psi_{\bf p} \,,
\end{equation}
where $\Lambda$ is a Lorentz boost matrix.  Each of these color-octet operators scales as ${\cal O}(\lambda)$ in SCET. The NRQCD power-counting has the ${}^1S_0$ octet operators scaling as ${\cal O}(v^3)$; however, this operator has an overlap with the $\Upsilon$ state beginning at ${\cal O}(v^2)$. (See Appendix~\ref{sec:states}.) Thus the $^1S_0$ operator contributes at order $v^5 \lambda$ to the $\Upsilon$ radiative decay rate. The ${}^3P_J$ octet operator has NRQCD scaling ${\cal O}(v^4)$, but overlaps with the $\Upsilon$ at order $v$. Thus the total power-counting of the $^3P_J$ contribution is ${\cal O}(v^5 \lambda)$, which is the same as the ${}^1S_0$ octet operators. The leading order matching coefficients for both are ${\cal O}(\sqrt{\alpha_s(M)})$. 

The color-singlet operators are 
\begin{equation}\label{3s1op}
J_\mu(1,{}^3S_1) =
\sum_i \Gamma^i_{\alpha \beta \delta \mu}
 \chi^\dagger_{-{\bf p}} \Lambda\cdot\bsigma^\delta \psi_{\bf p}
{\rm Tr} \big\{ B^\alpha_\perp \, 
C^{(1,{}^3S_1)}_i ( M,\bnP_{+} ) \, 
B^\beta_\perp \big\} \,,
\end{equation}
where $\bnP_+ = \bnP^\dagger + \bnP$, with $\bnP \equiv \bn\cdot\cP$. 
These operators scale as ${\cal O}(\lambda^2)$ in SCET and ${\cal O}(v^3)$ in NRQCD. The leading matching coefficients are ${\cal O}(\alpha_s(M))$. Thus the ratio of color-octet to color-singlet contributions in inclusive radiative $\Upsilon$ decay scales as:
\begin{equation}\label{incrat}
\frac{\textrm{octet}}{\textrm{singlet}} \sim 
\frac{v^2}{\lambda \sqrt{\alpha_s(M)}} \sim \frac{v}{\sqrt{\alpha_s(M)}} \,.
\end{equation}

\subsection{Exclusive Decays}

The situation changes when one considers exclusive decays. The currents we just discussed are $\textrm{SCET}_{\rm I}$ currents where the typical invariant mass of the collinear degrees of freedom is of order $\mu_c = \sqrt{M \Lambda}$. In order to have overlap with the meson state we must match onto $\textrm{SCET}_{\rm II}$ currents where the typical invariant mass of collinear particles is ${\cal O}(\Lambda)$. Furthermore, the interpolating field which annihilates the meson state in $\textrm{SCET}_{\rm II}$ is defined to consist only of collinear fields in a color-singlet configuration~\cite{Bauer:2002aj}. Given these considerations it is simple to match the color-singlet operator in $\textrm{SCET}_{\rm I}$ to an operator of identical form in $\textrm{SCET}_{\rm II}$.  However, the matching of the octet contributions from $\textrm{SCET}_{\rm I}$ onto $\textrm{SCET}_{\rm II}$ is more involved.

Before we consider the matching of the octet contributions from $\textrm{SCET}_{\rm I}$ onto $\textrm{SCET}_{\rm II}$ we turn our attention to the scaling of these contributions in $\textrm{SCET}_{\rm I}$. In order to produce a final state consisting only of collinear fields in a color-singlet configuration we need an interaction which changes the ultrasoft (usoft) gluon into a collinear gluon (as shown on the left-hand side of Fig.~\ref{octetmatch}).  This term in the SCET Lagrangian is power suppressed by $\lambda$ so that the time-ordered product of the octet current with the collinear-collinear-usoft vertex scales as ${\cal O}(\lambda^2)$ in the $\textrm{SCET}_{\rm I}$ power counting. In addition, the exchanged gluon introduces an extra factor of the coupling constant at the matching scale: $\alpha_s(\mu_c)$. Including these factors, the time-ordered product of octet currents with the subleading Lagrangian scales as ${\cal O}(\alpha_s(\mu_c) \sqrt{\alpha_s(M)} \lambda^2 v^5)$, and the ratio of time-ordered products to the singlet contribution is
\begin{equation}\label{excrat}
\frac{\textrm{octet}}{\textrm{singlet}} \sim
\frac{v^2 \alpha_s(\mu_c) }{\sqrt{\alpha_s(M)}} \approx 0.05 \,,
\end{equation}
for the bottomonium system. For charmonium the above ratio is about 0.2.
This result is very different from the result for the inclusive decay given in Eq.~(\ref{incrat}). In $\textrm{SCET}_{\rm I}$ the octet contribution to exclusive radiative $\Upsilon$ decay is not only suppressed in the limit $v,\lambda \to 0$, but numerically suppressed by a factor of $\sim 10$ for typical values of the parameters. This is the same order of suppression we expect from higher order SCET and NRQCD corrections; thus, we should be able to safely neglect the color-octet contribution in $\textrm{SCET}_{\rm I}$. However, before we can neglect the octet contribution in our analysis we must show that the suppression of the octet piece holds after matching onto $\textrm{SCET}_{\rm II}$.

We first turn our attention to the simpler calculation: matching the color-singlet operator.  In $\textrm{SCET}_{\rm I}$ we perform the field redefinition \cite{SCET3}:
\begin{equation}\label{fieldredef}
A_n \rightarrow Y A^{(0)}_n Y^\dag \,,
\end{equation}
which decouples usoft from collinear degrees of freedom. Under this field redefinition 
\begin{equation}\label{fieldredef2}
B^\alpha_\perp \rightarrow Y B^{(0)\alpha}_\perp Y^\dag\,,
\end{equation}
and the $Y$'s cancel in the trace of the color singlet operator given in Eq.~(\ref{3s1op}). Thus we match the $\textrm{SCET}_{\rm I}$ operator after the field redefinition onto an operator in $\textrm{SCET}_{\rm II}$ of a form identical to that in Eq.~(\ref{3s1op}):
\begin{equation}\label{3s1opII}
J_\mu^{(II)}(1,{}^3S_1) = \sum_i \Gamma^i_{\alpha \beta \delta \mu}
 \chi^\dagger_{-{\bf p}} \Lambda\cdot\bsigma^\delta \psi_{\bf p}
{\rm Tr} \big\{ B^\alpha_{\textrm{II} \perp} \, 
C^{(1,{}^3S_1)}_i ( M,\bnP_{+} ; \mu_c) \, 
B^\beta_{\textrm{II} \perp} \big\} \ ,
\end{equation}
where $\mu_c = \sqrt{ M \Lambda}$ is the $\textrm{SCET}_{\rm I}$--$\textrm{SCET}_{\rm II}$ matching scale, and the subscripts indicate $\textrm{SCET}_{\rm II}$ fields. From now on we drop the subscripts.   In $\textrm{SCET}_{\rm II}$ the power-counting parameter is $\eta \sim  \lambda^2$, and the $\textrm{SCET}_{\rm II}$ color-singlet operator in Eq.~(\ref{3s1opII}) is ${\cal O}(v^3 \eta^2)$. The short-distance coefficient is inherited from $\textrm{SCET}_{\rm I}$ and is ${\cal O}(\alpha_s(M))$. 

The matching of the color-octet current is more complicated. In order to match onto an $\textrm{SCET}_{\rm II}$ operator with color-singlet collinear degrees of freedom we must consider time-ordered products where a usoft gluon radiated from the $b\bar{b}$ pair is turned into a final state collinear degree of freedom. An example of such a diagram is given in Fig.~\ref{octetmatch}.
\begin{figure}[t]
\centerline{\includegraphics[width=4.0in]{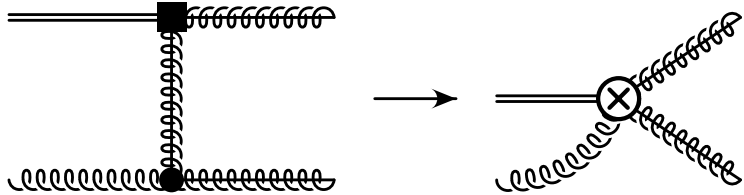}}
\vspace{2em}
\caption[Matching from \SCETa\ to \SCETb .]{Matching from \SCETa\ to \SCETb . This Feynman diagram is an example of a time-ordered product in $\textrm{SCET}_{\rm I}$ that matches onto an operator in $\textrm{SCET}_{\rm II}$ that has a nonzero overlap with the final-state collinear meson. }
\label{octetmatch}
\end{figure}
Two collinear gluons are required for the collinear final state to be color-singlet. One of the collinear gluons comes from the octet current, and the other can be produced by pulling a gluon out of the $b\bar b g$ Fock state of the $\Upsilon$, and kicking it with a collinear gluon from the current. This requires a collinear-collinear-ultrasoft coupling which first appears  at order $\lambda$ in the $\textrm{SCET}_{\rm I}$ Lagrangian~\cite{Chay:2002vy,Beneke:2002ph,Pirjol:2002km}:
\begin{equation}
\mathcal{L}_{cg}^{(1)} = \frac{2}{g^2}\Tr\left\{[i\mathcal{D}^\mu, iD_{us}^{\perp\nu}][i\mathcal{D}_\mu, iD_{c\nu}^\perp]\right\},
\end{equation}
where $\mathcal{D}^\mu = D_c^\mu + n\mcdot D_{us}\bar n^\mu/2$. The decay amplitude comes 
from a time-ordered product of the color-octet current and $\mathcal{L}_{cg}^{(1)}$, or a time-ordered product of the color-octet current, $\mathcal{L}_{cg}^{(1)}$, and a leading order gluon interaction. Though our result will hold for either type of time-ordered product we will, for the sake of concreteness, only consider the former:
\begin{equation}\label{top1}
T_8 = \int d^4 x \, T\left\{J(8,\cdot)(0),\mathcal{L}_{cg}^{(1)}(x)\right\},
\end{equation}
where the dot stands for $^1S_0$ or $^3P_J$. In the time-ordered product two gluon fields are contracted to form the internal propagator in Fig.~\ref{octetmatch}, which scales as $1/\lambda^2$. We require two uncontracted $A_{c\nu}^\perp$ fields (in a color-singlet configuration) so that we can match onto an $\textrm{SCET}_{\rm II}$ operator in the form ${\rm Tr}[A_{c\mu}^\perp A_{c\nu}^\perp]$ which annihilates the final state collinear meson. In this example the leading contribution is an $\bar n\mcdot A_n$ gluon field in one of the Wilson lines in $J(8,\cdot)$ contracted with an $n\mcdot A_n$ field in $\mathcal{L}_{cg}^{(1)}$. After the contraction what remains in $\mathcal{L}_{cg}^{(1)}$ is 
\begin{equation}\label{leftover}
\frac{2}{g^2}\Tr\left\{[gT^A, A_{us}^{\perp\nu}][\bnP, A_{c\nu}^\perp]\right\},
\end{equation}
which scales as $(\lambda^2)(1)(\lambda) = \lambda^3$. Note we now have the correct field content for the operator shown on the right-hand side of Fig.~\ref{octetmatch}: there are two outgoing $A^c_\perp$ fields, one from $\mathcal{L}_{cg}^{(1)}$ and one from $J(8,\cdot )$, in a color-singlet configuration, and an incoming soft gluon field also from $\mathcal{L}_{cg}^{(1)}$.

Next we  decouple collinear and usoft in $\textrm{SCET}_{\rm I}$ through the field redefinition in Eq.~(\ref{fieldredef}). This introduces factors of $Y$ and $Y^\dagger$ into our expressions. When matching onto $\textrm{SCET}_{\rm II}$ these become soft Wilson lines $S$ and $S^\dagger$. Since these Wilson lines do not affect the power counting we ignore them. Now we can match onto a convolution of $\textrm{SCET}_{\rm II}$ operators with $\textrm{SCET}_{\rm I}$--$\textrm{SCET}_{\rm II}$ matching coefficients. Since these arise from integrating out the internal collinear propagators they scale as $\lambda^{-2}$ for each propagator. In our example there is one propagator so the matching coefficient scales as ${\cal O}(\lambda^{-2})$, which is ${\cal O}(\eta^{-1})$ in the $\textrm{SCET}_{\rm II}$ power counting (remember $\eta\sim\lambda^2$). Since the $\textrm{SCET}_{\rm II}$ operator has two $A_{c\nu}^\perp$ fields each scaling as $\eta$, and a soft field scaling as $\eta$, it scales as $\eta^3$. Combining the scaling of the $\textrm{SCET}_{\rm II}$ operator with the scaling of the $\textrm{SCET}_{\rm I}$--$\textrm{SCET}_{\rm II}$ matching coefficient gives an ${\cal O}(\alpha_s(\mu_c) \eta^2)$ contribution. If we include the order $v^5$ NRQCD scaling from the heavy sector, and the ${\cal O}(\sqrt{\alpha_s(M)})$ contribution from the QCD--$\textrm{SCET}_{\rm I}$ matching coefficient the color-octet contribution to exclusive decays scales as ${\cal O}(v^5 \eta^2 \sqrt{\alpha_s(M)} \alpha_s(\mu_c))$ in $\textrm{SCET}_{\rm II}$. Taking the ratio of the color-octet to the color-singlet contribution to exclusive $\Upsilon$ decay in  $\textrm{SCET}_{\rm II}$ we find: 
\begin{equation}
\frac{\textrm{octet}}{\textrm{singlet}} \sim \frac{v^2 \alpha_s(\mu_c)}{\sqrt{\alpha_s(M)}} \,,
\end{equation}
which is the same scaling we found in $\textrm{SCET}_{\rm I}$. Thus we can safely neglect the color-octet contributions at this order.

\section{Complete Basis of Color-Singlet Matching Coefficients}

Now that the color-octet contribution has been eliminated we determine a complete basis of Lorentz structures $\Gamma^i_{\alpha\beta\mu\nu}$  that can appear in the color-singlet matching coefficient in Eq.~(\ref{3s1op}). At leading order in $\alpha_s(M)$ only one Lorentz structure was found to be non-zero~\cite{Fleming:2002rv}:
\begin{equation}\label{matchcoeff1}
C_1^{(1,{}^3S_1)}(M,\omega) \Gamma^1_{\alpha\beta\delta\mu} = 
\frac{4 g^2_s e e_b}{3 M} g^\perp_{\alpha\beta} g_{\mu\delta} \,.
\end{equation}
However, at higher order other Lorentz structures may appear. These coefficients can be constructed from the set:
\begin{equation}\label{set}
\{ g_{\mu\nu}, n_\mu, \bar n_\mu, v_\mu \} \,,
\end{equation}
where $v$ is the four-velocity of the $\Upsilon$, under the restriction that $\Gamma^i_{\alpha\beta\delta\mu}$ satisfies the appropriate symmetries. For example, the full theory amplitude is parity even, as is the effective theory operator, meaning that the matching coefficient must also be parity even. As a result the epsilon tensor is not included in Eq.~(\ref{set}). \

We treat $v$ as an object independent of $n,\bar n$~\cite{Pirjol:2002km}, and use $n^2 = \bar n^2 = 0$, $n\cdot\bar n = 2$, and $v^2 = 1$. Before we write down all the possible operators which can appear we note some simple properties that will make our task more manageable. First we note that $\Gamma_{\alpha\beta\delta\mu}$ must be symmetric in $\alpha$ and $\beta$. To see this consider the object:
\begin{equation}\label{blah}
\sum_\omega C_i^{(1,{}^3S_1)}(M,\omega) \Gamma^i_{\alpha\beta\delta\mu} 
\Tr(B_\perp^\alpha \delta_{\omega, \bnP_{+}}  B_\perp^\beta) \, ,
\end{equation}
which is the collinear part of the color-singlet operator where a sum over $\omega$ has been introduced. First interchange $\alpha$ and $\beta$, and then use the cyclic nature of the trace to switch the two $B_\perp$ fields. Note, however there is a projection on these fields from the operator in the Kronecker delta involving $\bnP_{+}$. Since $\bnP_{+} = \bnP^\dagger + \bnP$ this operator projects out minus the label on $B_\perp^\alpha$ and projects out the label on $B_\perp^\beta$~\cite{Bauer:2001ct}. To preserve this relationship when the order of the fields is switched we must let $\delta_{\omega, \bnP_{+}} \to  \delta_{\omega, -\bnP_{+}}$. By letting $\omega \to -\omega$ we  have $\delta_{-\omega, -\bnP_{+}} =  \delta_{\omega, \bnP_{+}}$, and the operator goes into itself. However, the Wilson coefficient is now $C_i^{(1,{}^3S_1)}(M,-\omega)$. To demonstrate that Eq.~(\ref{blah}) is symmetric under $\alpha \leftrightarrow \beta$ we must show that $C_i^{(1,{}^3S_1)}(M,\omega)$ is even in $\omega$. We use charge conjugation for this. The heavy quark sector of the operator has charge conjugation $C=-1$ as does the photon. As noted in Ref.~\cite{Novikov:1977dq} two gluons in a color-singlet configuration must have $C$ even. Since QCD is charge conjugation conserving the product of operator and coefficient in Eq.~(\ref{blah}) must also be $C$ even. This is the case if the matching coefficient $ C_i^{(1,{}^3S_1)}(M,\omega) \Gamma^i_{\alpha\beta\delta\mu}$ is $C$ even. Following Ref.~\cite{Bauer:2002nz}, under charge conjugation the above product of operator and coefficient goes to itself with $\omega \to -\omega$ in the coefficient function. Thus charge conjugation implies $C_i^{(1,{}^3S_1)}(M,-\omega) = C_i^{(1,{}^3S_1)}(M,\omega)$, and as a result Eq.~(\ref{blah}) is symmetric in $\alpha$ and $\beta$.  
Second, any $v_\delta$ appearing in $\Gamma_{\alpha\beta\delta\mu}$ gives zero contribution to the operator, since $v\cdot\Lambda = 0$. Third, $n_\alpha,\bar n_\alpha$ (and by symmetry $n_\beta, \bar n_\beta$) appearing in the operator also gives a zero contribution since these indices contract with indices on the $B_\perp$ field. Finally, we use reparameterization invariance (RPI) of SCET~\cite{Manohar:2002fd,Chay:2002vy}. The terms satisfying these requirements are
\begin{align}
\label{coeff1}
\sum_i C_i^{(1,{}^3S_1)}(M,\omega) \Gamma^i_{\alpha\beta\delta\mu}  & =
c_1 g_{\alpha\beta}g_{\delta\mu} + c_2 g_{\alpha\beta}\frac{n_\delta n_\mu}{(n\cdot v)^2} + c_3 g_{\alpha\beta}\frac{n_\delta v_\mu}{n\cdot v} \\
& \hspace{-15ex}
+c_4\biggl[\left(g_{\alpha\mu}-\frac{v_\alpha n_\mu}{n\cdot v}\right)\left(g_{\beta\delta}-\frac{v_\beta n_\delta}{n\cdot v}\right) + \left(g_{\alpha\delta}-\frac{v_\alpha n_\delta}{n\cdot v}\right)\left(g_{\beta\mu}-\frac{v_\beta n_\mu}{n\cdot v}\right)\biggr] \,.
\nonumber
\end{align}

So far we have allowed $v$ to be an arbitrary vector. Now we restrict ourselves to a frame where $v_\perp^\mu$ = 0.  Furthermore we are interested in the case where the photon is real, so we can restrict the photon to have transverse polarizations. This leaves only two linearly independent terms:
\begin{equation}
\label{coefftrans}
\sum_i C_i^{(1,{}^3S_1)}(M,\omega) \Gamma^i_{\alpha\beta\delta\mu} = a^g_1 g^\perp_{\alpha\beta}g^\perp_{\delta\mu} + a^g_2\left( g^\perp_{\alpha\delta}g^\perp_{\beta\mu} + g^\perp_{\alpha\mu}g^\perp_{\beta\delta} - g^\perp_{\alpha\beta}g^\perp_{\delta\mu}\right)\,,
\end{equation}
where $g^{\mu\nu}_\perp = g^{\mu\nu}-(n^\mu \bn^\nu + \bn^\mu n^\nu)/2$. The first term projects out the trace part of the $\Tr(B_\perp^\alpha \delta_{\omega, \bnP_{+}}  B_\perp^\beta)$ operator, while the second term projects out the symmetric traceless component. Since the Lorentz symmetry in the perpendicular components is not broken in SCET these two terms do not mix under renormalization. The leading-order matching fixes the coefficients $a_1$ and $a_2$ to order $\alpha_s$:
\begin{equation}\label{matchcoeff}
a^g_1(\bnP_{+} ; \mu = M)  = \frac{4g^2_s e e_b}{3M}, \hspace{10ex}
a^g_2(\bnP_{+} ; \mu = M)  = 0 \,.
\end{equation}
Since there is no mixing, $a_2=0 + {\cal O}(\alpha_s^2(M))$ at all scales. Note this matching assumes the $\Upsilon$ states are non-relativistically normalized: $\langle \Upsilon (P') | \Upsilon (P) \rangle = \delta^3(P-P')$. 

In addition to gluon operators we must consider the basis of all possible quark operators which can appear in radiative $\Upsilon$ decays
\begin{equation}\label{quarkcur}
J_\mu^q (1,{}^3S_1) = \sum_{i} \chi^\dagger_{-{\bf p}} \Lambda\cdot\bsigma^\delta \psi_{\bf p}
\bar{\xi}_{n,p_1} W \Gamma^i_{\mu \delta} (M,\bnP_+) W^\dagger \xi_{n,p_2} \,.
\end{equation}
The basis of Dirac structures, $\{ \bnslash , \bnslash \gamma_5 , \bnslash \gamma^\mu_\perp \}$, was given in Ref.~\cite{Bauer:2002nz}, and the most general basis of quark operators can then be constructed out of these Dirac structures and the set $\{ \epsilon_{\alpha\beta\mu\nu}, g_{\mu\nu}, n_\mu, \bar n_\mu, v_\mu \}$. Using the symmetries of SCET and RPI we find
\begin{equation}
\sum_{i} \Gamma^i_{\mu \delta}  = 
a^q_1 \frac{\bnslash}{2} g^\perp_{\mu\delta} + 
a^q_2  \frac{\bnslash}{2} \gamma_5 \epsilon^\perp_{\mu\delta} + 
a^q_3  \frac{\bnslash}{2} \gamma^\perp_{\mu} n_{\delta} \,.
\end{equation}
The first term transforms as a scalar, the second term transforms as a pseudoscalar, and the third as a vector. 
The matching coefficients at the scale $\mu = M$ for the quark operators in radiative $\Upsilon$ decay are all zero at leading order in perturbation theory, but the scalar quark operator mixes with the scalar gluon operator through renormalization group running and can be generated in this manner. The pseudoscalar and vector term do not mix with the scalar gluon operator due to Lorentz symmetry and will not be generated at this order in the perturbative matching.

\section{Decay Rates \& Phenomenology}

We now consider the phenomenological implications of our analysis for exclusive radiative decays of quarkonium into either a single meson or a pair of mesons which are collinear. The $(n+1)$-body decay rate is given by:
\begin{equation}
\begin{split}
\Gamma (\Upsilon \to \gamma F_n ) &= \frac{1}{2\mups} \int \frac{d^3\vect{q}}{2 E_{\gamma} (2\pi)^3}
\prod^n_i \frac{d^3\vect{p}_i}{2 E_i (2\pi)^3} (2 \pi )^4 \delta^4(P - q - \sum^n_i p_i)  \\
& \qquad\qquad \times
| \langle \gamma(q) p_1 ... p_n | J^\mu\mathcal{A}_\mu | \Upsilon(P) \rangle |^2,
\end{split}
\end{equation}
where $J$ is the QCD current, $\mathcal{A}$ is the photon field, and $F_n$ denotes an exclusive final state consisting of $n$ collinear particles. We consider only decay rates where the final state momenta $p_i$ are all collinear with combined invariant mass $m^2_n = (\sum^n_i p_i)^2 \sim \Lambda$. The  effective theory decay rate is obtained by matching the current $J$ onto the $\textrm{SCET}_{\rm II}$ current given in Eq.~(\ref{3s1opII}) plus a quark operator:
\begin{equation}\label{currmatch}
\begin{split}
\langle \gamma(q) &p_1 ... p_n | J^\mu\mathcal{A}_\mu | \Upsilon(P) \rangle
\\
& \to 
\sum_i \Gamma^i_{\alpha \beta \delta \mu} \langle  \gamma(q)  p_1 ... p_n  | 
 \chi^\dagger_{-{\bf p}} \Lambda\mcdot\bsigma^\delta \psi_{\bf p}
{\rm Tr} \big\{ B^\alpha_{\perp} \, 
C^{(1,{}^3S_1)}_i ( \bnP_{+} ; \mu) \, 
B^\beta_{ \perp} \big\} \mathcal{A}^\mu |  \Upsilon(P) \rangle
\\
&\quad + \sum_i  \langle  \gamma(q)  p_1 ... p_n  | 
 \chi^\dagger_{-{\bf p}} \Lambda\mcdot\bsigma^\delta \psi_{\bf p}
\bar{\xi}_{n,p_1} W \Gamma^i_{\mu \delta} (M,\bnP_+) W^\dagger \xi_{n,p_2} 
\mathcal{A}^\mu|  \Upsilon(P) \rangle
 \\
& =  
\bra{0}\chi_{-\vect{p}}^\dag\Lambda\mcdot\bsigma^\delta\psi_{\vect{p}}\ket{\Upsilon(P)}
\bra{\gamma(q)}\mathcal{A}^\mu\ket{0}g^\perp_{\delta\mu} \\
&  \quad \times\bigg[ \langle p_1 ... p_n  |  {\rm Tr} \big\{ B^\alpha_{\perp} \, 
a^g_1( \bnP_{+} ; \mu) \, 
B_\alpha^{ \perp} \big\}  | 0 \rangle  + \langle p_1 ... p_n  |  \bar{\xi}_{n,p_1} W \frac{\bnslash}{2} \, \frac{a^q_1( \bnP_{+} ; \mu)}M \,
W^\dagger \xi_{n,p_2} | 0 \rangle 
\bigg] \,.
\end{split}
\end{equation}
In obtaining the second line we make use of the results in Eqs.~(\ref{coefftrans}) and~(\ref{matchcoeff}), and use the properties of $\textrm{SCET}_{\rm II}$ to factor soft and collinear degrees of freedom.   In the last line we  changed to a nonrelativistic normalization for the $\Upsilon$ state.

Next we define the light-cone wave functions
\begin{subequations}
\label{wfdefs}
\begin{align}
\bra{p_1\dots p_n} \mathcal{\bar P}\Tr[B_\perp^\alpha \delta_{\omega,\mathcal{P}_+} B^\perp_\alpha]\ket{0} &= M^{3-n} \phi_g^{F_n}(x),  \\
\bra{p_1\dots p_n}\bar\xi_{n,\omega_1} W \frac{\bnslash}{2}\delta_{\omega,\bnP_+} W^\dag\xi_{n,\omega_2} \ket{0} &= M^{3-n}\phi_q^{F_n}(x) \,,
\end{align}
\end{subequations}
where states are relativistically normalized, and the discrete label $\omega$ is converted to a continuous one, $x = \omega/\bar{n}\cdot p$, as explained in Ref.~\cite{Fleming:2004rk}. The wave functions $\phi_{q,g}^{F_n}$ are dimensionless. See Appendix~\ref{app:lc} for the relation of these SCET light-cone wave functions to those conventionally defined in QCD.  Then the collinear matrix elements in brackets in Eq.~(\ref{currmatch}) can be written as the convolution:
\begin{equation}
\begin{split}
\langle p_1 ... p_n  |&\Tr \big\{ B^\alpha_{\perp} \, 
a^g_1( \bnP_{+} ; \mu) \, 
B_\alpha^{ \perp} \big\}  | 0 \rangle 
 + \langle p_1 ... p_n  |  \bar{\xi}_{n,p_1} W \frac{\bnslash}{2} \, a^q_1( \bnP_{+} ; \mu) \,
W^\dagger \xi_{n,p_2} | 0 \rangle 
\\
& =  
M^{2-n}\int^1_{-1} dx \, \big( a_1^g(x;\mu)  \phi^{F_n}_g(x;\mu) + a_1^q(x;\mu)  \phi^{F_n}_q(x;\mu) \big)\,.
\end{split}
\end{equation}
The dependence on the scale $\mu$ cancels between the long-distance matching coefficients and the wave function. We will elaborate on this point in a moment. First we expand both $a^{g/q}_1(x;\mu)$ and $\phi^{F_n}_{g/q}(x;\mu)$ in Gegenbauer polynomials:
\begin{subequations}
\begin{align}
a_1^q(x;\mu) &= \sum_{n\textrm{ odd}}a_q^{(n)}(\mu)C_n^{3/2}(x) \,,
\\
a_1^g(x;\mu) &= \sum_{n\textrm{ odd}}a_g^{(n)}(\mu)(1-x^2)C_{n-1}^{5/2}(x) \,,
\\
\phi_q^{F_n}(x;\mu) &= \sum_{n\textrm{ odd}}b_q^{(n)}(\mu)(1-x^2)C_n^{3/2}(x) \,, 
 \\
\phi_g^{F_n}(x;\mu) &= \sum_{n\textrm{ odd}}b_g^{(n)}(\mu)(1-x^2)C_{n-1}^{5/2}(x) \,.
\end{align}
\end{subequations}
Then the convolution becomes an infinite sum of products of Gegenbauer coefficients:
\begin{equation}\label{gegenbauer}
\begin{split}
\int^1_{-1} dx \big( a_1^g(x;\mu)  \phi^{F_n}_g(x;\mu) &+ a_1^q(x;\mu)  \phi^{F_n}_q(x;\mu) \big) 
 \\
& = \sum_{n\textrm{ odd}} \big( f^{(n)}_{5/2} a_g^{(n)}(\mu) b_g^{(n)}(\mu) + 
f^{(n)}_{3/2} a_q^{(n)}(\mu) b_q^{(n)}(\mu)\big) \,,
\end{split}
\end{equation}
where 
\begin{equation}
f^{(n)}_{5/2} = \frac{n(n+1)(n+2)(n+3)}{9(n+3/2)} \,, \hspace{5ex} 
f^{(n)}_{3/2} = \frac{(n+1)(n+2)}{n+3/2} \,.
\end{equation}

We now return to the question of the scale. Here we pick $\mu \sim \Lambda$ which minimizes logarithms in the wave function; however, large logarithms of $M/\Lambda$ then appear in the Wilson coefficients. These large logarithms are summed using the renormalization group equations in SCET. This calculation was carried out in Ref.~\cite{Fleming:2004rk}, and we only quote the results here. We find:
\begin{equation}\label{resum}
\begin{split}
\int^1_{-1} dx &\big( a_1^g(x;\mu)  \phi^{F_n}_g(x;\mu) + a_1^q(x;\mu)  \phi^{F_n}_q(x;\mu) \big) 
\\
& = \frac{4}{3} a_1^g(M) \sum_{n\textrm{ odd}} \bigg\{ \bigg[ 
\gamma^{(n)}_+ \bigg( \frac{\alpha_s(\mu)}{\alpha_s(M)} \bigg)^{2 \lambda^{(n)}_+/\beta_0} -
\gamma^{(n)}_- \bigg( \frac{\alpha_s(\mu)}{\alpha_s(M)} \bigg)^{2 \lambda^{(n)}_-/\beta_0} \bigg]
b_g^{(n)}(\mu)
 \\
& \qquad\qquad + 
\frac{f^{(n)}_{3/2}}{f^{(n)}_{5/2}} \frac{\gamma^{(n)}_{gq}}{\Delta^{(n)}} \bigg[ 
 \bigg( \frac{\alpha_s(\mu)}{\alpha_s(M)} \bigg)^{2 \lambda^{(n)}_+/\beta_0} -
 \bigg( \frac{\alpha_s(\mu)}{\alpha_s(M)} \bigg)^{2 \lambda^{(n)}_-/\beta_0} \bigg]
b_q^{(n)}(\mu) \bigg\} \,, 
\end{split}
\end{equation}
where 
\begin{subequations}
\begin{align}
\beta_0 &= 11 - \frac{2n_f}{3} \,, \\
\gamma_\pm^{(n)} &= \frac{\gamma_{gg}^{(n)} - \lambda_\mp^{(n)}}{\Delta^{(n)}} \,,
\\
\lambda_\pm^{(n)} &= \frac{1}{2}\left[\gamma_{gg}^{(n)} + \gamma_{q\bar q}^{(n)} \pm \Delta^{(n)}\right] \,,
\\
\Delta^{(n)} &= \sqrt{\left(\gamma_{gg}^{(n)} - \gamma_{q\bar q}^{(n)}\right)^2 + 4\gamma_{gq}^{(n)}\gamma_{qg}^{(n)}} \,,
\\
\gamma_{q\bar q}^{(n)} &= C_F\left[\frac{1}{(n+1)(n+2)} - \frac{1}{2} - 2\sum_{i=2}^{n+1}\frac{1}{i}\right]  \,,
\\
\gamma_{gq}^{(n)} &= \frac{C_F}{3}\frac{n^2+3n+4}{(n+1)(n+2)}  \,,
\\
\gamma_{qg}^{(n)} &= 3n_f\frac{n^2+3n+4}{n(n+1)(n+2)(n+3)}  \,,
\\
\gamma_{gg}^{(n)} &= C_A\left[\frac{2}{n(n+1)} + \frac{2}{(n+2)(n+3)} - \frac{1}{6} - 2\sum_{i=2}^{n+1}\frac{1}{i}\right] - \frac{n_f}{3}.
\end{align}
\end{subequations}
The quantities $\lambda^{(n)}_\pm$ which appear in the exponents in  Eq.~(\ref{resum}) are negative for any $n >1$. Furthermore $\lambda^{(1)}_- < 0$, while $\lambda^{(1)}_+ = 0$. This property allows us to consider the asymptotic limit $M \gg\Lambda$, where $\alpha_s(M) \to 0$. Then
\begin{equation}
\begin{split}
\lim_{M \to \infty}  \int_{-1}^1 dx
&\left[a_1^g(x;\mu)\phi_g^{F_n}(x;\mu) + a_1^q(x;\mu)\phi_q^{F_n}(x;\mu)\right] \\
&\longrightarrow   
\frac{16}{3} \frac{C_F}{4C_F + n_f} 
a_1^g(M)\left[b_g^{(1)}(\Lambda) + \frac{3}{4}b_q^{(1)}(\Lambda)\right] \equiv B^{F_{n}} a_1^g(M), 
\end{split} 
\end{equation}
which defines a nonperturbative parameter $B^{F_n}$. However, for values of $M$ around the $\Upsilon$ mass this is not a very good approximation, and for values around the $J/\psi$ mass a much better approximation is to assume no running at all.

\subsection{Two-Body Decay: $\Upsilon\rightarrow\gamma f_2$}

Having taken care of the technical details we can now use the above results to study the two-body radiative decay $\Upsilon \to \gamma F_1$. The decay rate is
\begin{equation}
\begin{split}
\label{2body}
\Gamma (\Upsilon \to \gamma F_1 )_{\textrm{SCET}_{\rm II}} & =  \frac{1}{16\pi} 
\bra{\Upsilon } \psi_{{\bf p}'}^\dagger \sigma_\perp^i \chi_{-{\bf p}'}  
\chi^\dagger_{-{\bf p}} \sigma_\perp^i \psi_{\bf p} \ket{ \Upsilon}  \\ 
& \quad \times  
\bigg[ \int^1_{-1} dx \big( a_1^g(x;\mu)  \phi^{F_1}_g(x;\mu) + a_1^q(x;\mu)  \phi^{F_1}_q(x;\mu) \big) \bigg]^2,
\end{split}
\end{equation}
where the full expression for the term in brackets is given in Eq.~(\ref{resum}). After factoring, the soft matrix element involving the heavy quark fields was further simplified using the vacuum insertion approximation for the quarkonium sector, which holds up to corrections of order $v^4$~\cite{Bodwin:1995jh}. Note that the operator above overlaps only with the $\lambda=\pm 1$ helicities of the $\Upsilon$. Then using the rotation symmetries of NRQCD~\cite{Braaten:1996jt} we can relate the non-relativistic matrix element above to those conventionally used:
\begin{equation}
\bra{\Upsilon } \psi_{{\bf p}'}^\dagger \sigma_\perp^i \chi_{-{\bf p}'}  
\chi^\dagger_{-{\bf p}} \sigma_\perp^i \psi_{\bf p} \ket{ \Upsilon} =
\frac{2}{3} \bra{\Upsilon } {\cal O}(1,{}^3S_1)  \ket{ \Upsilon} \,.
\end{equation}
For the final state meson $F_1$ to have nonzero overlap with the operators in Eq.~(\ref{wfdefs}) it must be flavor singlet, parity even and charge conjugation even. One candidate with the correct quantum numbers is the $f_2(1270)$. Furthermore this decay has been measured both in $\Upsilon$ and $J/\psi$ radiative decay, which is why we consider it. An interesting point is that only the helicity $\lambda =0$ component of the $f_2$ contributes at the order to which we are working. To see this begin
by considering the decomposition of the following gluon matrix element into all possible light-cone form-factors:
\begin{equation}
\label{f2formfact}
\bra{f_2}\Tr[B_\perp^\alpha B_\perp^\beta]\ket{0} = 
A(e(\lambda))g_\perp^{\alpha\beta} + B_\lambda e_\perp^{\alpha\beta}(\lambda),
\end{equation}
where $e^{\alpha\beta}$ is the symmetric-traceless polarization tensor of the $f_2$. We give the explicit form in Appendix~\ref{sec:pol}.  There are only two form factors above since the matrix element must be decomposed into tensors that have non-zero perpendicular components. The only structures available are $g_\perp^{\alpha\beta}$ and $e_\perp^{\alpha\beta}$. For $\lambda=\pm 1$, $e_\perp^{\mu\nu}(\lambda=\pm 1) = 0$, so this helicity component does not appear at this order. The coefficient $A(e(\lambda))$ is a scalar function which can be constructed from $\Tr(e_\perp)$ and $\bn_\alpha n_\beta e^{\alpha\beta}$. Because the helicity-zero polarization tensor has the property that $e_\perp^{\mu\nu}(\lambda = 0)\propto g_\perp^{\mu\nu}$, and the helicity-two polarization tensor has $\Tr(e_\perp) = 0$ and $\bn_\alpha n_\beta e^{\alpha\beta}= 0$, we can fix the normalization of  the coefficient $A(e(\lambda))$ so  that the first  term on the right-hand side of Eq.~(\ref{f2formfact}) parameterizes the $\lambda = 0$ contribution while the second term parameterizes the $\lambda = \pm 2$ contributions. The helicity-zero piece is picked out by the $a_1 g_\perp^{\alpha\beta}$ term in Eq.~(\ref{coefftrans}), while the helicity-two piece is picked out by the $a_2$ term. Thus at leading order in perturbation theory, the dominant decay should be to the helicity-zero component of the $f_2$.

The NRQCD matrix element in Eq.~(\ref{2body}) can be expressed in terms of the leptonic decay width of the $\Upsilon$. At leading order,
\begin{equation}
\Gamma(\Upsilon\rightarrow e^+ e^-) = \frac{8\pi\alpha^2 e_b^2}{3 M^2}\bra{\Upsilon} {\cal O}(1,{}^3S_1)   \ket{ \Upsilon} \,,
\end{equation}
and the decay rate for $\Upsilon\rightarrow\gamma f_2$ can be expressed as:
\begin{equation}
\Gamma(\Upsilon\rightarrow\gamma f_2) = \frac{16\pi\alpha_s(M)^2}{9\alpha}(B^{f_2})^2\Gamma(\Upsilon\rightarrow e^+ e^-)\,.
\end{equation}
We can repeat the same analysis for the decay rate $\Gamma(J/\psi\rightarrow\gamma f_2)$ and form a ratio of branching fractions, which in the asymptotic limit is:
\begin{equation}
\frac{B(\Upsilon\rightarrow\gamma f_2)}{B(J/\psi\rightarrow\gamma f_2)} = \left[\frac{\alpha_s(M_{b\bar{b}})}{\alpha_s(M_{c\bar{c}})}\right]^2\left(\frac{4 C_F +  3}{4C_F+4}\right)^2\frac{B(\Upsilon\rightarrow e^+ e^-)}{B(J/\psi\rightarrow e^+ e^-)}\,,
\end{equation}
where $M_{Q\bar{Q} } = 2 m_Q$.  Using $m_b = 4.1-4.4$ GeV, $m_c = 1.15-1.35$ GeV, $B(\Upsilon\rightarrow e^+e^-) = (2.38\pm 0.11)\times 10^{-2}$, and $B(J/\psi\rightarrow e^+e^-) = (5.93\pm 0.10)\times 10^{-2}$ \cite{Eidelman:2004wy}, we predict the ratio of branching fractions to be in the range:
\begin{equation}
\label{ratioasymptotic}
\left[\frac{B(\Upsilon\rightarrow\gamma f_2)}{B(J/\psi\rightarrow\gamma f_2)}\right]_{M\rightarrow\infty} = 0.14-0.19\,.
\end{equation}
As was mentioned earlier the asymptotic limit is not a particularly good approximation for the $\Upsilon$, and quite bad for the $J/\psi$. We can improve this approximation by keeping more terms in the resummed formula in Eq.~(\ref{resum}). The dominant term is the part of the $n=1$ term proportional to $b_g^{(1)}(\mu)$, and in this approximation:
\begin{equation}
\label{ratioimproved}
\begin{split}
\frac{B(\Upsilon\rightarrow\gamma f_2)}{B(J/\psi\rightarrow\gamma f_2)} &= \left[\frac{\alpha_s(M_{b\bar{b}})}{\alpha_s(M_{c\bar{c}})}\right]^2
\left[\frac{ \gamma_+^{(1)}  - 
                 \gamma_-^{(1)} \big( \alpha_s(\mu ) / \alpha_s(M_{b\bar{b}}) \big)^{2 \lambda_-^{(1)}/\beta_0^{n_f=4}} }
               { \gamma_+^{(1)}  - 
                 \gamma_-^{(1)} \big( \alpha_s(\mu ) / \alpha_s(M_{c\bar{c}}) \big)^{2 \lambda_-^{(1)}/\beta_0^{n_f=3}} }
\right]^2 \\
&\qquad\times
\frac{B(\Upsilon\rightarrow e^+ e^-)}{B(J/\psi\rightarrow e^+ e^-)}
\\
&= 0.13 - 0.18\,,
\end{split}
\end{equation}
where $\mu\sim 1$ GeV. The range of values has not changed much from Eq.~(\ref{ratioasymptotic}), however, theoretical errors are reduced: corrections to Eq.~(\ref{ratioimproved}) from the $b_q^{(1)}$ and higher-order terms in Eq.~(\ref{resum}) are estimated to be roughly $40\%$, while corrections to the infinite mass limit from higher order terms are estimated to be roughly $80\%$. In addition there are theory errors from neglecting higher-order terms in the perturbative expansion, as well as in the expansions in $v$ and $\eta$.
Our prediction can be compared to the measured value of $0.06\pm 0.03$, using the measurements $B(\Upsilon\rightarrow\gamma f_2) = (8\pm 4)\times 10^{-5}$ and $B(J/\psi\rightarrow\gamma f_2) = (1.38\pm 0.14)\times 10^{-3}$ \cite{Eidelman:2004wy}. Given the theoretical errors we can only conclude that our prediction does not disagree with data.

Our predictions for the ratios of $\Upsilon$ and $J/\psi$ branching fractions to $\gamma f_2$ are consistent with those derived in Refs.~\cite{Baier:1,Baier:1985wv,Ma:2001tt}, which use an expansion in twist. In particular, we reproduce the suppression of the helicities $\abs{\lambda} = 1,2$ in the final state relative to $\lambda = 0$. In contrast with Ref.~\cite{Ma:2001tt}, we extract the NRQCD color-singlet matrix elements from the leptonic decay widths of $\Upsilon$ and $J/\psi$ instead of the decay widths to light hadrons, for which corrections from color-octet contributions must be taken into account for a reliable calculation \cite{Gremm:1997dq}. The leptonic decay width, however, receives large corrections at NNLO in perturbation theory \cite{Beneke:1997jm, Czarnecki:1997vz}. In either case, one hopes that the uncertainties are mitigated in taking the ratios of branching fractions.

We can also compare the decay rates of $J/\psi$ and $\psi'$ to $\gamma f_2$ predicted by Eq.~(\ref{2body}) at the matching scale $\mu = M$, where $a_1^q(x;M) = 0$ and $a_1^g(x;M)$ is a constant. Dependence on the integral of the wave function $\phi_g^{f_2}(x;M)$ cancels out in the ratio of branching fractions:
\begin{equation}
\frac{B(J/\psi\rightarrow\gamma f_2)}{B(\psi'\rightarrow\gamma f_2)} = \frac{B(J/\psi\rightarrow e^+ e^-)}{B(\psi'\rightarrow e^+ e^-)} = 7.85\pm 0.35\,,
\end{equation}
while the measured value is $6.57\pm 1.42$. We used $B(\psi'\rightarrow e^+ e^-) = (7.55\pm 0.31)\times 10^{-3}$ and $B(\psi'\rightarrow \gamma f_2) = (2.1\pm 0.4)\times 10^{-4}$ \cite{Eidelman:2004wy}.

\subsection{Three-Body Decay: $\Upsilon\rightarrow\gamma\pi\pi$}

Next we consider a two pion final state in the kinematic region where the pions are collinear to each other with large energy and small total invariant mass $m_{\pi\pi} \sim \Lambda$. In this case we we have a three-body final state where the two pions are collinear. It is convenient to define the variables:
\begin{equation}
\mpipi = (p_1 + p_2)^2 \,, \qquad z = \frac{\bar n\mcdot p_1}{\mups} \,.
\end{equation}
In terms of these variables the differential decay rate is
\begin{equation}\label{temp}
\begin{split}
\frac{d\Gamma}{d\mpipi\,dz} &= \frac{1}{512\pi^3\mups^2}
\bra{\Upsilon } \psi_{{\bf p}'}^\dagger \sigma_\perp^i \chi_{-{\bf p}'}  
\chi^\dagger_{-{\bf p}} \sigma_\perp^i \psi_{\bf p} \ket{ \Upsilon}  \\ 
& \quad\times  
\bigg[ \int^1_{-1} dx \big( a_1^g(x;\mu)  \phi^{\pi\pi}_g(x;\mu) + a_1^q(x;\mu)  \phi^{\pi\pi}_q(x;\mu) \big) \bigg]^2
\end{split}
\end{equation}
to leading order in $\mpipi/\mups^2$. The properties of the meson pair light-cone wave function $\phi^{\pi\pi}$ have been investigated in Refs.~\cite{Grozin:1986at,Grozin:1983tt}, which interestingly find that in the region where $\Lambda \ll m_{\pi\pi} \ll \mups$ they are given by an integral over two single-particle wave functions. The ratio of the $\Upsilon$ and $J/\psi$ rates to $\gamma\pi\pi$ in the kinematic region of low $\mpipi$ is numerically the ratio in Eq.~(\ref{ratioimproved}) times an extra factor of $m_c^2/m_b^2 \sim 0.07 - 0.1$, that is, $0.01 - 0.02$. This suppression is due to the much larger total phase space available in $\Upsilon \to \gamma\pi\pi$ relative to that in $\Upsilon \to \gamma f_2$. No $\Upsilon\to\gamma\pi\pi$ events have yet been observed in the region $\mpipi < (1.0\text{ GeV})^2$ \cite{Anastassov:1998vs}.

\section{Conclusions}

We have systematically analyzed the exclusive radiative decays of quarkonium to energetic light mesons within the framework of soft-collinear effective theory and non-relativistic QCD to leading order in the effective theory power counting, as well as to leading order in the strong coupling. We show that color-octet contributions are suppressed by a factor of $v^2 \alpha_s(\mu_c) / \sqrt{\alpha_s(M)} \approx 0.05$ in exclusive $\Upsilon$ decays,  and can therefore be safely neglected. This is different from the situation in inclusive radiative decays in the endpoint region where octet contributions must be kept. 

We then turn to the color-singlet contribution. The tree-level matching onto this operator is carried out in Refs.~\cite{Fleming:2002rv, Fleming:2002sr}; however, the authors do not consider the complete set of operators that could appear in this decay. We use the symmetries of SCET and NRQCD, including RPI, to  show that the operator which is matched onto in Refs.~\cite{Fleming:2002rv, Fleming:2002sr} is the only operator that can appear for the decays in question. We also consider the set of possible quark  operators which can arise. Again only one of the possible quark operators can contribute to  the decays we are interested in. This operator has zero matching coefficient, but it can be generated through running.  We use the results of Ref.~\cite{Fleming:2004rk} for the renormalization group mixing of the quark and gluon operators, thus resumming large logarithms. Our results agree with an analysis in twist carried out in Refs.~\cite{Baier:1,Baier:1985wv,Ma:2001tt}.

Finally we study the phenomenology of quarkonium radiative decay to the $f_2$, as well as to $\pi\pi$ where the pions are collinear. We make predictions for the ratios of branching fractions $B(\Upsilon\to\gamma f_2)/B(J/\psi\to\gamma f_2)$ and $B(J/\psi\to\gamma f_2)/B(\psi'\to\gamma f_2)$, as well as for the differential decay rates of $\Upsilon$ and $J/\psi$ to $\gamma\pi\pi$ in the kinematic region of two collinear pions. Our predictions for the decays to $\gamma f_2$ are consistent with experimental data, but with large theoretical uncertainties, while there is insufficient data for $\gamma\pi\pi$ with which to compare. Further theoretical work and more experimental data, especially for the light-cone wave functions of $f_2$, will improve the precision of these predictions greatly.

\setcounter{section}{0}
\renewcommand{\thesection}{\arabic{chapter}.\Alph{section}}

\section{Appendix: Power Counting of States in NRQCD}
\label{sec:states}

The color-octet operators introduced in Eqs.~(\ref{1s0op2}) and (\ref{3pjop}) do not overlap with the $\Upsilon$ state until subleading order in $v$. We choose to represent the $\Upsilon$ state---a color-singlet spin-1 state---as being created from the vacuum with an interpolating field:
\begin{equation}
\ket{\Upsilon(\vect{p})} \sim \psi_{\vect{p}}^\dag\Lambda\cdot\boldsigma^\delta\chi_{-\vect{p}}\ket{0}.
\end{equation}
Thus, the NRQCD parts of the color-singlet operator $\mathcal{O}(1,^3S_1)$ in Eq.~(\ref{3s1op}) overlap with the $\Upsilon$ state at leading order in $v$:
\begin{equation}
\bra{0}\chi_{-\vect{p}}^\dag\Lambda\cdot\boldsigma^\delta\psi_{\vect{p}}\ket{\Upsilon}\sim\mathcal{O}(v^3),
\end{equation}
where the $v^3$ is the scaling of the operator itself. However, the color-octet operators do not overlap with the $\Upsilon$ until we put in insertions from the subleading part of the NRQCD Lagrangian. For example, for the $^1S_0$ operator,
\begin{equation}
\bra{0}\mathcal{O}(8,{}^1S_0)\ket{\Upsilon} \sim \bra{0}T\bigl\{\chi_{-\vect{p}}^\dag T^A\psi_{\vect{p}},\psi_{\vect{q}}^\dag(\boldsigma\cdot\vect{B})\psi_{\vect{q}}\bigr\}\ket{\Upsilon} \sim \mathcal{O}(v^5)\, .
\end{equation}
The field $\vect{B}$ scales as $v^4$ while the $\psi$ propagator in the time-ordered product scales as $1/v^2$, giving an overall $v^2$ suppression in the matrix element relative to the color-singlet matrix element. Thus the total scaling is $\mathcal{O}(v^5)$. The matrix element of the color-octet $^3P_J$ operator is the same as the $^1S_0$:
\begin{equation}
\bra{0}\mathcal{O}(8,^3P_J)\ket{\Upsilon}\sim\bra{0}T\{\chi_{-\vect{p}}^\dag T^A\Lambda\mcdot\vect{\hat p}^\sigma\Lambda\mcdot\boldsigma^\delta\psi_{\vect{p}},\psi_{\vect{q}}^\dag\frac{\vect{p}\cdot\vect{D}}{2M}\psi_{\vect{q}}\}\ket{\Upsilon}\sim\mathcal{O}(v^5),
\end{equation}
since the operator $\mathcal{O}(8,^3P_J)$ itself scales as $v^4$, $\vect{p}\cdot\vect{D}$ scales as $v^3$, and the $\psi$ propagator scales as $1/v^2$. 

Thus these two color-octet operators contribute at the same order in $v$ to the $\Upsilon$-to-vacuum matrix element, suppressed by $v^2$ relative to the contribution of the color-singlet operator.

\section{Appendix: Nonperturbative Matrix Elements and Light-Cone Wave Functions \label{app:lc}}

The matrix elements in Eq.~(\ref{wfdefs}) defining the SCET wave functions $\phi_{g,q}^{F_n}$ can be related to conventional QCD wave functions for flavor-singlet mesons. The two-gluon wave functions for a meson with momentum $q$ and net helicity $\lambda=0,\pm 2$ are defined as~\cite{Chernyak:1983ej}:
\begin{subequations}
\begin{align}
\bra{0}\Tr G_{\mu\nu}(z)&Y(z,-z)G_{\nu\lambda}(-z)\ket{q,\lambda=0}_{\mu_0} = f_S^L q_\mu q_\lambda\int_{-1}^1 d\zeta\,e^{iz\cdot q \zeta}\phi_S^L(\zeta,\mu_0) \,, \\
\bra{0}\Tr G_{\mu\nu}(z)&Y(z,-z)G_{\nu\lambda}(-z)\ket{q,\lambda=\pm 2}_{\mu_0}  \\
&= f_S^\perp[(q_\mu e^\perp_{\nu\beta} - q_\nu e^\perp_{\mu\beta})q_\alpha - (q_\mu e^\perp_{\nu\alpha} - q_\nu e^\perp_{\mu\alpha})q_\beta] 
 \int_{-1}^1 d\zeta\,e^{iz\cdot q \zeta}\phi_S^\perp(\zeta,\mu) \,, \nonumber
\end{align}
\end{subequations}
where $Y(z,-z)$ is the path-ordered exponential of gluon fields:
\begin{equation}
Y(z,-z) = P\exp\biggl[ig\int_{-z}^z d\sigma\cdot A(\sigma)\biggr].
\end{equation}
Going to the light-cone frame where $q_\mu = \frac{\bar n\cdot q}{2}n_\mu$ and $z^\mu = \frac{n\cdot z}{2}\bar n^\mu$, we invert these formulas to find 
\begin{subequations}
\label{lightconewfs}
\begin{align}
\phi_S^L(\zeta;\mu_0) &= \frac{\bar n^\mu \bar n^\lambda}{4\pi f_S^L q^-}\int_{-\infty}^\infty dz^+\,e^{-i\zeta q^- z^+/2}\bra{0}\Tr G_{\mu\nu}(z^+)Y(z^+,-z^+){G^\nu}_\lambda(-z^+)\ket{q,\lambda=0}  \,,\\
\phi_S^\perp(\zeta;\mu_0) &= \frac{\bar n^\mu \bar n^\alpha e_\perp^{*\nu\beta}}{4\pi f_S^\perp q^-}\int_{-\infty}^\infty dz^+\,e^{-i\zeta q^- z^+/2} \\
&\qquad\qquad\qquad\qquad\times\bra{0}\Tr G_{\mu\nu}(z^+)Y(z^+,-z^+)G_{\alpha\beta}(-z^+)\ket{q,\lambda=\pm 2}, \nonumber
\end{align}
\end{subequations}
where $z^+ = n\mcdot z$, and $q^-=\bar n\mcdot q$. 
Now we match the QCD fields on the right-hand side to fields in SCET:
\begin{subequations}
\begin{align}
\bar n^\mu G_{\mu\nu}(z^+) &\rightarrow \left[e^{-i\bnP z^+/2}\bar n_\mu G_n^{\mu\nu}\right ]  \,,\\
Y(z^+,-z^+) &\rightarrow \left[W_n e^{i(\bnP^\dag + \bnP)z^+/2} W_n^\dag\right],
\end{align}
\end{subequations}
where
\begin{equation}
G_n^{\mu\nu} = \frac{i}{g}[\mathcal{D}^\mu - ig A_{n,q}^\mu,\mathcal{D}^\nu - ig A_{n,q'}^\nu],
\end{equation}
with
\begin{equation}
i\mathcal{D}^\mu = \frac{n^\mu}{2}\bnP + \ppP^\mu + \frac{\bar n^\mu}{2}in\mcdot D_{us}.
\end{equation}
Therefore, for example, the matching between the QCD light-cone wave-function $\phi_S^L$ and the SCET operator is
\begin{equation}
\begin{split}
\phi_S^L&(\zeta ; \mu_0 )  \to  
 \frac{\bar n^\mu \bar n^\lambda}{4\pi f_S^L q^-}\int_{-\infty}^\infty dz^+\,e^{-i\zeta q^- z^+/2}
 \bra{0} \Tr G^n_{\mu\nu}(0) W_n e^{i \bnP_+ z^+ /2} 
 W^\dagger_n {G^{n \nu}}_\lambda(0)\ket{q,\lambda=0} \\
&  
 = 
\frac{-1}{16 \pi f^L_s q^- } \sum_\omega \int_{-\infty}^\infty dz^+\, e^{-i(\zeta q^-  - \omega)z^+/2}
(q^{- } - \omega)(q^{- } + \omega)
\bra{0} \Tr B_\perp^\nu \delta_{\bnP_+ , \omega} B^\perp_\nu \ket{q,\lambda=0} \,.\end{split}
\end{equation}
Integrating over $z^+$, and converting from the discrete index $\omega$ to the a continuous $\omega_c$ where $\zeta \equiv \omega_c / q^-$ we obtain the matching relation between the QCD and SCET light-cone wave functions
\begin{equation}
\label{wfdictionary}
\begin{split}
\phi_S^{L,\perp}(\zeta ;\mu) &\to -\frac{ q^-}{4f_S^L}(1-\zeta^2)\phi_g^{M(L,\perp)}(\zeta;\mu) \,.
\end{split}
\end{equation}
The SCET wave functions on the right-hand side are given by [cf. Eq.~(\ref{wfdefs})]:
\begin{subequations}
\begin{align}
\bra{0}\bnP\Tr[B_\perp^\alpha\delta_{\omega,\bnP_+}B^\perp_\alpha]\ket{M(q)}&= (q^-)^2\phi_g^{M(L)} \,, \\
e_{\perp\alpha\beta}^*\bra{0}\bnP\Tr[B_\perp^\alpha\delta_{\omega,\bnP_+}B_\perp^\beta]\ket{M(q)}&= ( q^-)^2\phi_g^{M(\perp)} \,.
\end{align}
\end{subequations}
Relations between the wave functions for the quark operator in QCD and SCET can be derived as in Ref.~\cite{Bauer:2002nz}.

\section{Appendix: Spin-2 Polarization Tensors \label{sec:pol}}

The spin-2 polarization tensor for a particle of mass $m$ and momentum $k$ can be built from spin-1 polarization vectors using Clebsch-Gordan coefficients to arrive at:
\begin{subequations}
\label{spin2pol}
\begin{align}
e^{\mu\nu}(\lambda=\pm 2) &= \frac{1}{2}
\begin{pmatrix}
0 & 0 & 0 & 0 \\
0 & 1 & \pm i & 0 \\
0 & \pm i & -1 & 0 \\
0 & 0 & 0 & 0
\end{pmatrix}  \,,\\
e^{\mu\nu}(\lambda = \pm 1) &= \mp\frac{1}{2m}
\begin{pmatrix}
0 & \abs{\vect{k}} & \pm i\abs{\vect{k}} & 0 \\
\abs{\vect{k}} & 0 & 0 & E_{\vect{k}} \\
\pm i\abs{\vect{k}} & 0 & 0 & \pm iE_{\vect{k}} \\
0 & E_{\vect{k}} & \pm iE_{\vect{k}} & 0
\end{pmatrix}  \,,\\
e^{\mu\nu}(\lambda = 0) &= \frac{1}{m^2}\sqrt{\frac{2}{3}}
\begin{pmatrix}
\vect{k}^2 & 0 & 0 & \abs{\vect{k}}E_{\vect{k}} \\
0 & -\frac{m^2}{2} & 0 & 0 \\
0 & 0 & -\frac{m^2}{2} & 0 \\
\abs{\vect{k}}E_{\vect{k}} & 0 & 0 & E_{\vect{k}}^2
\end{pmatrix}.
\end{align}
\end{subequations}

\renewcommand{\thesection}{\arabic{chapter}.\arabic{section}}

\chapter{Enhanced Nonperturbative Effects in $Z$ Decays to Hadrons \label{chap:jet}}

\begin{quotation}
\small\singlespace\noindent ``White!'' he sneered. ``It serves as a beginning. White cloth may be dyed. The white page can be overwritten; and the white light can be broken.''

``In which case it is no longer white,'' said I. ``And he that breaks a thing to find out what it is has left the path of wisdom.''

\flushright\emph{Gandalf, in \emph{The Fellowship of the Ring} by J.R.R. Tolkien}
\end{quotation}
In this chapter, we apply soft-collinear effective theory for inclusive decays---\SCETa--- to the hadronic decays of $Z$ bosons. The material in this chapter is based on Ref.~\cite{Bauer:2003di}.

\section{Introduction \label{sec:intro}}

Some of the most successful applications of perturbative QCD are to processes such as $Z$ decay to hadrons or  $e^+ e^-$ annihilation at large center-of-mass energy, in which a state with no strong interactions decays into final hadronic states. Here we will discuss the case of $Z$ decay, but the results apply equally well to the other cases.  Not only is the total hadronic $Z$ decay width calculable, but so are less inclusive infrared-safe quantities like the $Z$ decay rate into 2-jet and 3-jet events, the thrust distribution and jet mass distributions. Comparison of perturbative predictions for these and other quantities with experimental data on $Z$ decays from LEP and SLD has led to a remarkably accurate extraction of the strong coupling constant $\alpha_s(M_Z)$~\cite{compare,expdata1,expdata2,expdata3,expdata4,expdata5,alphasreview}.
Although the extraction of the strong coupling from event shape variables is less accurate than from the total hadronic $Z$ width, it is more model-independent since (neglecting quark mass effects) it does not depend on the values of the quark couplings to the $Z$.

For the totally inclusive hadronic $Z$ decay width, the operator product expansion allows one to include in theoretical predictions nonperturbative strong interaction effects that are characterized by vacuum expectation values of local operators. The effects of higher-dimension operators are suppressed by powers of the strong interaction scale  $\lqcd$ divided by the center-of-mass energy $M_Z$.  Since the $Z$ mass is large, these effects are very small. For example, if quark masses are neglected, the leading nonperturbative effects in the $Z$ decay width come from the vacuum expectation value of the gluon field strength tensor squared, $\vev{G_{\mu \nu} G^{\mu\nu}}$. This dimension-four operator gives rise to corrections to the total hadronic width suppressed by $\lqcd^4/M_Z^4\sim10^{-9}$.

Less inclusive variables that characterize $Z$ decay to hadrons give rise to nonperturbative effects suppressed by smaller powers of $\lqcd/M_Z$~\cite{DokWeb95,Webber,MW,KS95,KS99,KT00,Doktalk}. Furthermore, these corrections often become even more important in corners of phase space where hadronization effects are significant, such as in the thrust distribution very near $T=1$. It has been conjectured that the enhanced nonperturbative effects to many event shape distributions have a universal form with a single nonperturbative parameter \cite{DokWeb95,KT00,Doktalk,CataniWebber,DokMarSal,DokMarWeb}. These arguments are based on analysis of renormalon ambiguities in the QCD perturbation series and on the behavior of resummed perturbation theory. The conjectured relationship between the nonperturbative corrections to event shape distributions has recently been tested experimentally \cite{compare}. 

In Ref.~\cite{shape}, the enhanced nonperturbative effects that occur for the jet energy distribution in corners of phase space were studied using effective field theory methods. This approach uses the fact that very low momentum degrees of freedom which contain the nonperturbative physics couple to the degrees of freedom with energies of order $M_Z$ via Wilson lines.  Nonperturbative effects have been extensively studied previously~\cite{KS99} using factorization methods to divide the process into hard, jet-like  and soft subprocesses~\cite{StermanTasi,factorization}. Nonperturbative effects are computed from the soft subprocess. The effective field theory approach is similar to the one based on factorization methods. In this chapter we elaborate on the work in~\cite{shape} and extend it to other shape variables. The enhanced nonperturbative effects are expressed in terms of weighted matrix elements of operators involving Wilson lines, where the weighting depends on the event variable being considered. Our hope in this chapter is to make the results of Ref.~\cite{KS99} more accessible to the community of high energy theorists who are most familiar with effective field theory methods. 

We study smeared distributions, allowing us to expand the nonperturbative effects in powers of $\lqcd$, and write them as matrix elements of Wilson line operators and their derivatives. The computations are similar to those of smeared distributions in the endpoint region in $B$ decay---the point-by-point computation requires knowing the nonperturbative shape function, whereas nonperturbative effects in the smeared distributions can be written in terms of $\lambda_{1,2}$ provided the smearing region is large enough. 
 
For pedagogical reasons we start with a detailed treatment of the jet energy $E_J$ in $Z$ decay to two jets, where the jets are defined as Sterman and Weinberg did in their original work on jets in QCD~\cite{SW}. We spend considerable effort on this variable because the theoretical expression for its enhanced nonperturbative corrections is simpler than for other more phenomenologically interesting variables like thrust.  At lowest order in perturbation theory, the $Z$ boson creates a quark and an antiquark, each with energy $M_{Z}/2$, and so the jet energy distribution is equal to
\begin{equation}
{{\rm d}\Gamma_{\text{2-jet}}\over {\rm d} E_J}=\Gamma_{\text{2-jet}}^{(0)}\ \delta(E_J-M_{Z}/2),
\end{equation}
where $\Gamma_{\text{2-jet}}^{(0)}$ is the total two-jet rate at lowest order in perturbation theory. This leading order theoretical expression for the jet energy distribution is singular at $E_J=M_Z/2$. Furthermore, the leading perturbative and nonperturbative corrections are also singular at that kinematic point.  However, a non-singular quantity that can be compared with experiment without any resummation of singular terms is obtained by smearing the jet energy distribution over a region of size $\Delta$ that contains the lowest order partonic endpoint at $E_J=M_{Z}/2$. The leading nonperturbative correction to this smeared energy distribution is suppressed by $\lqcd/\Delta$. So, for example, with $\Delta \sim 10~ {\rm GeV}$ the nonperturbative corrections are expected to be of order $10\%$, roughly the same size as perturbative corrections, and an order of magnitude larger than the order $\lqcd/M_Z$ correction expected in the complete two jet rate. We argue that for $E_J$ very near $M_Z/2$ it is not possible to capture the dominant nonperturbative effects simply by shifting, $E_J \rightarrow E_J-\mu_{\rm np}$, in the perturbative expression for ${\rm d }\Gamma_{\rm 2-jet}/{\rm d}E_J$ (where $\mu_{\rm np}$ is a nonperturbative parameter of order $\lqcd$).

In the next section, we derive an expression for the leading enhanced nonperturbative correction to the smeared jet energy distribution for two jet events using methods from soft-collinear effective field theory (SCET)~\cite{SCET1a,SCET1b,SCET3,Bauer:2001ct}. This correction is given by the vacuum expectation value of a nonlocal operator involving Wilson lines. Perturbative order $\alpha_s$ corrections to this variable are derived in Appendix~\ref{sec:pert}.  

Section~\ref{sec:other} discusses the leading nonperturbative corrections for thrust, jet masses, the jet broadening variables, the $C$ parameter and energy-energy correlations. In agreement with Ref.~\cite{KS99} we find that the correction to jet mass sum and thrust are related. However, without additional model-dependent assumptions we do not find that the enhanced nonperturbative corrections to the $C$ parameter and jet broadening variables can be related to those for thrust and the jet masses. We compare the level of our understanding of the enhanced nonperturbative effects in these variables.

\section{Operator Product Expansion For The Two Jet Energy Distribution \label{sec:ope}}

The nonperturbative corrections to the jet energy distribution for $Z$ decay to two jets, ${\rm d}\Gamma_{\text{2-jet}}/ {\rm d} E_J$ near $E_J=M_Z/2$ are computed in this section. The perturbative corrections will be discussed in Sec.~\ref{sec:pert}. The results are given for the Sterman-Weinberg jet definition, where a cone of half-angle $\delta$ contains a jet if the energy contained in the cone is more than $E_{\text{cut}}=\beta M_Z$, as illustrated in Fig.~\ref{cones}. We take the cone half-angle $\delta$ and the dimensionless energy cut variable $\beta$ to be of order a small parameter $\lambda$, and compute in a systematic expansion in powers of $\lambda$. We are interested in the jet energy distribution within a region $\Delta$ of $M_Z/2$, where $M_Z\gg \Delta \gg \lambda^2 M_Z$. For example,  $\Delta \sim \lambda M_Z$. 
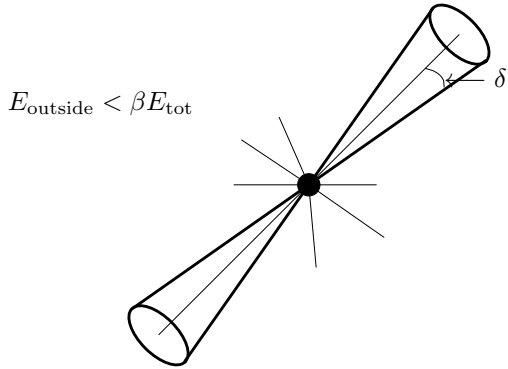
\begin{figure}
\begin{center}
\setlength{\unitlength}{1mm}
\begin{picture}(20,45)(-15,-20)
\put(0,0){\circle*{3}}
\drawline(-20,-20)(20,20)
\drawline(-4,9)(0,0)(10,-7)
\drawline(-9,6)(0,0)(1,-11)
\drawline(-10,0)(9,0)
\spline(15.5,15.5)(17,15)(18,14)(18,13)
\Thicklines
\put(20,20){\rotatebox{-45}{\ellipse{10}{5}}}
\put(-20,-20){\rotatebox{-45}{\ellipse{10}{5}}}
\drawline(-16.5,-23.5)(16.5,23.5)
\drawline(-23.5,-16.5)(23.5,16.5)
\put(18,13){\small$\longleftarrow\delta$}
\put(-40,10){\small$E_{\text{outside}}<\beta E_{\text{tot}}$}
\end{picture}
\end{center}
\caption[Sterman-Weinberg jets.]{Sterman-Weinberg jets. An event is characterized as a two-jet event for specified cuts $\beta,\delta$ if two cones of half-angle $\delta$ contain most of the energy in the event, with no more than $\beta E_{\text{tot}}$ of energy outside the cones.}
\label{cones}
\end{figure}

\SCETa\ is the appropriate effective field theory for the kinematic region of interest, and will be used for the derivation of the nonperturbative corrections to ${\rm d}\Gamma_{\text{2-jet}}/ {\rm d} E_J$ near $E_J=M_Z/2$. It is convenient  to introduce two lightlike vectors $n$ and $\bar n$ which satisfy $n^0=\bar n^0=1$ and ${\bf n}=-{\bf \bar n}$.   Four-vectors are decomposed along the $n$, $\bar n$ and perpendicular directions: $V=(V^+,V^-,V_{\perp})$ where $V^+ = n \cdot V$, $V^- = \bar n \cdot V$ and $ V_{\perp}^{\mu}=  V^{\mu}-V^+ {\bar n}^{\mu}/2-V^- n^{\mu}/2 $. For the problem of interest, \SCETa\ contains $n$-collinear, $\bar n $-collinear  and ultrasoft degrees of freedom~\cite{Bauer:2002nz}. The $n$-collinear and $\bar n$-collinear degrees of freedom have typical momenta that scale as 
\begin{equation}
\label{moment}
p_c^{(n)} \sim M_Z(\lambda^2,1, \lambda),\qquad p_c^{(\bar n)} \sim M_Z(1,\lambda^2, \lambda),
\end{equation}
 and the ultrasoft degrees of freedom have momenta that scale as
\begin{equation}
p_{u} \sim M_Z(\lambda^2,\lambda^2,\lambda^2).
\end{equation}
 We take $\lambda \sim \sqrt{\lqcd /M_Z}$ which implies that the typical ``off-shellness" of the ultrasoft degrees of freedom, $p_{u}^2 \sim M_Z^2 \lambda ^4 \sim \lqcd^2$, is set by the QCD scale while the typical ``off-shellness" of the collinear degrees of freedom, $p_c^2 \sim M_Z^2 \lambda^2 \sim M_Z \lqcd$, is much larger than $\lqcd^2$.
Hence the collinear degrees of freedom can be treated in perturbation theory.

Cone algorithms for jets, like that of Sterman and Weinberg, are ambiguous at higher orders in perturbation theory~\cite{Seymour,RUNIIJET}. This arises when there is more than one way to assign a particle to a particular jet. However, in this section we work to lowest order in perturbation theory, where the events consist of two almost back-to-back jets plus ultrasoft degrees of freedom. Since the cones are well separated, there is no ambiguity in assigning partons to the jets.

The nonperturbative effects we are after are characterized by matrix elements of operators composed from the ultrasoft degrees of freedom. In $Z$ decay into two jets, the jets are almost back-to-back, and $\mathbf{n}$ is  chosen along one of the jet directions. The degrees of freedom in the two jets are then represented by $n$-collinear (for the antiquark jet) and $\bar n$-collinear fields (for the quark jet). In this section we work to lowest order in perturbation theory in the collinear fields. Hence we match the weak neutral current in full QCD onto the effective theory at tree level,
\begin{equation}
\label{current}
j^{\mu}=[\bar \xi_{\bar n}W_{\bar n}]\Gamma^{\mu}[W_n^{\dagger}\xi_n],
\end{equation}
where $\Gamma^{\mu}=g_V\gamma^{\mu}_{\perp}+g_A\gamma^{\mu}_{\perp}\gamma_5$ involves the vector and axial couplings of the $Z$ boson. The fields $\xi_{\bar n}$ and $\xi_{n}$ are collinear quark fields in the $\bar n$ and $n$ directions and we have adopted the convention
\begin{equation}
\label{sumlabels}
\xi_n(x)=\sum_{\tilde p} e^{-i \tilde{p} \cdot x}\xi_{n, \tilde p}(x),
\end{equation}
where the label momentum $\widetilde p$ contains the components of order $1$ and $\lambda$, $\bar n \cdot p$ and $\mathbf{p}_\perp$, and the order $\lambda^2$ components are associated with the space-time dependence of the fields. The Wilson lines $W_{n, \bar n}$ are required to ensure collinear gauge invariance~\cite{Bauer:2001ct}. Since in this section we work to lowest order in QCD perturbation theory, they play no role in the analysis and can be set to unity.

The typical momenta of the partons in the jets are of the order of the collinear momenta, Eq.~(\ref{moment}), where the overall scale of their momentum is set by $M_Z$. However, it is possible for the jets to contain partons with momenta that have an overall scale that is much less than $M_Z$. Because of the sum over all values of $\widetilde p$ in Eq.~(\ref{sumlabels}), such partons can still be represented by collinear fields. The interaction of $n$-collinear fields among themselves is given by the full QCD Lagrangian, and so the hadronization of $n$-collinear partons into a jet is described by the full theory.

The Lagrangian of the effective theory does not contain any direct couplings between collinear particles moving in the two different lightlike directions labeled by $\bar n$ and $n$; however, they can interact via the exchange of ultrasoft gluons. It is convenient to remove the couplings of the collinear degrees of freedom to the ultrasoft ones via the field redefinition~\cite{SCET3}:
\begin{equation}
\label{def1}
\xi_{n} \rightarrow Y_n^{\dagger}\xi_n,\qquad A_n \rightarrow Y_n^{\dagger} A_n Y_n,
\end{equation}
where $A_n$ is an $n$-collinear gluon field and
\begin{equation}
\label{def2}
Y_n(z)=P \exp\left[ig\int_0^\infty {\rm d}s\ n \cdot A_{u}(ns+z)\right]
\end{equation}
denotes a path-ordered Wilson line of ultrasoft gluons in the $n$ direction from $s=0$ to $s=\infty$. This is the appropriate field redefinition for outgoing collinear fields, since if a factor of $\exp(-\epsilon s)$ is inserted in the integrand to decouple the interactions at late times, one reproduces the correct $i\epsilon$ prescription for the collinear quark propagator. For annihilation which contains incoming collinear particles $Y_n$ is from $s=-\infty$ to $s=0$ and the daggers are reversed in Eq.~(\ref{def1}). An analogous field redefinition with $n \to \bar n$ removes the couplings in the Lagrangian of ultrasoft fields to the $\bar n$-collinear fields. 

The differential decay rate for $Z$ decay to two jets is
\begin{equation}
\begin{split}
{\rm d}\Gamma_{\text{2-jet}}={1 \over 2 M_Z}\sum_{\rm {final ~states}}{1 \over 3}\sum_{\rm {\epsilon}}&\abs{ \me{ J_n J_{\bar n}X_{u} } { j^{\mu}(0)\epsilon_{\mu}} { 0 } }^2 \\
&\times (2\pi)^4\delta^4(p_Z-p_{J_{n}}-p_{J_{\bar n}}-k_{u}),
\label{1.01}
\end{split}
\end{equation}
where the sum over final states includes the usual phase space integrations and $\epsilon$ is the polarization vector of the decaying $Z$ boson. 
Since after the field redefinitions shown in Eq.~(\ref{def1}), there are no interactions between the ultrasoft and collinear degrees of freedom, the matrix element factorizes, and at lowest order in perturbation theory in the collinear degrees of freedom,
\begin{equation}
\begin{split}
\label{main}
{\rm d}\Gamma_{\text{2-jet}} =& {1 \over 2 M_Z}{{\rm d}^3{\bf p}_q \over (2 \pi)^3 2 p_q^0}{{\rm d}^3{\bf p}_{\bar q} \over (2 \pi)^3 2 p_{ \bar q}^0}\abs{ {\cal M}^{(0)}_{if} }^2  \sum_{X_{u}}(2\pi)^4\delta^4(p_Z-p_{q}-p_{\bar q}-k_{u})\\
& \times  {1 \over N_C}\me{ 0 }{\bar T [Y_{n d}~^{e} Y^{\dagger}_{{\bar n}e}~^a](0)}{X_{u}(k_{u})}
\me{X_{u}(k_{u})}{\vphantom{\Bigr|}T [Y_{\bar n a}~^{c} Y^{\dagger}_{{ n}c}~^d](0)}{0}.
\end{split}
\end{equation}
In Eq.~(\ref{main}), $|{\cal M}^{(0)}_{if}|^2$ is the square of the $Z \rightarrow q \bar q $ decay amplitude averaged over $Z$ polarizations and summed over the quark and antiquark spins and colors, $T $(${\bar T}$) denotes time- (anti-time-) ordering, $N_C$ is the number of colors, and we have explicitly displayed the color indices on the ultrasoft Wilson lines.
 
The derivation of Eq.~(\ref{main}) in many ways parallels the use of the operator product expansion to compute the deep inelastic scattering cross-section, or the rate for inclusive semileptonic $B$ decay. There is, however, one important distinction. The sum over final states in deep inelastic scattering and $B$ decay is a sum over a complete set of color-singlet hadron states. In Eq.~(\ref{1.01}), one is summing over a complete set of jet and ultrasoft states. These are a complete set of partonic states, and are not necessarily color-singlet states. In fact, unitarity would be violated if one separately imposed the color-singlet condition on each of $\ket{J_n}$, $\ket{J_{\bar n}}$ and $\ket{X_u}$. The derivation of Eq.~(\ref{main}) is valid to the extent that the sums over partonic and hadronic states are equivalent. In jet production, the color of the fast quark that turns into a jet is eventually transferred to low-energy partons during the fragmentation process. The low-energy partons communicate between the different jets, and make sure the whole process is color-singlet. The assumption is that this color recombination does not affect the decay rate at order $\lqcd/M_Z$.

To calculate ${\rm d}\Gamma_{\text{2-jet}}/ {\rm d} E_J$ we integrate Eq.~(\ref{main}) over the allowed values of the quark and antiquark three-momentum with the factor $\delta(E_J-p_q^0)$ inserted. This corresponds to choosing the quark jet as the ``observed'' jet. If one does not distinguish between quark and antiquark jets then Eq.~(\ref{main}) still applies since the value of ${\rm d}\Gamma_{\text{2-jet}}/ {\rm d} E_J$ when the ``observed'' jet is an antiquark jet is the same. It is convenient to work in the rest frame of the decaying $Z$, $p_Z=(M_Z,M_Z,{\bf 0}_{\perp})$, and align $\bf{\bar n}$ with the quark three-momentum ${\bf p}_{q}$.  The decomposition of the quark's four-momentum in terms of label and residual momentum, $p_q=\widetilde p_q +k_q$, has the form $p_q^+=\widetilde p_q^++k_q^+$ with $p_q^-=0$, ${\bf p}_{q \perp}=0$. (Note this means that $\widetilde {\bf p}_{q \perp}={\bf k}_{q \perp}=0$ and $k_q^-=0$.) Hence the phase space integration over quark three-momentum becomes
\begin{equation}
\int {{\rm d}^3 {\bf p}_q \over (2 \pi)^3 2 p_q^0}={1 \over 4(2\pi)^2}\sum_{\tilde p_q^+} \widetilde p_q^+ \int {\rm d} k_q^+.
\end{equation}
For the antiquark's four-momentum the decomposition into residual and label momentum is $p_{\bar q}^+=k_{\bar q}^+$, $p_{\bar q}^-={\widetilde p}_{\bar q}^-+k_{\bar q}^-$ and  ${\bf p}_{\bar q \perp}=\widetilde {\bf p}_{\bar q \perp}+{\bf k}_{\bar q \perp}$. One cannot set ${\bf p}_{\bar q \perp}=0$ by a choice of $\mathbf{n}$, since $\mathbf{n}=-\mathbf{\bar n}$, and $\mathbf{\bar n}$ has already been fixed by the direction of the quark jet.

Expressed in terms of label and residual momenta the phase space integration over antiquark three-momentum is
\begin{eqnarray}
\int {{\rm d}^3 {\bf p}_{\bar q} \over (2 \pi)^3 2 p_{\bar q}^0}&=&\sum_{\tilde p_{\bar q}}\int {{\rm d}^4 {k}_{\bar q} \over (2 \pi)^3 }\delta\bigl(({\widetilde p_{\bar q}}+k_{\bar q})^2\bigr) \nn
&=& \sum_{\tilde p_{\bar q}}\int {{\rm d}k_{\bar q}^- {\rm d^2}{\bf k}_{\bar q \perp} \over 2(2\pi)^3}{1 \over \widetilde p_{\bar q^-}}.
\end{eqnarray}
Here the delta function $\delta\bigl(({\widetilde p_{\bar q}}+k_{\bar q})^2\bigr)=\delta(\widetilde p_{\bar q}^- k_{\bar q}^+ -\widetilde {\bf p}_{\bar q \perp}^2)$ was used to do the $k_{\bar q}^+$ integration setting $k_{\bar q}^+=\widetilde {\bf p}_{\bar q \perp}^2/\widetilde p_{\bar q}^-$. 
At leading order in the SCET expansion parameter $\lambda$ the invariant matrix element ${\cal M}^{(0)}_{if}$ only depends on the label momenta $\widetilde p_q^+$ and $ \widetilde p_{\bar q}^-$. In terms of label and residual momentum the energy-momentum conserving delta function becomes:
\begin{equation}
\begin{split}
\label{delta1}
\delta^4(p_Z-p_q-p_{\bar q}-k_{u})&=2\delta(p_Z^--p_{\bar q}^--k_{u}^-)\delta(p_Z^+-p_q^+-p_{\bar q}^+-k_{u}^+) \delta^2({\bf p}_{\bar q \perp}+{\bf k}_{u \perp})\\
&=2\delta_{M_Z,\tilde p_{\bar q}^-}\delta_{M_Z,\tilde p_{q}^+} \delta^2_{{\tilde {\bf p}}_{\bar q \perp},{\bf 0}}\delta(k_{\bar q}^-+k_{u}^-)\delta(k_q^+ + k_{u}^+)  \delta^2({\bf k}_{\bar q \perp}+{\bf k}_{u \perp}).
\end{split}
\end{equation}
The relation  $k_{\bar q}^+=\widetilde {\bf p}_{\bar q \perp}^2/\widetilde p_{\bar q}^-$ and the Kronecker delta that sets $\widetilde{\bf p}_{\bar q \perp}$ to zero imply that $k_{\bar q}^+=0$, and so this variable does not appear in the penultimate delta function in Eq.~(\ref{delta1}).

Using these results gives:
\begin{equation}
\begin{split}
\label{almost}
{{\rm d} \Gamma_{\text{2-jet}} \over {\rm d } E_J}&={|{\cal M}^{(0)}_{if}|^2 \over 8M_Z (2\pi)}\int {\rm d}k_q^+\int {\rm d}k_{\bar q}^-{\rm d^2}{\bf k}_{ \bar q \perp}\sum_{X_{u}} \delta\left(\frac{M_Z}{2}-E_J+\frac{k_q^+}{2}\right) \\
&\quad\times\delta(k_{\bar q}^-+k_{u}^-)\delta(k_q^+ + k_{u}^+) \delta^2({\bf k}_{\bar q \perp}+{\bf k}_{u \perp}) \\
&\quad\times {1 \over N_C}\me {0 } { \bar T [Y_{n d}~^{e} Y^{\dagger}_{{\bar n}e}~^a](0)}{X_{u}(k_{u})}
\me{ X_{u}(k_{u})}{ \vphantom{\Bigr|} T [Y_{\bar n a}~^{c} Y^{\dagger}_{{ n}c}~^d](0)}{0}\\
&= {|{\cal M}^{(0)}_{if}|^2 \over 16 \pi M_Z} \sum_{X_{u}} \delta\left(\frac{M_Z}{2}-E_J-\frac{k_u^+}{2}\right) \\
&\quad\times \frac{1}{N_C}\me {0 } { \bar T [Y_{n d}~^{e} Y^{\dagger}_{{\bar n}e}~^a](0)}{X_{u}(k_{u})}
\me{ X_{u}(k_{u})}{ \vphantom{\Bigr|} T [Y_{\bar n a}~^{c} Y^{\dagger}_{{ n}c}~^d](0)}{0}.
\end{split}
\end{equation}
We write the remaining delta function as the integral:
\begin{equation}
\delta\left(\frac{M_Z}{2}-E_J-\frac{k_u^+}{2}\right)=\int \frac{{{\rm d}u }}{2\pi}\exp \left[-i\left(\frac{M_Z}{2}-E_J-\frac{k_u^+}{2}\right){u}\right]\,.
\end{equation}

At this stage the collinear degrees of freedom have been integrated out, and the matrix elements above, which involve only ultrasoft degrees of freedom, are evaluated at leading order in the SCET expansion parameter (i.e. $\lambda \rightarrow 0$). Recall that the Sterman-Weinberg jet criteria restrict particles outside the cones used to define the two jets associated with the quark and antiquark to have energy less than $E_{\text{cut}}$ which we are taking to be order $\lambda M_Z$. In the limit $\lambda \rightarrow 0$ this energy cut becomes much larger than  a typical component of an ultrasoft four-momentum. Hence, for the matrix elements of these operators, $E_{\text{cut}}$ should be taken to infinity and does not restrict these matrix elements. Similarly the cone angle is taken to be of order $\lambda$ while the typical angle between components of ultrasoft momenta is order unity. Thus the cone angle should be taken to zero in the effective theory that contains only ultrasoft degrees of freedom  and so  there is no restriction on the ultrasoft states that are summed over in Eq.~(\ref{almost}).

Using the exponential dependence on $k_{u}$ to translate the anti-time ordered product to the space-time point $u n/2$, and then using the completeness relation to perform the sum over all ultrasoft intermediate states, 
we find for the jet energy distribution:
\begin{equation}
\label{doneit}
{{\rm d} \Gamma_{\text{2-jet}} \over {\rm d } E_J}=\Gamma_{\text{2-jet}}^{(0)}\ S(M_Z/2-E_J),
\end{equation}
where the shape function $S$ is defined by \cite{KS95}
\begin{equation}
\label{shapefunction}
S(k)={1 \over N_C}\int {{\rm d}u \over 2 \pi}{\rm e}^{i k u}
\me{ 0 }{ \bar T [Y_{n d}~^{e} Y^{\dagger}_{{\bar n}e}~^a](un/2)T [Y_{\bar n a}~^{c} Y^{\dagger}_{{ n}c}~^d](0)}{ 0},
\end{equation}
and the total two jet Z-decay width at lowest order in perturbation theory is
\begin{equation}
\Gamma_{\text{2-jet}}^{(0)}={ \abs{ {\cal M}^{(0)}_{if}}^2  \over 16 \pi M_Z }={N_C M_Z \over 12 \pi}(g_V^2+g_A^2),
\end{equation}
having implicitly summed over spins and colors.
The $n$-directed and $\bar n$-directed ultrasoft Wilson lines commute since $(s_1n-s_2\bar n)^2=-4s_1s_2<0$, and the gauge fields in the Wilson lines are space-like separated. 

In this derivation we chose the jets to be composed entirely of collinear degrees of freedom. This is appropriate since jets are confined to narrow cones. For example, the momentum of any massless particle in the quark jet satisfies $p^- \ll p^+$, which is the appropriate scaling for collinear particles in the $\bar n$ direction. However, it is possible to repeat the above derivation allowing ultrasoft degrees of freedom to be inside a jet. Then instead of inserting $\delta(E_J - p_q^0)$ into Eq.~(\ref{main}), one inserts $\delta(E_J - p_q^0 - k^0_{uJ})$, where $k^0_{uJ} = (k^+_{uJ}/2)[1+{\cal O}(\lambda)]$ denotes the total ultrasoft energy inside the quark jet. Using the delta functions in Eq.~(\ref{delta1}) we obtain again Eq.~(\ref{almost}), with $k_u^+$ in the final delta function now denoting the total ultrasoft momentum \emph{outside} the quark jet. However, as mentioned previously, at leading order in $\lambda$ the cone angle of the jet shrinks to zero, and one recovers the previous result. 

It is possible to remove the time- and anti-time-ordering completely in the definition of the shape function $S$.  
Using the results from Appendix~\ref{A1} our expression for the shape function becomes
\begin{equation}
\label{notimeorder}
\begin{split}
S(k)&={1 \over N_C}\int {{\rm d}u \over 2 \pi}{\rm e
}^{i k u}\me{ 0} { [{\overline Y^{\dagger}_n}^e~_d Y^{\dagger}_{{\bar n}e}~^a](un / 2)[Y_{\bar n a}~^{c} {\overline Y_n}^d~_c](0)}{0} \\
&={1\over N_C}\me{0}{[{\overline Y^{\dagger}_n}^e~_d Y^{\dagger}_{{\bar n}e}~^a]\delta(k+in \cdot \partial/2)[Y_{\bar n a}~^{c} {\overline Y_n}^d~_c]}{0},
\end{split}
\end{equation}
where the overline denotes an anti-triplet Wilson line.  

Since in the kinematic region of interest $M_Z/2-E_J$ is much larger than $n\cdot \partial$ acting on ultrasoft gauge fields it is appropriate to expand the delta function above which gives
\begin{equation}
\begin{split}
\label{expand}
S(M_Z/2-E_J)&=\delta(M_Z/2-E_J)+\delta^{\prime}(M_Z/2-E_J)\me{0}{O_1}{0} \\
&\quad +{1 \over 2}\delta^{\prime \prime}(M_Z/2-E_J)\me{ 0}{O_2}{0}+\cdots
\end{split}
\end{equation}
where
\begin{equation}
\label{operators}
O_m={1\over N_C}\left[{\overline Y^{\dagger}_n}^e~_d Y^{\dagger}_{{\bar n}e}~^a\right]\left({i n \cdot \partial \over 2 }\right)^m\left[Y_{\bar n a}~^{c} {\overline Y_n}^d~_c\right]={1\over N_C}{\rm Tr}\,\left[Y^{\dagger}_{\bar n}\left( {in \cdot D \over 2}\right)^mY_{\bar n }\right].
\end{equation}
The simple form for the operators $O_m$ arises because the variable $E_J$ is totally inclusive on the ``unobserved'' antiquark jet. 

The formula for ${\rm d}\Gamma_{\text{2-jet}}/{\rm d}E_J$ is
\begin{equation}
{{\rm d}\Gamma_{\text{2-jet}}\over {\rm d}E_J}=\Gamma^{(0)}_{\text{2-jet}}\ \Bigl[ \delta(M_Z/2-E_J) +\delta^{\prime}(M_Z/2-E_J)\me{0}{O_1}{0} + \cdots \Bigr]  .
\label{pert}
\end{equation}
The delta function term in Eq.~(\ref{expand}) simply reproduces the leading perturbative formula for ${\rm d}\Gamma^{(0)}_{\text{2-jet}}/{\rm d}E_J$ while the higher-order terms contain the effects of nonperturbative physics. 
The derivation presented here assumes the observed jet is the quark jet. A similar derivation in the case where the antiquark jet is observed gives operators
\begin{equation}
{\overline  O_m}={1 \over N_C}{\rm Tr} \left[{\overline Y_n}^{\dagger}\left({i{\bar n} \cdot D \over 2}\right)^m{\overline Y_{ n }}\right].
\end{equation}
Since the vacuum expectation values of $O_m$ and ${\overline O_m}$ are equal by charge conjugation, our results also hold in the case where one does not distinguish between quark and antiquark jets.

We define the matrix elements using dimensional regularization with $\overline{{\rm MS}}$ subtraction so that in perturbation theory the vacuum expectation values $\me{ 0 }{O_m}{0 }$ are zero. 

Note that $O_2$ is a very different operator than $O_1$ so it is not possible to capture the effects of nonperturbative physics for $|E_J -M_Z/2|\sim \lambda^2 M_Z$ \footnote{More correctly the differential cross section ${\rm d}\Gamma_{\text{2-jet}}/{\rm d}E_J$ smeared over a region $\Delta$ of energy (that contains $E_J=M_Z/2$) with $\Delta$ of order $\lambda^2M_Z$.} simply by taking the lowest order perturbative formula in Eq.~(\ref{pert}) and shifting $E_J$ by a nonperturbative parameter $\mu_{\text{np}}$, that is,  $E_J\rightarrow E_J-\mu_{\text{np}}$.  This ansatz results in the shape function
\begin{equation}
S(M_Z/2-E_J)=\delta(M_Z/2-E_J)+\delta^{\prime}(M_Z/2-E_J)\mu_{\text{np}}
+{1 \over 2}\delta^{\prime \prime}(M_Z/2-E_J)\mu_{\text{np}}^2+\cdots
\label{3.01}
\end{equation}
where the series of derivatives of delta functions has coefficients that are simply related by $\me{ 0}{O_m}{0} =\me{ 0}{O_1}{0}^m$, which is not correct. 

For $|E_J -M_Z/2| \sim \lambda^2 M_Z$ all terms in the series of Eq.~(\ref{expand}) are equally important. However for $|E_J -M_Z/2|\sim \Delta\gg \lambda^2 M_Z$ the vacuum expectation value of $O_1$ provides the leading  order $\lqcd/\Delta$ nonperturbative correction. In this kinematic region the shift $E_J \rightarrow E_J-\mu_{\text{np}}$, with $\mu_{\text{np}}=\langle 0|O_1|0\rangle$, correctly captures the most important effects of nonperturbative physics.

We have focused on nonperturbative effects that are enhanced in the region near $E_J=M_Z/2$. If one considers a variable like the average value of the jet energy over the entire allowed phase space, then there are sources of nonperturbative corrections that we have not considered.

Using the results of Appendix~\ref{A2}, the operator $O_1$ in Eq.~(\ref{operators}) can be expressed in terms of the gluon field strength tensor \cite{KS95}:
\begin{equation}
\label{fancy}
\begin{split}
O_1&={1 \over 2}{\rm Tr} [Y^{\dagger}_{\bar n}(in~ \cdot D)Y_{\bar n }] \\
&=\frac{1}{2}{\rm Tr}\left[ig\int_0^{\infty}{\rm d}s\ Y_{\bar n}^{\dagger}(z;s,0)n^{\mu}{\bar n}^{\nu}G_{\mu \nu} Y_{\bar n}(z;s,0)\right].
\end{split}
\end{equation}
$O_1$ in Eq.~(\ref{fancy}) vanishes if the ultrasoft gauge field is treated as a classical degree of freedom. Then the Wilson lines in Eq. (\ref{fancy}) are unitary matrices and the trace vanishes since the gluon field strength tensor is in the adjoint representation. Note that  the  vacuum expectation value of $O_1$ can still be nonzero because of quantum effects. Usually operators involving products of gluon fields require renormalization. However, it is straightforward  to show that $O_1$ is not renormalized at one loop. For example, we have shown that the two-gluon matrix element of the operator $O_1$ shown in Fig.~\ref{O1loop} is identically zero, even before performing the loop integration. That is,
\begin{equation}
\bra{0}O_1\lvert A_\alpha^a(\epsilon_1,p_1)A_\beta^b(\epsilon,p_2)\rangle = 0,
\end{equation}
after extensive algebra.
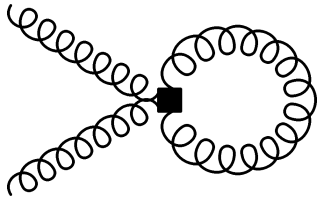
\begin{figure}
\vspace{-5cm}
\begin{equation*}
\raisebox{-5cm}{\rotatebox{-90}{
\begin{fmffile}{O1}
\begin{fmfgraph*}(100,120)
\fmfleft{A1}
\fmfright{A2}
\fmf{gluon}{A1,v,v,A2}
\fmfv{decor.shape=square,decor.filled=1,decor.size=70}{v}
\fmfforce{(.5w,.5h)}{v}
\fmfforce{(.142857w,0)}{A1}
\fmfforce{(.857142w,0)}{A2}
\end{fmfgraph*}
\end{fmffile}}}
\end{equation*}
\vspace{-5mm}
\caption[Two-gluon matrix element of $O_1$ at one-loop order.]{Two-gluon matrix element of $O_1$ at one-loop order. Matrix elements of $O_1$ such as this are identically zero in perturbation theory; hence, $O_1$ is not renormalized.}
\label{O1loop}
\end{figure}

\section{Enhanced Nonperturbative Corrections to Event Shape Variables \label{sec:other}}

There are a number of event shape distributions that are commonly studied in the literature.  Conventionally, one defines a general event shape distribution ${\rm d} \sigma/{\rm d}e$, where $e$ is an event shape variable defined such that the region $e \to 0$ corresponds to the two jet limit. Examples are $e=1-T$ for thrust, $e=B$ for jet broadening and $e=C$ for the $C$ parameter. Any event shape distribution in $Z$ decay contains both perturbative and nonperturbative contributions. The perturbative effects can be computed as a perturbation series in $\alpha_s(M_Z)$. At leading order, only two-jet (i.e.\ $q \bar q$) events contribute. Events with more hard partons are suppressed by powers of $\alpha_s(M_Z)$. In general, nonperturbative effects are suppressed by  powers of $\Lambda_{\rm QCD}/M_Z$, but in corners of phase space where $e \ll 1$ these nonperturbative effects become enhanced. Here we consider the region $\Lambda_{\rm QCD} \ll M_Z e \ll M_Z$ and focus on the enhanced nonperturbative contribution suppressed only by a single power of $\Lambda_{\rm QCD}/(M_Z e)$. 

Perturbative expressions for the jet variables considered in this section have been extensively studied in the literature \cite{pertThrust,pertJetMass,pertEvent,pertB,pertB2,pertC}. Our main interest is in nonperturbative physics. Working to leading order in $\alpha_s(M_Z)$, the dominant nonperturbative effects are corrections to the two-jet distribution. Nonperturbative corrections to higher-order processes are suppressed by additional powers of $\alpha_s(M_Z)$. We will compute the enhanced nonperturbative corrections to some commonly measured event shape distributions, just as we did for the jet energy distribution in Sec.~\ref{sec:ope}. Recall for the jet energy distribution the dominant nonperturbative correction came from expanding 
\begin{equation}
\label{EJdelta}
\delta\left(\frac{M_Z}{2}-E_J+ \frac{k_q^+}{2}\right) = \delta\left(\frac{M_Z}{2}-E_J\right) +\delta^{\prime}\left(\frac{M_Z}{2}-E_J\right)\frac{k_q^+}{2}+\cdots
\end{equation}
in Eq.~(\ref{almost}) to linear order in $k_q^+$. The delta function from Eq.~(\ref{delta1}) sets $k_q^+=-k_{u}^+$, and we therefore find
\begin{equation}
\label{poor1}
{{\rm d}\Gamma^{(0)}_{\text{2-jet}}\over {\rm d}E_J}=\Gamma^{(0)}_{\text{2-jet}}\left[\delta\left(\!\frac{M_Z}{2}-E_J\!\right)\!-\!\delta^{\prime}\left(\!\frac{M_Z}{2}-E_J\!\right)\!\frac{\vev{ k_{u}^+ }}{2} \right],
\end{equation}
where
\begin{equation}
\label{poor2}
\vev{ k_{u}^+ } = \sum_{X_{u}} {1 \over N_C}\me{ 0}{\bar T \left[Y_{n d}~^{e} Y^{\dagger}_{{\bar n}e}~^a \right] (0)}{X_{u}(k_{u})}\me{ X_{u}(k_{u})}{\vphantom{\Bigr|} T \left[Y_{\bar n a}~^{c} Y^{\dagger}_{{ n}c}~^d \right](0) }{0}  k_{u}^+.
\end{equation}
The jet energy distribution has the nice property that one can write $\vev{ k_{u}^+ }$ as the vacuum expectation value of an operator involving Wilson lines of ultrasoft gauge fields [namely, Eq.~(\ref{fancy})]. For some shape variables this is not possible. However, expressions analogous to Eqs.~(\ref{poor1}--\ref{poor2})  can be derived.

\subsection{Thrust \label{ssec:thrust}}

First we consider the thrust distribution ${\rm d } \Gamma/{\rm d} T$ where the thrust $T$ is defined by
\begin{equation}
M_Z\, T=\max_{\mathbf{\hat t}} \sum_{i} \abs{\mathbf{\hat t}\cdot{\bf p_i}},
\end{equation}
where $\mathbf{\hat t}$ is a unit vector that defines the thrust axis. The maximum is taken over all possible directions of $\mathbf{\hat t}$, and the sum is over all final state particles.
To the order we are working the thrust axis $\mathbf{\hat t}$ can be set equal to the spatial part of the lightlike 
four-vector $n$ used to define the collinear antiquark field. It is convenient to call this direction the $z$-axis. The thrust distribution is calculated analogously to the two jet distribution except that the delta function $\delta(E_J-p_q^0)$  is replaced by  $\delta\left(M_ZT-\abs{p_q^z}- \abs{p_{ \bar q}^z}-\sum_\alpha \abs{k_{u\alpha}^z}\right)$, where the sum is over all ultrasoft particles. We adopt the same conventions as in the jet energy distribution so that the phase space integrals are again done using the delta function in Eq.~(\ref{delta1}). Decomposing the total ultrasoft four-momentum, $k_{u}=k_{u}^{(a)}+k_{u}^{(b)}$, into the sum of the ultrasoft momentum from particles in the same hemisphere as the antiquark (type $a$) and the same hemisphere as the quark (type $b$) the thrust $T$ can be written as
\begin{equation}
\label{Tdelta}
\begin{split}
M_ZT &= \abs{p_q^z} +  \abs{p_{ \bar q}^z}+ \sum_\alpha \abs{k_{u\alpha}^z} \\
&= {1 \over 2}\left(p_{\bar q}^- -p_{\bar q}^+\right)- {1 \over 2}\left(p_{q}^- -p_{q}^+\right) +{1 \over 2}\left(k_{u}^{(a)-}- k_{u}^{(a)+}\right) -{1 \over 2}\left(k_{u}^{(b)-}- k_{u}^{(b)+}\right) \\
&={1 \over 2}\widetilde p_{\bar q}^- + {1 \over 2}\widetilde p_{q}^+ +{1 \over 2}\left(k_{\bar q}^- - k_{\bar q}^+\right)+{1 \over 2}k_q^+ + {1 \over 2}\left(k_{u}^{(a)-}- k_{u}^{(a)+}\right) -{1 \over 2}\left(k_{u}^{(b)-}- k_{u}^{(b)+}\right).
\end{split}
\end{equation}
Now the delta functions in Eq.~(\ref{delta1}) set $\widetilde p_{\bar q}^- = \widetilde p_q^+ = M_Z$, $k_{\bar q}^- = -k_u^-$, $k_q^+ = -k_u^+$, and $k_{\bar q}^+ = 0$.
Thus we find
\begin{equation}
\label{thrustdef}
T = 1 - \frac{1}{M_Z} \left(k_{u}^{(a)+} + k_{u}^{(b)-}\right),
\end{equation}
where we have also used  $k_u = k_u^{(a)} + k_u^{(b)}$. Thus,
\begin{equation}
\label{thrust}
\begin{split}
{{\rm d} \Gamma \over {\rm d} T} &= \Gamma^{(0)}_{\text{2-jet}}\left[\delta(1-T) 
- \delta^{\prime}(1-T)\frac{\vev{ k_{u}^{(a)+} +k_{u}^{(b)-}}}{M_Z} \right]\\
&\equiv \Gamma^{(0)}_{\text{2-jet}}\left[\delta(1-T) 
- \delta^{\prime}(1-T)\frac{\vev{O_1^T}}{M_Z} \right].
\end{split}
\end{equation}
The thrust axis and the hemispheres are determined by the jet directions, and can be defined in terms of the label momenta of the quark and antiquark. Thus $\mathbf{\hat t}$ and the hemispheres $a$ and $b$ are label variables. Nevertheless, because of the hemisphere condition on the ultrasoft momentum in Eq.~(\ref{thrust}), there isn't a simple formula expressing the correction in terms of the vacuum expectation value of an operator involving Wilson lines like the one in Eq.~(\ref{fancy}).  

In a region $|1-T|\sim \lambda ^2$ the higher order terms in the ultrasoft momentum that were neglected in Eq.~(\ref{thrust}) are important. Eq.~(\ref{thrust}) is appropriate for a region $\delta T$ near $T=1$  that satisfies $1 \gg \delta T \gg \lambda ^2$,  for example, $\delta T  \sim \lambda$.

\subsection{Jet Masses \label{ssec:jmass}}

The squared jet masses $M^2_{a,b}$ are the squares of the invariant mass of all the particles in the two hemispheres $a$ and $b$, defined by the plane perpendicular to the thrust axis. Two commonly used variables are the sum of jet masses, $\widehat{M}_S^2=(M_a^2+M_b^2)/M_Z^2$, and the heavy jet mass $\widehat{M}^2_H=\max(M_a^2,M_b^2)/M_Z^2$. The jet masses are $M^2_a=(p_{\bar q}+k_{u}^{(a)})^2$ and $M^2_b=(p_{ q}+k_{u}^{(b)})^2$. More explicitly,
\begin{subequations}
\label{MaMb}
\begin{align}
M_a^2 &= (p_{\bar q}^+ + k_u^{(a)+})(p_{\bar q}^- + k_u^{(a)-}) - (\vect{p}_{\bar q\perp} + \vect{k}_{u\perp}^{(a)})^2 \\
M_b^2 &= (p_q^+ + k_u^{(b)+})(p_q^- + k_u^{(b)-}) - (\vect{p}_{q\perp} + \vect{k}_{u\perp}^{(b)})^2.
\end{align}
\end{subequations}
Recall that $\vect{p}_q$ is aligned along $\vect{\bar n}$ so that $\vect{p}_{q\perp}=0$. Also, the delta function in Eq.~(\ref{delta1}) sets $\vect{\widetilde p}_{\bar q\perp} = 0$ and $\widetilde p_{\bar q}^- = \widetilde p_q^+ = M_Z$. Then, working to linear order in the ultrasoft momenta, $M^2_a = M_Z k_{u}^{(a)+}$ and $M^2_b = M_Z k_{u}^{(b)-}$, so
\begin{subequations}
\begin{align}
{{\rm d} \Gamma \over {\rm d} \widehat{M}_S^2}&=\Gamma^{(0)}_{\text{2-jet}}\left[\delta(\widehat{M}^2_S)- \delta^{\prime}(\widehat{M}_S^2)\frac{\vev{O_1^{M_S}}}{M_Z} \right]\\
{{\rm d} \Gamma \over {\rm d} \widehat{M}_H^2}&=\Gamma^{(0)}_{\text{2-jet}}\Bigl[\delta(\widehat{M}^2_H)e
-  \delta^{\prime}(\widehat{M}_H^2)\frac{\vev{O_1^{M_H}}}{M_Z} \Bigr],
\end{align}
\end{subequations}
where
\begin{subequations}
\begin{align}
\vev{O_1^{M_S}} &= \vev{ k_{u}^{(a)+}+k_{u}^{(b)-}}\\
\vev{O_1^{M_H}} &= \vev{ \max\left( k_{u}^{(a)+},k_{u}^{(b)-}\right)}.
\end{align}
\end{subequations}
Note that in the kinematic region where expanding to linear order in ultrasoft and residual momentum is appropriate, the nonperturbative corrections to the $M_S^2$ and $1-T$ distributions are given by the same nonperturbative matrix element. The nonperturbative corrections to the $M_S^2$ and $M_H^2$ distributions are different.

Working to higher orders in $k_u/M_Z$, the definitions of thrust in Eq.~(\ref{thrustdef}) and of jet masses in Eq.~(\ref{MaMb}) become different beyond linear order. However, the corrections to event shape distributions at higher orders in $\Lambda_{\text{QCD}}/(M_Ze)$ come not from expanding the argument of the delta functions used to define these variables to higher orders in $k_u/M_Z$, but rather from expanding these delta functions as power series in the ultrasoft momentum, as in Eq.~(\ref{EJdelta}) for the jet energy. So even at higher orders, the enhanced nonperturbative corrections, i.e. of order $[\Lambda_{\text{QCD}}/(M_Ze)]^n$, $n>1$, come from the leading-order correction to the argument of the delta function, which are the same for thrust and jet mass sum. So the enhanced nonperturbative corrections to thrust and jet mass sum are related to all orders in $\Lambda_{\text{QCD}}/(M_Ze)$.
 
\subsection{Jet Broadening}

Jet broadening variables $B_{a,b}$ are  defined by
\begin{eqnarray}
B_{a,b} = {1 \over 2 M_Z} \sum_{i \in a,b} \abs{ \mathbf{p}_i \times \mathbf{\hat t} },
\label{3.16}
\end{eqnarray}
where the hemispheres $a$ and $b$ are defined as before, and $\mathbf{\hat t}$ is the thrust axis. The jet broadening variables at order $k_u/M_Z$ require knowing the thrust axis to order $k_u/M_Z$. The thrust axis $\vect{\hat t}$ maximizes $\sum_i\abs{\vect{\hat t}\cdot\vect{p}_i}$.

The angle between $\vect{p}_{\bar q}$ and the $z$-axis is given by 
\begin{equation}
\label{thetaqbar}
\theta_{\bar q} = \frac{\abs{\vect{k}_{\bar q\perp}}}{M_Z/2},
\end{equation}
and the thrust axis $\vect{\hat t}$ can be written as
\begin{equation}
\vect{\hat t} = (0,-\sin\theta_t,\cos\theta_t).
\end{equation}
By symmetry,
\begin{equation}
\label{thetat}
\theta_t = \frac{\abs{\vect{k}_{\bar q\perp}}}{M_Z}\,,
\end{equation}
which is half the size of $\theta_{\bar q}$ (see Fig.~\ref{thrustaxis}).
\begin{figure}
\begin{center}
\includegraphics[width=10cm]{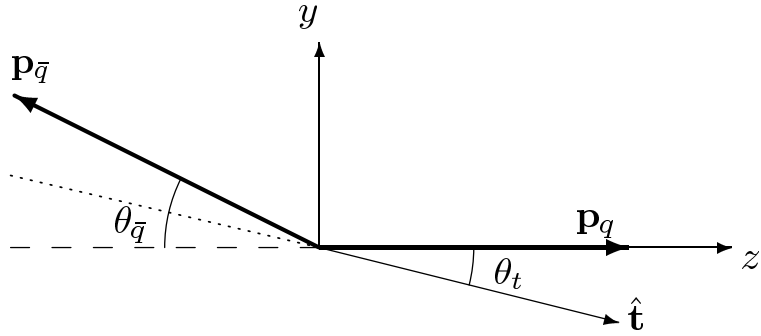}
\end{center}
\caption[Determination of the thrust axis.]{Determination of the thrust axis. To the order we are working, the quark and antiquark have momenta $\abs{\vect{p}_q}=\abs{\vect{p}_{\bar q}}= M_Z/2$. The antiquark then makes an angle $\theta_{\bar q}= 2\abs{\vect{k}_{\bar q\perp}}/M_Z$ with the $z$-axis, and the thrust axis $\vect{\hat t}$ makes an angle $\theta_t = \abs{\vect{k}_{\bar q\perp}}/M_Z$ with both the quark and antiquark.}
\label{thrustaxis}
\end{figure}

Now calculate $\abs{\vect{p}_i\times\vect{\hat t}}$ for each particle. To linear order in $k_u/M_Z$ we find for the quark,
\begin{equation}
\label{quarkcross}
\abs{\vect{p}_q\times\vect{\hat t}} = \frac{M_Z}{2}\sin\theta_t= \frac{\abs{\vect{k}_{\bar q\perp}}}{2},
\end{equation}
and for the antiquark,
\begin{equation}
\label{antiquarkcross}
\abs{\vect{p}_{\bar q}\times\vect{\hat t}} = \abs{\vect{k}_{\bar q\perp}}\cos\theta_t - \frac{M_Z}{2}\sin\theta_t  = \frac{\abs{\vect{k}_{\bar q\perp}}}{2}.
\end{equation}
For each ultrasoft particle $\alpha$, the cross product $\vect{k}_\alpha\times\vect{\hat t}$ is given by the determinant
\begin{equation}
\left|
\begin{array}{ccc}
\vect{\hat x} & \vect{\hat y} & \vect{\hat z} \\
k_\alpha^x & k_\alpha^y & k_\alpha^z \\
0 & -\sin\theta_t & \cos\theta_t
\end{array}
\right|.
\end{equation}
Since $\sin\theta_t$ is already of order $k_u/M_Z$ the cross product is, to linear order in $k_u/M_Z$,
\begin{equation}
\vect{k}_\alpha\times\vect{\hat t} = (k_\alpha^y,-k_\alpha^x, 0),
\end{equation}
so $\abs{\vect{k}_\alpha\times\vect{\hat t}} = \abs{{\vect{k}}_{\alpha\perp}}$. Combining the contributions of each particle to the sum in Eq.~(\ref{3.16}), and using the delta function in Eq.~(\ref{delta1}) to set $\vect{k}_{\bar q\perp} = -\vect{k}_{u\perp}$, we obtain for the jet broadening variables (to linear order in $k_u/M_Z$):
\begin{subequations}
\begin{align}
B_a &= {1 \over 2M_Z} \left(\frac{\abs{ \mathbf{k}_{u\perp} }}{2} + \sum_{\alpha \in a} \abs{\mathbf{k}_{\alpha \perp} } \right) ,\\
B_b &= {1 \over 2M_Z} \left(\frac{\abs{ \mathbf{k}_{u\perp} }}{2} + \sum_{\alpha \in b} \abs{\mathbf{k}_{\alpha \perp} } \right), 
\end{align}
\label{BaBb}
\end{subequations}
where the sum on $\alpha$ is over the ultrasoft particles in hemisphere $a$ or $b$.

One conventionally defines two other broadening variables as 
\begin{subequations}
\label{Bsum}
\begin{align}
B_{\text{max}}&=\max\left(B_a,B_b\right),\\
B_{\text{sum}}&=B_a+B_b.
\end{align}
\end{subequations}
The jet broadening distribution is
\begin{eqnarray}
{{\rm d} \Gamma \over {\rm d}B} &=&  \Gamma^{(0)}_{\text{2-jet}} \Bigl[\delta(B) - \delta^{\prime}(B)\frac{\vev{ O_1^B }}{M_Z} \Bigr],
\end{eqnarray}
for $B_{a,b,\text{sum},\text{max}}$, where $\vev{ O_1^B } \equiv M_Z\vev{B}$ is the matrix element of the appropriate quantity in Eqs.~(\ref{BaBb},\ref{Bsum}). Nonperturbative effects in the jet broadening measures are not related to the jet energy or thrust.

In this paper, we have assumed that the nonperturbative physics is completely described by ultrasoft degrees of freedom. It is possible that some of the subtleties associated with nonperturbative corrections to the jet broadening variables that have been discussed in the literature \cite{DokMarSal} can be attributed to nonperturbative effects in the collinear sector, which we have not included.

\subsection{$C$ Parameter}

The $C$ parameter is defined as
\begin{eqnarray}
C &=& 3 \left( \lambda_1 \lambda_2 + \lambda_2 \lambda_3 + \lambda_3 \lambda_1 \right),
\label{3.19}
\end{eqnarray}
where $\lambda_i$ are the eigenvalues of
\begin{eqnarray}
\theta^{rs} &=& {1 \over M_Z} \sum_i {\mathbf{p}_i^r \mathbf{p}_i^s \over \abs{\mathbf{p}_i}},
\label{3.20}
\end{eqnarray}
and $r,s=1,2,3$ are the space components of the momentum $\mathbf{p}_i$ of the $i^{\text{th}}$ particle. 

The largest component of $\theta^{rs}$ is $\theta^{zz}$. The quark and antiquark in the jets have $z$-momentum $p_q^z=-{p}_{\bar q}^z = M_Z/2$ to the order we are working. Then, to linear order in $k_u/M_Z$, the eigenvalues of $\theta^{rs}$ are given by:
\begin{eqnarray}
\label{determinant}
\det(\theta-\lambda I) = (1-\lambda) \, {\rm det} (X - \lambda I),
\end{eqnarray}
where $I$ is the identity matrix and 
\begin{subequations}
\begin{align}
X_{11}&=\sum_\alpha\frac{k_\alpha^{x2}}{M_Z\abs{\vect{k}_\alpha}},\\
X_{22}&=\sum_\alpha\frac{k_\alpha^{y2}}{M_Z\abs{\vect{k}_\alpha}},\\
X_{12} = X_{21} &= \sum_\alpha\frac{k_\alpha^x k_\alpha^y}{M_Z\abs{\vect{k}_\alpha}}.
\end{align}
\end{subequations}
Here the sums over $\alpha$ are only over ultrasoft particles. (The contributions from the quark and antiquark to these components of $\theta^{rs}$ are suppressed by another factor of $1/M_Z$, since $\abs{\vect{p}_q}=\abs{\vect{p}_{\bar q}} = M_Z/2$.)

The largest eigenvalue is $\lambda_1 = 1$, and the other two eigenvalues satisfy
\begin{equation}
\lambda_2 + \lambda_3 = \frac{1}{M_Z}\sum_\alpha\frac{(k_\alpha^{x})^2 + (k_\alpha^{y})^2}{\abs{\vect{k}_\alpha}}.
\end{equation}
Thus,
\begin{equation}
C = \frac{3}{M_Z}  \sum_{\alpha}  { \abs{\mathbf{k}_{\alpha \perp} }^2 \over
\abs{\mathbf{k}_{\alpha} } }.
\label{3.21}
\end{equation}
The $C$ distribution is then:
\begin{eqnarray}
{{\rm d} \Gamma \over {\rm d}C} &=&  \Gamma^{(0)}_{\text{2-jet}} \left[\delta(C) -\delta^{\prime}(C)
\frac{\vev{O_1^C}}{M_Z}\right],
\end{eqnarray}
where $\vev{O_1^C} \equiv M_Z\vev{C}$ defined in Eq.~(\ref{3.21}). Like jet broadening, the $C$ parameter distribution is not local on the ultrasoft fields, and the nonperturbative correction is not related to that for any of the above distributions.

\subsection{Energy-Energy Correlation and Jet-Cone Energy Fraction}

The angular correlations of radiated energy can be characterized by the one-point and two-point correlations~\cite{BBEL},
\begin{subequations}
\begin{align}
{ {\rm d} \Sigma \over {\rm d} \Omega} &= \int {\rm d}\Gamma \sum_i {E_i \over M_Z} \delta\left(\Omega-\Omega_i\right) ,\\
{ {\rm d} ^2 \Sigma \over {\rm d} \Omega{\rm d} \Omega^\prime} &= \int {\rm d}\Gamma \sum_{i,j} {E_iE_j \over M_Z^2} \delta\left(\Omega-\Omega_i\right) \delta\left(\Omega'-\Omega_j\right),
\label{5.01}
\end{align}
\end{subequations}
where the sum is over all particles, and includes the terms with $i=j$. They are normalized so that
\begin{subequations}
\begin{align}
\int {\rm d}\Omega { {\rm d} \Sigma \over {\rm d} \Omega} &=\Gamma , \\
\int {\rm d}\Omega^\prime { {\rm d} ^2 \Sigma \over {\rm d} \Omega{\rm d} \Omega^\prime} &= { {\rm d} \Sigma \over {\rm d} \Omega}.
\end{align}
\end{subequations}

The energy-energy correlation function $P(\cos \chi)$ is defined by
\begin{eqnarray}
P(\cos \chi) &=& \int {\rm d}\Omega^\prime {\rm d}\Omega{ {\rm d} ^2 \Sigma \over {\rm d} \Omega{\rm d} \Omega^\prime} \delta(\cos \chi - \cos \theta_{\Omega\Omega^\prime}),
\end{eqnarray}
where $\theta_{\Omega\Omega^\prime}$ is the angle between vectors in the $\Omega$ and $\Omega^\prime$ directions.

The angular energy correlations Eq.~(\ref{5.01}) were defined in Ref.~\cite{BBEL} for $e^+e^-$ annihilation, and the solid angle was defined with respect to the beam direction. For unpolarized $Z$ decay, there is no preferred direction, so ${\rm d}\Sigma/{\rm d}\Omega$ is a constant, and ${\rm d}^2\Sigma/{\rm d}\Omega{\rm d}\Omega^\prime$ contains the same information as the energy-energy correlation function $P(\cos\chi)$. One can, however, define distributions analogous to Eq.~(\ref{5.01}) where the solid angle is measured with respect to the thrust axis $\mathbf{\hat t}$. The one-point function is called the jet cone energy fraction $J$.

The energy-energy correlation and the jet cone energy fraction both are proportional to $\delta$ functions if one considers the leading order process of $Z$ decay into a quark-antiquark pair:
\begin{equation}
\begin{split}
P(\cos \chi) &= J(\cos \chi) \\
&= \frac12\Gamma_0\bigl[ \delta\left(\cos \chi - 1 \right)+ \delta\left(\cos \chi + 1 \right) \bigr].
\end{split}
\end{equation}
Ultrasoft emission (in two-jet events) changes the distribution in two ways: (a) by changing the energy or (b) by changing the solid angle of the emitted particles. At order $k_u/M_Z$, the change in energy can be neglected, because it does not shift the angles of the partons; thus there is no contribution proportional to $\delta^\prime(\cos \chi\pm1)$, as for variables such as thrust. The angle between the quark and antiquark is [compare Eq.~(\ref{thetaqbar})]:
\begin{equation}
\cos \theta_{q \bar q} = -1 + 2 {\mathbf{k}_\perp^2 \over M_Z^2},
\end{equation}
and the angle of the quark or antiquark with respect to the thrust axis is [compare Eq.~(\ref{thetat})]:
\begin{equation}
\cos \theta_{q \mathbf{\hat t}} = -\cos \theta_{\bar q \mathbf{\hat t}} = 1 -  {\mathbf{k}_\perp^2 \over2 M_Z^2},
\end{equation}
where $\mathbf{k}_\perp$ is the total $\perp$ momentum of the ultrasoft particles. The shift in angle is second order in $k_u/M_Z$, and so to first order, there is no enhanced contribution near $\cos \chi = \pm 1$. There are nonperturbative contributions at second order.

\subsection{Classes of Observables}

The different observables we have discussed can be divided into classes, based on the extent to which their nonperturbative corrections are inclusive on the ultrasoft degrees of freedom. 

A class I observable is the jet energy distribution. The nonperturbative correction to the jet energy depends on $\vev{k_u^+}$, where $k_u$ is the total ultrasoft momentum, so the jet energy distribution is totally inclusive on the ultrasoft fields. The derivation of nonperturbative corrections to the two jet energy distribution is not quite on the same footing as the derivation of nonperturbative corrections to the $B$ meson semileptonic decay rate, because of the additional assumption about the equivalence of sums over partonic and hadronic states discussed after Eq.~(\ref{main}).

Class II observables are thrust and the jet masses $M^2_{S,H}$. The nonperturbative corrections to these variables require the ultrasoft momentum to be broken up into two parts, $k_u=k_u^{(a)}+k_u^{(b)}$, corresponding to the contributions from ultrasoft partons in the two hemispheres. The hemispheres are chosen based on the jet directions, i.e., based on the collinear degrees of freedom. The momentum in each hemisphere can then be defined by integrating the ultrasoft energy-momentum tensor over the hemisphere at infinity \cite{KS99,infinity1,infinity2,infinity3,infinity4}. The class II  variables are not totally inclusive on the ultrasoft variables, but require them to be divided globally into two parts. Whether our derivation of the nonperturbative corrections for class II observables (e.g., the relation between jet mass and thrust distributions) is valid  depends on the nature of hadronization in QCD. The ultrasoft fields end up inside final state hadrons. The final hadron can contain ultrasoft partons from different hemispheres, so the hadronic energy flow in each hemisphere does not have to equal the parton energy flow in each hemisphere. If the hadronic and partonic  energy flows differ by order unity,  the derivation of nonperturbative effects in class II  observables is invalid. If, for a smearing region of size $\Delta$ the mixing of ultrasoft momenta between the two hemispheres during hadronization is an effect of order $\lqcd/\Delta$, then its impact on class II observables is the same size as $k_u^2$ effects, which are one higher order than the terms we have computed.

Class III observables are the jet broadening measures $B_{a,b,\text{sum},\text{max}}$ and the $C$ parameter. These depend on knowing the individual ultrasoft momenta of each parton. This appears to be a notion that cannot be made rigorous in field theory.

\subsection{Model-Dependent Relations Among Event Shape Variables}
\label{ssec:comparison}

Nonperturbative corrections to event shape distributions have been considered extensively in the literature in the past.  For example, in the work of Ref.~\cite{KS99}, nonperturbative shape functions were derived for thrust and jet mass distributions.  The enhanced nonperturbative corrections to these distributions are given by first moments of these shape functions, and the results in sections \ref{ssec:thrust} and \ref{ssec:jmass} are in agreement with Ref.~\cite{KS99}.

The derivations of the enhanced non-perturbative corrections in this section have only relied on the fact that they arise from matrix elements of ultrasoft operators. It is insightful to understand what further conditions have to be imposed to reproduce other proposed relations amongst nonperturbative parameters for event shape distributions~\cite{DokWeb95,KT00,Doktalk,CataniWebber}. 

As an example, consider the $C$ parameter, for which the nonperturbative matrix element was defined as
\begin{eqnarray}
\langle O_1^C \rangle = 3 \left \langle \sum_\alpha \frac{|{\bf k}_{\alpha\perp}|^2}{|{\bf k}_\alpha|} \right \rangle .
\end{eqnarray}
For on-shell soft gluons collinear to the antiquark or quark jet (i.e., in hemisphere $a$ or $b$, respectively), $k^{(a)+} \ll k^{(a)-}$ and $k^{(b)-} \ll k^{(b)+}$. This implies that
\begin{equation}
\begin{split}
\left \langle \sum_\alpha \frac{|{\bf k}_{\alpha\perp}|^2}{|{\bf k}_\alpha|} \right \rangle_{\rm coll} 
&=
2 \left\langle \sum_\alpha \frac{|k_\alpha^+\, k_\alpha^-|}{|k_\alpha^+ + k_\alpha^-|} \right\rangle\\
&=
 2 \left\langle \sum_\alpha |k^{(a)+}_\alpha| + \sum_\beta |k^{(b)-}_\beta| \right \rangle\\
&=
2 \left\langle k_u^{(a)}+k_u^{(b)-}\right\rangle.
\end{split}
\end{equation}
This leads to
\begin{eqnarray}
\langle O_1^C \rangle_{\rm coll} = 6 \, \langle O_1^T \rangle .
\end{eqnarray}

To take into account that ultrasoft gluons can also be radiated at a finite angle, one can impose the condition that the matrix elements of $O_1^C$ and $O_1^T$ are given by the one-gluon contribution in perturbation theory, performing the angular integrals in the phase space at a fixed value of $\abs{\vect{k}_\perp}$. Under this assumption, the matrix element of $O_1^C$ is given by:
\begin{equation}
\begin{split}
\left \langle \frac{|{\bf k}_{\alpha\perp}|^2}{|{\bf k}_\alpha|} \right \rangle _{\rm 1-gluon}
&=
2\left\langle \int_0^{\frac{\pi}{2}} \!\!{\rm d}\theta\sin\theta\frac{\abs{\bf k_\perp}^2}{(\abs{\bf k_\perp}/\sin\theta)}\frac{1}{\sin^2\theta}\right\rangle\\
&=
\pi \langle \abs{\bf k_\perp} \rangle, 
\end{split}
\end{equation}
where the factors of $\sin\theta$ from the phase space, from the relation $\abs{\vect{k}_\perp} = \abs{\vect{k}}\sin\theta$, and from the squared amplitude for one gluon emission have all canceled out to give the final result.
For the matrix element of $O_1^T$, we calculate:
\begin{equation}
\begin{split}
\left \langle k^{(a)+} + k^{(b)-} \right \rangle_{\rm 1-gluon} 
&=
2\left\langle \int_0^{\frac{\pi}{2}}\!\!{\rm d}\theta\sin\theta\frac{\abs{\bf k} (1-\cos\theta)}{\sin^2\theta} \right\rangle \\
&=
2\left\langle \int_0^{\frac{\pi}{2}}\!\!{\rm d}\theta\frac{\abs{\bf k_\perp} (1-\cos\theta)}{\sin^2\theta} \right\rangle \\
&= 2\langle \abs{\bf k_\perp} \rangle.
\end{split}
\end{equation}
This leads to the result
\begin{eqnarray}
\label{3pi/2}
\langle O_1^C \rangle_{\rm 1-gluon} = \frac{3 \pi}{2} \langle O_1^T \rangle_{\rm 1-gluon} .
\end{eqnarray}
Given the assumptions that have to be made to obtain Eq.~(\ref{3pi/2}) (or analogous relations based on higher orders in perturbation theory), it does not seem likely to us that there is a simple analytic nonperturbative relation  between $\vev{O_1^C}$ and $\vev{O_1^T}$.

\subsection{Comparison with the Data}

Predictions for event shape variables have been compared with  experimental data in Refs.~\cite{compare,expdata1}. Nonperturbative corrections have been included using the ansatz that their effect on distributions for shape variables is described by shifting the variable by $c_f \mu_{\text{np}}/E_{{\rm cm}}$ in the perturbative formula for the distribution:
\begin{equation}
\der{\Gamma}{f}(f) \rightarrow \der{\Gamma}{f}\Bigl(f - \frac{c_f\mu_{\text{np}}}{E_{\text{cm}}}\Bigr).
\end{equation}
Here $c_f$ is a constant that depends on the kinematic variable $f$,  $\mu_{\text{np}}$ is a universal nonperturbative parameter, and $E_{{\rm cm}}$ is the center-of-mass energy. An analysis in perturbation theory (similar to what was done in section  \ref{ssec:comparison}) provides simple relations between the $c$'s for some of the event shape variables. We have found that, provided one is not in a  kinematic region that is  extremely close to the partonic endpoint (i.e., the shape function region), $c$ for $1-T$ and $M_S^2$ are the same. However, we argued that $c$ for other parameters like the heavy jet mass and $C$ are not connected to $c$ for thrust. Some experimental evidence for this can be found in the analysis of Ref.~\cite{compare}. For $1-T$ and the jet mass sum\footnote{Ref.~\cite{compare} advocates the use of a modified $E$-scheme jet mass to reduce sensitivity to hadronic masses.} a simultaneous fit for $\alpha_s$ and $\mu_{\text{np}}$ under the assumption that $c$ takes on its conjectured values (see Fig.~\ref{delphi}) yields values of $\mu_{\text{np}}$ that are close to each other, and values of $\alpha_s$ that are consistent with other extractions of the strong coupling. However, Ref.~\cite{compare} finds that $\mu_{\text{np}}$ for the heavy jet mass, $C$ parameter, and jet broadenings are not related to $\mu_{\text{np}}$ for thrust in the way that the analysis based on perturbation theory suggests, and, furthermore, a fit to these variables does not yield a value of $\alpha_s$ that is consistent with other extractions.
\begin{figure}
\begin{center}
\includegraphics[width=11cm]{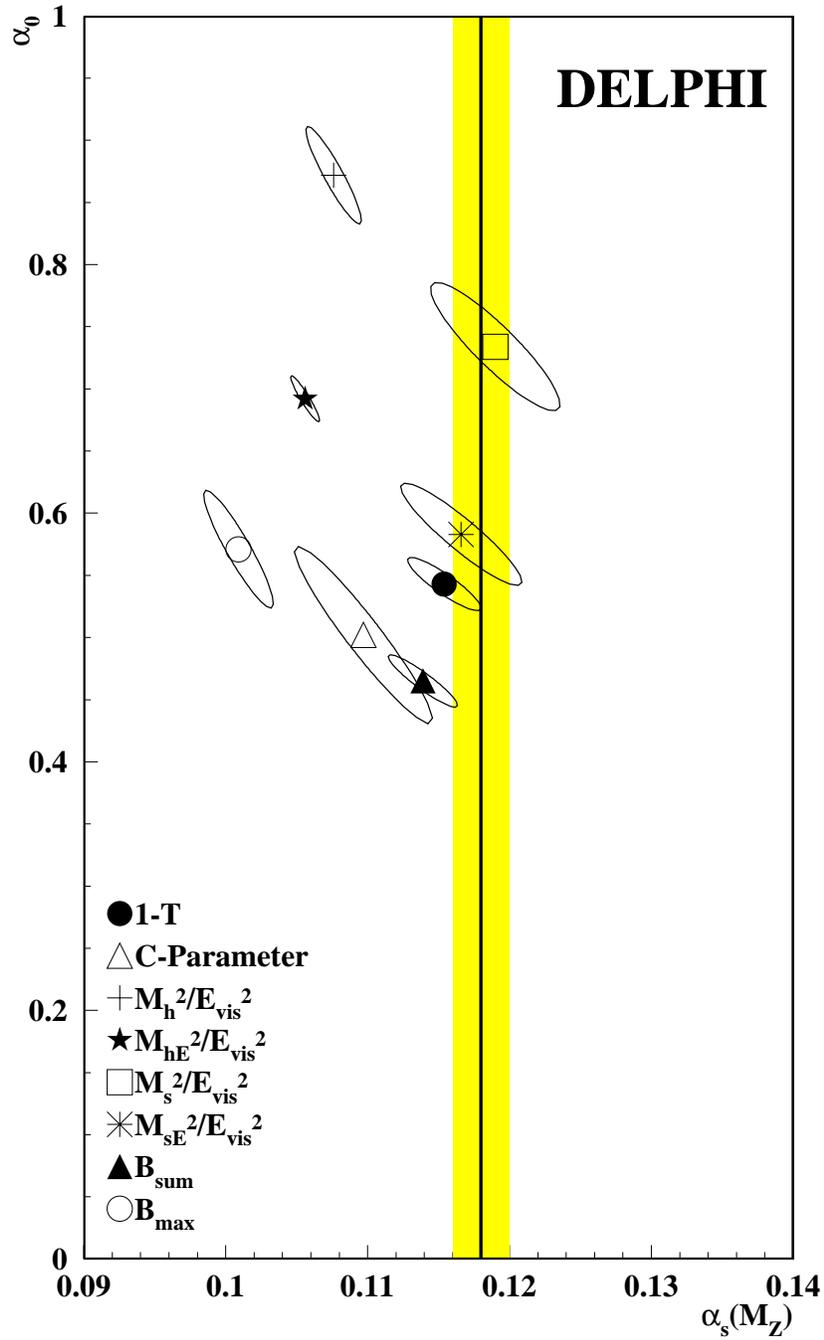}
\end{center}
\caption[Experimental fit for model of power corrections to event shape variables.]{Experimental fit for model of nonperturbative power corrections to event shape variables. The DELPHI collaboration \cite{compare} tested the Dokshitzer-Webber model for nonperturbative power corrections to event shape variables in $e^+e^-$-annihilation, fitting to the two model parameters $\alpha_0$ and $\alpha_S(M_Z)$. The data suggest a fairly poor fit for most variables, except for the thrust and jet mass sum, which are most nearly consistent with each other while also agreeing with independent extractions of $\alpha_s(M_Z)$. These findings are consitent with the theoretical predictions in Sec.~\ref{sec:other}.}
 \label{delphi}
\end{figure}

\section{Perturbative corrections to $d \Gamma/dE_J$ \label{sec:pert}}

Neglecting  order $\alpha_s$ corrections to the nonperturbative effects proportional to $O_1$, perturbative corrections to ${\rm d}\Gamma_{\text{2-jet}}/{\rm d}E_J$ in Eq.~(\ref{main}) can be calculated in full QCD using standard methods. In this section we first review the computation of perturbative $\mathcal{O}(\alpha_s)$ corrections to the total two-jet rate $\Gamma_{\text{2-jet}}$ and then compute the jet energy distribution ${\rm d}\Gamma_{\text{2-jet}}/{\rm d}E_J$ at order $\alpha_s$. We work in $d=4-\epsilon$ dimensions to regulate infrared, collinear and ultraviolet divergences that occur in contributions to the differential decay rate. The jets are defined using the Sterman-Weinberg criteria which involve  an energy cut $\beta M_Z$ and a cone half-angle $\delta$. Corrections suppressed by $\alpha_s \beta $ and $\alpha_s \delta$ are neglected.

\subsection{Two Jet Decay Rate \label{ssec:2jet}}

Using the Sterman-Weinberg definition of jets, there are three contributions to the two-jet rate at $\mathcal{O}(\alpha_s)$:
\begin{enumerate}
\item[(a)] One quark and one antiquark each creating a jet;
\item[(b)] One quark and one antiquark each creating a jet, plus a gluon with energy $E_g<\beta M_Z$;
\item[(c)] One quark and one antiquark each creating a jet, plus a gluon with energy $E_g>\beta M_Z$ inside one of the jets (within an angle $\delta$ of the quark or antiquark).
\end{enumerate}

Contribution (a) is simply the rate $\Gamma(Z\rightarrow q\bar q)$. The tree and virtual gluon graphs give the amplitude:
\begin{eqnarray}
\label{ampA}
\mathcal{M}_{Z\rightarrow q\bar q} = \epsilon_\mu(p_Z)\bar u^a(p_q)\Gamma^\mu v_a(p_{\bar q})\left(1 + \frac{\alpha_s C_F}{2\pi}X\right),
\end{eqnarray}
where the color index $a$ is summed over values $a=1,\dots,N_C$, and $C_F$ is the Casimir of the fundamental representation. Explicit computation of the one-loop vertex correction gives,
\begin{equation}
\label{X}
X = -\frac{4}{\epsilon^2} - \frac{3}{\epsilon} + \frac{2}{\epsilon}\ln\left(\frac{-2p_q\cdot p_{\bar q}}{\mu^2}\right) - 4 + \frac{\pi^2}{12} - \frac{1}{2}\ln^2\left(\frac{-2p_q\cdot p_{\bar q}}{\mu^2}\right) + \frac{3}{2}\ln\left(\frac{-2p_q\cdot p_{\bar q}}{\mu^2}\right).
\end{equation}
Integrating the square of the amplitude over the $d$ dimensional two body phase space gives:
\begin{equation}
\label{Zqq}
\begin{split}
\Gamma_{Z\rightarrow q\bar q} &= \frac{N_C}{32\pi^2}(g_V^2+g_A^2)\left[M_Z^{1-\epsilon}(4\pi)^\epsilon\frac{2-\epsilon}{3-\epsilon}\Omega_{3-\epsilon}\right] \\
&\times\Biggl[1 + \frac{\alpha_s C_F}{\pi}\biggl(-\frac{4}{\epsilon^2} - \frac{3}{\epsilon} + \frac{2}{\epsilon}\ln\frac{M_Z^2}{\mu^2} - 4 + \frac{7\pi^2}{12}  - \frac{1}{2}\ln^2\frac{M_Z^2}{\mu^2} + \frac{3}{2}\ln\frac{M_Z^2}{\mu^2}\biggr)\Biggr],
\end{split}
\end{equation}
where $\Omega_{d}$ is the total solid angle in $d$ dimensions. The $1/\epsilon$ poles will cancel out against divergences from the real gluon emission graphs. We do not need to expand the bracketed prefactor in Eq.~(\ref{Zqq}) in powers of $\epsilon$ because the identical factor will appear in the real gluon graphs.

Contributions (b) and (c) come from integrating the square of the amplitude for real gluon emission, $Z\rightarrow q\bar q g$, over the three-body phase space in $d$ dimensions. We find for the terms that do not vanish as $\beta$ and $\delta$ go to zero,
\begin{eqnarray}
\Gamma_{Z\rightarrow q\bar q g}^{(b)} = \frac{g_s^2 M_Z^{1-\epsilon}N_C C_F}{256\pi^5}(2\pi)^{2\epsilon}\Bigl(\frac{\mu}{M_Z}\Bigr)^\epsilon\Omega_{2-\epsilon}\Omega_{3-\epsilon}(g_V^2 + g_A^2)
\Bigl(-\frac{1}{\epsilon}\Bigr)\frac{\Gamma(-\frac{\epsilon}{2})^2}{\Gamma(-\epsilon)}\beta^{-\epsilon},
\end{eqnarray}
and
\begin{equation}
\begin{split}
\Gamma^{(c)}_{Z\rightarrow q\bar q g} &= \frac{g_s^2 M_Z^{1-\epsilon}N_C C_F}{256\pi^5}(4\pi)^{2\epsilon}\Bigl(\frac{\mu}{M_Z}\Bigr)^\epsilon\Omega_{2-\epsilon}\Omega_{3-\epsilon}\frac{2-\epsilon}{3-\epsilon}(g_V^2+g_A^2)\frac{(2\delta )^{-\epsilon}}{-\epsilon} \\
&\qquad\times\biggl[\frac{4}{\epsilon}(2\beta)^{-\epsilon} + 2\biggl(1+\frac{3\epsilon}{4}+\frac{13\epsilon^2}{8}\biggr)\frac{\Gamma(-\epsilon)^2}{\Gamma(-2\epsilon)}\biggr].
\end{split}
\end{equation}
Adding together contributions (b) and (c), expanding in powers of $\epsilon$ and converting to the $\overline{\text{MS}}$ scheme yields for the total rate for $Z\rightarrow q\bar q g$ in the two-jet region
\begin{equation}
\begin{split}
\label{Zqqg}
\Gamma_{Z\rightarrow q\bar q g} &= \frac{N_C C_F\alpha_s}{32\pi^3}(g_V^2+g_A^2)\left[M_Z^{1-\epsilon}(4\pi)^\epsilon\frac{2-\epsilon}{3-\epsilon}\Omega_{3-\epsilon}\right] \\
&\quad\times\biggl(\frac{4}{\epsilon^2} + \frac{3}{\epsilon} - \frac{2}{\epsilon}\ln\frac{M_Z^2}{\mu^2}  - \frac{3}{2}\ln\frac{M_Z^2}{\mu^2} + \frac{1}{2}\ln^2\frac{M_Z^2}{\mu^2} \\
&\qquad\quad - 4\ln 2\beta\ln\delta - 3\ln \delta + \frac{13}{2} - \frac{11\pi^2}{12}\biggr).
\end{split}
\end{equation}
Finally, we  add together the rates $\Gamma_{Z\rightarrow q\bar q}$ and $\Gamma_{Z\rightarrow q\bar q g}$ from Eqs.~(\ref{Zqq}) and (\ref{Zqqg}). The $\epsilon$-dependent prefactors in brackets in the two equations are identical, as promised. The $1/\epsilon$-poles in the remainder of the expressions cancel out exactly (as do all the logarithms of $M_Z/\mu$), so we can set $\epsilon=0$ in the remaining finite parts, leaving
\begin{eqnarray}
\label{SW}
\Gamma_{\text{2-jet}} = \frac{N_C M_Z}{12\pi}(g_V^2+g_A^2)\Biggl[1
+ \frac{\alpha_s C_F}{\pi}\biggl(\frac{5}{2} - \frac{\pi^2}{3}- 3\ln\delta - 4\ln 2\beta\ln\delta \biggr)\Biggr],
\end{eqnarray}
which agrees with Sterman and Weinberg's original result~\cite{SW}.

\subsection{Differential Decay Rate \protect{${\rm d}\Gamma_{\text{2-jet}}/{\rm d} E_J$} \label{ssec:diff}}

We now turn our attention to the differential decay rate ${\rm d}\Gamma_{\text{2-jet}}/{\rm d} E_J$. The contribution of $\Gamma_{Z\rightarrow q\bar q}$ to this rate is simply
\begin{equation}
\frac{{\rm d}\Gamma_{Z\rightarrow q\bar q}}{{\rm d}E_J} = \Gamma_{Z\rightarrow q\bar q}\ \delta\left(E_J-\frac{M_Z}{2}\right),
\end{equation}
where $\Gamma_{Z\rightarrow q\bar q}$ is the total rate for $Z\rightarrow q\bar q$ calculated to $\mathcal{O}(\alpha_s)$, which is given by Eq.~(\ref{Zqq}).

For the contribution of real gluon emission processes, we write the three-body phase space for this rate:
\begin{equation}
\begin{split}
\frac{{\rm d}\Gamma_{\text{2-jet}}}{{\rm d}E_J} &= \frac{1}{16M_Z}\frac{1}{(2\pi)^{2d-3}}\Omega_{d-2}\Omega_{d-1}{\rm d}E_1 E_1^{d-4} {\rm d}E_2 E_2^{d-4} {\rm d}\cos\theta\sin^{d-4}\theta \\
&\quad\times\delta\biggl[\frac{M_Z^2-2M_Z(E_1+E_2)}{2E_1 E_2} + 1-\cos\theta\biggr]\delta(E_J - \cdots)|\mathcal{M}|^2,
\end{split}
\label{3body}
\end{equation}
where the $\delta(E_J - \cdots)$ defines $E_J$ according to which partons actually go inside the jet. It is useful to split up the phase space slightly differently than for the case of the total rate:
\begin{enumerate}
\item[(a)] Gluon with energy $E_g>\beta M_Z$ inside unobserved jet;
\item[(b)] Gluon with any energy inside observed jet;
\item[(c)] Gluon with energy $E_g<\beta M_Z$ outside observed jet.
\end{enumerate}
These three regions exhaust the possible gluon energies and locations with respect to the jets. It is convenient to introduce the variable
\begin{equation}
e_J= {M_Z \over 2}-E_J
\end{equation}
and focus on a region of $e_J$ near the origin with size of order $\beta M_Z$.

For case (a), where a gluon with $E_g>\beta M_Z$ is inside the unobserved jet, take $E_1 = E_g$, $E_2 = E_{\bar q}$, so $\theta$ is the angle between the gluon and antiquark, and $E_J=E_q$. Integrating over $\theta$ and $E_{\bar q}$ using the delta functions leaves an integral over $E_g$ running between the limits
\begin{equation}
E_g^\pm = \frac{M_Z}{4}\biggl(1\pm\sqrt{1-\frac{8e_J}{M_Z\delta^2}}\biggr),
\end{equation}
and restricts $e_J$ to lie between
\begin{equation}
\delta^2\beta M_Z<e_J<\frac{M_Z\delta^2}{8}.
\end{equation}

Similarly, for case (b), where a gluon with any energy lies inside the observed jet, $E_1 = E_q$, $E_2 = E_g$, and $E_J=E_g +E_q$. Integrate over $\theta$ and $E_q$ using the delta functions. Then the limits of the $E_g$ integral are
\begin{equation}
E_g^\pm = \frac{M_Z}{4}\biggl(1\pm\sqrt{1+\frac{8e_J}{M_Z\delta^2}}\biggr),
\end{equation}
and $e_J$ is restricted to the region
\begin{equation}
-\frac{M_Z\delta^2}{8}<e_J<0.
\end{equation}

Before we proceed to case (c), note that the physical observable we actually want to calculate is the smeared distribution
\begin{equation}
\frac{{\rm d}\Gamma}{{\rm d}E_J}\biggr|_{\Delta} = \int {\rm d}E_J w_\Delta(E_J)\frac{{\rm d}\Gamma}{{\rm d}E_J},
\end{equation}
where $w_\Delta$ is a smooth function which smears the differential rate over a region of jet energy whose size is of order $\beta M_Z$. But the contributions to the rate from cases (a) and (b) have support only over a region of size $\delta^2 M_Z\ll \beta M_Z$ near $e_J = 0$. Consider smearing ${{\rm d}\Gamma}/{{\rm d}E_J}$, or equivalently ${{\rm d}\Gamma}/{{\rm d}e_J}$, over a region near $E_J = M_Z/2$ ($e_J=0$) of size of order $\beta M_Z$. Then $w(0)\sim 1/\beta M_Z$, $w'(0)\sim 1/(\beta M_Z)^2$, etc.\ Expanding,
\begin{equation}
\int {\rm d}e_J w(e_J)\frac{{\rm d}\Gamma}{{\rm d}e_J} = \int {\rm d}e_J[w(0)+w'(0)e_J+\cdots]\frac{{\rm d}\Gamma}{{\rm d}e_J}.
\end{equation}
Since $w'(0)/w(0)\sim 1/\beta M_Z$, and, for the contributions in cases (a) and (b), $e_J \sim\delta^2 M_Z$ in the region where ${{\rm d}\Gamma}/{{\rm d}e_J}$ is nonzero, the second term is suppressed by a power of $\delta^2/\beta\ll 1$. Thus only the first term is relevant.\footnote{This argument assumes that the integral $\int {\rm {\rm d}}e_J\,e_J {\rm d}\Gamma/{\rm d}e_J$ is finite, which can easily be shown.} Keeping only the first term amounts to replacing the full ${{\rm d}\Gamma}/{{\rm d}e_J}$ by
\begin{equation}
\frac{{\rm d}\Gamma}{{\rm d}e_J}\rightarrow\delta (e_J)\int {\rm d}e_J^\prime \frac{{\rm d}\Gamma}{{\rm d}e_J^\prime}.
\end{equation}
However, integrating the contributions of (a) and (b) to ${{\rm d}\Gamma}/{{\rm d}e_J}$ over all allowed values of $e_J$ simply gives their contribution to the total Sterman-Weinberg jet rate, that is, they will build up part of the term $\Gamma_{Z\rightarrow q\bar q g}\delta(E_J-M_Z/2)$ in ${\rm d}\Gamma_{\text{2-jet}}/{\rm d}E_J$. Since we have already calculated the total rate, we need not analyze cases (a) and (b) any further, as long as we can get the remaining contribution to the total rate from case (c).

In case (c) we have a gluon with $E_g<\beta M$ anywhere outside the observed jet. Here $E_1=E_q$, $E_2=E_g$, and  $E_J = E_q$. Writing out the formula for the rate explicitly,
\begin{equation}
\begin{split}
\frac{{\rm d}\Gamma^{(c)}}{{\rm d}E_J} &= \frac{1}{16M_Z}\frac{1}{(2\pi)^{2d-3}}\Omega_{d-2}\Omega_{d-1}\theta(e_J)\theta[\beta M_Z(1-\delta^2)-e_J] \\
&\quad\times 
\int_{e_J (1+\delta^2)}^{\beta M_Z}\!\!{\rm d}E_g\,E_g^{d-4}E_q^{d-4}\sin^{d-4}\theta|\mathcal{M}^{(c)}_{Z\rightarrow q\bar q g}|^2.
\end{split}
\end{equation}
The part of the amplitude that gives a contribution that survives as $\beta\rightarrow 0$  is
\begin{equation}
\label{ampB}
\abs{ \mathcal{M}^{(c)}_{Z\rightarrow q\bar q g} }^2 \!\! = 4N_C C_F g_s^2\mu^\epsilon\frac{d-2}{d-1}(g_V^2+g_A^2)\frac{M_Z^2 p_q\cdot p_{\bar q}}{(k\cdot p_q)(k\cdot p_{\bar q})}.
\end{equation}
Substituting  Eq.~(\ref{ampB}) into the phase space,
\begin{equation}
\label{plus}
\begin{split}
\frac{{\rm d}\Gamma^{(c)}}{{\rm d}E_J} &= \frac{M_Z g_s^2 N_C C_F}{256\pi^5}\Bigl(\frac{\mu}{M_Z}\Bigr)^\epsilon(2\pi)^{2\epsilon}\frac{d-2}{d-1}
\Omega_{d-2}\Omega_{d-1} \\
&\quad\times(g_V^2+g_A^2)e_J^{-1-\frac{\epsilon}{2}} \theta(e_J)\theta[\beta M_Z(1-\delta^2) - e_J]\ln\left[\frac{\beta M_Z - e_J}{e_J\delta^2}\right].
\end{split}
\end{equation}
The factor, $(1/e_J)\ln [(\beta M_Z - e_J)/(e_J\delta^2)]$,
is singular as $e_J\to 0$, and must be rewritten in terms of an integrable quantity. Use the ``plus distribution'':
\begin{equation}
\int_0^{\beta M_Z} \!\!\!\!{\rm d}e_J \,f(e_J)_+\ g(e_J) \equiv \int_0^{\beta M_Z} \!\!\!\!{\rm d}e_J\,f(e_J)[g(e_J)-g(0)],
\end{equation}
where $f$ diverges at $e_J=0$ and $g$ is a test function finite at $e_J =0$. To replace $f$ by $f_+$, we would write
\begin{equation}
\int_0^{\beta M_Z} \!\!\!\!{\rm d}e_J\,f(e_J)g(e_J) = \int_0^{\beta M_Z}\!\!\!\! {\rm d}e_J\,f(e_J)_+\ g(e_J)
 + g(0)\int_0^{\beta M_Z}\!\!\!\! {\rm d}e_J\,f(e_J).
\end{equation}
The second term amounts to replacing
\begin{equation}
f(e_J)\rightarrow \delta(e_J)\int {\rm d}e_J'\,f(e_J').
\end{equation}
But making this replacement in Eq. (\ref{plus}) means writing a delta function $\delta(E_J-M_Z/2)$ and integrating the differential rate over all allowed values of $e_J$, which again just gives its contribution to the total Sterman-Weinberg jet rate. Together with the contributions from (a) and (b) this gives the one loop contribution to $\delta(E_J-M_Z/2)\Gamma_{\rm 2-jet}$. Only the plus function piece gives a deviation of the jet energy distribution away from $E_J=M_Z/2$. The final result for the differential rate to $\mathcal{O}(\alpha_s)$ is:
\begin{equation}
\begin{split}
\frac{{\rm d}\Gamma_{\text{2-jet}}}{{\rm d}E_J} &= \delta\Bigl(E_J - \frac{M_Z}{2}\Bigr)\Gamma_{\text{2-jet}} \label{59}\\ 
&\quad+\frac{M_Z\alpha_s N_C C_F}{12\pi^2}(g_V^2+g_A^2) \theta(e_J)\theta(\beta M_Z-e_J)\biggl[\frac{1}{e_J}\ln \biggl(\frac{\beta M_Z - e_J}{e_J\delta^2}\biggr)\biggr]_+,
\end{split}
\end{equation}
where the total rate $\Gamma_{\text{2-jet}}$ is given by Eq.~(\ref{SW}).

\subsection{First Moment of the Jet Energy Distribution}

As an application of the above  result consider the first moment of the jet
energy distribution, defined by
\begin{equation}
M_1(f) =\int_{{M_Z \over 2}-f \beta M_Z}^{ { M_Z }} 
\hspace{-2em}{\rm d}
E_J \left( {1 \over\Gamma_{{\rm 2-jet}}} {{\rm d} \Gamma_{{\rm 2-jet}}\over
{\rm d} E_J}\right)\left( {1\over 2}-{E_J\over M_Z} \right),
\end{equation}
Using the expression in Eq.~(\ref{pert}) for the nonperturbative correction and in Eq.~(\ref{59}) for the order $\alpha_s$
perturbative correction to the jet energy distribution gives
\begin{equation}
M_1(f)={\alpha_s  C_F \beta  \over \pi}\left[f \log \left({1 \over f
\delta^2}\right)-(1-f)\log(1-f)\right] +\frac{\langle 0|O_1|0\rangle}{M_Z},
\end{equation}
for $f<1$ and
\begin{equation}
M_1(f)={\alpha_s C_F \beta  \over \pi}\log \left({1 \over
\delta^2}\right)+\frac{\langle 0|O_1|0\rangle}{M_Z},
\end{equation}
for $f>1$.
Note that the order $\alpha_s$ contribution to  $M_1(f)$ is independent of $f$ for $f>1$. This occurs because the perturbative correction vanishes for $E_J <M_Z/2-\beta M_Z$. 

In Fig.~\ref{momentfig}  we plot $M_1(f)$, for $f<1$. For this figure the value of the
energy cut is $\beta=0.15$ and the cone half-angle is $\delta=15^\circ$ and the vacuum expectation value of $O_1$ is set equal to $500~{\rm MeV}$. We evaluate $\alpha_s$ at the scale $\beta M_Z$ and find with
these parameters that the order $\alpha_s$ corrections reduce the two jet
rate by about $16\%$ from its tree level value lending support to the validity of perturbation theory for the values of the cone angle and energy cut used in Fig.~\ref{momentfig}.

\begin{figure}
\begin{center}
\includegraphics[width=12cm]{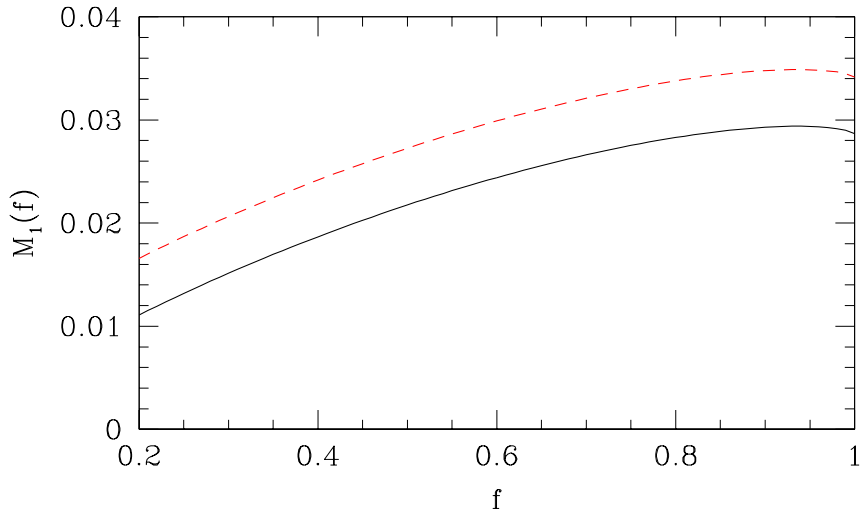}
\end{center}
\vspace{-1.5cm}
\caption[First moment $M_1(f)$ of the jet energy distribution.]{First moment $M_1(f)$ of the jet energy distribution. The black solid curve shows the perturbative contributions only, while the red dashed line represents the moment including the nonperturbative contribution. The figure corresponds to $\beta=0.15$, $\delta=\pi/12$, $\langle 0|O_1|0\rangle = 0.5~{\rm GeV}$, and we have evaluated the strong coupling constant at the scale $\mu = \beta M_Z$.}
\label{momentfig}
\end{figure}

\subsection{Perturbative Corrections in the Effective Theory \label{ssec:match}}

Although we have used full QCD to calculate the jet energy distribution it is possible to do the computation in the effective theory. Here we briefly discuss how that computation would proceed.

The full theory amplitude for $Z\rightarrow q\bar q$ is reproduced in SCET by the Wilson coefficient in the current matching:
\begin{equation}
j_{\rm QCD}^\mu = [\bar\xi_{\bar n} W_{\bar n}]C(\mu,\widetilde p_q\cdot \widetilde p_{\bar q})\Gamma^\mu[W_n^\dag\xi_{ n}],
\end{equation}
where there is an implicit sum over label momenta  and the matching coefficient $C(\mu,\widetilde p_q\cdot \widetilde p_{\bar q})$ can be read off\footnote{The matching coefficient is just given by the finite part of the full theory matrix element $\langle q\bar q | j^\mu | 0\rangle$ because the full theory current has no anomalous dimension, so the $1/\epsilon$ poles are pure IR divergences, which must cancel out in the matching condition. The loop graphs in the effective theory contributing to this matrix element are zero in dimensional regularization, so the finite part of the matching coefficient is just the finite part of the QCD matrix element, given by Eqs.~(\ref{ampA}) and (\ref{X}), while the infinite parts become the UV counterterm in the effective theory~\cite{EFTnotes,EFT}.} from  Eqs.~(\ref{ampA}) and (\ref{X}):
\begin{equation}
C(\mu,\widetilde p_q\!\cdot\!\widetilde p_{\bar q}) = 1 + \frac{\alpha_s C_F}{2\pi}\!\Biggl[-4 + \frac{\pi^2}{12} - \frac{1}{2}\ln^2\!\left(\!\frac{-2\widetilde p_q\cdot \widetilde p_{\bar q}}{\mu^2}\!\right) + \frac{3}{2}\ln\left(\frac{-2\widetilde p_q\cdot \widetilde p_{\bar q}}{\mu^2}\right)\Biggr],
\end{equation}
and the UV renormalization factor for the current in the effective theory is
\begin{equation}
Z_V = 1 + \frac{\alpha_s C_F}{2\pi}\left[-\frac{4}{\epsilon^2} - \frac{3}{\epsilon} + \frac{2}{\epsilon}\ln\left(\frac{-2\widetilde p_q\cdot \widetilde p_{\bar q}}{\mu^2}\right)\right].
\end{equation}
Note that both the renormalization factor and the matching coefficient depend on the label momenta for the quark and antiquark.
For outgoing particles, the collinear Wilson lines are  defined as
\begin{equation}
W_n(z) = P\,\exp\biggl[ig\int_0^\infty ds\,\bar n\cdot A_n(\bar n s + z)\biggr],\end{equation}
and one  must include collinear gluons produced by a Wilson line in real gluon emission to get the correct $Z \rightarrow q \bar q g$ amplitude. We find that the perturbative expressions for the two jet rate presented in the previous sections are reproduced by the effective theory if we call any particles inside the observed quark jet $\bar n$-collinear particles and all other particles $ n$-collinear. In the effective theory, ultrasoft gluons in the final state contribute zero in perturbation theory and appear only in the nonperturbative shape function. A similar result holds for deep inelastic scattering~\cite{dis}.

\section{Concluding Remarks \label{sec:conc}}
We have studied nonperturbative effects in $Z$ decay to hadrons using soft-collinear effective theory (SCET). The jet energy distribution for two jet events has enhanced nonperturbative effects when the jet energy is near $M_Z/2$. These nonperturbative effects can be expressed in terms of the vacuum expectation value of operators involving Wilson lines. The Wilson lines arise from the coupling of ultrasoft gluons to collinear degrees of freedom in the jet. In Sec.~\ref{sec:pert} we derive the order $\alpha_s$ perturbative corrections to
the jet energy distribution and discuss the implications of perturbative and nonperturbative physics on the first moment of this distribution.

For a region of $|E_J-M_Z/2|$ that is of size $\Delta$, the leading nonperturbative corrections to the jet energy distribution are of order $\Lambda_{{\rm QCD}}/\Delta$ when $\Delta$ is large compared to $\Lambda_{{\rm QCD}}$.  In this region they can be characterized by the vacuum expectation value of a single operator involving ultrasoft fields which provides a contribution to the jet energy spectrum that is proportional to $\delta'(M_Z/2-E_J)$. For multijet events, a similar analysis holds; however,  an additional operator analogous to $O_1$ but involving adjoint Wilson lines occurs for a gluon jet~\cite{shape}.

When $\Delta \sim \Lambda_{{\rm QCD}}$, one is in the shape function region, and the functional dependence on $E_J$ is much more complicated. While we focused mostly on the kinematic region where $M_Z \gg \Delta \gg \Lambda_{{\rm QCD}}$, it was shown that in the shape function region, it is not possible to capture the effects of nonperturbative physics by introducing a single nonperturbative parameter $\mu_{np}$ and shifting $E_J \rightarrow E_J-\mu_{\rm np}$ in the perturbative formula for the jet energy distribution.

The jet energy distribution has the special property that it is totally inclusive in one of the jets, and hence expressions for nonperturbative effects can be derived using operator methods that are similar to those used for the endpoint region in inclusive semileptonic $B$ decay. Other event shape variables (e.g. thrust, jet mass, jet broadening) have nonperturbative effects that are enhanced in the partonic endpoint region. We discussed the extent to which these effects can be understood using field theoretic methods in QCD.

\setcounter{section}{0}
\renewcommand{\thesection}{\arabic{chapter}.\Alph{section}}

\section{Appendix: Properties of Wilson Lines}

In this section we derive some useful properties of the ultrasoft Wilson lines introduced in Eq.~(\ref{def2}).

\subsection{Relations Between Triplet and Anti-triplet Wilson Lines}
\label{A1}

Consider the time- and anti-time-ordering of the Wilson lines in the shape function $S$ defined in Eq.~(\ref{shapefunction}). For $Y_n$, the path ordering is the same as time ordering and so $T\left[Y_n\right]=Y_n$. Consider writing $Y_n$ as the product of $N$ infinitesimal integrals over path segments of length ${\rm ds}$,
\begin{equation}
 Y_{n a}~^b= \left({\rm e}^{ig A_N ds}\right)_a~^{b_{N-1}}\left({\rm e}^{ig A_{N-1} ds}\right)_{b_{N-1}}~^{b_{N-2}}\ldots \left({\rm e}^{ig A_{1} ds}\right)_{b_{1}}~^{b},
\end{equation}
with the subscripts on the ultrasoft gauge fields denoting their space-time location along the path of integration. Taking its adjoint
\begin{equation}
Y^{\dagger}_{n a}~^b=\left({\rm e}^{-ig A_{1} ds}\right)_a~^{b_{1}}\ldots \left({\rm e}^{-ig A_{N-1} ds}\right)_{b_{N-2}}~^{b_{N-1}} \left({\rm e}^{ig A_N ds}\right)_{b_{N-1}}~^{b}.
\end{equation}
Time ordering this expression,
\begin{equation}
\begin{split}
T\left [Y^{\dagger}_{n a}~^b \right]&= \left({\rm e}^{ig A_N ds}\right)_{b_{N-1}}~^{b}\left({\rm e}^{-ig A_{N-1} ds}\right)_{b_{N-2}}~^{b_{N-1}}\ldots \left({\rm e}^{-ig A_{1} ds}\right)_a~^{b_{1}} \\
&=\left({\rm e}^{-ig A^T_N ds}\right)^{b}~_{b_{N-1}}\left({\rm e}^{-ig A^T_{N-1} ds}\right)^{b_{N-1}}~_{b_{N-2}}\cdots
\left({\rm e}^{-ig A^T_{1} ds}\right)^{b_{1}}~_a={\overline Y_n}^b~_a,
\end{split}
\end{equation}
where the overline denotes an anti-triplet Wilson line. (Recall that the generators  in the ${\overline {\bf 3}}$ representation are minus the transpose of those in the ${\bf 3}$.) Similarly,
\begin{equation}
 {\bar T} \left[Y_{{ n} a}~^b \right]={\overline Y^{\dagger}_n}^b~_a,\qquad {\bar T}\left[Y^{\dagger}_{na}~^b\right]=Y^{\dagger}_{na}~^b.
\end{equation}
From these results, Eq.~(\ref{notimeorder}) follows.

\subsection{$O_1$ in Terms of the Gluon Field Strength}
\label{A2}

We can express the operator $O_1$ in terms of the gluon field strength tensor as written in Eq.~(\ref{fancy}).
It is convenient for this purpose to generalize the expression for the ultrasoft Wilson line to
\begin{equation}
Y_{\bar n}(z;b,a)=P \exp \left[ig\int_a^b {\rm d}s\ {\bar n}\cdot A(z+\bar ns)\right]
\end{equation}
so that with $a=0$ and $b=\infty$ we recover the standard Wilson line used above, $Y_{\bar n}(z)=Y_{\bar n}(z;0,\infty)$. 
Differentiating along the $n$ direction,
\begin{equation}
\begin{split}
\label{derivative}
n \cdot \partial\ Y_{\bar n}(z) &= ig\int_0^{\infty}{\rm ds}\ Y_{\bar n}(z;\infty,s) \left[n \cdot \partial_z \bar n\cdot
 A \right](z+{\bar n} s)Y_{\bar n}(z;s,0) \\
&=ig\int_0^{\infty}{\rm ds}\ Y_{\bar n}(z;\infty,s) \left[n \cdot \partial_z \bar n\cdot A-\bar n \cdot \partial _z n \cdot A+\bar n \cdot \partial _z n \cdot A \right](z+{\bar n} s) \\
&\qquad\qquad\times Y_{\bar n}(z;s,0) \\
&=ig\int_0^{\infty}{\rm ds}\ Y_{\bar n}(z;\infty,s) \left[n \cdot \partial_z \bar n\cdot A-\bar n \cdot \partial _z n \cdot A \right](z+{\bar n} s)Y_{\bar n}(z;s,0) \\
&\quad+ig\int_0^{\infty}{\rm ds}\ Y_{\bar n}(z;\infty,s) \left[{{\rm d} (n\cdot A)\over {\rm d}s}\right](z+{\bar n} s)Y_{\bar n}(z;s,0).
\end{split}
\end{equation}
Using the chain rule,
\begin{equation}
\begin{split}
\int_0^\infty {\rm d}s\ &{{\rm d} \over {\rm d}s} \left[Y_{\bar n}(z;\infty,s)\left[n\cdot A\right](z+{\bar n} s ) Y_{\bar n}(z;s,0)\right] \\
&= \int_0^{\infty}{\rm d}s\  \left[{{\rm d} \over {\rm d}s}Y_{\bar n}(z;\infty,s)\right]\left[n\cdot A\right](z+{\bar n} s ) Y_{\bar n}(z;s,0) \\
&\quad+\int_0^\infty {\rm d}s\  Y_{\bar n}(z;\infty,s)\left[{{\rm d} (n\cdot A)\over {\rm d}s}\right](z+{\bar n} s)Y_{\bar n}(z;s,0) \\ 
&\quad+ \int_0^\infty {\rm d}s\  Y_{\bar n}(z;\infty,s)  \left[n\cdot A \right](z+{\bar n} s )\left[{{\rm d} \over {\rm d}s}Y_{\bar n}(z;s,0)\right]\\
&=-ig\int_0^\infty{\rm d}s\ Y_{\bar n}(z;\infty,s)[{\bar n}\cdot A(z+\bar n s),n \cdot A(z+n s)]Y_{\bar n}(z;s,0) \\
&\quad+ \int_0^\infty {\rm d}s\  Y_{\bar n}(z;\infty,s)\left[{{\rm d} (n\cdot A)\over {\rm d}s}\right](z+{\bar n} s)Y_{\bar n}(z;s,0).
\end{split}
\end{equation}
Using the above equation to eliminate the last term in Eq.~(\ref{derivative}) yields,
\begin{equation}
n\cdot D\,Y_{\bar n}(z)=ig\int_0^\infty {\rm d}s\ Y_{\bar n}(z;\infty ,s)n^{\mu} {\bar n}^{\nu} G_{\mu \nu}(z+\bar n s) Y_{\bar n}(z;s,0),
\end{equation}
where
\begin{equation}
n\cdot D\,Y_{\bar n}(z)=n\cdot \partial\,Y_{\bar n}(z)-ig n \cdot A(\infty)Y_{\bar n}(z)+igY_{\bar n}(z)n \cdot A(z),
\end{equation}
and the gluon field strength tensor is defined by,
\begin{equation}
G_{\mu \nu}= \partial_{\mu} A_{\nu}-\partial_{\nu}A_{\mu}-ig[A_{\mu},A_{\nu}].
\end{equation}
Hence
\begin{equation}
O_1={1 \over 2}{\rm Tr} [Y^{\dagger}_{\bar n}(in~ \cdot D)Y_{\bar n }]={1 \over 2}{\rm Tr}\left[ig\int_0^{\infty}{\rm d}s\ Y_{\bar n}^{\dagger}(z;s,0)n^{\mu}{\bar n}^{\nu}G_{\mu \nu} Y_{\bar n}(z;s,0)\right],
\end{equation}
which is Eq.~(\ref{fancy}).

\renewcommand{\thesection}{\arabic{chapter}.\arabic{section}}
\chapter{Summary and Outlook}
\label{fin}

\begin{quote}
\small\singlespace``There is no supersymmetry.'' \\ \flushright \vspace{-12pt} \emph{Carlos Wagner, in the middle of his 2002 TASI Lecture,\\ \emph{Introduction to Supersymmetry.}}
\end{quote}
Experiments to search with unprecedented sensitivity for particles and phenomena beyond the Standard Model of particle physics lie just over the horizon. These experimental advances make urgent the task of theorists to prepare to extract evidence for new physics from the empirical data by improving our ability to calculate reliably within the Standard Model, and then to be ready to deduce the implications of the new discoveries for our existing models of physical phenomena.

In this work we have chipped modestly away at these complementary tasks. First, taking note of the already-existing evidence of the inadequacy of the Standard Model in the baryon asymmetry of the Universe, we took steps toward a more reliable calculation of the BAU predicted by the scenario of electroweak baryogenesis in the MSSM, and comparing the sizes of $CP$-violating phases required by this scenario with the limits already placed on them by the null results of searches for electric dipole moments of electrons, neutrons, and $^{199}$Hg atoms. Making use of the closed-time path formalism for quantum field theory incorporating the effects of finite-temperature and nonequilbrium physics, we found regions of parameter space in the MSSM for which the baryon asymmetry would be enhanced over the predictions of semi-classical calculations. In these regions, electroweak baryogenesis is found to account successfully for the BAU with $CP$-violating phases in the MSSM as small as $10^{-2-3}$, which are still consistent with current bounds from EDM searches. Future electron and neutron EDM searches, however, should be sensitive enough to rule out this scenario for baryogenesis if no EDMs (or too small EDMs) are found, or to detect positively nonzero EDMs of this size. Such a discovery, combined with discoveries of superpartners with masses consistent with our predictions, would constitute strong evidence that the baryon asymmetry could indeed have been produced during the electroweak phase transition. More likely, EDM searches and LHC results will further restrict the parameter space in the MSSM that could allow for successful electroweak baryogenesis, and point our investigations in the direction of other models (either extending the MSSM or favoring a different mechanism for baryogenesis) that could be the correct account of the particles which actually exist in Nature and the origin of those which ended up as the matter that makes up all of our observable (and improbable \cite{Graesser:2004ng}) Universe today.

On the Standard Model front, we employed effective field theories in QCD to untangle perturbatively-calculable and unknown nonperturbative effects in processes involving the strong interactions by organizing the theory in powers of some small parameter. We applied soft-collinear effective theory combined with non-relativistic QCD to the exclusive radiative decay of $\Upsilon$ or other quarkonia to light hadrons. It has been known that in inclusive radiative $\Upsilon$ decays the decay channel involving the $b\bar b$ pair in a color-octet configuration can be as important as the color-singlet channel in the kinematic region of near-maximal photon energy, $E_\gamma\sim M_\Upsilon/2$. In exclusive decays, however, we showed that the color-octet channel could be safely ignored, thanks to the power counting in the effective theory of the color-singlet and color-octet contributions to the $\Upsilon\rightarrow\gamma H$ decay rate. The power counting and perturbative matching produced an operator which dominantly contributes to the decay rate, leading us to predict that the dominant decay product in radiative $\Upsilon$ decay should be the $f_2(1270)$, which should be produced in a helicity-zero state. The limited experimental evidence so far lends support to this prediction, although the helicity of the produced $f_2$ has not yet been positively identified. Our analysis also suggests a larger branching fraction for $\Upsilon\rightarrow\gamma f_2$ than the data so far indicate. Further data on this decay and other radiative decays, if consistent with the effective theory predictions, would lend powerful support to the validity of the effective theory expansion, establishing its reliability for use in other processes, especially those which relate more directly to searches for new particles.

Such processes are the production of hadronic jets in lepton collisions or, in our formulation, $Z$ decays. SCET isolated for us the soft gluon matrix elements giving the leading nonperturbative corrections to event shape distributions in hadronic $Z$ decays or, equivalently, $e^+e^-$ annhilation. Reliable calculation of these variables in QCD is essential to separate the Standard Model background in such processes in searches for new physics. Universalities among nonperturbative contributions to different event shapes would reduce the uncertainties in these calcualtions greatly, improving our ability to find deviations from Standard Model predictions while also revealing new information about QCD itself. We attempted to find such relations in the $Z$ decay distributions in jet energy, thrust, jet masses, jet broadenings, $C$ parameter, and other variables. The theory, however, did not fully acquiesce to our hopes, leaving us only one relation between the thrust and jet mass sum distributions. Experimental tests of models proposing more extensive relations among these nonperturbative corrections seem to confirm our findings. This suggests a future direction of research to define  new event shape variables, other than those standardly used in the past, which may receive universal nonperturbative corrections. This task was begun in perturbative QCD in Refs.~\cite{Berger:2003iw,Berger:2003pk}, and could be analyzed also in SCET. Similar methods could also be applied to other classes of events with QCD-jet backgrounds, or to the study of processes with collinear hadrons in the initial state, as occurs at any hadron collider.

Precision tests at low energy which constrain parameters of proposed models of physics beyond the Standard Model together with reduction of uncertainties in calculations of strong interaction phenomena prepare the way for the further scrutiny of the Standard Model to ever higher precision and perhaps the discovery of new particles and phenomena which lie beyond it. It is the task of the theorist to develop reliable methods to predict the observables in these phenomena and be prepared to understand the implications thereof on our models for the particles which make up our Universe and how they were generated in the very beginning.

\vfill
\pagebreak

\begin{center}
\raisebox{-3in}{\large\textsc{Quod scripsi scripsi.} \emph{Jn 19:22}}
\end{center}

\end{document}